\input epsf
\documentstyle{amsppt}
\pagewidth{5.4truein}\hcorrection{0.55in}
\pageheight{7.5truein}\vcorrection{0.75in}
\TagsOnRight
\NoRunningHeads
\catcode`\@=11
\def\logo@{}
\footline={\ifnum\pageno>1 \hfil\folio\hfil\else\hfil\fi}
\topmatter
\title A random tiling model for two dimensional electrostatics
\endtitle
\author Mihai Ciucu\endauthor
\thanks Research supported in part by NSF grants DMS 9802390 and DMS 0100950.
\endthanks
\affil
  School of Mathematics, Georgia Institute of Technology\\
  Atlanta, Georgia 30332-0160
\\
\\
\centerline{\it Dedicated to Richard Stanley on his sixtieth birthday}
\endaffil
\abstract We consider triangular holes on the hexagonal lattice and we study their
interaction when the rest of the lattice is covered by dimers. More precisely, we analyze 
the joint correlation of these triangular holes in a ``sea'' of dimers. We determine 
the asymptotics of the joint correlation (for large separations between the holes) 
in the case when one of the holes has side 1, all remaining
holes have side 2, and the holes are distributed symmetrically 
with respect to a symmetry axis. Our result has a striking physical
interpretation. If we regard the holes as electrical charges, with charge equal to
the difference between the number of down-pointing and up-pointing unit triangles in a
hole, the logarithm of the joint correlation behaves exactly like the
electrostatic potential energy of this two-dimensional electrostatic system: it is
obtained by a Superposition Principle from the interaction of all pairs, and the pair
interactions are according to Coulomb's law. 
\endabstract
\endtopmatter
\document

\def\mysec#1{\bigskip\centerline{\bf #1}\message{ * }\nopagebreak\par\bigskip}

\def\myref#1{\item"{[{\bf #1}]}"} 
 
\def\pf{{\it Proof.\ }} 
\def\endpf{\hbox{\qed}\bigskip}
\def\epf{\hbox{\qed}}
\def\cite#1{\relaxnext@
  \def\nextiii@##1,##2\end@{[{\bf##1},\,##2]}%
  \in@,{#1}\ifin@\def\next{\nextiii@#1\end@}\else
  \def\next{[{\bf#1}]}\fi\next}
\def\proclaimheadfont@{\smc}

\def\pf{{\it Proof.\ }}

\define\Z{{\Bbb Z}}
\define\Q{{\Bbb Q}}
\define\R{{\Bbb R}}
\define\C{{\Bbb C}}
\define\M{\operatorname{M}}
\define\Rep{\operatorname{Re}}
\define\wt{\operatorname{wt}}

\define\ch{\operatorname{ch}}
\define\ph{\operatorname{ph}}
\define\sign{\operatorname{sgn}}
\define\de{\operatorname{d}}
\define\mo{\operatorname{mod}}
\define\twoline#1#2{\line{\hfill{\smc #1}\hfill{\smc #2}\hfill}}

\def\mypic#1{\epsffile{figs/#1}}



\define\A{1}
\define\Ca{2}
\define\Cone{3}
\define\Ctwo{4}
\define\Cthree{5}
\define\Cfour{6} 

\define\CEP{7}
\define\CLP{8}
\define\FMS{9}
\define\Fone{10}
\define\Ftwo{11}

\define\FS{12}
\define\GV{13}
\define\GR{14}
\define\GKP{15}
\define\Ham{16}
\define\Har{17}
\define\HG{18}
\define\K{K}
\define\Kastel{19}
\define\Kone{20}
\define\Ktwo{21}

\define\KM{22}
\define\LP{23}
\define\Lu{24}
\define\MPW{25}
\define\Myers{26}
\define\Nien{27}
\define\Ol{28}
\define\Ste{29}

\define\WMTB{30}

\define\TwoZero{2.1}
\define\TwoOne{2.2}
\define\TwoTwo{2.3}
\define\TwoThree{2.4}
\define\TwoFour{2.5}
\define\TwoFourFive{2.6}
\define\TwoFive{2.7}
\define\TwoSix{2.8}
\define\TwoSeven{2.9}
\define\TwoEight{2.10}
\define\TwoNine{2.11}
\define\TwoTen{2.12}
\define\TwoEleven{2.14}
\define\TwoTwelve{2.13}


\mysec{1. Introduction}

Monomer-monomer and especially dimer-dimer correlations\footnote{The monomers, 
respectively the dimers, are interacting via a sea of dimers that cover all lattice 
sites not occupied by them.} on a plane bipartite lattice 
(especially the square and hexagonal lattice) 
have been studied quite extensively (see for instance \cite{\FS}, 
\cite{\Har}, \cite{\K1} and \cite{\K2}). Color the vertices
of the lattice white and black so that each edge has one white and one black endpoint. 
From the point of
view of this paper, there is a fundamental difference between studying dimer-dimer
and monomer-monomer correlations: the former have the same number of vertices of each
color, while the latter have an excess of either a white or a black vertex. 

In this paper we consider correlations of triangular holes on the hexagonal
lattice.  

This type of hole has the convenient feature that the difference between the
number of its white and black constituent vertices is equal to the length of its side. 
We will be lead by our results to interpret the white vertices as elementary negative
charges (``electrons''), and the black vertices as elementary positive charges (``positrons''),
so that the triangular plurimers become charges of magnitude given by their
side-length, and sign given by their orientation (up-pointing or down-pointing). The
main result of this paper, Theorem 2.1 (see also its much simpler restatement (2.6)),
implies that in the fairly general situation in which it applies 
(namely, when the holes are symmetrically distributed 
about an axis, and all holes have side 2, except for one of side 1, on the symmetry axis)
the logarithm of the joint correlation of 
triangular holes behaves exactly like the two-dimensional electrostatic potential
energy of the corresponding system of charges: it is obtained by a Superposition Principle from the 
interaction of all pairs, and the pair interactions are according to Coulomb's law. 
(It is now clear why dimer-dimer and monomer-monomer 
correlations are fundamentally different: dimers are neutral!) 

To present our results in the background of previous related results in the literature, we 
point out the following facts. 

First, we mention that there is an alternative approach for expressing joint correlations of holes
on the hexagonal (or square) lattice due to Kenyon (\cite{\Ktwo,Theorem\,2.3}; see also \cite{\Kone}). 
When it applies, it provides an expression for the joint correlation as a $k\times k$ determinant (where
$2k$ is the total number of vertices in the holes).
However, while not requiring symmetry, the set-up of Kenyon's approach limits its applicability to the 
case when all holes have even side (thus not accommodating our hole of side 1 on the symmetry axis), and,
more restrictively, to the case when the total ``charge'' of the holes is zero\footnote{Indeed, 
Kenyon \cite{\Kone}
defines the correlation of holes as the limit of $M(\hat{H}_{m,n})/M(H_{m,n})$ when $m,n\to\infty$, 
where $H_{m,n}$ is a toroidal hexagonal grid graph, and $\hat{H}_{m,n}$ is its subgraph obtained by
removing the holes ($M(G)$ denotes the number of perfect matchings of the graph $G$). 
In order for $M(H_{m,n})$
to be non-zero one must have the same number of vertices in the two bipartition classes of the vertices
of $H_{m,n}$. Thus, since $M(\hat{H}_{m,n})$ also has to be non-zero in order for Kenyon's correlation
to be non-zero, the number of vertices in the two bipartition classes that fall in the holes must also
be the same; i.e., the total charge of the holes must be zero.}. 
The main advantage of
our approach is that it sets no restriction on the the total charge. 

Furthermore, our result (14.9) gives
the joint correlation of a general distribution of collinear {\it monomers} on the square lattice, 
a situation in which none of the holes satisfies the even-sidedness required by Kenyon's approach.


Second, there are other discrete models in the physics literature (see e.g. the survey \cite{\Nien} 
by Nienhuis)
believed to behave like a Coulomb gas. However, there are several important differences between
them and our model: (1) we do not require that the total charge is 0, a fact built into the definition
of the Coulomb gas model in the survey by Nienhuis \cite{(2.7),\Nien}; (2) by studying correlation
of holes, our discrete model seems quite different from the others in the literature; 
(3) for those models surveyed in \cite{\Nien} for which the (believed) Coulomb behavior is only 
asymptotic (as it is the case for our model), 
the arguments 
for their equivalence with the Coulomb gas model are only heuristic, while our results are proved 
rigorously; (4) in our model all states have the same
energy, so the emergence of the Coulomb interaction is entirely due to the number of different 
geometrical configurations (unit rhombus tilings) compatible with the holes---in the language of 
physicists, our model is {\it stabilized by entropy}; by contrast, in all models surveyed in 
\cite{\Nien} different configurations have different energies, specified by a Hamiltonian---they 
are {\it stabilized by energy}. Furthermore, in our model the electrical charge
has a purely geometric origin: the charges are holes in the lattice, and their magnitude is
the difference between the number of right-pointing and left-pointing unit triangles in the hole.

Third, in a recent paper \cite{\KM} Krauth and Moessner study numerically (using
Monte Carlo algorithms) a very special case of the problem considered in this paper, namely just
the two-point correlations of monomers on the square lattice. Their data leads them to 
conjecture that the two-point correlations on monomers behave like a Coulomb potential (the case of
monomers on different sublattices appeared already in \cite{\FS}; the new part is the simulation data 
for monomers on the same sublattice). Krauth and Moessner also state that they could not find these
correlations worked out in the literature. Since in the current paper we address the case of 
$(2m+2n+1)$-point correlations, showing that they satisfy the much more general Superposition Principle,
their remark suggests that our results are also new.

And fourth, shortly after the current paper was posted on the preprint archive 
(web address arxiv.org/abs/math-ph/0303067, March 2003), 
physicists D. A. Huse, W. Krauth, R. Moessner and S. L. Sondhi
posted an article (arxiv.org/abs/cond-mat/0305318, May 2003) 
presenting numerical simulations that suggest a
positive answer to Question 15.1 (which concerns a three dimensional analog of the model presented 
in this paper) for the case of two monomers.

\mysec{2. Definitions, statement of results and physical interpretation}

We will find it more convenient to present our results on dimer coverings of
portions of the hexagonal lattice in the equivalent, dual language of lozenge tilings 
of lattice
regions on the triangular lattice. In this language a monomer is just a unit triangle
of the lattice. 
A lattice triangle of side two will be called a {\it quadromer}. 
The results proved in this paper involve only monomers and quadromers.
More generally, triangular plurimers are lattice triangles of arbitrary side. 
A dimer becomes a lozenge---a unit rhombus covering precisely two unit triangles. We 
will often refer to lozenges as dimers.

Let $m$ and $n$, and $N$ be nonnegative integers. Consider also the nonnegative integers 
$R_i$, $v_i$, $i=1,\dotsc,m$ and $R'_i$, $v'_i$, $i=1,\dotsc,n$.
Define the region 
$$
H_N\left(\matrix{R_1}\\{v_1}\endmatrix
\cdots\matrix{R_m}\\{v_m}\endmatrix;
\matrix{R'_1}\\{v'_1}\endmatrix\cdots\matrix{R'_n}\\{v'_n}\endmatrix\right)\tag\TwoZero
$$ 
as follows. 

Consider a lattice hexagon $H$ whose sides alternate between $2N+4n+1$ and $2N+4m$,
starting with the base.
Denote by $\ell$ the vertical symmetry axis of $H$. Let $u$ be the up-pointing unit 
triangle on $\ell$ whose base is $(N+2m)\sqrt{3}$ units above the base of $H$ (that
is, the base of $u$ is along the horizontal diagonal of $H$). 

Let $D(R,v)$ be the down-pointing quadromer 
(i.e., down-pointing lattice 
triangle of side 2) whose base is centered $R$ units to the left of $\ell$, and 
$(2v+1)\sqrt{3}/2$ units below the base of $u$ (two instances of this appear in
Figure 2.1). Let $U(R,v)$ be the up-pointing 
quadromer whose base is centered $R$ units to the left of $\ell$ 
and $(2v+1)\sqrt{3}/2$ units {\it above} the base of $u$ (see Figure~2.1 for three
instances of this).

Finally, let $\bar{D}(R,v)$ and $\bar{U}(R,v)$ be the mirror images in $\ell$ of 
the above quadromers. We define 
$H_N\left(\matrix{R_1}\\{v_1}\endmatrix
\cdots\matrix{R_m}\\{v_m}\endmatrix;
\matrix{R'_1}\\{v'_1}\endmatrix\cdots\matrix{R'_n}\\{v'_n}\endmatrix\right)$ 
to be the region obtained from the hexagon $H$ by removing $u$
(a monomer), $D(R_i,v_i)$, $\bar{D}(R_i,v_i)$, $i=1,\dotsc,m$ and 
$U(R'_j,v'_j)$, $\bar{U}(R'_j,v'_j)$, $j=1,\dotsc,n$ (a total 
of $2m+2n+1$ holes). 

We assume throughout the paper that $R_i\geq 1$, $i=1,\dotsc,m$, and 
$R'_j\geq 1$, $j=1,\dotsc,n$, so that $\ell$ separates the original plurimers from
their mirror images.
Figure 2.1 shows the region 
$H_2\left(\matrix{5}&2\\{0}&1\endmatrix;
\matrix4&2&3\\1&2&4\endmatrix\right)$.

\topinsert
\centerline{\mypic{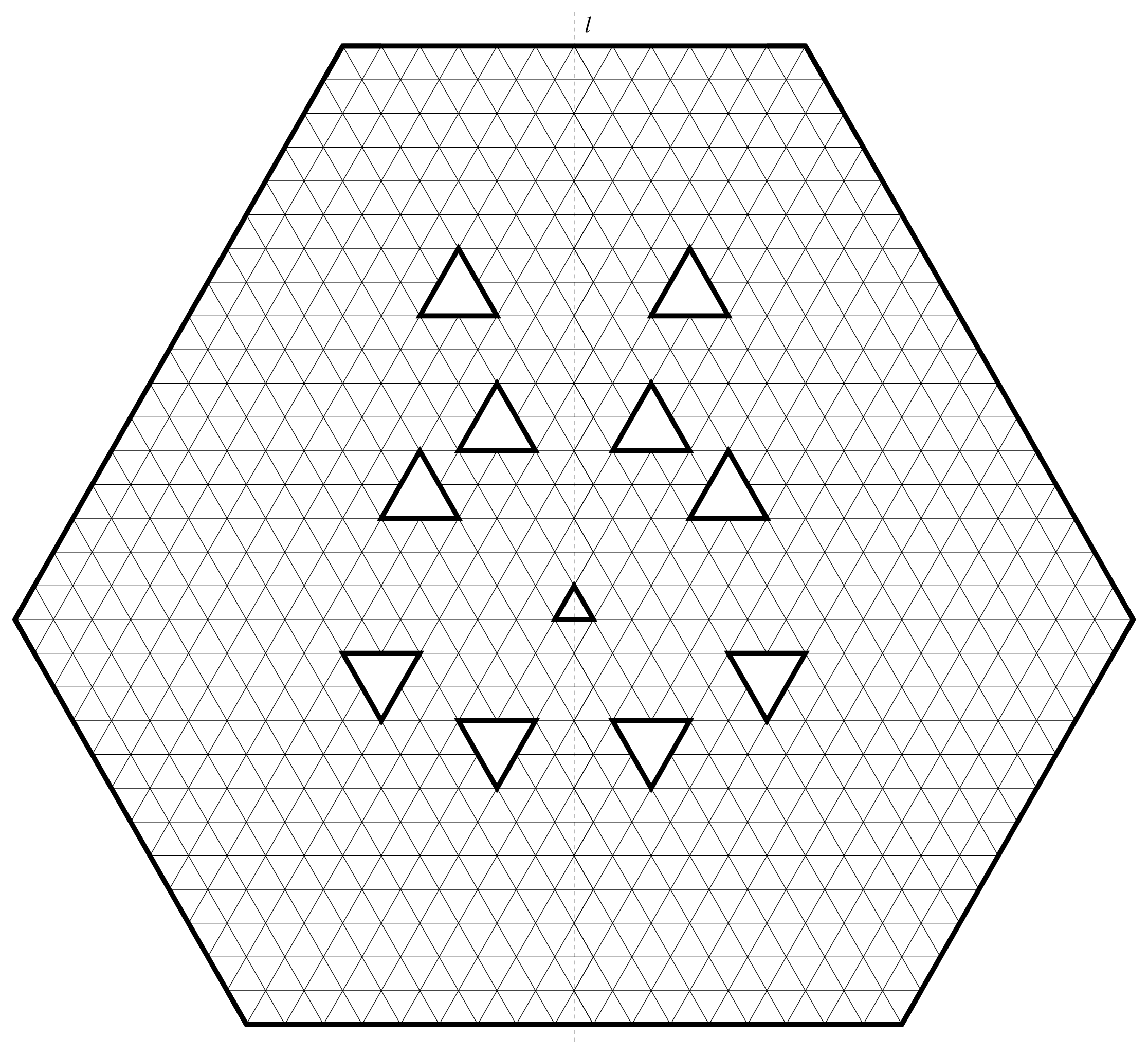}}
\centerline{{\smc Figure~2.1.} 
{\rm $H_2\left(\matrix{5}&2\\{0}&1\endmatrix;
\matrix4&2&3\\1&2&4\endmatrix\right)$.}}

\centerline{\mypic{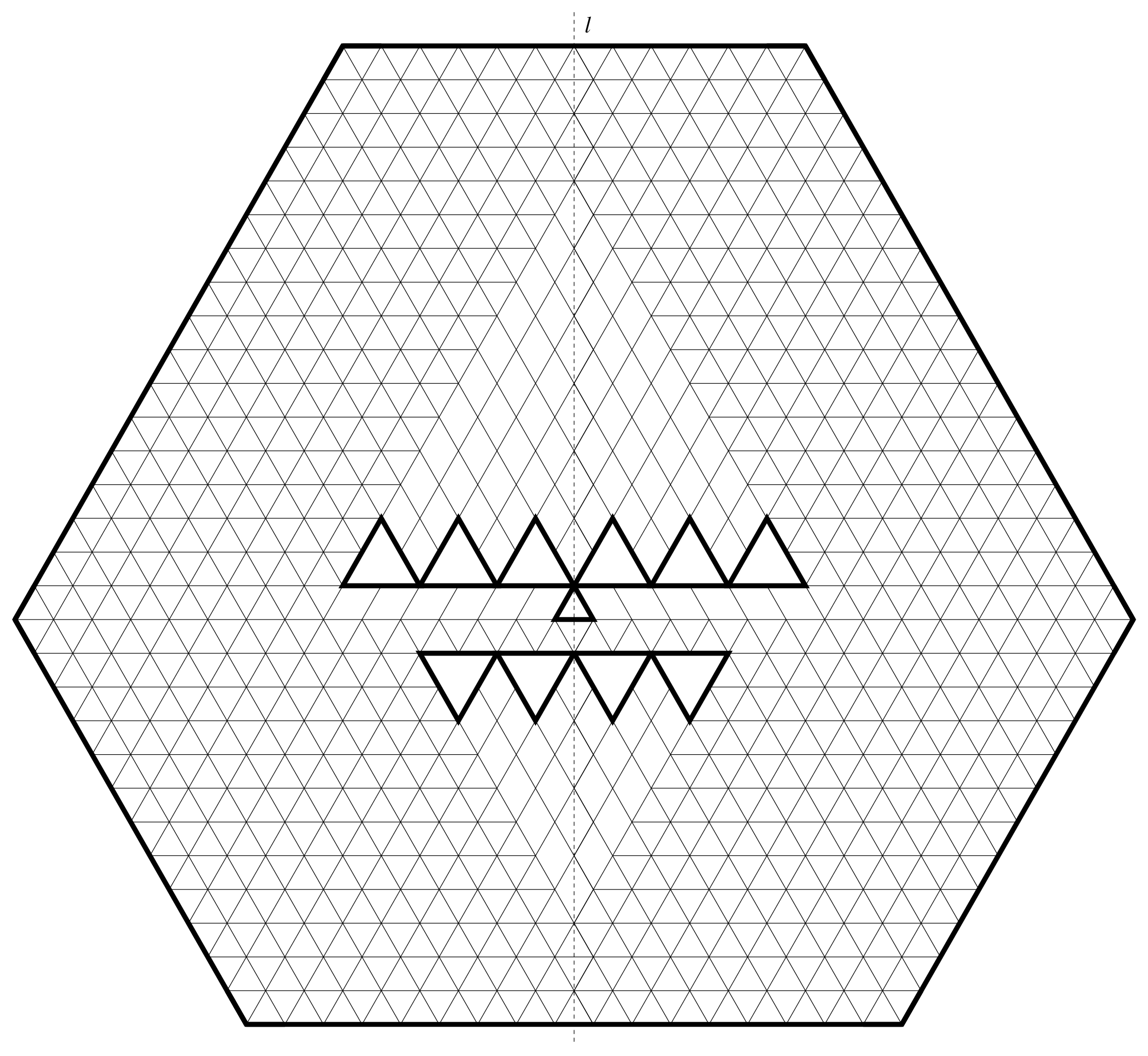}}
\centerline{{\smc Figure~2.2.} 
{\rm $H_2\left(\matrix{1}&3\\{0}&0\endmatrix;
\matrix1&3&5\\0&0&0\endmatrix\right)$.}}
\endinsert

Define the {\it correlation at the center} (or simply correlation) of these $2m+2n+1$
plurimers by
$$
\align
\!\!\!\!\!\!\!\!
\omega\left(\matrix{R_1}\\{v_1}\endmatrix
\cdots\matrix{R_m}\\{v_m}\endmatrix;
\matrix{R'_1}\\{v'_1}\endmatrix\cdots\matrix{R'_n}\\{v'_n}\endmatrix\right)&:=\\
&\!\!\!\!\!\!\!\
\lim_{N\to\infty}
\frac{\M\left(H_N\left(\matrix{R_1}&{R_2}\\{v_1}&{v_2}\endmatrix
\cdots\matrix{R_m}\\{v_m}\endmatrix;
\matrix{R'_1}&{R'_2}\\{v'_1}&{v'_2}\endmatrix\cdots\matrix{R'_n}\\{v'_n}\endmatrix\right) \right)}
{\M\left(H_N\left(\matrix 1&3\\0&0
\endmatrix
\cdots\matrix 2m-1\\0\endmatrix;
\matrix 1&3\\0&0\endmatrix\cdots\matrix
2n-1\\0\endmatrix\right)\right)},\tag\TwoOne\\
\endalign
$$
where $\M(D)$ is the number of dimer coverings (i.e, lozenge tilings---tilings by 
unit rhombuses each
covering precisely two unit triangles) of the lattice region $D$. The existence of the
limit on the right hand side of (\TwoOne) follows from Proposition 3.2.

In the above definition, 
the region at the denominator is obtained from the one in the numerator by packing 
the quadromers as tightly as possible around $u$---for the example of Figure 2.1 this
is illustrated in Figure 2.2. As indicated by the latter figure, because of forced
lozenges, the situation is equivalent to removing just two large plurimers from $H$,
one down-pointing of side $2m$ and one up-pointing of side $2n+1$.

Assume that the midpoints of the bases of the quadromers have coordinates
$$
\align
R_i&=A_iR\\
v_i&=q_iR_i+c_i=q_iA_iR+c_i\\
R'_j&=B_jR\\
v'_j&=q'_jR'_j+c'_j=q'_jB_jR+c'_j,\tag\TwoTwo
\endalign
$$
where: $A_i>0$, $q_i>0$, $i=1,\dotsc,m$ are fixed rational numbers chosen so that the pairs 
$(A_1,q_1),\dotsc,(A_m,q_m)$ are distinct;
$B_j>0$, $q'_j>0$, $j=1,\dotsc,n$ are fixed rational numbers so that the pairs
$(B_1,q'_1),\dotsc,(B_n,q'_n)$ are distinct;
$c_i\geq0$, $i=1,\dotsc,m$ and $c'_j\geq0$, $j=1,\dotsc,n$ are fixed integers; 
and $R$ is an integer parameter.
 
The main result of this paper is the following.

\proclaim{Theorem 2.1} If the coordinates of the quadromers are 
as in $(\TwoTwo)$ and $R\to\infty$, their joint correlation with the fixed monomer $u$
is given asymptotically by
$$
\align
\omega&\left(\matrix{R_1}\\{v_1}\endmatrix
\cdots\matrix{R_m}\\{v_m}\endmatrix;
\matrix{R'_1}\\{v'_1}\endmatrix\cdots\matrix{R'_n}\\{v'_n}\endmatrix\right)=\\
&\ \ \ \ \ \ \ \ 
c_{2m,2n}
\prod_{i=1}^m(2R_i)^{2}\prod_{j=1}^n(2R'_j)^{2}
\frac{\prod_{j=1}^n(R'_j)^2+3(v'_j)^2}{\prod_{i=1}^m R_i^2+3v_i^2}\\
&\ \ \ \ \ \ \ 
\times
\prod_{1\leq i<j\leq m}[(R_j-R_i)^2+3(v_j-v_i)^2]^{2}\,
[(R_j+R_i)^2+3(v_j-v_i)^2]^{2}\\
&\ \ \ \ \ \ \ 
\times
\frac{\prod_{1\leq i<j\leq n}[(R'_j-R'_i)^2+3(v'_j-v'_i)^2]^{2}\,
[(R'_j+R'_i)^2+3(v'_j-v'_i)^2]^{2}}
{\prod_{i=1}^m\prod_{j=1}^n[(R'_j-R_i)^2+3(v'_j+v_i)^2]^{2}\,
[(R'_j+R_i)^2+3(v'_j+v_i)^2]^{2}}\\
&\ \ \ \ \ \ \ 
+O(R^{4(m-n)^2-4m-1}),\tag\TwoThree
\endalign
$$
where 
$$
c_{k,l}=\frac{2^{k+l}3^{k-(k-l)^2/2}}{\pi^{k+l}}
\left(\prod_{j=0}^{k-1}\frac{(2)_j}{(1)_j(3/2)_j}
\prod_{j=0}^{l-1}\frac{(j+2)_k}{(3/2)_j}\right)^2.\tag\TwoFour
$$
\endproclaim

Remarkably, each factor in (\TwoThree), except for the multiplicative constant $c_{2m,2n}$,
is exactly equal to the Euclidean distance between the midpoints of
the bases\footnote{For the monomer $u$---involved in the factors of the fraction on the second line
of (\TwoThree)---instead of the midpoint of its base we need to consider
the closest lattice point above it or below it, according as the factor in question is at the 
numerator or denominator, respectively.} of some pair of our $2m+2n+1$ plurimers (one monomer and
$2m+2n$ quadromers). Indeed:

\smallpagebreak
$(i)$ $2R_i$ is the distance between $D(R_i,v_i)$ and $\bar{D}(R_i,v_i)$;

$(ii)$ $2R'_j$ is the distance between $U(R'_j,v'_j)$ and $\bar{U}(R'_j,v'_j)$;

$(iii)$ $[(R'_j)^2+3(v'_j)^2]^{1/2}$ is both the distance between $U(R'_j,v'_j)$ and monomer $u$,
and 

the distance between $\bar{U}(R'_j,v'_j)$ and monomer $u$;

$(iv)$ $[R_i^2+3v_i^2]^{1/2}$ is both the distance between $D(R_i,v_i)$ and monomer $u$,
and the 

distance between $\bar{D}(R_i,v_i)$ and monomer $u$;

$(v)$ $[(R_j-R_i)^2+3(v_j-v_i)^2]^{1/2}$ is both the distance between $D(R_i,v_i)$ and $D(R_j,v_j)$,

and the distance between $\bar{D}(R_i,v_i)$ and $\bar{D}(R_j,v_j)$;

$(vi)$ $[(R_j+R_i)^2+3(v_j-v_i)^2]^{1/2}$ is both 
the distance between $D(R_i,v_i)$ and $\bar{D}(R_j,v_j)$,

and the distance between $\bar{D}(R_i,v_i)$ and $D(R_j,v_j)$;

$(vii)$ $[(R'_j-R'_i)^2+3(v'_j-v'_i)^2]^{1/2}$ is both 
the distance between $U(R'_i,v'_i)$ and $U(R'_j,v'_j)$,

and the distance between $\bar{U}(R'_i,v'_i)$ and $\bar{U}(R'_j,v'_j)$;

$(viii)$ $[(R'_j+R'_i)^2+3(v'_j-v'_i)^2]^{1/2}$ is both 
the distance between $U(R'_i,v'_i)$ and $\bar{U}(R'_j,v'_j)$,

and the distance between $\bar{U}(R'_i,v'_i)$ and $U(R'_j,v'_j)$;

$(ix)$ $[(R'_j-R_i)^2+3(v'_j+v_i)^2]^{1/2}$ is both 
the distance between $D(R_i,v_i)$ and $U(R'_j,v'_j)$,

and the distance between $\bar{D}(R_i,v_i)$ and $\bar{U}(R'_j,v'_j)$;

$(x)$ $[(R'_j+R_i)^2+3(v'_j+v_i)^2]^{1/2}$ is both 
the distance between $D(R_i,v_i)$ and $\bar{U}(R'_j,v'_j)$,

and the distance between $\bar{D}(R_i,v_i)$ and $U(R'_j,v'_j)$.

\smallpagebreak
This remarkable phenomenon goes even further: if one defines the {\it charge} 
$\ch(Q)$ of a hole to be the number of its up-pointing unit lattice triangles 
minus the number of its down-pointing ones (this clearly gives 1 for the monomer $u$, and
$-2$ and $2$ for the quadromers of type $D$ and $U$, respectively), the
exponent with which each such factor occurs in (\TwoThree) is precisely half the product of the charges
of the corresponding two plurimers. Therefore, the statement of Theorem 2.1 can be rewritten as
follows.


\proclaim{Theorem 2.1 (equivalent restatement)} Denote the $(2m+2n+1)$ triangular plu\-ri\-mers 
removed in region $(\TwoZero)$ by $Q_1,\dotsc,Q_{2m+2n+1}$, and define the charge $\ch(Q)$ of $Q$ 
to be the number 
of up-pointing unit lattice triangles of $Q$ minus the number of its down-pointing unit triangles.
Let $\de(Q_i,Q_j)$ denote the Euclidean distance between the midpoints of the bases of plurimers
$Q_i$ and $Q_j$.

Then as the coordinates of the midpoints of the bases of the $2m+2n$ quadromers
approach infinity as specified by $(\TwoThree)$, the 
asymptotics of their joint correlation with the monomer $u$ 
(which by definition is kept fixed at the origin) is given by 
$$
\align
\omega(Q_1,\dotsc,Q_{2m+2n+1})=
c_{2m,2n}\prod_{1\leq i<j\leq 2m+2n+1}&\de(Q_i,Q_j)^{\ch(Q_i)\ch(Q_j)/2}\\
+&O(R^{4(m-n)^2-4m-1}),\tag\TwoFourFive
\endalign
$$
where the constant $c_{k,l}$ is given by $(\TwoFour)$.
\endproclaim

Our result has the following striking physical interpretation. 

Take the logarithm of both sides of (\TwoFourFive), and assume $4(m-n)^2-4m\neq0$. 
Since then the main term on the right hand side of 
(\TwoThree) approaches either $0$ or infinity as $R\to\infty$, the resulting contribution 
from the multiplicative constant is negligible. We obtain
$$
\log\omega(Q_1,\dotsc,Q_{2m+2n+1})\sim\sum_{1\leq i<j\leq 2m+2n+1}
\frac{\ch(Q_i)\ch(Q_j)}{2}\log\de(Q_i,Q_j),\tag\TwoFive
$$
where $\de(Q_i,Q_j)$ is the Euclidean distance between the 
plu\-ri\-mers $Q_i$ and $Q_j$.

But the logarithm of the distance is, up to a negative multiplicative constant, 
just the potential in two dimensional electrostatics!
So if we view the plurimers as {\it electrical charges} of (signed) magnitude given by 
the operator $\ch$ (in this case, just their side-length if they point upward, or minus
their side-length if they point downward),
the logarithm of the joint correlation (multiplied by a negative constant) 
acts precisely like the electrostatic 
potential energy of this two dimensional electrostatic system: it is
obtained by a Superposition Principle from the interaction of all pairs, and the pair
interactions are according to Coulomb's law. 

\smallpagebreak
Both for our proof of Theorem 2.1 and from the point of view of physical
interpretation we find it useful to define another kind of plurimer correlation, in
which the boundary of the defining regions makes its presence felt. To this
end, define another family of regions as follows. 

Cut the region
$H_N\left(\matrix{R_1}\\{v_1}\endmatrix
\cdots\matrix{R_m}\\{v_m}\endmatrix;
\matrix{R'_1}\\{v'_1}\endmatrix\cdots\matrix{R'_n}\\{v'_n}\endmatrix\right)$
along a zig-zag line that starts at the midpoint of its top side, follows $\ell$ as
closely as possible on its right until it reaches the central monomer $u$, crosses
$\ell$ along the western edge of $u$ and then follows $\ell$ as closely as possible on
its {\it left}, ending at the midpoint of the base (this cut is pictured in Figure
2.3). We define
$$
W_N\left(\matrix{R_1}\\{v_1}\endmatrix
\cdots\matrix{R_m}\\{v_m}\endmatrix;
\matrix{R'_1}\\{v'_1}\endmatrix\cdots\matrix{R'_n}\\{v'_n}\endmatrix\right)
$$
to be the region obtained this way to the west of our cut, in which in addition the
$N+2n$ lozenge positions above $u$ and immediately to the left of the cut are
distinguished and given
weight 1/2 (i.e., each lozenge tiling $T$ of the regions $W_N$ is weighted by $1/2^k$,
where $k$ is the number of distinguished lozenge positions occupied by a lozenge in 
$T$). An example is illustrated in Figure 2.3 (the distinguished lozenge positions are
marked by shaded ellipses).

\topinsert
\centerline{\mypic{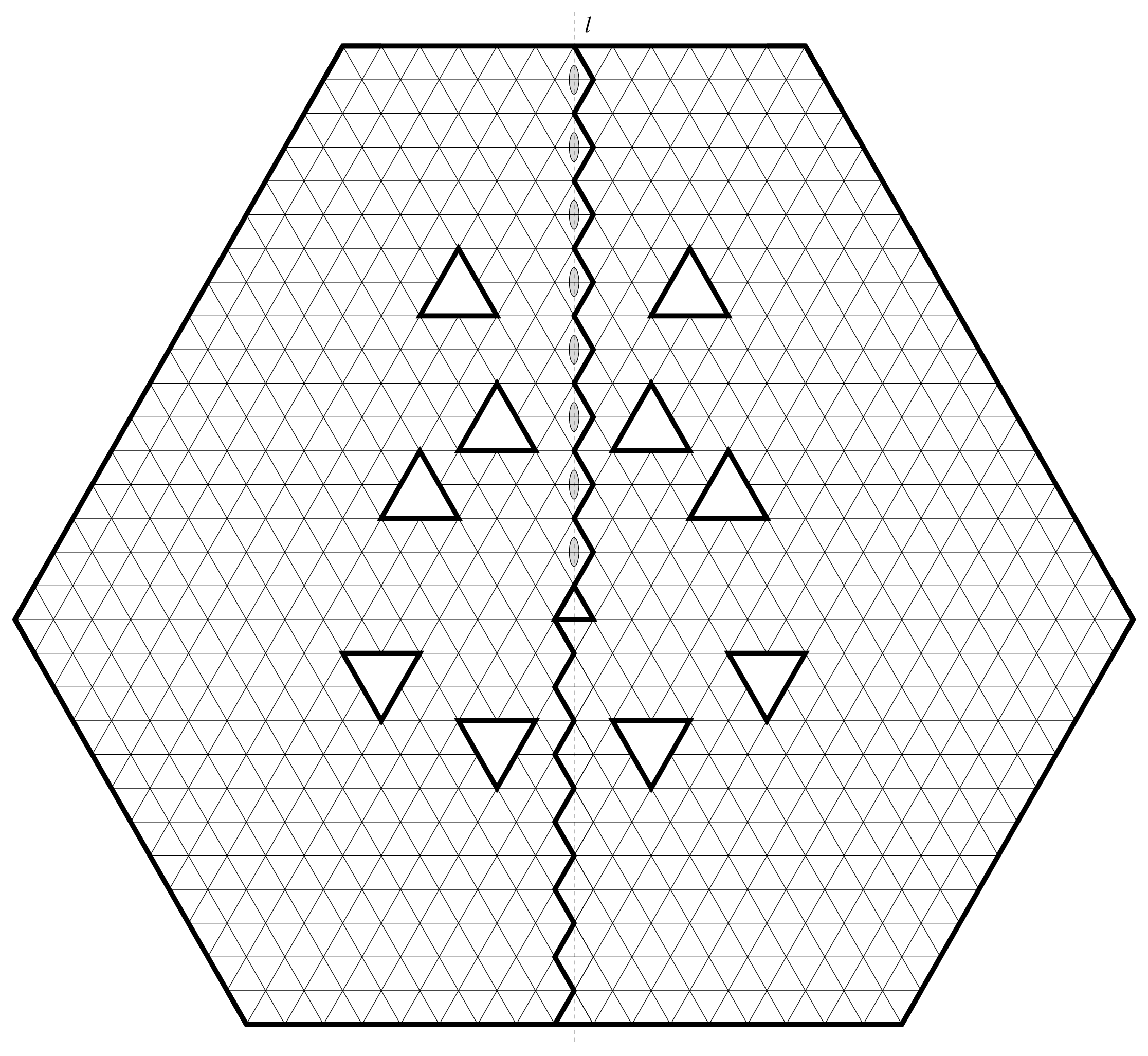}}
\centerline{{\smc Figure~2.3.} 
{\rm $W_2\left(\matrix{5}&2\\{0}&1\endmatrix;
\matrix4&2&3\\1&2&4\endmatrix\right)$.}}
\endinsert

When $N\to\infty$ and the coordinates $R_i$, $v_i$, $R'_i$, $v'_i$ of the quadromers
are fixed, they maintain a fixed relative position with respect to the right boundary of the
regions $W_N$. We define the {\it boundary-influenced correlation} $\omega_b$ by
$$
\align
\!\!\!\!\!\!\!\!
\omega_b\left(\matrix{R_1}\\{v_1}\endmatrix
\cdots\matrix{R_m}\\{v_m}\endmatrix;
\matrix{R'_1}\\{v'_1}\endmatrix\cdots\matrix{R'_n}\\{v'_n}\endmatrix\right)&:=\\
&\!\!\!\!\!\!\!\!
\lim_{N\to\infty}
\frac{\M\left(W_N\left(\matrix{R_1}&{R_2}\\{v_1}&{v_2}\endmatrix
\cdots\matrix{R_m}\\{v_m}\endmatrix;
\matrix{R'_1}&{R'_2}\\{v'_1}&{v'_2}\endmatrix\cdots\matrix{R'_n}\\{v'_n}\endmatrix\right) \right)}
{\M\left(W_N\left(\matrix 1&3\\0&0
\endmatrix
\cdots\matrix 2m-1\\0\endmatrix;
\matrix 1&3\\0&0\endmatrix\cdots\matrix
2n-1\\0\endmatrix\right)\right)}\tag\TwoSix
\endalign
$$
(the fact that this limit exists follows by Lemma 5.1).

An important part of the proof of Theorem 2.1---and a result interesting on its 
own---is the determination of the asymptotics of 
$\omega_b$. 

\proclaim{Theorem 2.2} If the coordinates of the quadromers $D(R_i,v_i)$, $i=1,\dotsc,m$ and
$U(R_j,v_j)$, $j=1,\dotsc,n$ are given by $(\TwoTwo)$, the
large $R$ asymptotics of their boundary-influenced correlation is
$$
\align
\omega_b&\left(\matrix{R_1}\\{v_1}\endmatrix
\cdots\matrix{R_m}\\{v_m}\endmatrix;
\matrix{R'_1}\\{v'_1}\endmatrix\cdots\matrix{R'_n}\\{v'_n}\endmatrix\right)=\\
&\ \ \ \ \ 
\phi_{2m,2n}
\prod_{i=1}^m(2R_i)\prod_{j=1}^n(2R'_j)
\frac{\prod_{j=1}^n[(R'_j)^2+3(v'_j)^2]^{1/2}}{\prod_{i=1}^m[R_i^2+3v_i^2]^{1/2}}\\
&\ \ \ \ \ \ \ 
\times
\prod_{1\leq i<j\leq m}[(R_j-R_i)^2+3(v_j-v_i)^2]\,
[(R_j+R_i)^2+3(v_j-v_i)^2]\\
&\ \ \ \ \ \ \ 
\times
\frac{\prod_{1\leq i<j\leq n}[(R'_j-R'_i)^2+3(v'_j-v'_i)^2]\,
[(R'_j+R'_i)^2+3(v'_j-v'_i)^2]}
{\prod_{i=1}^m\prod_{j=1}^n[(R'_j-R_i)^2+3(v'_j+v_i)^2]\,
[(R'_j+R_i)^2+3(v'_j+v_i)^2]}\\
&\ \ \ \ \ \ \ 
+O(R^{2(m-n)^2-2m-1}),\tag\TwoSeven
\endalign
$$
where 
$$
\phi_{k,l}=\frac{2^{k}3^{(k+l)/4-(k-l)^2/4}}{\pi^{(k+l)/2}}
\prod_{j=0}^{k-1}\frac{(2)_j}{(1)_j(3/2)_j}
\prod_{j=0}^{l-1}\frac{(j+2)_k}{(3/2)_j}.\tag\TwoEight
$$
\endproclaim

This result also has an interesting physical interpretation. 
Relabel the $m+n$ quadromers by $Q_1,\dotsc,Q_{m+n}$. Assuming 
$2(m-n)^2-2m\neq0$, one deduces from (\TwoSeven)---just as (\TwoFive) was deduced
from (\TwoThree)---that
$$
\log\omega_b(Q_1,\dotsc,Q_{m+n})\sim\sum_{Q,Q'\in\Cal Q,Q\neq Q'}
\frac{\ch(Q)\ch(Q')}{4}\log\de(Q,Q'),\tag\TwoNine
$$
where $\ch$ denotes the charge, $\de$ the Euclidean distance, and
$\Cal Q$ contains, {\it in addition} to our $m+n$ plurimers, {\it their mirror images 
in the vertical line} touching all zig-zags on the right boundary of $W_N$, and also an
{\it up-pointing monomer} just outside the lower of the two
aligned consecutive segments of this right boundary.

Relation (\TwoNine) also has a strong electrostatic reminiscence. It shows that 
the quad\-ro\-mers near a zig-zag boundary can be interpreted as electrical charges near a straight 
line whose
effect is to bring in the mirror images of our charges. A well-known situation in
which this happens is for electrical
charges near a conductor---but in that case the image charges need to be taken of {\it
opposite} sign. In fact, there is another physical situation in which the images are
taken with the same sign: that when we have fixed charges inside a dielectric and near
a straight boundary separating this dielectric from another one, of a dielectric
constant negligible in comparison with the first (see e.g. \cite{\Ftwo,\S12-2}). 
Our relation (\TwoNine) models this
situation---with the additional specification that the slight irregularity of the right
boundary near its middle behaves like an extra charge of magnitude $+1$.


We conclude this section by presenting a view of our results as a possible random
tiling model for two dimensional electrostatics.

Suppose one pictures the two dimensional ``universe'' being spatially quantized,
consisting of a 
very fine lattice of triangular ``quanta of space'', each of them responsible for 
creating a unit charge (of sign given by its orientation) when displaced by a
``body'' (which, since we are considering space to be quantized, can be placed only in
such a way that it consists of whole unit triangles). Suppose also that once a number
of bodies are placed in this space---thus acquiring the charge determined by the 
above interpretation---the unoccupied quanta of space (i.e.,
unit triangles) have the tendency to pair up with their neighbors to form a sea of 
dimers, according to the uniform distribution on all such possible pairings. 
(Assuming
such a tendency for unoccupied unit triangles to pair up with a neighbor seems natural from
the point of view of quantum mechanics, as the quantum fluctuations of the vacuum ceaselessly cause
particle-antimatter-companion pairs to erupt into existence only to be annihilated after an instant. 
One particular rhombus tiling corresponds to one particular way for these virtual particles to annihilate.
To consider averaging over all such possible tilings is analogous to Feynman's ``sum-over-paths'' approach
to quantum mechanics; see the remark at the end of Section 15.)
Then (\TwoFive)
shows---under the assumption that the charges are symmetrically distributed and
are all of magnitude $\pm2$, except for a single $+1$ on the symmetry axis---that these
charged bodies would interact precisely like charged bodies in two dimensional
electrostatics. We would get a {\it proof} of the Superposition Principle (for such
charge distributions), and very strong evidence for Coulomb's law!\footnote{The reason
we don't obtain a proof for Coulomb's law from Theorem 2.1 is that we are assuming
there a
symmetric charge distribution, and therefore the case of two charged bodies in general
position is not covered. This case is addressed in \cite{\Cfour} for opposite sign charges 
of magnitude 2. The general even-magnitude case will be presented in a sequel of the 
present paper.} 

To make this parallelism more explicit, let $\Cal S$ be the family of all $(2m+2n+1)$-element 
sets of triangular plurimers (one monomer, and the rest quadromers) 
satisfying the hypothesis of Theorem 2.1, and being contained in some fixed disk $D$ 
centered at the center of the regions (\TwoZero) (the radius of $D$ is some large absolute 
constant, the ``diameter'' of our system of plurimers). 
  
Let $\Cal A=\{A_1,\dotsc,A_{2m+2n+1}\}$ and $\Cal B=\{B_1,\dotsc,B_{2m+2n+1}\}$
be two members of $\Cal S$. Re-denote the charges of their elements by 
$\ch(A_i)=\ch(B_i)=q_i$, $i=1,\dotsc,2m+2n+1$, and denote their corresponding regions
(\TwoZero) by $H_N(\Cal A)$ and $H_N(\Cal B)$, respectively. Denote also their corresponding
correlations (\TwoOne) by $\omega(\Cal A)$ and $\omega(\Cal B)$, respectively.

Define a probability distribution on $\Cal S$ by requiring the ratio of the probabilities
$P_{\Cal A}$ and $P_{\Cal B}$ to be
$$
\frac{P_{\Cal A}}{P_{\Cal B}}:=\lim_{N\to\infty}\frac{\M(H_N(\Cal A))}{\M(H_N(\Cal B))}
=\frac{\omega(\Cal A)}{\omega(\Cal B)},\tag\TwoTen
$$
the second equality being valid by (\TwoOne). 

By (\TwoFourFive) we obtain from (\TwoTen) that
$$
\align
&\frac{P_{\Cal A}}{P_{\Cal B}}\sim\\
&\ 
\exp\left(-\frac12\left[\sum_{1\leq i<j\leq 2m+2n+1}q_iq_j(-\ln\de(A_i,A_j))
-\sum_{1\leq i<j\leq 2m+2n+1}q_iq_j(-\ln\de(B_i,B_j))\right]\right),\\
\tag\TwoTwelve
\endalign
$$
for large mutual distances between the elements of $\Cal A$ and $\Cal B$. 

Consider, on the other hand, a two dimensional physical system of $2m+2n+1$ charges 
$Q_1,\dotsc,Q_{2m+2n+1}$
of magnitudes $q_1q_e,\dotsc,q_{2m+2n+1}q_e$ (expressed as integer multiples of the elementary 
charge $q_e$). Then assuming that there are only electrical forces between them
(governed by the two dimensional electrostatic potential, $-1/(2\pi\epsilon_0)\ln(d)$, where 
$d$ is the distance and $\epsilon_0$ is the permittivity of empty space 
\cite{\Ftwo,\S4-2,\,\S5-5}), we obtain 
by the Fundamental Theorem of Statistical Physics (see e.g. \cite{\Fone,\S 40-3}) that
the probability $P_e({\bold d})$ of finding the charges at mutual distances 
${\bold d}=(d_{ij})_{1\leq i<j\leq 2m+2n+1}$, 
relative to the probability $P_e({\bold d'})$ of finding them at 
mutual distances ${\bold d'}=(d'_{ij})_{1\leq i<j\leq 2m+2n+1}$, is
$$
\align
&\!\!\!\!\!\!\!\!\!\!\!
\frac{P_e({\bold d})}{P_e({\bold d'})}=\\
&\!\!\!\!\!\!
\exp\left(-\frac{q_e^2}{2\pi\epsilon_0 kT}\left[\sum_{1\leq i<j\leq 2m+2n+1}q_iq_j(-\ln d_{ij})
-\sum_{1\leq i<j\leq 2m+2n+1}q_iq_j(-\ln d'_{ij})\right]\right),\tag\TwoEleven
\endalign
$$
where $k$ is Boltzmann's constant (see e.g. \cite{\Fone,\S39-4}) and $T$ is absolute temperature.
%

Relations (\TwoTwelve) and (\TwoEleven) show that, at least in the case when the two dimensional 
physical system of electrical charges satisfies the magnitude and geometrical 
distribution requirements in the hypothesis of Theorem 2.1,
their physical electrostatic interaction at temperature $T=q_e^2/(\pi\epsilon_0k)$ 
is correctly modeled by our 
model\footnote{In fact, if one accepts Conjecture 14.1,
our set-up allows one to introduce an extra
parameter in (\TwoTwelve), a positive integer $x$, which makes the parallelism between (\TwoTwelve) and 
(\TwoEleven) go through for any temperature.
To this end, refine our triangular lattice 
$\Cal T$ so that each unit triangle is subdivided into equilateral triangles of side $1/x$; denote
the new lattice by ${\Cal T}_x$. 
The 
hypotheses of Conjecture 14.1 are clearly satisfied when the plurimers in $\Cal A$ and $\Cal B$ are 
regarded
as plurimers on ${\Cal T}_x$. Both their charges and their mutual distances get multiplied by $x$.
By (\TwoTwelve) one readily sees that the effects of the change in distance cancel out, while the
effect of changing the charges is that the fraction $-1/2$ in front of the square brackets in
(\TwoTwelve) changes to $-x^2/2$. 
\newline
\phantom{aaa} 
A second ``calibration'' parameter $a$, a given positive integer, can be introduced 
as follows. Arrange, by $a$-fold lattice refinement, for all $q_i$'s in (\TwoTwelve)
to be divisible by $a$. Factor out the common factor of $a^2$ in front of the
square parenthesis in (\TwoTwelve) and re-denote each leftover $q_i/a$ by $q_i$. The formula
obtained this way from (\TwoTwelve) still parallels (\TwoEleven), but now the elementary
physical charge $q_e$ corresponds to a plurimer of charge $a$. In particular, $a$ should have
a fixed value.
\newline
\phantom{aaa}
The overall effect of $x$-fold lattice refinement and ``calibration'' by $a$ is 
that the fraction $-1/2$ in front of the square brackets in (\TwoTwelve) changes to $-(ax)^2/2$.
}. 
We present in Section 15 a possible three dimensional
analog of this parallelism that allows any temperature $T$.

Moreover,
what plays the role of {\it electrostatic potential} in our model is just the {\it
entropy}---the logarithm of the number of dimer coverings---which had been considered
before, in a different context \cite{\Ca}\cite{\LP} (it involved
the number of dimer coverings of the {\it inside} structure of a molecule, not of the
{\it outside}, as in our situation), as a possible measure for chemical
energy (one of the original motivations was simply its satisfying the addition 
principle, for isolated systems).

Our interpretation has the advantage that Coulomb's law emerges as an asymptotic law 
valid when the distances between the charges are large in comparison with the size of
the charges. This would explain why Coulomb's law seems to break down at very small
distances (cf. \cite{\Ftwo,\S5-8}, at distances lower than about $10^{-14}$). 

Furthermore, in this context (\TwoNine) can be interpreted as describing the interaction of
electrical charges with an ``edge'' of this two dimensional ``universe.'' The same sign
for the image charges ensures a repelling effect, and this helps keeping the charges
inside the ``universe''!

An appealing feature of such a model is that it is discrete, and therefore
models the physical electrostatic interaction by considering a discrete ambient space.

The possibility of a discrete ``machinery'' underlying the electrostatics of the real
physical world is mentioned by Feynman in \cite{\Ftwo, \S12-7}. When
indicating how several equations in physics, like for instance that for neutron
diffusion, are true only as approximations when the distance over which one looks is
large (for neutron diffusion, large in comparison to the mean free
path), Feynman goes on to ask: 

\eightpoint
\medskip
``Is the same statement perhaps also true for the {\it electrostatic} equations? Are
they also correct only as a smoothed-out imitation of a really much more complicated
microscopic world? Could it be that the real world consists of little X-ons which can
be seen only at {\it very} tiny distances? And that in our measurements we are always
observing on such a large scale that we can't see these little X-ons, and that is why
we get the differential equations?

Our currently most complete theory of electrodynamics does indeed have its
difficulties at very short distances. So it is possible, in principle, that these
equations are smoothed-out versions of something. They appear to be correct at
distances down to about $10^{-14}$ cm, but then they begin to look wrong. It is
possible that there is some as yet undiscovered underlying `machinery,' and that the
details of an underlying complexity are hidden in the smooth-looking equations---as is
so in the `smooth' diffusion of neutrons. But no one has yet formulated a successful
theory that works that way.''

%

\medskip
\tenpoint
The random tiling model presented in this paper (and pursued further in a subsequent paper) 
seems to be a possible such
``machinery'' that would produce, in a two dimensional world, precisely the effects of
electrostatics. As presented in Section 14, we conjecture that the validity of our
model does not depend on the particular choice of the hexagonal lattice, but it holds
in fact for any plane bipartite lattice under a suitable embedding. This would imply
that the electrostatic effects we obtain depend only on the {\it space}---in
accordance with Feynman's suggestion \cite{\Ftwo, \S12-7} that it might be
the space itself, the common ``framework into which physics is put,'' that is responsible for
the emergence of such a simple equation governing electrostatics (as well as other
physical phenomena).

The model presented in this paper has natural analogs in higher dimensions---for
instance, by considering the cubic lattices. We believe that the three dimensional
analog presents the effects of electrostatics in the real three dimensional world.
Some details about these considerations are presented in Section 15.


\mysec{3. Reduction to boundary-influenced correlations}

We will find it convenient to express the number of dimer coverings of the region
$H_N\left(\matrix{R_1}\\{v_1}\endmatrix
\cdots\matrix{R_m}\\{v_m}\endmatrix;
\matrix{R'_1}\\{v'_1}\endmatrix\cdots\matrix{R'_n}\\{v'_n}\endmatrix\right)$
in terms of the number of tilings of two ``halves'' of it. One of them is the region
$W_N\left(\matrix{R_1}\\{v_1}\endmatrix
\cdots\matrix{R_m}\\{v_m}\endmatrix;
\matrix{R'_1}\\{v'_1}\endmatrix\cdots\matrix{R'_n}\\{v'_n}\endmatrix\right)$ defined
in Section 2---the portion to the west of the cut illustrated in Figure 2.3.
The other is the part of $H_N$ to the east of that cut---more precisely, what is left
from the region east of the cut after removing the forced dimers (see Figure 3.1 for
an example; the forced dimers are shaded), and weighting by $1/2$ the $N+2m-1$ dimer
positions below the monomer $u$ that are closest to the cut; denote it by
$E_N\left(\matrix{R_1}\\{v_1}\endmatrix
\cdots\matrix{R_m}\\{v_m}\endmatrix;
\matrix{R'_1}\\{v'_1}\endmatrix\cdots\matrix{R'_n}\\{v'_n}\endmatrix\right)$. 
Figure 3.1 illustrates an example 
(for clarity, the two regions are pictured after separating them
horizontally by one unit; the dimer positions weighted by $1/2$ are indicated by
shaded ellipses).

\topinsert
\centerline{\mypic{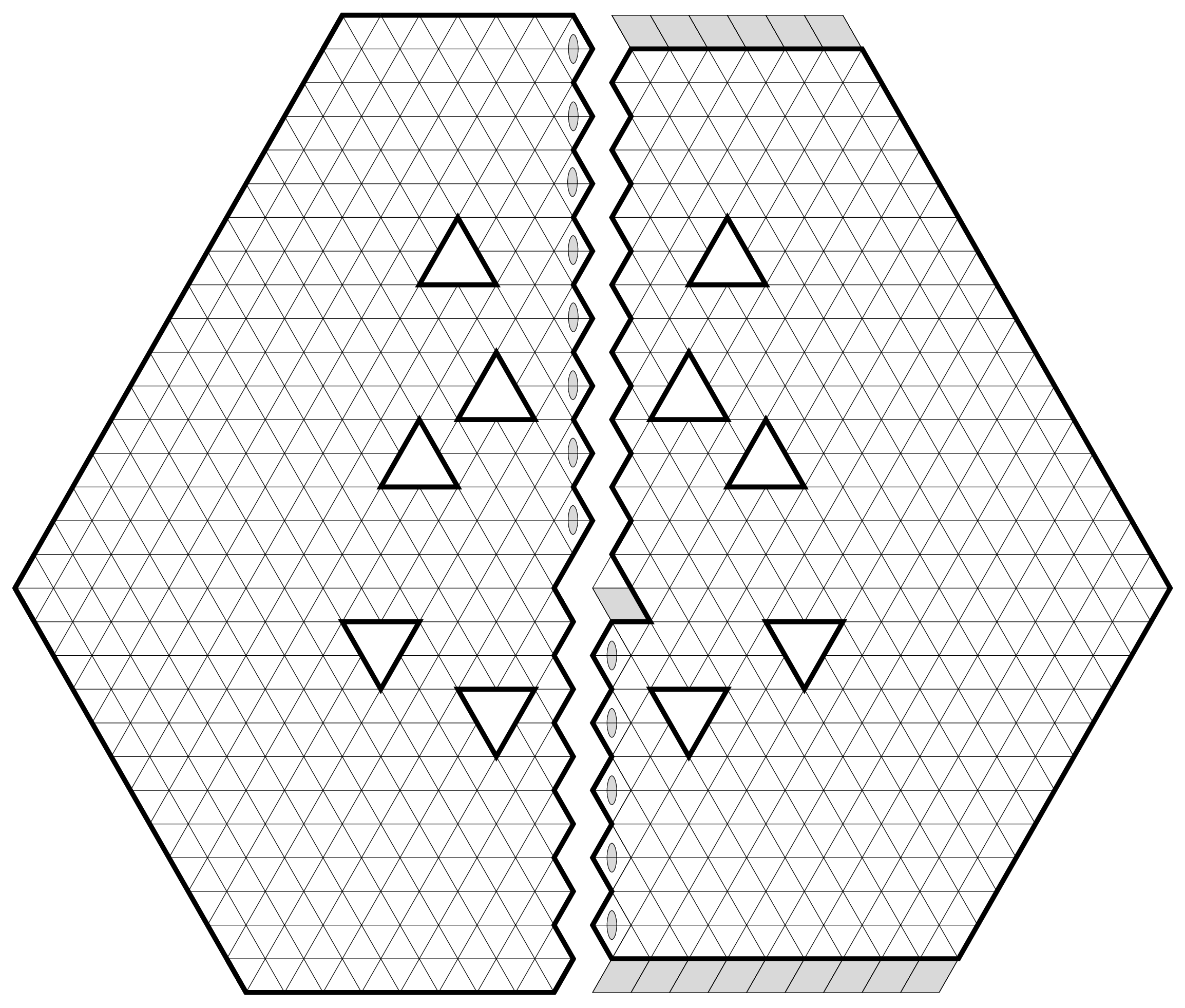}}
\centerline{{\smc Figure~3.1.} 
{\rm $W_2\left(\matrix{5}&2\\{0}&1\endmatrix;
\matrix4&2&3\\1&2&4\endmatrix\right)$ and 
$E_2\left(\matrix{5}&2\\{0}&1\endmatrix;
\matrix4&2&3\\1&2&4\endmatrix\right)$
.}}
\endinsert

The regions $W_N$ and $E_N$ were defined in such a way that they are precisely the ones that 
result from applying the
Factorization Theorem \cite{\Cone,Theorem 1.2} to the region $H_N$. Therefore, 
the quoted Factorization Theorem yields the following result.

\proclaim{Proposition 3.1} 
$$
\align
&\!\!\!\!\!\!\!\!\!
\M\left(H_N\left(\matrix{R_1}\\{v_1}\endmatrix
\cdots\matrix{R_m}\\{v_m}\endmatrix;
\matrix{R'_1}\\{v'_1}\endmatrix\cdots\matrix{R'_n}\\{v'_n}\endmatrix\right)\right)=\\
&
2^{N+m+n}
\M\left(W_N\left(\matrix{R_1}\\{v_1}\endmatrix
\cdots\matrix{R_m}\\{v_m}\endmatrix;
\matrix{R'_1}\\{v'_1}\endmatrix\cdots\matrix{R'_n}\\{v'_n}\endmatrix\right)\right)
\M\left(E_N\left(\matrix{R_1}\\{v_1}\endmatrix
\cdots\matrix{R_m}\\{v_m}\endmatrix;
\matrix{R'_1}\\{v'_1}\endmatrix\cdots\matrix{R'_n}\\{v'_n}\endmatrix\right)\right).
\tag3.1
\endalign
$$
\endproclaim
In particular, (3.1) holds for the reference position of the plurimers used in
definitions (\TwoOne) and (\TwoSix). We obtain
$$
\align
&
\M\left(H_N\left(\matrix 1\\0\endmatrix
\cdots\matrix 2m-1\\0\endmatrix;
\matrix 1\\0\endmatrix\cdots\matrix
2n-1\\0\endmatrix\right)\right)=\\
&
2^{N+m+n}\M\!\left(\!W_N\!\left(\matrix 1\\0\endmatrix
\cdots\matrix 2m-1\\0\endmatrix;
\matrix 1\\0\endmatrix\cdots\matrix
2n-1\\0\endmatrix\right)\right)
\M\!\left(\!E_N\!\left(\matrix 1\\0\endmatrix
\cdots\matrix 2m-1\\0\endmatrix;
\matrix 1\\0\endmatrix\cdots\matrix
2n-1\\0\endmatrix\right)\right).\\
\tag3.2
\endalign
$$
Dividing (3.1) and (3.2) side by side and letting $N\to\infty$, one sees by (\TwoOne) and
(\TwoSix) that $\omega$ and $\omega_b$ can be related by considering another
boundary-influenced correlation, defined by means of the regions $E_N$:
$$
\bar{\omega}_b
\left(\matrix{R_1}\\{v_1}\endmatrix
\cdots\matrix{R_m}\\{v_m}\endmatrix;
\matrix{R'_1}\\{v'_1}\endmatrix\cdots\matrix{R'_n}\\{v'_n}\endmatrix\right):=
\lim_{N\to\infty}
\frac{\M\left(E_N\left(\matrix{R_1}\\{v_1}\endmatrix
\cdots\matrix{R_m}\\{v_m}\endmatrix;
\matrix{R'_1}\\{v'_1}\endmatrix\cdots\matrix{R'_n}\\{v'_n}\endmatrix\right) \right)}
{\M\left(E_N\left(\matrix 1\\0
\endmatrix
\cdots\matrix 2m-1\\0\endmatrix;
\matrix 1\\0\endmatrix\cdots\matrix
2n-1\\0\endmatrix\right)\right)}\tag3.3
$$
(the fact that this limit exists follows by Lemma 13.1).

By (3.1)--(3.3), the definitions (\TwoOne) and (\TwoSix), and the existence of the limits
in the latter two definitions (guaranteed by Lemma 5.1 and Lemma 13.1),
one obtains the following result.

\proclaim{Proposition 3.2} The limit $(\TwoOne)$ defining the correlation at the center 
exists and its value is the product of the 
two boundary-influenced correlations:
$$
\align
&\!\!\!\!\!\!\!\!\!\!\!\!\!\!\!\!\!\!\!\!\!\!\!\!
\omega\left(\matrix{R_1}\\{v_1}\endmatrix
\cdots\matrix{R_m}\\{v_m}\endmatrix;
\matrix{R'_1}\\{v'_1}\endmatrix\cdots\matrix{R'_n}\\{v'_n}\endmatrix\right)=\\
&
\omega_b\left(\matrix{R_1}\\{v_1}\endmatrix
\cdots\matrix{R_m}\\{v_m}\endmatrix;
\matrix{R'_1}\\{v'_1}\endmatrix\cdots\matrix{R'_n}\\{v'_n}\endmatrix\right)
\bar{\omega}_b\left(\matrix{R_1}\\{v_1}\endmatrix
\cdots\matrix{R_m}\\{v_m}\endmatrix;
\matrix{R'_1}\\{v'_1}\endmatrix\cdots\matrix{R'_n}\\{v'_n}\endmatrix\right).\tag3.4
\endalign
$$
\endproclaim

Therefore, in order to obtain the asymptotic behavior of the correlation at the center
$\omega$, it is enough to study the boundary-influenced correlations $\omega_b$ and
$\bar{\omega}_b$.

\mysec{4. A simple product formula for correlations along the boundary}

Our calculations are built upon an explicit product formula that we present in this
section for certain correlations along the boundary of the regions $W_N$. To state
this we need to introduce a new family of regions, closely related to the $W_N$'s, and
in terms of which the boundary-influenced correlation $\omega_b$ turns out to be
expressible.

Let $W$ be the region determined by the common outside boundary of the regions
$W_N\left(\matrix{R_1}\\{v_1}\endmatrix
\cdots\matrix{R_m}\\{v_m}\endmatrix;
\matrix{R'_1}\\{v'_1}\endmatrix\cdots\matrix{R'_n}\\{v'_n}\endmatrix\right)$, for
fixed $N$, $m$ and $n$. Then $W$ is the half-hexagonal lattice region with  
four straight sides---the southern side of length $N+2n$, 
southwestern of length $2N+4m$, northwestern of length $2N+4n+1$, and northern 
of length $N+2m$---followed by
$N+2n$ descending zig-zags to the lattice point $O$, one extra unit step southwest of
$O$, and $N+2m$ more descending zig-zags to close up the boundary (the boundary of the
region $W$ corresponding 
to $m=4$, $n=6$, and $N=2$ can be seen in Figure 4.1). In addition, 
the $N+2n$ dimer positions weighted 1/2 in the regions $W_N$ are also weighted so in
$W$.

\topinsert
\centerline{\mypic{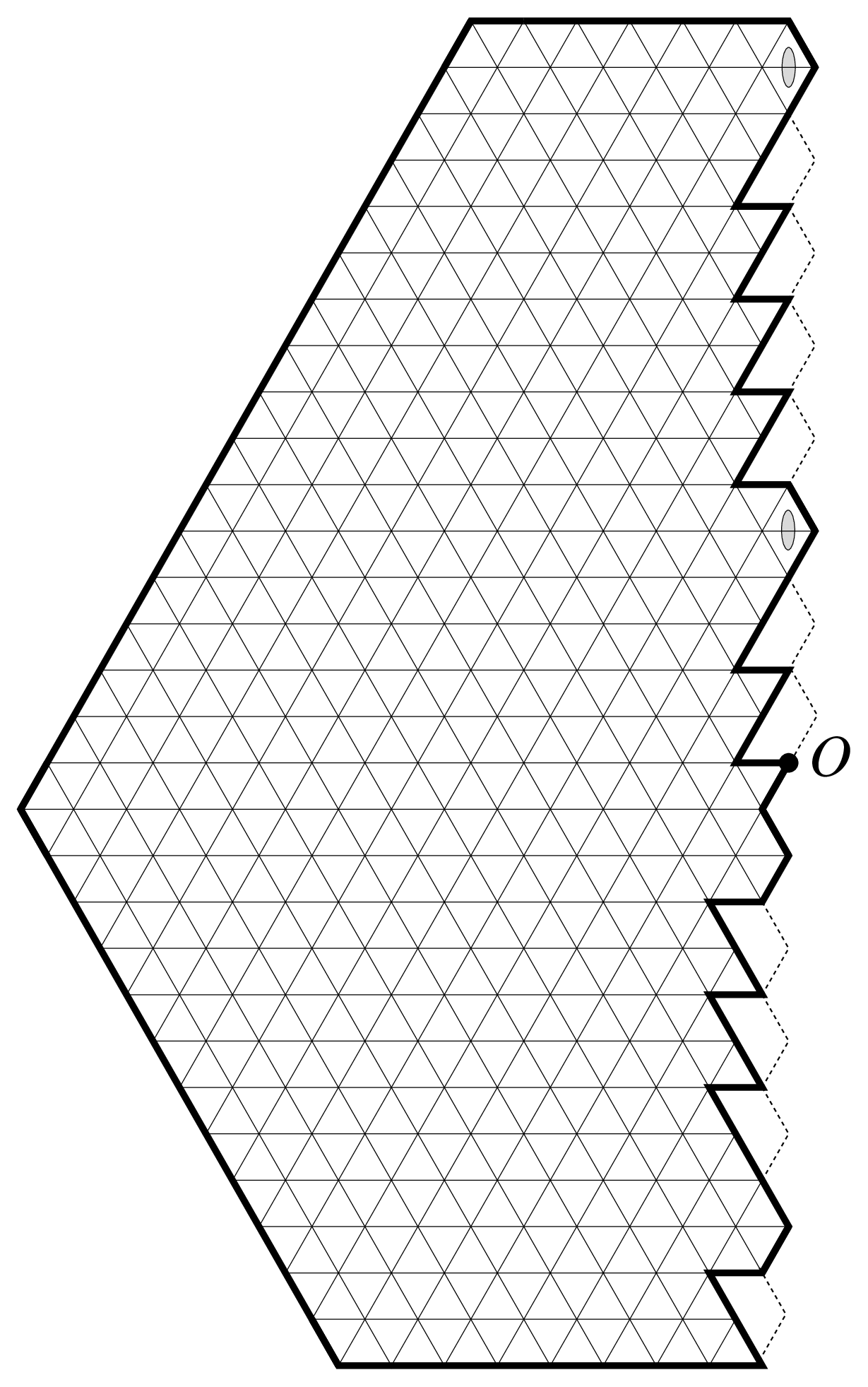}}
\centerline{{\smc Figure~4.1.} 
{\rm $W_2[1,2,3,5;0,1,3,4,5,6]$
.}}
\endinsert

The eastern side of $W$ can be viewed as consisting of {\it bumps}---pairs of adjacent
lattice segments forming an angle that opens to the west: $N+2m$ bumps below $O$, and
$N+2n$ above $O$. Label the former by $0,1,\dotsc,N+2m-1$ and the latter by 
$0,1,\dotsc,N+2n-1$, both labelings starting with the bumps closest to $O$ and then moving
successively outwards. 

Since everywhere in the description of $W$ the parameters $m$
and $n$ appear with even multiplicative coefficients, 
we re-denote, for notational simplicity, $2m$ by $m$
and $2n$ by $n$. Therefore we consider the four straight sides of $W$ to have lengths 
$N+n$, $2N+2m$, $2N+2n+1$ and $N+m$, while the number of bumps below and above $O$ 
is $N+m$ and $N+n$, respectively. The results of this section do not assume that $m$
and $n$ are even (even though we will only use them for even $m$ and $n$).

We allow any bump above $O$ to be ``removed'' by placing an up-pointing quadromer 
(lattice triangle of side two) across it and discarding the three monomers of $W$ it
covers. Similarly, a bump below $O$ can be removed by placing a {\it down}-pointing
quadromer across it and discarding the three monomers it covers.

We are now ready to introduce the family of regions mentioned in the first
paragraph of this section:
define $W_N[k_1,\dotsc,k_{m};l_1,\dotsc,l_{n}]$ to be the region 
obtained from $W$ by removing the bumps below $O$ with labels 
$0\leq k_1<k_2<\cdots<k_{m}\leq N+m-1$, and the bumps above $O$ with labels 
$0\leq l_1<l_2<\cdots<l_{n}\leq N+n-1$. Figure 4.1 shows $W_2[1,2,3,5;0,1,3,4,5,6]$.

The product formula referred to in the title of this section is stated in the following
result.

\proclaim{Proposition 4.1} For $m,n\geq0$ and fixed integers 
$0\leq k_1<k_2<\cdots<k_{m}$ and $0\leq l_1<l_2<\cdots<l_{n}$ we have
$$
\align
\lim_{N\to\infty}&\frac{\M\left(W_N[k_1,\dotsc,k_{m};l_1,\dotsc,l_{n}]\right)}
{\M\left(W_N[0,\dotsc,m-1;0,\dotsc,n-1]\right)}=\\
&\ \ \ \ \ \ \ 
\prod_{i=1}^{m}
\frac{\frac{{\displaystyle (3/2)_{k_i}}}{{\displaystyle (2)_{k_i}}}}
{\frac{{\displaystyle (3/2)_{i-1}}}{{\displaystyle (2)_{i-1}}}}
\prod_{i=1}^{n}
\frac{\frac{{\displaystyle (3/2)_{l_i}}}{{\displaystyle (1)_{l_i}}}}
{\frac{{\displaystyle (3/2)_{i-1}}}{{\displaystyle (1)_{i-1}}}}
\,\frac{
{\displaystyle \prod_{1\leq i<j\leq m}}
\frac{{\displaystyle k_j-k_i}}{{\displaystyle j-i}}
{\displaystyle \prod_{1\leq i<j\leq n}}
\frac{
{\displaystyle l_j-l_i}}{{\displaystyle j-i}}
}
{
{\displaystyle \prod_{i=1}^{m}\prod_{i=1}^{n}}
\frac{{\displaystyle k_i+l_j+2}}
{\displaystyle {i+j}}
}\tag4.1\\
=
&
\chi_{m,n}
\prod_{i=1}^{m}\frac{(3/2)_{k_i}}{(2)_{k_i}}
\prod_{i=1}^{n}\frac{(3/2)_{l_i}}{(1)_{l_i}}
\frac{
{\displaystyle \prod_{1\leq i<j\leq m}}(k_j-k_i)
{\displaystyle \prod_{1\leq i<j\leq n}}(l_j-l_i)}
{{\displaystyle \prod_{i=1}^{m}\prod_{i=1}^{n}}(k_i+l_j+2)},\tag4.2
\endalign
$$
where $(a)_k:=a(a+1)\cdots(a+k-1)$ $($by convention, $(a)_0:=1$$)$ and
$$
\chi_{m,n}=\prod_{i=0}^{m-1}\frac{(2)_i}{(1)_i(3/2)_i}
\prod_{i=0}^{n-1}\frac{(i+2)_m}{(3/2)_i}
.\tag4.3
$$
\endproclaim

\pf An explicit product formula for 
$\M\left(W_N[k_1,\dotsc,k_{m};l_1,\dotsc,l_{n}]\right)$ follows from the results 
of \cite{\Cthree}. Indeed, our regions
$W_N[k_1,\dotsc,k_{m};l_1,\dotsc,l_{n}]$ are just a different notation
for the regions $\bar{R}_{\bold k,\bold l}(x)$ defined in \cite{\Cthree, Section 2}: In
the notation of \cite{\Cthree}, $x$ is the length of the base---$N+n$ in the current
notation---while ${\bold k}$ and ${\bold l}$ are the (strictly increasing) lists of 
labels---incremented 
by 1, due to a shift of one unit in labeling bumps in the present paper compared to 
\cite{\Cthree}---of the bumps that {\it remain} in the region. Therefore, 
$$
\align
W_N[k_1,\dotsc,k_{m};&l_1,\dotsc,l_{n}]=\\
&\!\!\!\!\!\!\!\!\!\!\!\!
\bar{R}_{[1,\dotsc,N+m]\setminus[k_1+1,\dotsc,k_{m}+1],
[1,\dotsc,N+n]\setminus[l_1+1,\dotsc,l_{n}+1]}(N+n).\tag4.4
\endalign
$$

Relations \cite{\Cthree,(6.1)}, \cite{\Cthree,(6.2)} and \cite{\Cthree,(6.4)} state that for any 
pair of lists $\bold p=[p_1,\dotsc,p_s]$, 
$1\leq p_1<p_2<\dotsc<p_s$ and $\bold q=[q_1,\dotsc,q_t]$, 
$1\leq q_1<q_2<\dotsc<q_t$, and for any nonnegative integer 
$x\geq q_t-p_s-t+s$ (see \cite{\Cthree,p.10}), one has
$$
\M\left(\bar{R}_{\bold p,\bold q}(x)\right)=
\bar{c}_{\bold p,\bold q}F_{{\bold p},{\bold q}}(x),\tag4.5
$$ 
where
$$
\bar{c}_{\bold p,\bold q}=2^{{t-s\choose 2}-s}\prod_{i=1}^s\frac{1}{(2p_i-1)!}
\prod_{i=1}^t\frac{1}{(2q_i)!}
\frac{
{\prod_{1\leq i<j\leq s}}(p_j-p_i)
{\prod_{1\leq i<j\leq t}}(q_j-q_i)}
{{\prod_{i=1}^{s}\prod_{j=1}^{t}}(p_i+q_j)}\tag4.6
$$
and the polynomials $F_{{\bold p},{\bold q}}(x)$ satisfy 
$$
\align
\frac{F_{{\bold p}^{|i\rangle},{\bold q}}(x)}{F_{{\bold p},{\bold q}}(x)}
&=(x-p_i+p_s)(x+p_i+p_s-s+t+1),\ \ \ \ \text{\rm for $1\leq i<s$}\tag4.7\\
\frac{F_{{\bold p},{\bold q}^{|i\rangle}}(x)}{F_{{\bold p},{\bold q}}(x)}
&=(x+q_i+p_s+1)(x-q_i+p_s-s+t),\ \ \ \ \text{\rm for $1\leq i\leq t$}\tag4.8
\endalign
$$
(here ${\bold p}^{|i\rangle}$ is the list obtained from ${\bold p}$ by increasing its $i$-th
element by 1---in particular, it is defined only if $l_{i+1}-l_i\geq2$).

To deduce the limit (4.1) from these formulas, it will be convenient to consider
first the limit
$$
\lim_{N\to\infty}\frac{\M\left(W_N[k_1,\dotsc,k_{i-1},k_i+1,k_{i+1},\dotsc,k_{m};
l_1,\dotsc,l_{n}]\right)}
{\M\left(W_N[k_1,\dotsc,k_{i-1},k_i,k_{i+1},\dotsc,k_{m};
l_1,\dotsc,l_{n}]\right)},\tag4.9
$$
for $k_i+1<k_{i+1}$.

Use (4.4) to view the regions involved in this fraction as 
$\bar{R}_{{\bold p},{\bold q}}(x)$'s. 
The lists ${\bold p}$ and ${\bold q}$ 
corresponding to the regions at the numerator and denominator in (4.9) are 
illustrated in Figure 4.2 (the shaded squares indicate the entries removed at the
indices on the right hand side of (4.4)). Since the ${\bold q}$-lists are the same,
the limit (4.9) can be found applying formulas (4.6) and (4.7) to the lists
illustrated in Figure 4.2. 

Since the only difference between the ${\bold p}$-lists in Figure 4.2(a) is the one
indicated by the dot in that figure, we obtain from (4.6) that the contribution to the 
fraction in (4.9) coming from the ratio of the 
$\bar{c}_{{\bold p},{\bold q}}$'s is
$$
\align
&\frac{
\frac{{\displaystyle 1}}{{\displaystyle (2k_i+1)!}}
}
{
\frac{{\displaystyle 1}}{{\displaystyle (2k_i+3)!}}
}
\frac{
\frac{{\displaystyle k_i!}}{{\displaystyle (k_i-k_1)\cdots(k_i-k_{i-1})}}
\frac{{\displaystyle [N+m-(k_i+1)]!}}{{\displaystyle (k_{i+1}-k_i)\cdots(k_m-k_{i})}}
}
{
\frac{{\displaystyle (k_i+1)!}}{{\displaystyle (k_i-k_1+1)\cdots(k_i-k_{i-1}+1)}}
\frac{{\displaystyle [N+m-(k_i+2)]!}}{{\displaystyle(k_{i+1}-k_i-1)\cdots(k_m-k_{i}-1)}}
}\\
&\ \ \ \ \ \ \ \ \ \ \ \ \ \times
\frac{
\frac{{\displaystyle (k_i+3)\cdots(k_i+2+N+m)}}
{{\displaystyle (k_i+l_1+3)\cdots(k_i+l_n+3)}}
}
{
\frac{{\displaystyle (k_i+2)\cdots(k_i+1+N+m)}}
{{\displaystyle (k_i+l_1+2)\cdots(k_i+l_n+2)}}
}\\
&=(k_i+2+N+n)(N+m-(k_i+1))\frac{(2k_i+2)(2k_i+3)}{(k_i+1)(k_i+2)}\\
&\ \ \ \ \ \ \ \ \ \ \ \ \ \times
\frac{((k_i+1)-k_1)\cdots((k_i+1)-k_{i-1})(k_{i+1}-(k_i+1))\cdots(k_{m}-(k_i+1))}
{(k_i-k_1)\cdots(k_i-k_{i-1})(k_{i+1}-k_i)\cdots(k_{m}-k_i)}\\
&\ \ \ \ \ \ \ \ \ \ \ \ \ \times
\frac{(k_i+l_1+2)\cdots(k_i+l_n+2)}
{(k_i+l_1+3)\cdots(k_i+l_n+3)}.\tag4.10
\endalign
$$

\topinsert
\centerline{\mypic{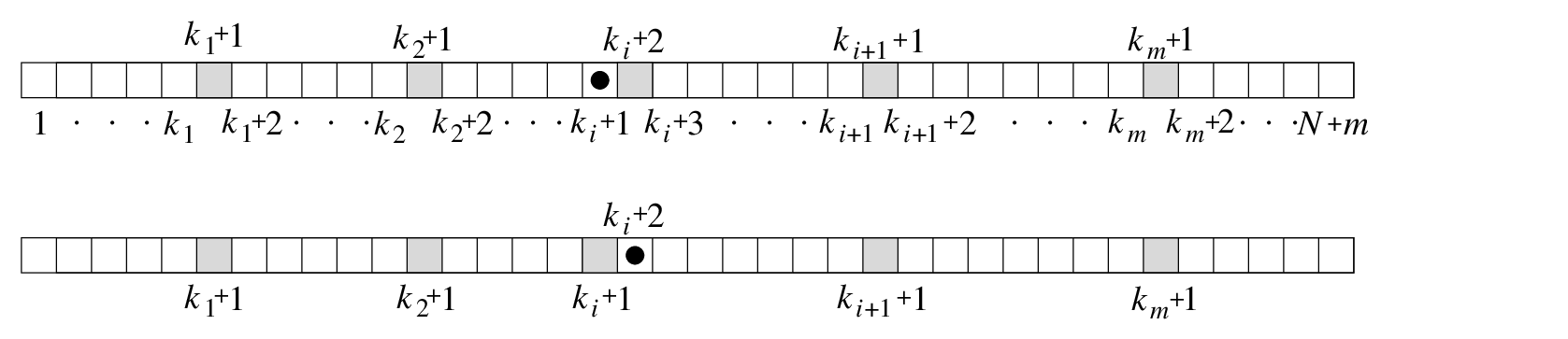}}
\centerline{{\rm (a). 
The ${\bold p}$'s.}}

\centerline{\mypic{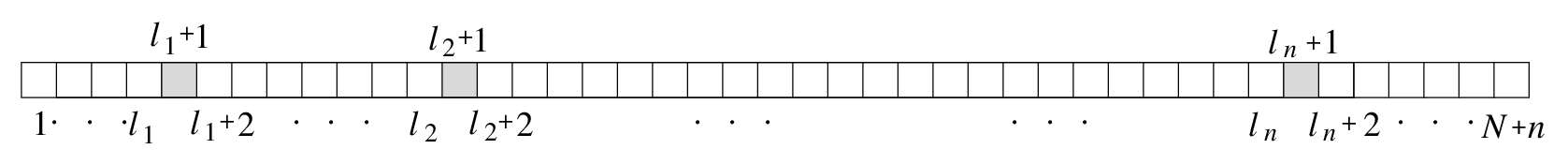}}
\centerline{{\rm (b). 
The ${\bold q}$'s.}}
\medskip
\centerline{{\smc Figure~4.2.} {\rm Incremental change in the first index list.}}
\endinsert

On the other hand, the contribution to the fraction in (4.9) coming from the
polynomials $F_{{\bold p},{\bold q}}(x)$ is readily seen from (4.7) to be
$$
\frac{1}{(2N+m+n-(k_i+1))(2N+m+n+(k_i+1)+1)}.\tag4.11
$$
By (4.10) and (4.11) one readily finds the limit (4.9) to be
$$
\align
\lim_{N\to\infty}&\frac{\M\left(W_N[k_1,\dotsc,k_{i-1},k_i+1,k_{i+1},\dotsc,k_{m};
l_1,\dotsc,l_{n}]\right)}
{\M\left(W_N[k_1,\dotsc,k_{i-1},k_i,k_{i+1},\dotsc,k_{m};
l_1,\dotsc,l_{n}]\right)}=\\
\ \ \ \ \ \ \ \ \ \ \ \ 
&\frac{k_i+3/2}{k_i+2}
\frac{((k_i+1)-k_1)\cdots((k_i+1)-k_{i-1})(k_{i+1}-(k_i+1))\cdots(k_{m}-(k_i+1))}
{(k_i-k_1)\cdots(k_i-k_{i-1})(k_{i+1}-k_i)\cdots(k_{m}-k_i)}\\
&\ \ \ \ \ \ \ 
\times
\frac{(k_i+l_1+2)\cdots(k_i+l_n+2)}
{((k_i+1)+l_1+2)\cdots((k_i+1)+l_n+2)}.\tag4.12
\endalign
$$

Regard (4.12) as giving the effect of decrementing the entry $k_i+1$ in the argument
to $k_i$. A further decrementation of this entry to $k_i-1$ will produce, by (4.12),
the factor
$$
\align
&\frac{k_i+1/2}{k_i+1}
\frac{(k_i-k_1)\cdots(k_i-k_{i-1})(k_{i+1}-k_i)\cdots(k_{m}-k_i))}
{((k_i-1)-k_1)\cdots((k_i-1)-k_{i-1})(k_{i+1}-(k_i-1))\cdots(k_{m}-(k_i-1))}\\
&\ \ \ \ \ \ \ 
\times
\frac{(k_i+l_1+1)\cdots(k_i+l_n+1)}
{(k_i+l_1+2)\cdots(k_i+l_n+2)}.\tag4.13
\endalign
$$
One notices that telescoping simplifications occur when multiplying together the
second and third fractions in (4.12) for successive decrementations of each
argument $k_i$. In particular, if the argument $k_1$ is decremented
all the way to $0$, we obtain by repeated application of (4.12) that
$$
\align
\lim_{N\to\infty}&\frac{\M\left(W_N[k_1,k_2,\dotsc,k_{m};
l_1,\dotsc,l_{n}]\right)}
{\M\left(W_N[0,k_2,\dotsc,k_{m};
l_1,\dotsc,l_{n}]\right)}=\\
&\ \ \ 
\frac{(3/2)_{k_1}}{(2)_{k_1}}
\frac{(k_2-k_1)(k_3-k_1)\cdots(k_{m}-k_1)}
{(k_2-0)(k_3-0)\cdots(k_{m}-0)}
\frac{(l_1+2)\cdots(l_n+2)}
{(k_1+l_1+2)\cdots(k_1+l_n+2)}.\tag4.14
\endalign
$$
Similarly, we obtain
$$
\align
\lim_{N\to\infty}&\frac{\M\left(W_N[0,k_2,k_3,\dotsc,k_{m};
l_1,\dotsc,l_{n}]\right)}
{\M\left(W_N[0,1,k_3,\dotsc,k_{m};
l_1,\dotsc,l_{n}]\right)}=\\
&\ \ \ 
\frac{(5/2)_{k_2-1}}{(3)_{k_2-1}}
\frac{(k_2-0)(k_3-k_2)\cdots(k_{m}-k_2)}
{(1-0)(k_3-1)\cdots(k_{m}-1)}
\frac{(l_1+3)\cdots(l_n+3)}
{(k_2+l_1+2)\cdots(k_2+l_n+2)},\\
\lim_{N\to\infty}&\frac{\M\left(W_N[0,1,k_3,k_4,\dotsc,k_{m};
l_1,\dotsc,l_{n}]\right)}
{\M\left(W_N[0,1,2,k_4,\dotsc,k_{m};
l_1,\dotsc,l_{n}]\right)}=\\
&\ \ \ 
\frac{(7/2)_{k_3-2}}{(4)_{k_3-2}}
\frac{(k_3-0)(k_3-1)(k_4-k_3)\cdots(k_{m}-k_3)}
{(2-0)(2-1)(k_4-2)\cdots(k_{m}-2)}
\frac{(l_1+4)\cdots(l_n+4)}
{(k_3+l_1+2)\cdots(k_3+l_n+2)},\\
\\
&\ \ \ \ \ \ \ \ \ \ \ \ \ \ \ \ \ \ \vdots\\
\\
\lim_{N\to\infty}&\frac{\M\left(W_N[0,1,\dotsc,m-2,k_{m};
l_1,\dotsc,l_{n}]\right)}
{\M\left(W_N[0,1,\dotsc,m-1;
l_1,\dotsc,l_{n}]\right)}=\\
&\ \ \ 
\frac{((2m+1)/2)_{k_m-m+1}}{(m+1)_{k_m-m+1}}
\frac{(k_m-0)(k_m-1)\cdots(k_{m}-(m-2))}
{((m-1)-0)((m-1)-1)\cdots((m-1)-(m-2))}\\
&\ \ \ \ \ \ \ \ \ \ \ \ \ \ \ \ \ \ 
\times
\frac{(l_1+m+1)\cdots(l_n+m+1)}
{(k_m+l_1+2)\cdots(k_m+l_n+2)}
.\tag4.15
\endalign
$$
Now multiply together the equalities in (4.14) and (4.15). The first fractions on
their right hand sides combine to give 
$\prod_{i=1}^m((3/2)_{k_i}/(2)_{k_i})/\prod_{i=1}^{m}((3/2)_{i-1}/(2)_{i-1})$. Due to
simplifications, the second fractions on the right hand sides yield
$\prod_{1\leq i<j\leq m}(k_j-k_i)/(j-i)$. The third fractions give 
$\prod_{i=1}^n(l_i+2)_m/\prod_{i=1}^m\prod_{j=1}^n(k_i+l_j+2)$. We obtain
$$
\align
\lim_{N\to\infty}\frac{\M\left(W_N[k_1,k_2,\dotsc,k_{m};
l_1,\dotsc,l_{n}]\right)}
{\M\left(W_N[0,1,\dotsc,m-1;
l_1,\dotsc,l_{n}]\right)}&=\prod_{i=1}^n(l_i+2)_m\\
&\!\!\!\!\!\!\!\!\!\!\!\!\!\!\!\!\!\!\!\!\!\!
\times
\frac{
\prod_{i=1}^m\frac{ {\displaystyle (3/2)_{k_i}} }{ {\displaystyle (2)_{k_i}} }
}
{
\prod_{i=1}^{m}\frac{ {\displaystyle (3/2)_{i-1}} }{ {\displaystyle (2)_{i-1}} }
}
\frac{
\prod_{1\leq i<j\leq m}\frac{ {\displaystyle k_j-k_i} }{ {\displaystyle j-i} }
}
{
\prod_{i=1}^m\prod_{j=1}^n(k_i+l_j+2)
}
.\tag4.16
\endalign
$$

The effect of decrementing arguments in the list $[l_1,\dotsc,l_n]$ can be analyzed in a
similar way. Indeed, compare the situation when the lists of $k_j$'s at the indices on
the left hand side of (4.1) are the same, and the lists of $l_j$'s are 
$[l_1,\dotsc,l_{i-1},l_i+1,l_{i+1},\dotsc,l_n]$ and 
$[l_1,\dotsc,l_{i-1},l_i,l_{i+1},\dotsc,l_n]$, respectively. These lists
are illustrated in Figure 4.3. We need to find
$$
\lim_{N\to\infty}\frac{\M\left(W_N[k_1,\dotsc,k_{m};
l_1,\dotsc,\dotsc,l_{i-1},l_i+1,l_{i+1},l_{n}]\right)}
{\M\left(W_N[k_1,\dotsc,k_{m};
l_1,\dotsc,l_{i-1},l_i,l_{i+1},\dotsc,l_{n}]\right)},\tag4.17
$$
for $l_i+1<l_{i+1}$.

\topinsert
\centerline{\mypic{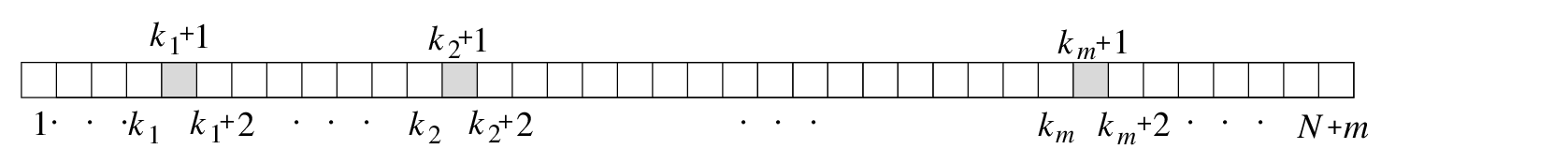}}
\centerline{{\rm (a). 
The ${\bold p}$'s.}}

\centerline{\mypic{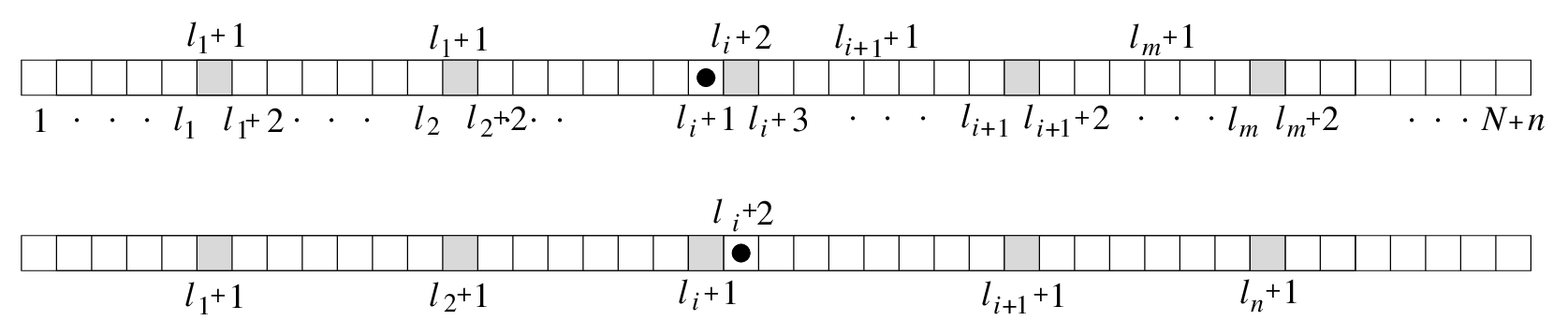}}
\centerline{{\rm (b). 
The ${\bold q}$'s.}}
\medskip
\centerline{{\smc Figure~4.3.} {\rm Incremental change in the second index list.}}
\endinsert

By (4.4), view the regions in (4.17) as $\bar{R}_{{\bold p},{\bold q}}(x)$'s and use
formulas (4.5), (4.6) and (4.8). Repeating the reasoning that proved (4.12), we see
that the formulas we obtain now are almost exactly (4.10) and (4.11), with the roles
of the lists $[k_1,\dotsc,k_m]$ and $[l_1,\dotsc,l_n]$ interchanged. 

Indeed, the only
difference from (4.10) of the contribution coming from the 
$\bar{c}_{{\bold p},{\bold q}}$'s of (4.6) is that the first fraction on the left hand
side of the analog of (4.10) is $(1/(2(l_i+1))!)/(1/(2(l_i+2))!)$, as opposed to the
$(1/(2l_i+1)!)/(1/(2l_i+3)!)$ that results from (4.10) by interchanging the lists
$[k_1,\dotsc,k_m]$ and $[l_1,\dotsc,l_n]$. The consequence of this difference is that
the numerator of the first fraction after 
the equality sign in the present situation analog of (4.10) is $(2l_i+3)(2l_i+4)$, and
therefore the first fraction on the right hand side of the analog of (4.12) is
$(l_i+3/2)/(l_i+1)$.

On the other hand, the contribution to the left hand side of (4.17) coming from the 
$F_{{\bold p},{\bold q}}$'s of (4.6) is readily seen to have, after letting
$N\to\infty$, {\it exactly} the same effect as (4.11). 

Therefore formulas (4.14) and (4.15) have perfect analogs when changing the arguments
of the second list, with the only difference that the integers between the round
parentheses at the denominators in the first
fractions on their right hand sides are now decremented by one unit: they are $(1)_{l_1}$,
$(2)_{l_2-1}$,$\,$$\dotsc$, $(n)_{l_n-n+1}$.

Just as we deduced (4.16) from (4.14) and (4.15), we obtain from the above analysis of
the differences between decrementing the $k_i$'s and the $l_j$'s that
$$
\align
\lim_{N\to\infty}\frac{\M\left(W_N[k_1,\dotsc,k_{m};
l_1,l_2,\dotsc,l_{n}]\right)}
{\M\left(W_N[k_1,\dotsc,k_{m};
0,1,\dotsc,n-1]\right)}&=\prod_{i=1}^m(k_i+2)_n\\
&\!\!\!\!\!\!\!\!\!\!\!\!\!\!\!\!\!\!\!\!\!\!\!\!\!\!\!\!\!
\times
\frac{
\prod_{i=1}^n\frac{ {\displaystyle (3/2)_{l_i}} }{ {\displaystyle (1)_{l_i}} }
}
{
\prod_{i=1}^{n}\frac{ {\displaystyle (3/2)_{i-1}} }{ {\displaystyle (1)_{i-1}} }
}
\frac{
\prod_{1\leq i<j\leq n}\frac{ {\displaystyle l_j-l_i} }{ {\displaystyle j-i} }
}
{
\prod_{i=1}^m\prod_{j=1}^n(k_i+l_j+2)
}
.\tag4.18
\endalign
$$
Combining (4.16) with the specialization of (4.18) for $k_i=i-1$, $i=1,\dotsc,m$, 
and using $\prod_{i=1}^m((i-1)+2)_n=\prod_{i=1}^{m}\prod_{j=1}^{n}(i+j)$, we obtain 
the first equality in (4.1). The second equality is just a repackaging of the first,
using $\prod_{1\leq i<j\leq m}(j-i)=\prod_{i=1}^{m}(1)_{i-1}$ and
$\prod_{i=1}^{m}\prod_{j=1}^{n}(i+j)=\prod_{j=0}^{n-1}(j+2)_n$.
\epf

\mysec{5. A $(2m+2n)$-fold sum for $\omega_b$}

%

In this section we present an expression for the 
boundary-influenced correlation $\omega_b$ as a $(2m+2n)$-fold sum. 
We deduce this by expressing the generating function for
dimer coverings of the region
$W_N\left(\matrix{R_1}\\{v_1}\endmatrix
\cdots\matrix{R_m}\\{v_m}\endmatrix;
\matrix{R'_1}\\{v'_1}\endmatrix\cdots\matrix{R'_n}\\{v'_n}\endmatrix\right)$
in terms of generating functions for the dimer coverings of the regions
$W_N[k_1,\dotsc,k_{m};l_1,\dotsc,l_{n}]$ introduced in the previous section, and
then using the explicit product formula (4.1).
The remaining part of the paper will consist mainly of analyzing the asymptotics
of this multiple sum.


The result mentioned in the title of this section is the following.

\proclaim{Lemma 5.1} For fixed $R_1,\dotsc,R_m,R'_1,\dotsc,R'_n\geq1$ and
$v_1,\dotsc,v_m,v'_1,\dotsc,v'_n\geq0$ we have
$$
\align
&\!\!\!\!\!\!\!\!\!\!\!\!\!\!\!\!\!\!\!\!\!\!\!\!
\omega_b\left(\matrix{R_1}\\{v_1}\endmatrix
\cdots\matrix{R_m}\\{v_m}\endmatrix;
\matrix{R'_1}\\{v'_1}\endmatrix\cdots\matrix{R'_n}\\{v'_n}\endmatrix\right)
=\chi_{2m,2n}\prod_{i=1}^mR_i
\prod_{i=1}^nR'_i(R'_i-1/2)(R'_i+1/2)
\\
&\!\!\!\!\!\!\!\!\!\!\!\!\!\!\!\!\!\!\!\!\!\!\!\!\!\!\!\!\!\!\!\!\!\!\!\!
\times
\left|
\sum_{a_1,b_1=0}^{R_1}\cdots\sum_{a_m,b_m=0}^{R_m}
\sum_{c_1,d_1=0}^{R'_1}\cdots\sum_{c_n,d_n=0}^{R'_n}
(-1)^{\sum_{i=1}^m (a_i+b_i) +\sum_{i=1}^n (c_i+d_i)}
\right.
\\
&\!\!\!\!\!\!
\times\prod_{i=1}^m\frac{(R_i+a_i-1)!\,(R_i+b_i-1)!}
{(2a_i)!\,(R_i-a_i)!\,(2b_i)!\,(R_i-b_i)!}
\frac{(3/2)_{v_i+a_i}\,(3/2)_{v_i+b_i}}{(2)_{v_i+a_i}\,(2)_{v_i+b_i}}
\\
&\!\!\!\!\!\!
\times\prod_{i=1}^n\frac{(R_i+c_i-1)!\,(R_i+d_i-1)!}
{(2c_i+1)!\,(R_i-c_i)!\,(2d_i+1)!\,(R_i-d_i)!}
\frac{(3/2)_{v'_i+c_i}\,(3/2)_{v'_i+d_i}}{(1)_{v'_i+c_i}\,(1)_{v'_i+d_i}}
\\
%
%
\times
\prod_{1\leq i<j\leq m}&(v_j-v_i+a_j-a_i)(v_j-v_i+a_j-b_i)(v_j-v_i+b_j-a_i)(v_j-v_i+b_j-b_i)\\
\times\prod_{1\leq i<j\leq
n}&(v'_j-v'_i+c_j-c_i)(v'_j-v'_i+c_j-d_i)(v'_j-v'_i+d_j-c_i)(v'_j-v'_i+d_j-d_i)\\
&\!\!\!\!\!\!\!\!\!\!\!\!\!\!\!\!\!\!\!\!\!\!\!\!\!\!\!\!
\left.
\times\frac{\prod_{i=1}^m(a_i-b_i)^2\prod_{i=1}^n(c_i-d_i)^2}
{\prod_{i=1}^m\prod_{j=1}^n(u_{ij}+a_i+c_j)(u_{ij}+a_i+d_j)
(u_{ij}+b_i+c_j)(u_{ij}+b_i+d_j)  }
\right|,
\tag5.1
\endalign
$$
where 
$$
u_{ij}=v_i+v'_j+2
$$
for $i=1,\dotsc,m$, $j=1,\dotsc,n$, and $\chi_{m,n}$ is given by $(4.3)$.
\endproclaim

To prove this Lemma we will need the following special case of the 
Lindstr\"om-Gessel-Viennot theorem on non-intersecting 
lattice paths (see e.g. \cite{\GV} or \cite{\Ste}). 

Consider lattice paths on the directed grid graph $\Z^2$, with edges 
oriented so that they point in the positive coordinate directions. We allow the edges of 
$\Z^2$ to be weighted, and define the weight of a lattice path to be the product of the 
weights on its steps. The weight of an $N$-tuple of lattice paths is the product of the
individual weights of its members. 
The weighted count of a set of $N$-tuples of 
lattice paths is the sum of the weights of its elements.

Let ${\bold u}=(u_1,\dotsc,u_N)$ and ${\bold v}=(v_1,\dotsc,v_N)$ be two fixed sets of
starting and ending points on $\Z^2$, and let  $\Cal N({\bold u},{\bold v})$ be the set
of non-intersecting $N$-tuples of lattice paths with these starting and ending points. For 
${\bold P}\in\Cal N({\bold u},{\bold v})$, let $\sigma_{\bold P}$ be the permutation 
induced by ${\bold P}$ on the set consisting of the $N$ indices of its starting and 
ending points.

\proclaim{Theorem 5.2 (Lindstr\"om-Gessel-Viennot)} 
$$\sum_{{\bold P}\in\Cal N({\bold u},{\bold v})}(-1)^{\sigma_{\bold P}}\wt({\bold P})=
\det\left((a_{ij})_{1\leq i,j\leq n}\right),$$
where $a_{ij}$ is the weighted count of the lattice paths from $u_i$ to $v_j$.
\endproclaim

What makes possible the use of this result in our setting is a well-known procedure of
encoding dimer coverings by families of
non-intersecting ``paths of dimers:'' given a dimer covering $T$ of a region $R$ on the 
triangular lattice and a lattice line direction $d$, the dimers of $T$ parallel to $d$
(i.e., having two sides parallel to $d$) can naturally be grouped into
non-intersecting paths joining the lattice segments on the boundary of $R$ that are
parallel to $d$, and conversely this family of paths determines the dimer covering
(see Figure 5.1 for an illustration of this, and e.g. \cite{\Cthree} for a more detailed 
account). 

Consider such a path of dimers $P$. Let $\Cal T$ denote our triangular lattice.
Clearly, $P$ can be identified with a lattice path on the 
lattice $\Cal L$ of rhombi formed by the midpoints of the segments of $\Cal T$ that are parallel to the
encoding direction $d$. In turn, by deforming the lattice of rhombi $\Cal L$ to a square lattice, the
path of dimers $P$ can be regarded as a lattice path on $\Z^2$. It is in this sense that we
will view the paths of dimers as lattice paths on $\Z^2$ in the remainder of this section.

\topinsert
\twoline{\mypic{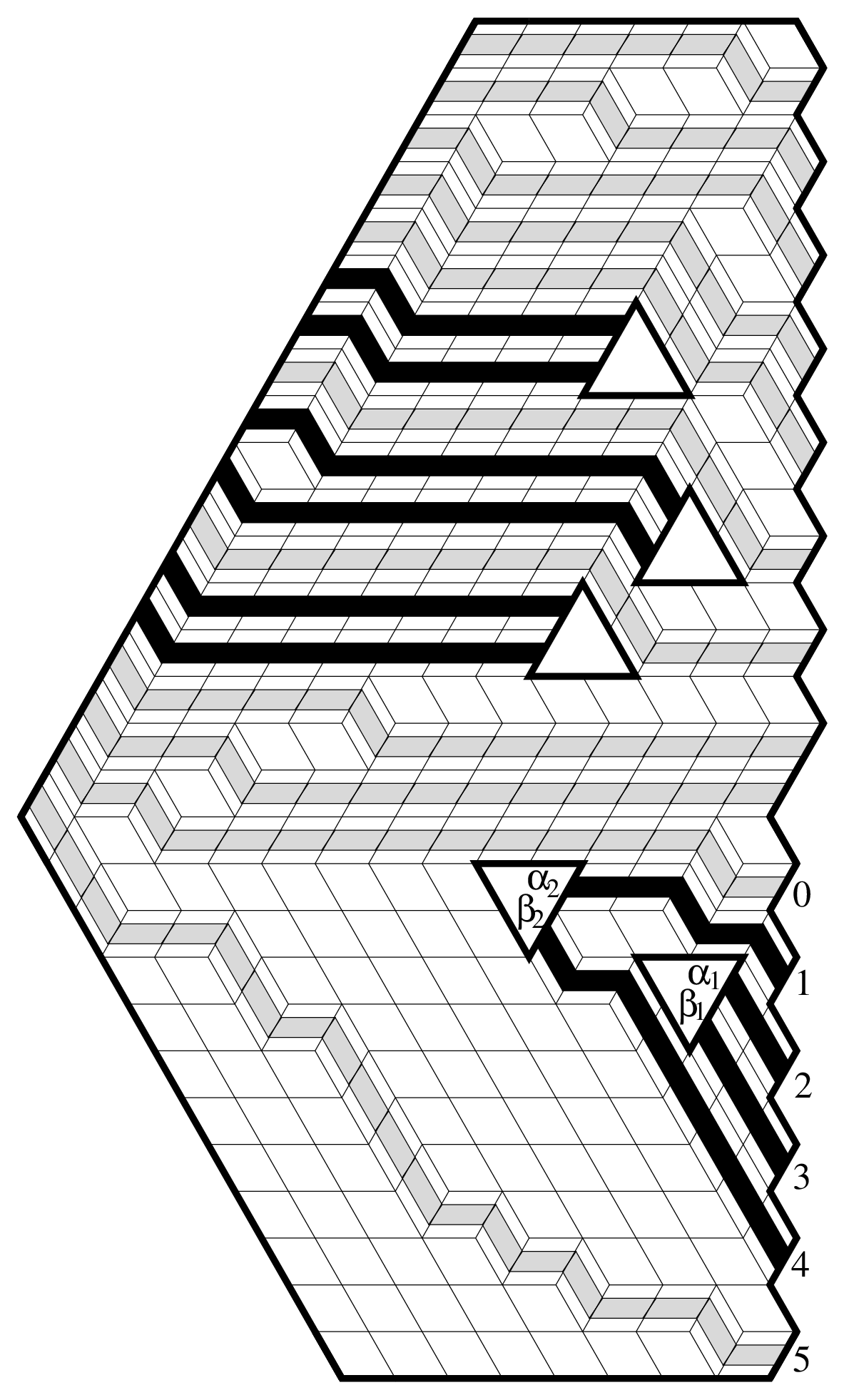}}{\mypic{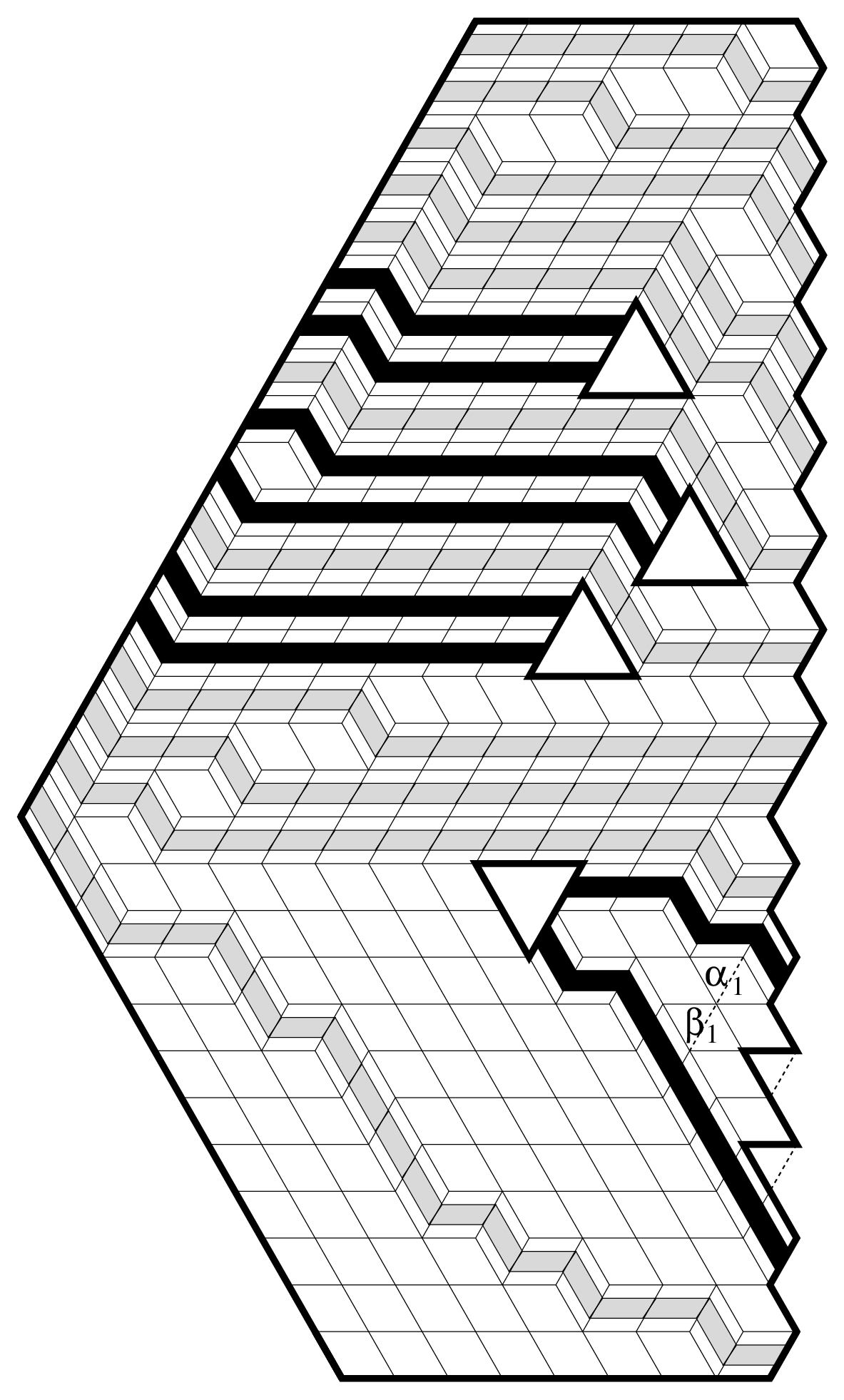}}
\twoline{Figure~5.1. {\rm Lattice path encoding of}}{Figure~5.2. {\rm The effect of Laplace expan-}}
\twoline{{\rm a tiling of $W_2\left(\matrix{5}&2\\{0}&1\endmatrix;
\matrix4&2&3\\1&2&4\endmatrix\right)$.\ \ \ }}{{\rm sion over the rows indexed by $\alpha$ and $\beta$. }}
\endinsert

\smallpagebreak
{\it Proof of Lemma 5.1.} 
Choose the lattice line direction $d$ in the above encoding procedure to be the
southwest-northeast direction of $\Cal T$. Encode the dimer coverings of
$W_N\left(\matrix{R_1}\\{v_1}\endmatrix
\cdots\matrix{R_m}\\{v_m}\endmatrix;
\matrix{R'_1}\\{v'_1}\endmatrix\cdots\matrix{R'_n}\\{v'_n}\endmatrix\right)$ 
by $(2N+2m+4n+1)$-tuples of non-intersecting paths---$2N+4n+1$ starting at the northwestern 
side, and $2m$ starting at the $m$ down-pointing removed quadromers---consisting of 
dimers parallel to $d$ (see Figure 5.1; there and in the following figures the 
$N+2n$ dimer positions weighted by 1/2 are not distinguished, but are understood to carry that
weight). 

Let ${\bold P}$ be such a $(2N+2m+4n+1)$-tuple. Consider the
permutation $\sigma_{\bold P}$ induced by ${\bold P}$ on the set of the $2N+2m+4n+1$ indices 
of its starting and ending points. We claim that the sign of $\sigma_{\bold P}$ is
independent of ${\bold P}$. Indeed, denote by $D_i$, $i=1,\dotsc,m$ and $U_j$, $j=1,\dotsc,n$, 
the removed down-pointing and up-pointing quadromers, respectively. While the way in
which the starting and ending points of ${\bold P}$---clearly independent of ${\bold
P}$---are matched up depends on ${\bold P}$, it is always the case that 
the two paths starting at $D_i$ end at consecutive ending points, and the two
ending on $U_i$ start at consecutive starting points. 
It is easy to see that this implies that all $\sigma_{\bold P}$'s have the same sign.

Consider the lattice $\Cal L$ defined just before the beginning of this
proof. Weight by 1/2 its edges corresponding to
dimer positions weighted by 1/2 in $W_N\left(\matrix{R_1}\\{v_1}\endmatrix
\cdots\matrix{R_m}\\{v_m}\endmatrix;
\matrix{R'_1}\\{v'_1}\endmatrix\cdots\matrix{R'_n}\\{v'_n}\endmatrix\right)$.  
Weight all other edges of $\Cal L$ by 1.
Then the weight of the dimer covering encoded by ${\bold P}$ is just $\wt({\bold P})$,
and we obtain by Theorem 5.2 and the constancy of the sign of $\sigma_{\bold P}$ that
$$
\M\left(W_N\left(\matrix{R_1}\\{v_1}\endmatrix\cdots\matrix{R_m}\\{v_m}\endmatrix;
\matrix{R'_1}\\{v'_1}\endmatrix\cdots\matrix{R'_n}\\{v'_n}\endmatrix\right)\right)=
\left|\det A\right|,\tag5.2
$$
where $A$ is the $(2N+2m+4n+1)\times(2N+2m+4n+1)$ matrix recording the weighted counts of the
lattice paths on $\Cal L$ with given starting and ending points (note that the right hand 
side of (5.2) is independent of the ordering of these starting and ending points).

We deduce (5.1) by applying a sequence of Laplace expansions to the determinant in (5.2). 
Recall that for any $m\times m$ matrix $M$ and any $s$-subset $S$ of $[m]:=\{1,\dotsc,m\}$, 
Laplace expansion along the rows with indices in $S$ states that
$$\det M=
\sum_K(-1)^{\epsilon(K)}\det M_S^K\det M_{[m]\setminus S}^{[m]\setminus K},\tag5.3$$
where $K$ ranges over all $s$-subsets of $[m]$, 
$\epsilon(K):=\sum_{k\in K}(k-1)$ and $M_I^J$ is the 
submatrix of $M$ with row-index set $I$ and column-index set $J$.

The rows and columns of the matrix $A$ in (5.2) are indexed by the starting and ending
points of the $(2N+2m+4n+1)$-tuples of non-intersecting lattice paths encoding the dimer
coverings of $W_N\left(\matrix{R_1}\\{v_1}\endmatrix
\cdots\matrix{R_m}\\{v_m}\endmatrix;
\matrix{R'_1}\\{v'_1}\endmatrix\cdots\matrix{R'_n}\\{v'_n}\endmatrix\right)$. 
The starting points are the $2N+4n+1$ unit segments
along the northwestern boundary of $W_N\left(\matrix{R_1}\\{v_1}\endmatrix
\cdots\matrix{R_m}\\{v_m}\endmatrix;
\matrix{R'_1}\\{v'_1}\endmatrix\cdots\matrix{R'_n}\\{v'_n}\endmatrix\right)$, together with 
the segments $\alpha_i$ and $\beta_i$ of $D_i$ that are parallel to $d$, $i=1,\dotsc,m$ 
(see Figure 5.1). The ending points are the $2N+2m+2n+1$ segments parallel to $d$ on the 
eastern boundary of $W_N\left(\matrix{R_1}\\{v_1}\endmatrix\cdots\matrix{R_m}\\{v_m}\endmatrix;
\matrix{R'_1}\\{v'_1}\endmatrix\cdots\matrix{R'_n}\\{v'_n}\endmatrix\right)$, together
with two more such segments on each $U_j$, $j=1,\dotsc,n$. 

Label the bottommost $N+2m$ ending segments on the eastern boundary, from top to bottom, by $0,
\,1,\,\dotsc,\,N+2m-1$ (this labeling is illustrated in Figure 5.1).

Apply Laplace expansion to the matrix $A$ of (5.2) along the two rows indexed by 
$\alpha_1$ and $\beta_1$ (see Figure 5.1). 
The first determinant in the summand in (5.3) is then just a two
by two determinant. Its entries are weighted counts of lattice paths on $\Cal L$ that
start at $\alpha_1$ or $\beta_1$ and end at some labeled segment on the eastern boundary. 
There are only $R_1+1$ labeled segments that can be reached this way: those with labels
ranging from $v_1$ to $v_1+R_1$. We can restrict summation in (5.3) to the
two-element subsets $K$ of this set of segments, since all other terms have at least
one zero column in the two by two determinant. Therefore we obtain from (5.2) that
$$
\align
&\M\left(W_N\left(\matrix{R_1}\\{v_1}\endmatrix\cdots\matrix{R_m}\\{v_m}\endmatrix;
\matrix{R'_1}\\{v'_1}\endmatrix\cdots\matrix{R'_n}\\{v'_n}\endmatrix\right)\right)=\\
&\ \ \ \ \ \ 
\left|\sum_{0\leq a_1<b_1\leq R_1}(-1)^{a_1+b_1}
\det A_{\{\alpha_1,\beta_1\}}^{\{v_1+a_1,v_1+b_1\}}
\det A_{[2N+2m+4n+1]\setminus\{\alpha_1,\beta_1\}}^{[2N+2m+4n+1]\setminus\{v_1+a_1,v_1+b_1\}}\right|
\tag5.4
\endalign
$$
(here $[2N+2m+4n+1]$ denotes the set of starting, respectively ending points
of the families of non-intersecting lattice paths encoding tilings of the region 
(5.2)).
Choosing the origin of the lattice $\Cal L$ to be at $\alpha_1$, and its positive axis 
directions to point east and southeast, one sees that $\beta_1$ has coordinates $(-1,1)$,
and the segment labeled $v_1+j$ on the eastern boundary has coordinates $(R_1-1-j,2j)$, 
$j=0,\dotsc,R_1$. 
Since the lattice paths counted by the entries of $A_{\{\alpha_1,\beta_1\}}^{\{v_1+a_1,v_1+b_1\}}$
have all steps weighted by 1, the determinant of this matrix is
$$
\align
\det A_{\{\alpha_1,\beta_1\}}^{\{v_1+a_1,v_1+b_1\}}&=
\det\left[\matrix
{R_1-1+a_1\choose 2a_1} {R_1-1+b_1\choose 2b_1}\\
{R_1-1+a_1\choose 2a_1-1} {R_1-1+b_1\choose 2b_1-1}
\endmatrix\right]\\
&=2R_1\frac{(b_1-a_1)(R_1+a_1-1)!\,(R_1+b_1-1)!}{(2a_1)!\,(R_1-a_1)!\,(2b_1)!\,(R_1-b_1)!}.
\tag5.5
\endalign
$$
On the other hand, the second determinant in the summand in (5.4) can be interpreted
as being the weighted count of dimer coverings of the region
$$
W_N\left(\matrix{R_2}\\{v_2}\endmatrix\cdots\matrix{R_m}\\{v_m}\endmatrix;
\matrix{R'_1}\\{v'_1}\endmatrix\cdots\matrix{R'_n}\\{v'_n}\endmatrix\right)
\left[\matrix v_1+a_1\\v_1+b_1\endmatrix\right]\tag5.6
$$
obtained from
$W_N\left(\matrix{R_1}\\{v_1}\endmatrix\cdots\matrix{R_m}\\{v_m}\endmatrix;
\matrix{R'_1}\\{v'_1}\endmatrix\cdots\matrix{R'_n}\\{v'_n}\endmatrix\right)$
by placing back quadromer $D_1$ and removing the two monomers that
contain the segments labeled $v_1+a_1$ and $v_1+b_1$ on the eastern boundary (see Figure 5.2 
for an illustration). Indeed, the
Lindstr\"om-Gessel-Viennot matrix of this region is precisely
$A_{[2N+2m+4n+1]\setminus\{\alpha_1,\beta_1\}}^{[2N+2m+4n+1]\setminus\{v_1+a_1,v_1+b_1\}}$,
and by the argument that proved (5.2) we obtain that 
$\M\left(W_N\left(\matrix{R_2}\\{v_2}\endmatrix\cdots\matrix{R_m}\\{v_m}\endmatrix;
\matrix{R'_1}\\{v'_1}\endmatrix\cdots\matrix{R'_n}\\{v'_n}\endmatrix\right)
\left[\matrix v_1+a_1\\v_1+b_1\endmatrix\right]\right)$ is equal to
$\det A_{[2N+2m+4n+1]\setminus\{\alpha_1,\beta_1\}}^{[2N+2m+4n+1]\setminus\{v_1+a_1,v_1+b_1\}}$, 
up to a sign that 
is independent of $a_1$ and $b_1$ 
(to check the latter statement, in the labeling of the starting and ending points of
${\bold P}$ that defines $\sigma_{\bold P}$, label the bottommost $N+2m-2$ northwest
facing unit segments on the boundary of (5.6), say from bottom to top, by $1,\dotsc,N+2m-2$, 
irrespective of the values of $a_1$ and $b_1$; the argument in the second paragraph of the
current proof shows that the sign of $\sigma_{\bold P}$ is independent of not just 
${\bold P}$, but also of $a_1$ and $b_1$).
Therefore, using (5.5) we can rewrite (5.4) as
$$
\align
&\M\left(W_N\left(\matrix{R_1}\\{v_1}\endmatrix\cdots\matrix{R_m}\\{v_m}\endmatrix;
\matrix{R'_1}\\{v'_1}\endmatrix\cdots\matrix{R'_n}\\{v'_n}\endmatrix\right)\right)=\\
&\ \ \ \ \ \ \ \ \ \ 
2R_1
\left|
\sum_{0\leq a_1<b_1\leq R_1}(-1)^{a_1+b_1}
\frac{(b_1-a_1)(R_1+a_1-1)!\,(R_1+b_1-1)!}{(2a_1)!\,(R_1-a_1)!\,(2b_1)!\,(R_1-b_1)!}
\right.
\\
&\ \ \ \ \ \ \ \ \ \ \ \ \ \ \ \ \ \ \ \ 
\left.
\times
\M\left(W_N\left(\matrix{R_2}\\{v_2}\endmatrix\cdots\matrix{R_m}\\{v_m}\endmatrix;
\matrix{R'_1}\\{v'_1}\endmatrix\cdots\matrix{R'_n}\\{v'_n}\endmatrix\right)
\left[\matrix v_1+a_1\\v_1+b_1\endmatrix\right]\right)
\right|.
\tag5.7
\endalign
$$
This way the matching generating function of 
$\M\left(W_N\left(\matrix{R_1}\\{v_1}\endmatrix\cdots\matrix{R_m}\\{v_m}\endmatrix;
\matrix{R'_1}\\{v'_1}\endmatrix\cdots\matrix{R'_n}\\{v'_n}\endmatrix\right)\right)$
is expressed as a double sum involving the matching generating functions of regions
having one less down-pointing triangular hole. By the same reasoning the second down-pointing
quadromer $D_2$ can be removed from the regions (5.6) and the last factor in the summand of
(5.7) is expressed as a double sum involving regions with {\it two} less down-pointing holes.
One obtains
$$
\align
&\M\left(W_N\left(\matrix{R_1}\\{v_1}\endmatrix\cdots\matrix{R_m}\\{v_m}\endmatrix;
\matrix{R'_1}\\{v'_1}\endmatrix\cdots\matrix{R'_n}\\{v'_n}\endmatrix\right)\right)=\\
&\ \ 
4R_1R_2
\left|
\sum_{0\leq a_1<b_1\leq R_1}\sum_{0\leq a_2<b_2\leq R_2}(-1)^{a_1+b_1+a_2+b_2}
\sign(v_1+a_1,v_1+b_1,v_2+a_2,v_2+b_2)
\right.
\\
&\ \ \ \ \ \ \ \ \ \ \ \ \ \ \ \ \ \ \ \ 
\times
\frac{(b_1-a_1)(R_1+a_1-1)!\,(R_1+b_1-1)!}{(2a_1)!\,(R_1-a_1)!\,(2b_1)!\,(R_1-b_1)!}\\
&\ \ \ \ \ \ \ \ \ \ \ \ \ \ \ \ \ \ \ \ 
\times
\frac{(b_2-a_2)(R_2+a_2-1)!\,(R_2+b_2-1)!}{(2a_2)!\,(R_2-a_2)!\,(2b_2)!\,(R_2-b_2)!}\\
&\ \ \ \ \ \ \ \ \ \ \ \ \ \ \ \ \ \ \ \ 
\left.
\times
\M\left(W_N\left(\matrix{R_3}\\{v_3}\endmatrix\cdots\matrix{R_m}\\{v_m}\endmatrix;
\matrix{R'_1}\\{v'_1}\endmatrix\cdots\matrix{R'_n}\\{v'_n}\endmatrix\right)
\left[\matrix v_1+a_1&v_2+a_2\\v_1+b_1&v_2+b_2\endmatrix\right]\right)
\right|,
\tag5.8
\endalign
$$
where the summation extends only over those indices for which
$(v_1+a_1,v_1+b_1,v_2+a_2,v_2+b_2)$ consists of distinct
components, $\sign$ denotes its sign when regarded as a permutation, and the region on the
right hand side of (5.8) is obtained from (5.6) by placing back the quadromer $D_2$ and
removing the two monomers containing the segments labeled $v_2+a_2$ and $v_2+b_2$ on its
eastern boundary.

Indeed, perform Laplace expansion for the region (5.6) over the rows indexed 
by $\alpha_2$ and $\beta_2$. Because of the geometry of the triangular lattice 
$\Cal T$, the paths starting at these segments can end on the
eastern boundary only at segments with labels in the range $[v_2,v_2+R_2]$; denote
these labels by $v_2+a_2$ and $v_2+b_2$.
Moreover, because the segments labeled $v_1+a_1$ and $v_1+b_1$ are not present
in the region (5.6), the labels $v_2+a_2$ and $v_2+b_2$ must in fact
also be different from both $v_1+a_1$ and $v_1+b_1$. This explains why the summation
range in (5.8) needs to be restricted to $v_1+a_1,v_1+b_1,v_2+a_2,v_2+b_2$ being
distinct.

The only other needed explanation for justifying (5.8) is the factor
$\sign(v_1+a_1,v_1+b_1,v_2+a_2,v_2+b_2)$. The reason it appears is the following.
In the summand of the Laplace expansion (5.3) one has the factor
$(-1)^{\epsilon(K)}$. In our situation this becomes $(-1)^{k_1+k_2}$, where the
``column indices'' $k_i$
indicate that the two path-ending segments on the eastern boundary are the $k_1$th and
$k_2$th from, say, the top, {\it in the current region}. 
This column index coincides---up to a
translation, which, pertaining to both $k_1$ and $k_2$, does not affect the sign 
$(-1)^{\epsilon(K)}$---in the case of 
$\M\left(W_N\left(\matrix{R_1}\\{v_1}\endmatrix\cdots\matrix{R_m}\\{v_m}\endmatrix;
\matrix{R'_1}\\{v'_1}\endmatrix\cdots\matrix{R'_n}\\{v'_n}\endmatrix\right)\right)$,
with the label of the path-ending segment.
However, when we do Laplace expansion for the region
(5.6), this column index is affected---unless $v_2+a_2<v_2+b_2<v_1+a_1<v_1+b_1$---
by the previous removal of the segments with labels $v_1+a_1$ and $v_1+b_1$. 

More precisely, say $v_2+a_2<v_1+a_1<v_2+b_2<v_1+b_1$. Then while $v_2+a_2$
correctly gives the column index of the segment where the path starting at $\alpha_2$
ends, $v_2+b_2$ is, due to the absence of the segment labeled $v_1+a_1$, 
{\it one unit more} than the column index of the ending segment of the path starting 
at $\beta_2$.
It is easy to see that in general the effect of this ``interference''
is precisely the multiplication of the summand by 
$\sign(v_2+a_2,v_2+b_2,v_1+a_1,v_1+b_1)$. Since clearly 
$\sign(v_2+a_2,v_2+b_2,v_1+a_1,v_1+b_1)=\sign(v_1+a_1,v_1+b_1,v_2+a_2,v_2+b_2)$, (5.8)
is completely justified.

Applying the same reasoning $m-2$ more times, we obtain
$$
\align
&\M\left(W_N\left(\matrix{R_1}\\{v_1}\endmatrix\cdots\matrix{R_m}\\{v_m}\endmatrix;
\matrix{R'_1}\\{v'_1}\endmatrix\cdots\matrix{R'_n}\\{v'_n}\endmatrix\right)\right)=
2^m\prod_{i=1}^mR_i\\
& 
\times
\left|
\sum_{0\leq a_1<b_1\leq R_1}\!\!\cdots\!\!\sum_{0\leq a_m<b_m\leq R_m}
(-1)^{\sum_{i=1}^m(a_i+b_i)}
\sign(v_1+a_1,v_1+b_1,\dotsc,v_m+a_m,v_m+b_m)
\right.
\\
&\ \ \ \ \ \ \ \ \ \ \ \ \ \ \ \ \ \ \ \ 
\times
\prod_{i=1}^m
\frac{(b_i-a_i)(R_i+a_i-1)!\,(R_i+b_i-1)!}{(2a_i)!\,(R_i-a_i)!\,(2b_i)!\,(R_i-b_i)!}\\
&\ \ \ \ \ \ \ \ \ \ \ \ \ \ \ \ \ \ \ \ 
\left.
\times
\M\left(W_N\left(\matrix\emptyset\\\emptyset\endmatrix;
\matrix{R'_1}\\{v'_1}\endmatrix\cdots\matrix{R'_n}\\{v'_n}\endmatrix\right)
\left[\matrix v_1+a_1\\v_1+b_1\endmatrix\cdots\matrix v_m+a_m\\v_m+b_m\endmatrix\right]
\right)
\right|,
\tag5.9
\endalign
$$
where the region on the right hand side of (5.9) is obtained from 
the region on the left
by placing back all $m$ down-pointing quadromers, and removing instead the $2m$ monomers
containing the segments with labels $v_i+a_i$ and $v_i+b_i$, $i=1,\dotsc,m$, on its eastern
boundary. 

We complete expressing the left hand side of (5.1) in terms of regions with no 
holes by repeating our hole-removing procedure for the 
regions 
$$
W_N\left(\matrix\emptyset\\\emptyset\endmatrix;
\matrix{R'_1}\\{v'_1}\endmatrix\cdots\matrix{R'_n}\\{v'_n}\endmatrix\right)
\left[\matrix v_1+a_1\\v_1+b_1\endmatrix\cdots\matrix v_m+a_m\\v_m+b_m\endmatrix
\right]\tag5.10
$$
on the right hand side of (5.9).

\topinsert
\twoline{\mypic{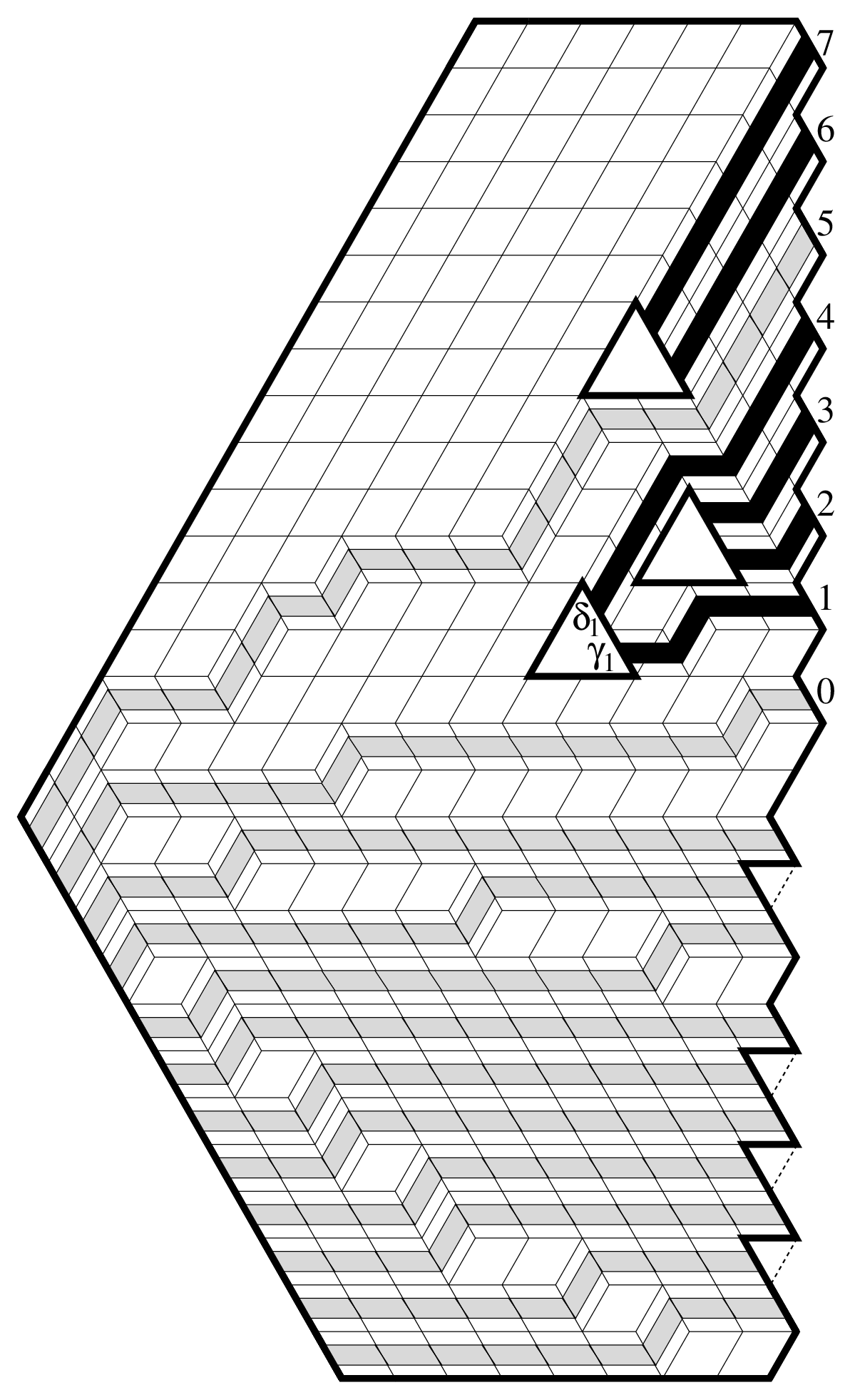}}{\mypic{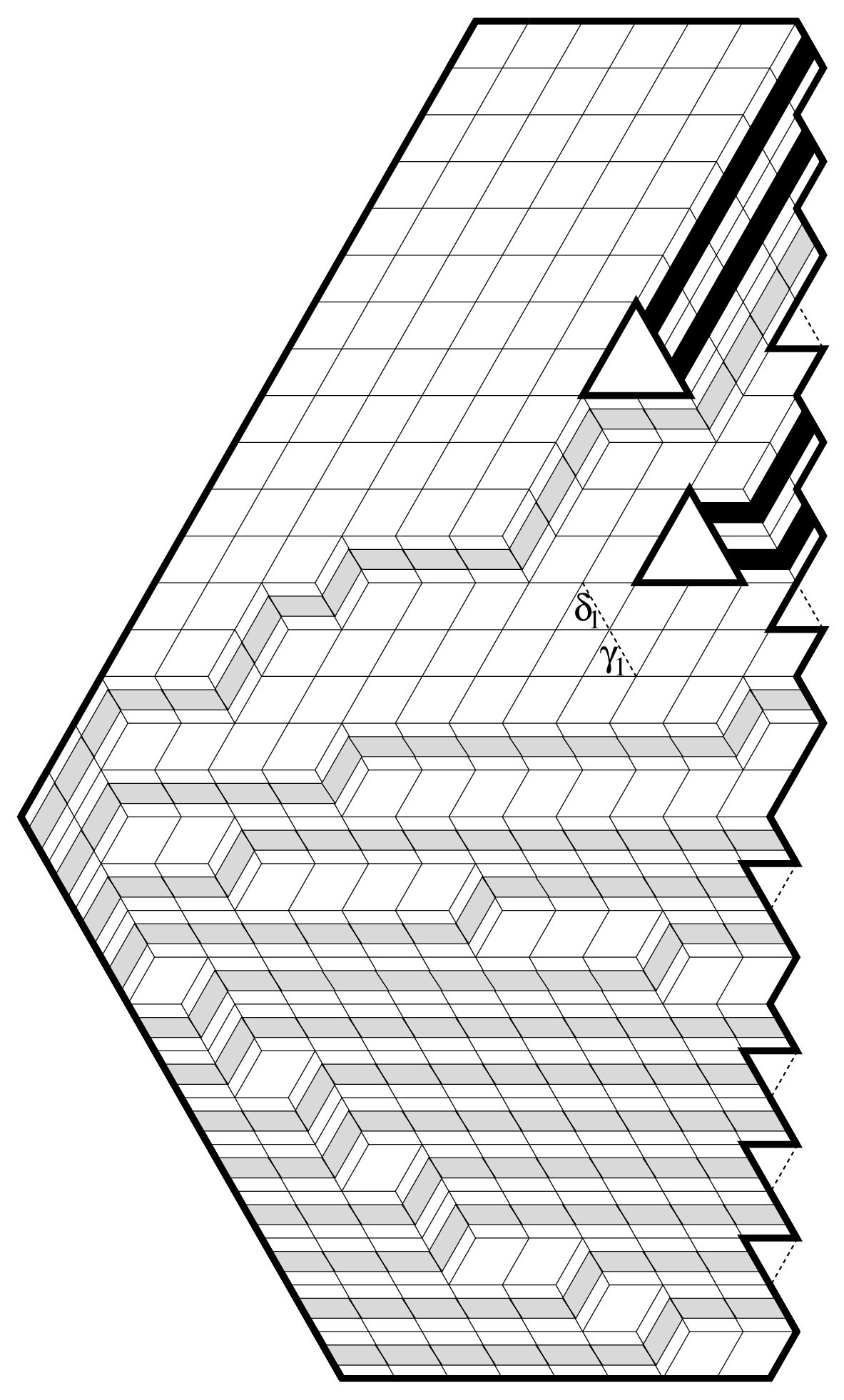}}
\twoline{Figure~5.3. {\rm Lattice path encoding of}}{\ \ \ \ \ Figure~5.4. {\rm The effect of Laplace expan-}}
\twoline{{\rm a tiling of $W_2\left(\matrix\emptyset\\\emptyset\endmatrix;
\matrix4&2&3\\1&2&4\endmatrix\right)\left[\matrix0&2\\4&3\endmatrix\right]$.}}
{{\rm sion over the rows indexed by $\gamma$ and $\delta$. }}
\endinsert

To this end, we encode the tilings of the regions (5.10) by paths of dimers 
parallel to the direction $d$ now chosen to be the southeast-northwest direction 
(an example is illustrated in 
Figure 5.3). As in the previous encoding, weight by 1/2 those segments of the
resulting lattice $\Cal L$ that correspond to dimer positions weighted 1/2 in 
the region (5.10), and weight all remaining segments of $\Cal L$ by 1.

Each tiling of the region (5.10) gets encoded this way by a $(2N+4m+2n)$-tuple 
${\bold P}$ of non-intersecting lattice paths on $\Cal L$, starting at
the unit segments on its southwestern boundary or at the unit segments $\gamma_i$
and $\delta_i$ of $U_i$ ($i=1,\dotsc,n$) that are parallel to $d$, and ending at the 
unit segments parallel to $d$ on its eastern boundary (see Figure 5.3).  

Label the topmost $N+2n$ ending segments on the eastern boundary, from bottom to top, 
by $0,\,1,\,\dotsc,\,N+2n-1$ (this labeling is illustrated in Figure 5.3).

As in the argument that proved (5.2), the sign of the permutation $\sigma_{\bold P}$
is independent of ${\bold P}$. Therefore, we obtain by Theorem 5.2 that
$$
\M\left(W_N\left(\matrix\emptyset\\\emptyset\endmatrix;
\matrix{R'_1}\\{v'_1}\endmatrix\cdots\matrix{R'_n}\\{v'_n}\endmatrix\right)
\left[\matrix v_1+a_1\\v_1+b_1\endmatrix\cdots\matrix v_m+a_m\\v_m+b_m\endmatrix
\right]\right)=\epsilon_1\det B,\tag5.11
$$
where $B$ is the $(2N+4m+2n)\times(2N+4m+2n)$ matrix recording the weighted counts 
of the lattice paths with specified starting and ending points, and the sign 
$\epsilon_1$ in front of the determinant is the same for all choices of $a_i$ and $b_i$,
$i=1,\dotsc,m$. 

Apply Laplace expansion in $\det B$ along the two rows indexed by $\gamma_1$ and
$\delta_1$. The first determinant in the summand of (5.3) is again two by two, and
records weighted counts of lattice paths starting at $\gamma_1$ or $\delta_1$ and 
ending at some segment on the eastern boundary (see Figure 5.3). 
There are $R'_1+1$ segments on the eastern boundary that can be
reached this way; in the labeling of the paragraph before (5.11), they are the ones
with labels $v'_1,v'_1+1,\dotsc,v'_1+R'_1$. 
As with our previous Laplace expansions, we can restrict the summation range in (5.3)
to obtain
$$
\align
&\M\left(W_N\left(\matrix\emptyset\\\emptyset\endmatrix;
\matrix{R'_1}\\{v'_1}\endmatrix\cdots\matrix{R'_n}\\{v'_n}\endmatrix\right)
\left[\matrix v_1+a_1\\v_1+b_1\endmatrix\cdots\matrix v_m+a_m\\v_m+b_m\endmatrix
\right]\right)=\\
&\ \ \ \ \ \ \ \ \ \ \ \ \ \ \ 
\epsilon_1\sum_{0\leq c_1<d_1\leq R'_1}(-1)^{c_1+d_1}
\det B_{\{\gamma_1,\delta_1\}}^{\{v'_1+c_1,v'_1+d_1\}}
\det B_{[2N+4m+2n]\setminus\{\gamma_1,\delta_1\}}^{[2N+4m+2n]\setminus\{v'_1+c_1,v'_1+d_1\}}.
\tag5.12
\endalign
$$
Centering $\Cal L$ at $\gamma_1$ and choosing the positive directions in the 
lattice $\Cal L$ to point east and northeast, $\delta_1$ has coordinates 
$(-1,1)$, and the segment labeled $v'_1+j$ on the eastern boundary has coordinates 
$(R'_1-1-j,2j+1)$, $j=0,1,\dotsc,R'_1$. The weighted counts involved in the entries 
of $B_{\{\gamma_1,\delta_1\}}^{\{v'_1+c_1,v'_1+d_1\}}$ (which involve this time some steps of 
weight 1/2) are easily calculated and one obtains
$$
\align
\det B_{\{\gamma_1,\delta_1\}}^{\{v'_1+c_1,v'_1+d_1\}}&=\det\left[\matrix
\frac{1}{2}{R'_1-1+c_1\choose 2c_1}+{R'_1-1+c_1\choose 2c_1+1} 
\ \ \frac{1}{2}{R'_1-1+d_1\choose 2d_1}+{R'_1-1+d_1\choose 2d_1+1}\\
\frac{1}{2}{R'_1-1+c_1\choose 2c_1-1}+{R'_1-1+c_1\choose 2c_1} 
\ \ \frac{1}{2}{R'_1-1+d_1\choose 2d_1-1}+{R'_1-1+d_1\choose 2d_1}
\endmatrix\right]\\
&\!\!\!\!\!\!\!\!\!\!\!\!\!\!\!\!\!\!\!\!\!\!\!\!\!\!\!\!\!\!\!\!\!\!\!\!
=2R'_1(R'_1-1/2)(R'_1+1/2)\frac{(d_1-c_1)(R'_1+c_1-1)!\,(R'_1+d_1-1)!}
{(2c_1+1)!\,(R'_1-c_1)!\,(2d_1+1)!\,(R'_1-d_1)!}.\tag5.13
\endalign
$$
On the other hand, by applying Theorem 5.2 one more time one sees that 
$$
\align
&\det B_{[2N+4m+2n]\setminus\{\gamma_1,\delta_1\}}^{[2N+4m+2n]\setminus\{v'_1+c_1,v'_1+d_1\}}=\\
&\ \ \ \ \ \ \ \ \ \ \ \ 
\epsilon'_1\M\left(W_N\left(\matrix\emptyset\\\emptyset\endmatrix;
\matrix{R'_2}\\{v'_2}\endmatrix\cdots\matrix{R'_n}\\{v'_n}\endmatrix\right)
\left[\matrix v_1+a_1\\v_1+b_1\endmatrix\cdots\matrix v_m+a_m\\v_m+b_m\endmatrix
\right]\left[\matrix v'_1+c_1\\v'_1+d_1\endmatrix\right]\right),\tag5.14
\endalign
$$ 
where 
$$
W_N\left(\matrix\emptyset\\\emptyset\endmatrix;
\matrix{R'_2}\\{v'_2}\endmatrix\cdots\matrix{R'_n}\\{v'_n}\endmatrix\right)
\left[\matrix v_1+a_1\\v_1+b_1\endmatrix\cdots\matrix v_m+a_m\\v_m+b_m\endmatrix
\right]\left[\matrix v'_1+c_1\\v'_1+d_1\endmatrix\right]\tag5.15
$$
is the region obtained from the region (5.10) by placing
back quadromer $U_1$ and removing the two monomers containing the segments labeled 
$v'_1+c_1$ and $v'_1+d_1$ on the eastern boundary, 
and the sign $\epsilon'_1$ is independent of $c_1$ and $d_1$. 
Therefore, by (5.12)--(5.14) we obtain that
$$
\align
&
\M\left(W_N\left(\matrix\emptyset\\\emptyset\endmatrix;
\matrix{R'_1}\\{v'_1}\endmatrix\cdots\matrix{R'_n}\\{v'_n}\endmatrix\right)
\left[\matrix v_1+a_1\\v_1+b_1\endmatrix\cdots\matrix v_m+a_m\\v_m+b_m\endmatrix
\right]\right)=2R'_1(R'_1-1/2)(R'_1+1/2)
\\&\ \ \ \ \ \ \ \ \ \ 
\times\epsilon_1\epsilon'_1\sum_{0\leq c_1<d_1\leq R'_1}(-1)^{c_1+d_1}
\frac{(d_1-c_1)(R'_1+c_1-1)!\,(R'_1+d_1-1)!}
{(2c_1+1)!\,(R'_1-c_1)!\,(2d_1+1)!\,(R'_1-d_1)!}\\
&\ \ \ \ \ \ \ \ \ \ \ \ \ \ \ \ \ \ \ \ 
\times
\M\left(W_N\left(\matrix\emptyset\\\emptyset\endmatrix;
\matrix{R'_2}\\{v'_2}\endmatrix\cdots\matrix{R'_n}\\{v'_n}\endmatrix\right)
\left[\matrix v_1+a_1\\v_1+b_1\endmatrix\cdots\matrix v_m+a_m\\v_m+b_m\endmatrix
;\matrix v'_1+c_1\\v'_1+d_1\endmatrix\right]\right).
\tag5.16
\endalign
$$
By further Laplace expansions applied on the pairs of rows indexed by 
$\gamma_i$ and $\delta_i$, $i=2,\dotsc,n$, one can successively remove all remaining
holes $U_2,\dotsc,U_n$. 
The argument that proved (5.9) yields now
$$
\align
&\M\left(W_N\left(\matrix\emptyset\\\emptyset\endmatrix;
\matrix{R'_1}\\{v'_1}\endmatrix\cdots\matrix{R'_n}\\{v'_n}\endmatrix\right)
\left[\matrix v_1+a_1\\v_1+b_1\endmatrix\cdots\matrix v_m+a_m\\v_m+b_m\endmatrix
\right]\right)=\prod_{i=1}^n\epsilon_i\epsilon'_i
\prod_{i=1}^n2R'_i(R'_i-1/2)(R'_i+1/2)\\
&\ \ 
\times
\sum_{0\leq c_1<d_1\leq R'_1}\cdots\sum_{0\leq c_n<d_n\leq R'_n}
(-1)^{\sum_{i=1}^n(c_i+d_i)}
\sign(v'_1+c_1,v'_1+d_1,\dotsc,v'_n+c_n,v'_n+d_n)\\
&\ \ \ \ \ \ \ \ \ \ \ \ \ \ \ \ \ \ \ \ 
\times
\prod_{i=1}^n
\frac{(d_i-c_i)(R'_i+c_i-1)!\,(R'_i+d_i-1)!}{(2c_i+1)!\,(R'_i-c_i)!\,(2d_i+1)!\,(R'_i-d_i)!}
\\
&\ \ \ \ \ \ \ \ \ \ \ \ \ \ \ \ \ \ \ \ 
\times
\M\left(W_N\left(\matrix\emptyset\\\emptyset\endmatrix;
\matrix\emptyset\\\emptyset\endmatrix\right)
\left[\matrix v_1+a_1\\v_1+b_1\endmatrix\cdots\matrix v_m+a_m\\v_m+b_m\endmatrix;
\matrix v'_1+c_1\\v'_1+d_1\endmatrix\cdots\matrix v'_n+c_n\\v'_n+d_n\endmatrix\right]
\right),
\tag5.17
\endalign
$$
where the multiple sum extends only to the summation indices for which 
$v'_1+c_1,v'_1+d_1,\dotsc,v'_n+c_n,v'_n+d_n$ are distinct, the signs $\epsilon_i$ and
$\epsilon'_i$, $i=1,\dotsc,n$, are independent of $a_1,b_1,\dotsc,a_m,b_m$,
and
$$
W_N\left(\matrix\emptyset\\\emptyset\endmatrix;
\matrix\emptyset\\\emptyset\endmatrix\right)
\left[\matrix v_1+a_1\\v_1+b_1\endmatrix\cdots\matrix v_m+a_m\\v_m+b_m\endmatrix;
\matrix v'_1+c_1\\v'_1+d_1\endmatrix\cdots\matrix v'_n+c_n\\v'_n+d_n\endmatrix\right]
\tag5.18
$$
is the region obtained from the region (5.10) by placing back all its removed quadromers
$U_i$, $i=1,\dotsc,n$, and removing the $2n$ monomers containing the segments labeled 
$\gamma_i$ and $\delta_i$, $i=1,\dotsc,n$, on its eastern boundary.

\topinsert
\centerline{\mypic{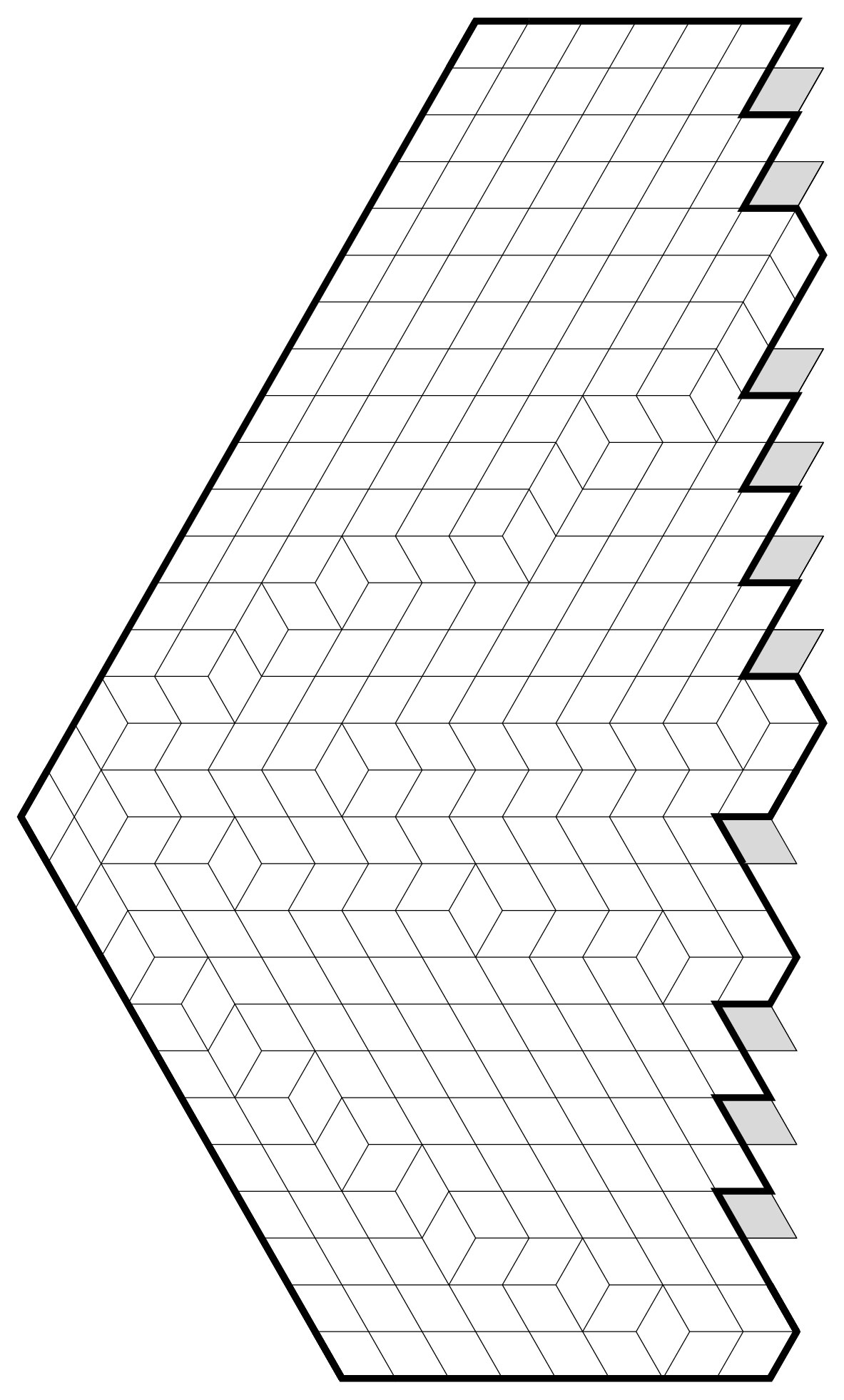}}
\centerline{{\smc Figure~5.5.} 
{\rm The regions
$W_N\left(\matrix\emptyset\\\emptyset\endmatrix;
\matrix\emptyset\\\emptyset\endmatrix\right)
\left[\matrix 0&2\\4&3\endmatrix;
\matrix 0&1&5\\3&2&6\endmatrix\right]$
and $W_N[0,2,3,4;0,1,2,3,5,6]$}} 
\centerline{{\rm have the same matching generating function.}}
\endinsert

However, the region (5.18) differs from the region
$W_N[\{v_1+a_1,v_1+b_1,\dotsc,v_m+a_m,v_m+b_m\}_{<};
\{v'_1+c_1,v'_1+d_1,\dotsc,v'_n+c_n,v'_n+d_n\}_{<}]$ (where $A_{<}$ denotes the list of 
increasingly sorted elements of the set $A$ of distinct nonnegative integers)
defined in Section 4 only in that the former contains $2m+2n$ more unit rhombi near the 
eastern boundary (see Figure 5.5; the additional rhombi are shaded). 
Moreover, all these additional dimer positions have 
weight 1 and are forced to be part of any dimer covering of the former region. Therefore
the two regions have the same matching generating function, and replacing (5.17) in the 
right hand side of (5.9) one obtains
$$
\align
&\M\left(W_N\left(\matrix{R_1}\\{v_1}\endmatrix\cdots\matrix{R_m}\\{v_m}\endmatrix;
\matrix{R'_1}\\{v'_1}\endmatrix\cdots\matrix{R'_n}\\{v'_n}\endmatrix\right)\right)=
2^{m+n}\prod_{i=1}^mR_i\prod_{i=1}^nR'_i(R'_i-1/2)(R'_i+1/2)\\
&\ \  
\times
\left|
\sum_{0\leq a_1<b_1\leq R_1}\cdots\sum_{0\leq a_m<b_m\leq R_m}
\sum_{0\leq c_1<d_1\leq R'_1}\cdots\sum_{0\leq c_n<d_n\leq R'_n}
(-1)^{\sum_{i=1}^m (a_i+b_i) +\sum_{i=1}^n (c_i+d_i)}
\right.
\\
&\ \ 
\times
\prod_{i=1}^m
\frac{(b_i-a_i)(R_i+a_i-1)!\,(R_i+b_i-1)!}{(2a_i)!\,(R_i-a_i)!\,(2b_i)!\,(R_i-b_i)!}\\
&\ \ 
\times
\prod_{i=1}^n
\frac{(d_i-c_i)(R'_i+c_i-1)!\,(R'_i+d_i-1)!}{(2c_i+1)!\,(R'_i-c_i)!\,(2d_i+1)!\,(R'_i-d_i)!}\\
&\ \ 
\times
\sign(v'_1+c_1,v'_1+d_1,\dotsc,v'_n+c_n,v'_n+d_n)
\sign(v_1+a_1,v_1+b_1,\dotsc,v_m+a_m,v_m+b_m)\\
&\ \ 
\left.
\times
\M(W_N[\{v_1+a_1,\dotsc,v_m+b_m\}_{<};
\{v'_1+c_1,\dotsc,v'_n+d_n\}_{<}])
\right|,
\tag5.19
\endalign
$$
the summation extending only over those summation indices for which both lists of 
arguments in the region on the right hand side have distinct elements.

After removing the forced dimers, the normalizing region at the denominator of (\TwoSix), 
$$
W_N\left(\matrix{1}&3\\{0}&0\endmatrix\cdots\matrix{2m-1}\\{0}\endmatrix;
\matrix{1}&3\\{0}&0\endmatrix\cdots\matrix{2n-1}\\{0}\endmatrix\right),\tag5.20
$$ 
is seen to be precisely $W_N[0,1,\dotsc,2m-1;0,1,\dotsc,2n-1]$.

Moreover, it follows from (4.2) that
$$
\align
\lim_{N\to\infty}&\frac{\M\left(W_N[\{v_1+a_1,\dotsc,v_m+b_m\}_{<};
\{v'_1+c_1,\dotsc,v'_n+d_n\}_{<}]\right)}
{\M\left(W_N[0,\dotsc,2m-1;0,\dotsc,2n-1]\right)}=\chi_{2m,2n}\\
&\!\!\!\!\!\!\!\!\!\!\!\!
\times
\sign(v'_1+c_1,v'_1+d_1,\dotsc,v'_n+c_n,v'_n+d_n)
\sign(v_1+a_1,v_1+b_1,\dotsc,v_m+a_m,v_m+b_m)\\
\times
&\prod_{i=1}^m\frac{(3/2)_{v_i+a_i}\,(3/2)_{v_i+b_i}}
{(2)_{v_i+a_i}\,(2)_{v_i+b_i}}
\prod_{i=1}^n\frac{(3/2)_{v'_i+c_i}\,(3/2)_{v'_i+d_i}}
{(1)_{v'_i+c_i}\,(1)_{v'_i+d_i}}\\
\times
&\prod_{1\leq i<j\leq m}(v_j-v_i+a_j-a_i)(v_j-v_i+a_j-b_i)(v_j-v_i+b_j-a_i)
(v_j-v_i+b_j-b_i)\\
&\times\prod_{1\leq i<j\leq n}(v'_j-v'_i+c_j-c_i)(v'_j-v'_i+c_j-d_i)(v'_j-v'_i+d_j-c_i)
(v'_j-v'_i+d_j-d_i)\\
&
\times\frac{\prod_{i=1}^m(a_i-b_i)\prod_{i=1}^n(c_i-d_i)}
{\prod_{i=1}^m\prod_{j=1}^n(u_{ij}+a_i+c_j)(u_{ij}+a_i+d_j)
(u_{ij}+b_i+c_j)(u_{ij}+b_i+d_j)  },\tag5.21
\endalign
$$
with $\chi$ given by (4.3) and $u_{ij}=v_i+v'_j+2$, $i=1,\dotsc,m$, $j=1,\dotsc,n$.
Indeed, (5.21) is a direct consequence of (4.2) 
when $v_1+a_1<v_1+b_1<\cdots<v_m+a_m<v_m+b_m$ and
$v'_1+c_1<v'_1+d_1<\cdots<v'_n+c_n<v'_n+d_n$. Since each ``elementary move'' changing the
relative order of just two consecutive elements in either of these two lists (except when
these two elements are $v_i+a_i$ and $v_i+b_i$, or $v'_i+c_i$ and $v'_i+d_i$, for some 
$i$, case in which their relative order is fixed by construction) has the 
effect of multiplying {\it both} the sign part and the product part of the right hand 
side of (5.21) by $-1$, it follows that (5.21) is valid in general.

Therefore, divide (5.19) by the matching generation function of the region (5.20) and
let $N\to\infty$. By (\TwoSix), the left hand side becomes the left hand side of (5.1), while
by (5.21) the summand on the right hand side becomes precisely the summand on the right
hand side of (5.1). (In particular, it follows that the limit (2.6) defining $\omega_b$ exists.)
We obtain
$$
\align
&\!\!\!\!\!\!\!\!\!\!\!\!\!\!\!\!\!\!\!\!\!\!\!\!
\omega_b\left(\matrix{R_1}\\{v_1}\endmatrix
\cdots\matrix{R_m}\\{v_m}\endmatrix;
\matrix{R'_1}\\{v'_1}\endmatrix\cdots\matrix{R'_n}\\{v'_n}\endmatrix\right)
=2^{m+n}\chi_{m,n}\prod_{i=1}^mR_i
\prod_{i=1}^nR'_i(R'_i-1/2)(R'_i+1/2)
\\
&\!\!\!\!\!\!\!\!\!\!\!\!\!\!\!\!\!\!\!\!\!\!\!\!\!\!\!\!\!\!\!\!\!\!\!\!
\times
\left|
\sum_{0\leq a_1<b_1\leq R_1}\cdots\sum_{0\leq a_m<b_m\leq R_m}
\sum_{0\leq c_1<d_1\leq R'_1}\cdots\sum_{0\leq c_n<d_n\leq R'_n}
(-1)^{\sum_{i=1}^m (a_i+b_i) +\sum_{i=1}^n (c_i+d_i)}
\right.
\\
&\!\!\!\!\!\!
\times\prod_{i=1}^m\frac{(R_i+a_i-1)!\,(R_i+b_i-1)!}
{(2a_i)!\,(R_i-a_i)!\,(2b_i)!\,(R_i-b_i)!}
\frac{(3/2)_{v_i+a_i}\,(3/2)_{v_i+b_i}}{(2)_{v_i+a_i}\,(2)_{v_i+b_i}}
\\
&\!\!\!\!\!\!
\times\prod_{i=1}^n\frac{(R_i+c_i-1)!\,(R_i+d_i-1)!}
{(2c_i+1)!\,(R_i-c_i)!\,(2d_i+1)!\,(R_i-d_i)!}
\frac{(3/2)_{v'_i+c_i}\,(3/2)_{v'_i+d_i}}{(1)_{v'_i+c_i}\,(1)_{v'_i+d_i}}
\\
%
%
\times
\prod_{1\leq i<j\leq m}&(v_j-v_i+a_j-a_i)(v_j-v_i+a_j-b_i)(v_j-v_i+b_j-a_i)(v_j-v_i+b_j-b_i)\\
\times\prod_{1\leq i<j\leq
n}&(v'_j-v'_i+c_j-c_i)(v'_j-v'_i+c_j-d_i)(v'_j-v'_i+d_j-c_i)(v'_j-v'_i+d_j-d_i)\\
&\!\!\!\!\!\!\!\!\!\!\!\!\!\!\!\!\!\!\!\!\!\!\!\!\!\!\!\!
\left.
\times\frac{\prod_{i=1}^m(a_i-b_i)^2\prod_{i=1}^n(c_i-d_i)^2}
{\prod_{i=1}^m\prod_{j=1}^n(u_{ij}+a_i+c_j)(u_{ij}+a_i+d_j)
(u_{ij}+b_i+c_j)(u_{ij}+b_i+d_j)  }
\right|,\tag5.22
\endalign
$$
where the summation range is restricted to those summation variables for which
$v_1+a_1,v_1+b_1,\dotsc,v_m+a_m,v_m+b_m$,
as well as $v'_1+c_1,v'_1+d_1,\dotsc,v'_n+c_n,v'_n+d_n$, are distinct.

The fortunate situation is that, on the one hand, when $a_i=b_i$, for some 
$i=1,\dotsc,m$, or when $c_i=d_i$, for some $i=1,\dotsc,n$, the summand in (5.22) 
becomes zero.
On the other, the summand in (5.22) is invariant under independently interchanging 
$a_i$ with $b_i$, $i=1,\dotsc,m$, and $c_i$ with $d_i$, $i=1,\dotsc,m$ (because the 
differences $b_i-a_i$ and $d_i-c_i$ end up, by the combination of (5.5), (5.13) and 
(5.21), appearing at the second power). Therefore the summation range may
be extended in (5.22), at the expense of a multiplicative factor of
$1/2^{m+n}$, to the summation range in (5.1). 
This leads precisely to the multiple sum given in the statement of the Lemma. \epf

\mysec{6. Separation of the $(2m+2n)$-fold sum for $\omega_b$ in terms of $4mn$-fold 
integrals}


If it weren't for the denominator in the last fraction on the right hand side of (5.1), 
one could expand the product at the numerator of the summand in terms of monomials in the 
summation variables, and the $(2m+2n)$-fold sum (5.1) would separate: one could express it
in terms of simple sums. 

We can get around this obstacle by expressing each factor at the denominator as an 
integral:
$$
\align
\frac{1}{u_{ij}+a_i+c_j}&=\int_0^1x_{ij}^{u_{ij}+a_i+c_j-1}dx_{ij}\\
\frac{1}{u_{ij}+a_i+d_j}&=\int_0^1y_{ij}^{u_{ij}+a_i+d_j-1}dy_{ij}\\
\frac{1}{u_{ij}+b_i+c_j}&=\int_0^1z_{ij}^{u_{ij}+b_i+c_j-1}dz_{ij}\\
\frac{1}{u_{ij}+b_i+d_j}&=\int_0^1w_{ij}^{u_{ij}+b_i+d_j-1}dw_{ij}
,\tag6.1
\endalign
$$
for $i=1,\dotsc,m$, $j=1,\dotsc,n$. Then the multiple sum inside the absolute value signs
in (5.1) can be expressed as a sum of $4mn$-fold integrals of products of simple sums. 
More precisely, expand the product of the numerators on the last three lines of (5.1) as
$$
\align
&\Cal E:=\prod_{i=1}^m(a_i-b_i)^2\prod_{i=1}^n(c_i-d_i)^2\\
&\times\!
\prod_{1\leq i<j\leq m}\!\!((v_j-v_i)+a_j-a_i)((v_j-v_i)+a_j-b_i)((v_j-v_i)+b_j-a_i)
((v_j-v_i)+b_j-b_i)\\
&\times\!
\prod_{1\leq i<j\leq n}\!\!
((v'_j-v'_i)+c_j-c_i)((v'_j-v'_i)+c_j-d_i)((v'_j-v'_i)+d_j-c_i)((v'_j-v'_i)+d_j-d_i)\\
&
=\sum_{C\in\Cal C} e(C)
a_1^{\alpha_1(C)}b_1^{\beta_1(C)}\cdots a_m^{\alpha_m(C)}b_m^{\beta_m(C)}
c_1^{\gamma_1(C)}d_1^{\delta_1(C)}\cdots c_n^{\gamma_n(C)}d_n^{\delta_n(C)},\tag6.2
\endalign
$$
where $\Cal C$ is the collection of all $2^{2m+2n}3^{4{m\choose2}+4{n\choose2}}$ 
signed monomials in the
$(v_j-v_i)$'s, $(v'_j-v'_i)$'s,  $a_i$'s, $b_i$'s, $c_j$'s and $d_j$'s 
obtained by expanding the left hand side of (6.2),
and for such a monomial $C\in\Cal C$, $\alpha_i(C)$, $\beta_i(C)$, $\gamma_j(C)$ and 
$\delta_j(C)$ are the exponents of $a_i$, $b_i$, $c_j$ and $d_j$, respectively, while 
$e(C)$ is the part of $C$ besides 
$\prod_{i=1}^m a_i^{\alpha_i(C)}b_i^{\beta_i(C)}
\prod_{j=1}^nc_j^{\gamma_j(C)}d_j^{\delta_j(C)}$ (so $e(C)$ is a signed monomial in the
$(v_j-v_i)$'s and $(v'_j-v'_i)$'s).

Define also
$$
\align
T^{(n)}(R,v;x)&:=\frac{1}{R}\sum_{a=0}^R \frac{(-R)_a\,(R)_a\,(3/2)_{v+a}}
{(1)_a\,(1/2)_a\,(2)_{v+a}}\left(\frac{x}{4}\right)^a a^n\tag6.3\\
{T'}^{(n)}(R,v;x)&:=\frac{1}{R}\sum_{c=0}^R \frac{(-R)_c\,(R)_c\,(3/2)_{v+c}}
{(1)_c\,(3/2)_c\,(1)_{v+c}}\left(\frac{x}{4}\right)^c c^n
.\tag6.4
\endalign
$$

We have the following result.

\proclaim{Proposition 6.1} The boundary-influenced correlation $\omega_b$ can be written as
$$
\align
&\omega_b\left(\matrix{R_1}\\{v_1}\endmatrix
\cdots\matrix{R_m}\\{v_m}\endmatrix;
\matrix{R'_1}\\{v'_1}\endmatrix\cdots\matrix{R'_n}\\{v'_n}\endmatrix\right)
=\chi_{2m,2n}\prod_{i=1}^mR_i
\prod_{i=1}^nR'_i(R'_i-1/2)(R'_i+1/2)
\\
&\ \ \ \ \ \ \ \ \ \ \ \ 
\times
\left|\sum_{C\in\Cal C} e(C)
M_{\alpha_1(C),\beta_1(C),\dotsc,\alpha_m(C),\beta_m(C);
\gamma_1(C),\delta_1(C),\dotsc,\gamma_n(C),\delta_n(C)}\right|,
\tag6.5
\endalign
$$
where $\chi$ is given by $(4.3)$, the collection $\Cal C$ and
$e(C)$, $\alpha_i(C)$, $\beta_i(C)$, $\gamma_j(C)$, $\delta_j(C)$
are as in $(6.2)$, and the ``moment'' 
$M_{\alpha_1,\beta_1,\dotsc,\alpha_m,\beta_m;\gamma_1,\delta_1,\dotsc,\gamma_n,\delta_n}$
equals the $4mn$-fold integral
$$
\align
&\!\!\!\!\!\!\!\!\!\!\!\
M_{\alpha_1,\beta_1,\dotsc,\alpha_m,\beta_m;\gamma_1,\delta_1,\dotsc,\gamma_n,\delta_n}=
\int_0^1\cdots\int_0^1 \prod_{i=1}^m\prod_{j=1}^n(x_{ij}y_{ij}z_{ij}w_{ij})^{v_i+v'_j+1}\\
&
\times
T^{(\alpha_1)}(R_1,v_1;\prod_{j=1}^n x_{1j}y_{1j})\,
T^{(\beta_1)}(R_1,v_1;\prod_{j=1}^n z_{1j}w_{1j})
\cdots\\
&\ \ \ \ \ \ \ \ \ \ \ \ 
\cdots
T^{(\alpha_m)}(R_m,v_m;\prod_{j=1}^n x_{mj}y_{mj})\,
T^{(\beta_m)}(R_m,v_m;\prod_{j=1}^n z_{mj}w_{mj})\\
&
\times
{T'}^{(\gamma_1)}(R'_1,v'_1;\prod_{i=1}^m x_{i1}z_{i1})\,
{T'}^{(\delta_1)}(R'_1,v'_1;\prod_{i=1}^m y_{i1}w_{i1})
\cdots\\
&\ \ \ \ \ \ \ \ \ \ \ \ 
\cdots
{T'}^{(\gamma_n)}(R'_n,v'_n;\prod_{i=1}^m x_{in}z_{in})\,
{T'}^{(\delta_n)}(R'_n,v'_n;\prod_{i=1}^m y_{in}w_{in})\,dx_{11}\cdots dw_{mn},\tag6.6
\endalign
$$
with $T^{(n)}(R,v;x)$ and ${T'}^{(n)}(R,v;x)$ defined by $(6.3)$--$(6.4)$.
\endproclaim

\pf We note first that, since
$$
\frac{(R+a-1)!}{(2a)!\,(R-a)!}=\frac{(-1)^a}{R}\frac{(-R)_a\,(R)_a}{4^a\, (1)_a\,(1/2)_a}
$$
and
$$
\frac{(R+c-1)!}{(2c+1)!\,(R-c)!}=\frac{(-1)^c}{R}\frac{(-R)_c\,(R)_c}{4^c\, (1)_c\,(3/2)_c},
$$
the sums in (6.3)--(6.4) can also be written as
$$
\align
T^{(n)}(R,v;x)&=\sum_{a=0}^R\, (-1)^a\frac{(R+a-1)!}{(2a)!\,(R-a)!}
\frac{(3/2)_{v+a}}{(2)_{v+a}}x^a a^n\tag6.7\\
{T'}^{(n)}(R,v;x)&=\sum_{c=0}^R\, (-1)^c\frac{(R+c-1)!}{(2c+1)!\,(R-c)!}
\frac{(3/2)_{v+c}}{(1)_{v+c}}x^c c^n.\tag6.8
\endalign
$$

Denote by $S$ the multiple sum in (5.1).
Express the factors at the denominator of the summand of $S$ as integrals, using (6.1). 
Expand the factors of the summand of $S$ contained in the left hand side of (6.2) as a sum 
of monomials in the summation variables $a_1,b_1,\dotsc,a_m,b_m$ and 
$c_1,d_1,\dotsc,c_n,d_n$ of $S$, as indicated by (6.2).
Bring all $4mn$ integration signs in front of $S$. We obtain
$$
\align
S=&\int_0^1\cdots\int_0^1
\left\{
\sum_{a_1,b_1=0}^{R_1}\cdots\sum_{a_m,b_m=0}^{R_m}
\sum_{c_1,d_1=0}^{R'_1}\cdots\sum_{c_n,d_n=0}^{R'_n}
(-1)^{\sum_{i=1}^m (a_i+b_i) +\sum_{i=1}^n (c_i+d_i)}
\right.
\\
&\times\prod_{i=1}^m\frac{(R_i+a_i-1)!\,(R_i+b_i-1)!}
{(2a_i)!\,(R_i-a_i)!\,(2b_i)!\,(R_i-b_i)!}
\frac{(3/2)_{v_i+a_i}\,(3/2)_{v_i+b_i}}{(2)_{v_i+a_i}\,(2)_{v_i+b_i}}
\\
&\times\prod_{i=1}^n\frac{(R_i+c_i-1)!\,(R_i+d_i-1)!}
{(2c_i+1)!\,(R_i-c_i)!\,(2d_i+1)!\,(R_i-d_i)!}
\frac{(3/2)_{v'_i+c_i}\,(3/2)_{v'_i+d_i}}{(1)_{v'_i+c_i}\,(1)_{v'_i+d_i}}
\\
&\times
\prod_{i=1}^m\prod_{j=1}^n
x_{ij}^{v_i+v'_j+a_i+c_j+1}y_{ij}^{v_i+v'_j+a_i+d_j+1}
z_{ij}^{v_i+v'_j+b_i+c_j+1}w_{ij}^{v_i+v'_j+b_i+d_j+1}\\
&
\left.
\times
\sum_{C\in\Cal C} e(C)
a_1^{\alpha_1(C)}b_1^{\beta_1(C)}\cdots a_m^{\alpha_m(C)}b_m^{\beta_m(C)}
c_1^{\gamma_1(C)}d_1^{\delta_1(C)}\cdots c_n^{\gamma_n(C)}d_n^{\delta_n(C)}
\right\}\,
dx_{11}\cdots dw_{mn}.
\endalign
$$
Reversing the order of the two multiple summations in the integrand above yields 
$$
\align
S=&\int_0^1\!\!\cdots\!\int_0^1
\sum_{C\in\Cal C}
\left\{
\sum_{a_1,b_1=0}^{R_1}\cdots\sum_{a_m,b_m=0}^{R_m}
\sum_{c_1,d_1=0}^{R'_1}\cdots\sum_{c_n,d_n=0}^{R'_n}
\!\!(-1)^{\sum_{i=1}^m (a_i+b_i) +\sum_{i=1}^n (c_i+d_i)}
\right.
\\
&\times\prod_{i=1}^m\frac{(R_i+a_i-1)!\,(R_i+b_i-1)!}
{(2a_i)!\,(R_i-a_i)!\,(2b_i)!\,(R_i-b_i)!}
\\
&\times\prod_{i=1}^n\frac{(R_i+c_i-1)!\,(R_i+d_i-1)!}
{(2c_i+1)!\,(R_i-c_i)!\,(2d_i+1)!\,(R_i-d_i)!}
\\
&\times\prod_{i=1}^m\frac{(3/2)_{v_i+a_i}\,(3/2)_{v_i+b_i}}
{(2)_{v_i+a_i}\,(2)_{v_i+b_i}}
\prod_{i=1}^n\frac{(3/2)_{v'_i+c_i}\,(3/2)_{v'_i+d_i}}
{(1)_{v'_i+c_i}\,(1)_{v'_i+d_i}}
\\
&\times
\prod_{i=1}^m\prod_{j=1}^n
x_{ij}^{v_i+v'_j+a_i+c_j+1}y_{ij}^{v_i+v'_j+a_i+d_j+1}
z_{ij}^{v_i+v'_j+b_i+c_j+1}w_{ij}^{v_i+v'_j+b_i+d_j+1}\\
&
\left.
\times
a_1^{\alpha_1(C)}b_1^{\beta_1(C)}\cdots a_m^{\alpha_m(C)}b_m^{\beta_m(C)}
c_1^{\gamma_1(C)}d_1^{\delta_1(C)}\cdots c_n^{\gamma_n(C)}d_n^{\delta_n(C)}
\right\}
\,e(C)
\,dx_{11}\cdots dw_{mn}.
\endalign
$$
Clearly, the inner multiple sum separates in terms of simple sums on 
$a_1,b_1,\dotsc,a_m,b_m$ and $c_1,d_1,\dotsc,c_n,d_n$. Moreover, all the simple sums arising 
this way have one of the forms (6.7) or (6.8).
Moving the $4mn$-fold integral sign 
inside the outer multiple sum above one obtains (6.6). \endpf

The sums (6.3) and (6.4) can conveniently be expressed in terms of hypergeometric functions
\footnote{The hypergeometric function of parameters
$a_1,\dotsc,a_p$ and $b_1,\dotsc,b_q$ is defined by
$${}_p F_q\!\left[\matrix a_1,\dotsc,a_p\\ b_1,\dotsc,b_q\endmatrix;
z\right]=\sum _{k=0} ^{\infty}\frac {(a_1)_k\cdots(a_p)_k}
{k!\,(b_1)_k\cdots(b_q)_k} z^k\ ,$$
where $(a)_0:=1$ and $(a)_k:=a(a+1)\cdots (a+k-1)$ for $k\geq1$.}.
Indeed, for integer $R$ the upper summation limits in (6.3) and (6.4) can be replaced by 
$\infty$ without affecting the definitions, due to the factors $(-R)_a$ and $(-R)_c$ in the 
summands. Using that $(x)_{v+a}=(x)_v(x+v)_a$, (6.3) becomes
$$
\align
&T^{(n)}(R,v;x)=\frac{1}{R}\frac{(3/2)_v}{(2)_v}
\sum_{a=0}^\infty \frac{(-R)_a\,(R)_a\,(v+3/2)_{a}}
{(1)_a\,(1/2)_a\,(v+2)_{a}}\left(\frac{x}{4}\right)^a a^n\\
&
=\frac{1}{R}\frac{(3/2)_v}{(2)_v}
\sum_{a=0}^\infty \frac{(-R)_a\,(R)_a\,(v+3/2)_{a}}
{(1)_a\,(1/2)_a\,(v+2)_{a}}\left(\frac{x}{4}\right)^a\\
&\ \ \ \ \ \ \ 
\times
\{f_na(a-1)\cdots(a-n+1)+f_{n-1}a(a-1)\cdots(a-n+2)+\cdots+f_1a+f_0\},\tag6.9
\endalign
$$
where the coefficients $f_i$ are defined so that the last factor in the summand equals $a^n$ (in particular, $f_n=1$ and $f_0$ is the Kronecker symbol $\delta_{n0}$).
Since $a(a-1)\cdots(a-k+1)/(1)_a=1/(1)_{a-k}$, for $a\geq k$, we have
$$
\align
&\sum_{a=0}^\infty\frac{(-R)_a\,(R)_a\,(v+3/2)_{a}}{(1)_a\,(1/2)_a\,(v+2)_{a}}
\left(\frac{x}{4}\right)^a a(a-1)\cdots(a-k+1)\\
&
=
\frac{(-R)_k\,(R)_k\,(v+3/2)_{k}}{(1/2)_k\,(v+2)_{k}}\left(\frac{x}{4}\right)^{k}
\sum_{a=k}^\infty\frac{(-R+k)_{a-k}\,(R+k)_{a-k}\,(v+k+3/2)_{a-k}}
{(1)_{a-k}\,(k+1/2)_{a-k}\,(v+k+2)_{a-k}}
\left(\frac{x}{4}\right)^{a-k}\\
&
=\frac{(-R)_k\,(R)_k\,(v+3/2)_{k}}{(1/2)_k\,(v+2)_{k}}\left(\frac{x}{4}\right)^{k}   
{}_3F_2\!\left[\matrix{-R+k,\,R+k,\,\frac{3}{2}+v+k}\\{\frac12+k,2+v+k}\endmatrix;\frac x4\right].
\endalign
$$
Substituting this into (6.9) we obtain the first part of the following result.

\proclaim{Lemma 6.2} We have
$$
\align
&\!\!\!\!\!\!\!\!\!\!
T^{(n)}(R,v;x)=\\
&\!\!\!\!\!
\frac{1}{R}\frac{(3/2)_v}{(2)_v}
\sum_{k=0}^n f_k\frac{(-R)_k\,(R)_k\,(v+3/2)_{k}}{(1/2)_k\,(v+2)_{k}}\left(\frac{x}{4}\right)^{k}
{}_3F_2\!\left[\matrix{-R+k,\,R+k,\,\frac{3}{2}+v+k}\\{\frac12+k,2+v+k}\endmatrix;\frac x4\right]
\tag6.10\\
&\!\!\!\!\!\!\!\!\!\!
{T'}^{(n)}(R,v;x)=\\
&\!\!\!\!\!
\frac{1}{R}\frac{(3/2)_v}{(1)_v}
\sum_{k=0}^n f_k\frac{(-R)_k\,(R)_k\,(v+3/2)_{k}}{(3/2)_k\,(v+1)_{k}}\left(\frac{x}{4}\right)^{k}
{}_3F_2\!\left[\matrix{-R+k,\,R+k,\,\frac{3}{2}+v+k}\\{\frac32+k,1+v+k}\endmatrix;\frac x4\right],
\tag6.11
\endalign
$$
where the $f_k$'s are as in $(6.9)$ $($in particular $f_n=1$$)$.
\endproclaim

\pf Starting from (6.4), we obtain (6.11) by nearly the same calculation that proved (6.10). \endpf

Since by Proposition 6.1 the boundary-influenced correlation $\omega_b$ is expressed in terms of
the moments
$M_{\alpha_1,\beta_1,\dotsc,\alpha_m,\beta_m;\gamma_1,\delta_1,\dotsc,\gamma_n,\delta_n}$
given by (6.6), which in turn depend on $T^{(n)}(R,v;x)$ and ${T'}^{(n)}(R,v;x)$, to determine the
asymptotics of $\omega_b$ we need to understand the asymptotics of the $T^{(n)}$'s and
${T'}^{(n)}$'s. In the next section we deduce these latter two asymptotics from a
result---stated in Proposition 7.2---whose technical proof we defer to Section 9. 
In Section 8 we show that the asymptotics of the $T^{(n)}$'s and ${T'}^{(n)}$'s found in
Section 7 can be used to obtain the asymptotics of 
$M_{\alpha_1,\beta_1,\dotsc,\alpha_m,\beta_m;\gamma_1,\delta_1,\dotsc,\gamma_n,\delta_n}$.

\mysec{7. The asymptotics of the $T^{(n)}$'s and ${T'}^{(n)}$'s}

Given that in Theorem 2.1 the coordinates of the quadromers approach infinity as indicated by
(\TwoTwo), we need more specifically to find the asymptotics of $T^{(n)}(R,qR+c;x)$ and
${T'}^{(n)}(R,qR+c;x)$, as $R\to\infty$. These are given by the following result.

\proclaim{Proposition 7.1} Let $q>0$ be fixed rational number, and let $n\geq0$ and $c$ be 
fixed integers. Then for any real number $x\in(0,1]$, we have
$$
\align
&\!\!\!\!\!\!\!\!\!
\left|
T^{(n)}(R,qR+c;x)-\frac{2}{\sqrt{\pi}}\frac{1}{\root 4\of {q^2+\frac{x}{4-x}}}
\frac{1}{R^{3/2}}\left(R\sqrt{\frac{x}{4-x}}\,\right)^n
\right.
\\
&\ \ \ \ \ \ 
\left.
\times
\cos\left[R\arccos\left(1-\frac x2\right)-\frac12\arctan\frac1q\sqrt{\frac{x}{4-x}}+\frac{n\pi}{2}
\right]
\right|
\leq MR^{n-5/2}\tag7.1\\
&\!\!\!\!\!\!\!\!\!
\left|
{T'}^{(n)}(R,qR+c;x)-\frac{1}{\sqrt{\pi}}\frac{\root 4\of{q^2+\frac{x}{4-x}}}
{\sqrt{\frac{x}{4-x}}}\frac{1}{R^{3/2}}\left(R\sqrt{\frac{x}{4-x}}\,\right)^n
\right.
\\
&
\left.
\times
\cos\left[R\arccos\left(1-\frac x2\right)+\frac12\arctan\frac1q\sqrt{\frac{x}{4-x}}
+\frac{(n-1)\pi}{2}
\right]
\right|
\leq \frac{1}{\sqrt{x}}M'R^{n-5/2},\tag7.2
\endalign
$$
for $R\geq R_0$, where $R_0$, $M$ and $M'$ are independent of $x\in(0,1]$.
\endproclaim

In our proof of the above statements we make use of the following result, whose proof is presented in
Section 9.

\proclaim{Proposition 7.2}
Let $p(t)$ and $Q(t)$ be functions depending on the parameter $x\in[0,1]$, defined on $(0,1)$ by 
$$
\align
p(t)&=-q\ln t - i\arccos\left(1-\frac{xt}{2}\right)\tag7.3\\
Q(t)&=\frac{t^l}{(1-t)^{1/2}}\frac{(4-2xt)^a}{(4-xt)^b},\tag7.4
\endalign
$$
where $0<q\in\Q$, $0\leq a\in\Z$, $-1/2\leq b\in\frac12\Z$ and $l\in\frac12\Z$ are all fixed.
Then
$$
\left|\int_0^1 e^{-Rp(t)}Q(t) dt - F(R,x)\right|\leq MR^{-3/2},\tag7.5
$$
for $R\geq R_0$, with $R_0$ and $M$ independent of $x\in[0,1]$, and $F(R,x)$ given by
$$
F(R,x)=\frac{\sqrt{\pi}}{\sqrt{R}}\frac{(4-2x)^a/(4-x)^b}{\root4\of{q^2+\frac{x}{4-x}}}
e^{i\left[R\arccos\left(1-\frac{x}{2}\right)-\frac12\arctan\frac1q\sqrt{\frac{x}{4-x}}\,\right]}.
\tag7.6
$$
\endproclaim

\proclaim{Corollary 7.3} Under the assumptions of Proposition $7.2$, we have
$$
\left|\int_0^1 e^{-Rp(t)}Q(t) dt\right|\leq MR^{-1/2},\tag7.7
$$
where $M$ is independent of $x\in[0,1]$.
\endproclaim

\pf The absolute value of the part of the right hand side of (7.6) not containing $R$ can 
clearly be majorized, for $x\in[0,1]$, by a constant independent of $x$. The statement of the 
Corollary follows by combining this observation with (7.5). \endpf

{\it Proof of Proposition 7.1.} By \cite{\Lu,(10),\,p.58}, a ${}_{p+1} F_{q+1}$ hypergeometric
function can be written as an integral of a ${}_{p} F_{q}$ as 
$$
{}_{p+1} F_{q+1}\left[\matrix{\beta,\,\alpha_p}\\{\beta+\sigma,\,\rho_q}\endmatrix;z\right]=
\frac{\Gamma(\beta+\sigma)}{\Gamma(\beta)\Gamma(\sigma)}
\int_0^1t^{\beta-1}(1-t)^{\sigma-1}{}_pF_q\left[\matrix \alpha_p\\\rho_q\endmatrix;zt\right]dt,
\tag7.8
$$
provided $p\leq q+1$, $\Rep\beta>0$, $\Rep\sigma>0$, and $|z|<1$ if $p=q+1$ (here $\alpha_p$
stands for a $p$-tuple, $\rho_q$ for a $q$-tuple of parameters).

Taking $\beta=v+k+3/2$ and $\sigma=1/2$, (7.8) yields
$$
\align
&{}_3F_2\!\left[\matrix{-R+k,\,R+k,\,\frac{3}{2}+v+k}\\{\frac12+k,2+v+k}\endmatrix;
\frac x4\right]
=\frac{\Gamma(v+k+2)}{\Gamma(v+k+3/2)\Gamma(1/2)}\\
&\ \ \ \ \ \ \ \ \ \ \ \ \ \ \ \ \ \ 
\times
\int_0^1t^{v+k+1/2}(1-t)^{-1/2}{}_2F_1\left[\matrix{-R+k,\,R+k}\\{\frac12+k}\endmatrix;
\frac{xt}{4}\right]dt.\tag7.9
\endalign
$$
On the other hand, from the definition of a ${}_2F_1$ hypergeometric function it readily follows
that
$$
\frac{d}{dz}\,{}_2F_1\left[\matrix{a_1,\,a_2}\\{b}\endmatrix;z\right]
=\frac{a_1a_2}{b}{}_2F_1\left[\matrix{a_1+1,\,a_2+1}\\{b+1}\endmatrix;z\right].
$$
Repeated application of this shows that
$$
{}_2F_1\left[\matrix{-R+k,\,R+k}\\{\frac12+k}\endmatrix;\frac{xt}{4}\right]
=\frac{(1/2)_k}{(-R)_k\,(R)_k\,(x/4)^k}
\frac{d^k}{dt^k}\,{}_2F_1\left[\matrix{-R,\,R}\\{\ \ \ \frac12}\endmatrix;\frac{xt}{4}\right]
.\tag7.10
$$
However, the latter ${}_2F_1$ evaluates exactly (see for instance \cite{\GR, p.1055,\,\#1}):
$$
{}_2F_1\left[\matrix{-R,\,R}\\{\ \ \ \frac12}\endmatrix;z\right]=
\cos \left[R\arccos(1-2z)\right].\tag7.11
$$
Replacing (7.9) and (7.10) into (6.10) and using (7.11) we obtain
$$
\align
T^{(n)}(R,v;x)=\frac{1}{R}\sum_{k=0}^n& f_k
\frac
{
\frac{{\displaystyle (-R)_k\,(R)_k\,(3/2)_{v+k}}}{{\displaystyle (1/2)_k\,(2)_{v+k}}}
\left(\frac {{\displaystyle x}}{{\displaystyle 4}}\right)^k
\frac{{\displaystyle \Gamma(v+k+2)}}{{\displaystyle \Gamma(v+k+3/2)\Gamma(1/2)}}
}
{
\left(\frac {{\displaystyle x}}{{\displaystyle 4}}\right)^k
\frac{{\displaystyle (-R)_k\,(R)_k}}{{\displaystyle (1/2)_k}}
}\\
&\ \ 
\times
\int_0^1 t^{v+k+1/2}(1-t)^{-1/2}\frac{d^k}{dt^k}
\cos\!\left[R\arccos\left(1-\frac{xt}{2}\right)\right]dt\\
&\!\!\!\!\!\!\!\!\!\!\!\!\!\!\!\!\!\!\!\!\!\!\!\!
=
\frac{2}{\pi R}\sum_{k=0}^n f_k
\int_0^1 t^{v+k+1/2}(1-t)^{-1/2}\frac{d^k}{dt^k}
\cos\!\left[R\arccos\left(1-\frac{xt}{2}\right)\right]dt
\tag7.12
\endalign
$$
(for the second equality we also used $\Gamma(1/2)=\sqrt{\pi}$ and the recurrence 
$\Gamma(x+1)=x\Gamma(x)$).

By Lemma 7.4, the successive derivatives with respect to $t$ of 
$\cos(R\arccos(1-xt/2))$ can be written as  
$$
\align
\frac{d^{k}}{dt^k}\cos(R\arccos(1-xt/2))=
\cos&\left[R\arccos\left(1-\frac{xt}{2}\right)+\frac{n\pi}{2}\right]
\left(R\sqrt{\frac{x}{4t-xt^2}}\,\right)^k\\
&+O(R^{k-1})
,\tag7.13
\endalign
$$
where the terms of the omitted linear combination are 
of the form $R^jx^s\cos(R\arccos(1-xt/2))$ or $R^jx^s\sin(R\arccos(1-xt/2))$
times a function of type (7.4), with $0\leq j\leq k-1$ and $s\geq0$. 
(In fact, this is how 
the family of functions (7.4) was chosen, to contain all functions arising this way from 
successive derivatives of $\cos(R\arccos(1-xt/2))$, and the analogous functions resulting 
when doing the same analysis to the ${T'}^{(n)}$'s.) 

Since clearly $|x|\leq1$ throughout the range $x\in(0,1]$,
one sees by Corollary 7.3 that for $v=qR+c$, $q>0$ and for any fixed $k\in\{0,\dotsc,n\}$, 
the total contribution of the lower 
order terms in (7.13) to the sum on the right hand side of (7.12) 
is bounded in absolute value by $L_kR^{k-3/2}$ for $R\geq R_k$, where $R_k$ and 
$L_k$ are independent of $x\in[0,1]$, $k=0,\dotsc,n$. Therefore, the combined contribution to 
$T^{(n)}(R,qR+c;x)$ of all lower order terms in (7.13), for $k=0,\dotsc,n$, is bounded in 
absolute value by $LR^{k-5/2}$ for $R\geq \rho_0$, where $\rho_0$ and 
$L$ are constants independent of $x\in[0,1]$.

On the other hand, by using Corollary 7.3 as in the previous paragraph, 
one sees that the combined
contribution of the leading terms of (7.13) for $k=0,\dotsc,n-1$ to ${T}^{(n)}(R,qR+c;x)$ 
is again bounded in absolute value by $K{R}^{n-5/2}$ for all $R\geq \rho_1$ and
$x\in(0,1]$, for some constants $K$ and $\rho_1$ independent of $x\in(0,1]$. 

Taking into account the leading term of (7.13) for $k=n$, we obtain by (7.12) and the previous
two paragraphs that
$$
\align
&\!\!\!\!\!\!\!\!\!\!\!\!\!\!\!\!\!\!\!\!\!\!
\left|
T^{(n)}(R,qR+c;x)-\frac{2}{\pi R}
\int_0^1 t^{qR}\frac{t^{n+c+1/2}}{(1-t)^{1/2}}\left(R\sqrt{\frac{x}{4t-xt^2}}\,\right)^n
\right.
\\
&\ \ \ \ \ \ \ \ \ \ \ \ \ \ \ \ \ \ \ \ \ \ \ \ \ 
\left.
\times
\cos\left[R\arccos\left(1-\frac{xt}{2}\right)+\frac{n\pi}{2}\right]dt
\right|
\leq M_0R^{n-5/2},\tag7.14
\endalign
$$
for $R\geq\rho_2$, with $\rho_2$ and $M_0$ independent of $x\in[0,1]$. 
However, it is readily seen by Proposition 7.2 that
$$
\align
&
\left|
\int_0^1 t^{qR}\frac{t^{n+c+1/2}}{(1-t)^{1/2}}\left(\sqrt{\frac{x}{4t-xt^2}}\,\right)^n
\cos\left[R\arccos\left(1-\frac{xt}{2}\right)+\frac{n\pi}{2}\right]dt
\right.
\\
&\ \ \ \ \ \ \ \ \ \ \ \ 
-\frac{\sqrt{\pi}}{\sqrt{R}}\left(q^2+\frac{x}{4-x}\right)^{-1/4}
\left(\sqrt{\frac{x}{4-x}}\,\right)^n\\
&\ \ \ \ \ \ \ \ \ \ \ \
\left.
\times
\cos\left[R\arccos\left(1-\frac{x}{2}\right)-\frac12\arctan\frac1q\sqrt{\frac{x}{4-x}}
+\frac{n\pi}{2}\right]
\right|
\\
&\ \ \ \ \ \ \ \ \ \ \ \ \ \ \ \ \ \ \ \ \ \ \ \ \
\leq  M_1R^{n-5/2},\tag7.15
\endalign
$$
for $R\geq \rho_3$, with $\rho_3$ and $M_1$ independent of $x\in[0,1]$.
Relations (7.14) and (7.15) imply (7.1).

We now turn to proving (7.2). Because of the requirement $\Rep \sigma>0$ we cannot
apply (7.8) directly to the ${}_3F_2$'s of (6.11), so we write first
$$
\align
&\!\!\!\!\!\!\!\!\!\!\!\!\!\!\!\!\!\!\!\!\!
{}_3F_2\left[\matrix{-R+k,\,R+k,\,\frac32+v+k}\\{\frac32+k,\,1+v+k}\endmatrix;\frac x4
\right]=\\
&
\sum_{a\geq0}\frac{(-R+k)_a\,(R+k)_a\,(1/2+v+k)_a
\left(1+\frac{{\displaystyle a}}{{\displaystyle 1/2+v+k}}\right)}
{a!\,(3/2+k)_a\,(1+v+k)_a}\left(\frac x4\right)^a\\
&\!\!\!\!\!\!\!\!\!\!\!\!\!\!
={}_3F_2\left[\matrix{-R+k,\,R+k,\,\frac12+v+k}\\{\frac32+k,\,1+v+k}\endmatrix;\frac x4
\right]+\\
&
\frac{1}{1/2+v+k}\frac{(-R+k)_1\,(R+k)_1\,(1/2+v+k)_1}{(3/2+k)_1\,(1+v+k)_1}
\left(\frac x4\right)\\
&
\times
\sum_{a\geq1}\frac{(-R+k+1)_{a-1}\,(R+k+1)_{a-1}\,(3/2+v+k)_{a-1}}
{(a-1)!\,(5/2+k)_{a-1}\,(2+v+k)_{a-1}}\left(\frac x4\right)^{a-1}\\
&\!\!\!\!\!\!\!\!\!\!\!\!\!\!
={}_3F_2\left[\matrix{-R+k,\,R+k,\,\frac12+v+k}\\{\frac32+k,\,1+v+k}\endmatrix;\frac x4
\right]+\\
&
\frac{x}{4}\frac{(-R+k)(R+k)}{(3/2+k)(1+v+k)}
{}_3F_2\left[\matrix{-R+k+1,\,R+k+1,\,\frac32+v+k}\\{\frac52+k,\,2+v+k}\endmatrix;\frac x4
\right].\tag7.16
\endalign
$$
The two ${}_3F_2$'s on the right hand side of (7.16) have the same form (the second is 
obtained from the first by replacing $k$ by $k+1$). Equality (7.8)
is applicable to them and yields
$$
\align
&
{}_3F_2\left[\matrix{-R+k,\,R+k,\,\frac12+v+k}\\{\frac32+k,\,1+v+k}\endmatrix;\frac x4
\right]=
\frac{\Gamma(v+k+1)}{\Gamma(v+k+1/2)\Gamma(1/2)}\\
&\ \ \ \ \ \ \ \ \ \ \ \ \ \ \ \ \ \ 
\times
\int_0^1t^{v+k-1/2}(1-t)^{-1/2}{}_2F_1\left[\matrix{-R+k,\,R+k}\\{\frac32+k}\endmatrix;
\frac{xt}{4}\right]dt.\tag7.17
\endalign
$$
Repeated application of the relation just before (7.10) shows that
$$
{}_2F_1\left[\matrix{-R+k,\,R+k}\\{\frac32+k}\endmatrix;\frac{xt}{4}\right]
=\frac{(3/2)_k}{(-R)_k\,(R)_k\,(x/4)^k}
\frac{d^k}{dt^k}\,{}_2F_1\left[\matrix{-R,\,R}\\{\ \ \ \frac32}\endmatrix;\frac{xt}{4}\right]
.\tag7.18
$$
To continue our analysis we need a closed formula for 
the ${}_2F_1$ on the right hand side of (7.18). We obtain it as follows. By
\cite{\GR,8.962,\,\#1} one has
$$
P_n^{(\alpha,\beta)}(x)=\frac{(-1)^n\Gamma(n+1+\beta)}{n!\,\Gamma(1+\beta)}
{}_2F_1\left[\matrix{-n,\,n+\alpha+\beta+1}\\{1+\beta}\endmatrix;\frac{1+x}{2}\right],
$$
where $P_n^{(\alpha,\beta)}(x)$ is the $n$th Jacobi polynomial of parameters 
$\alpha$ and $\beta$.
Taking $\alpha=-3/2$, $\beta=1/2$, this gives
$$
{}_2F_1\left[\matrix{-n,\,n}\\{\ \ \ \frac32}\endmatrix;\frac{1+x}{2}\right]=
\frac{n!\,\Gamma(3/2)}{(-1)^n\Gamma(n+3/2)}P_n^{(-\frac32,\frac12)}(x).\tag7.19
$$

On the other hand, from \cite{\GR,8.962,\,\#4} we obtain
$$
\align
P_n^{(\frac12,\frac12)}(x)&=\frac{\Gamma(2)\Gamma(n+3/2)}{\Gamma(n+2)\Gamma(3/2)}C_n^1(x)\\
&=\frac{\Gamma(2)}{\Gamma(3/2)}\frac{\Gamma(n+3/2)}{\Gamma(n+2)}
\frac{\sin[(n+1)\arccos x]}{\sin (\arccos x)},\tag7.20
\endalign
$$
where $C_n^\lambda$ is the $n$th ultraspherical polynomial of parameter $\lambda$, and at 
the last equality in (7.20) we used \cite{\GR,8.937,\,\#1}.

However, by \cite{\GR,8.961,\,\#8} the Jacobi polynomials satisfy the recurrence
$$
(2n+\alpha+\beta)P_n^{(\alpha-1,\beta)}(x)=
(n+\alpha+\beta)P_n^{(\alpha,\beta)}(x)-
(n+\beta)P_{n-1}^{(\alpha,\beta)}(x).
$$
By this the explicit formula (7.20) for $P_n^{(\frac12,\frac12)}(x)$ yields one for
$P_n^{(-\frac12,\frac12)}(x)$, which in turn, by another application of the above 
recurrence, yields an explicit formula for $P_n^{(-\frac32,\frac12)}(x)$. Substituting this
into (7.19) one obtains after simplifications that
$$
\align
{}_2F_1\left[\matrix{-n,\,n}\\{\ \ \ \frac32}\endmatrix;\frac{t}{4}\right]=
\frac{2n}{4n^2-1}\sqrt{\frac{4-t}{t}}\sin&\left[n\arccos\left(1-\frac t2\right)\right]\\
&
-\frac{1}{4n^2-1}\cos\left[n\arccos\left(1-\frac t2\right)\right].\tag7.21
\endalign
$$
Expressing the ${}_3F_2$'s in (6.11) with the use of (7.17), (7.18) and (7.21), (6.11) 
becomes
$$
\align
&
{T'}^{(n)}(R,v;x)=
\\
&
\frac{1}{R}\sum_{k=0}^nf_k
\frac
{
\frac{(-R)_k\,(R)_k\,(3/2)_{v+k}}{(3/2)_k\,(1)_{v+k}}
\left(\frac x4 \right)^k
\frac{\Gamma(v+k+1)}{\Gamma(v+k+1/2)\Gamma(1/2)}
}
{
\frac{(-R)_k\,(R)_k}{(3/2)_k}
\left(\frac x4 \right)^k
}
\\
&
\times
\int_0^1 t^{v+k-1/2}(1-t)^{-1/2}\frac{d^k}{dt^k}
\left\{
\frac{2R}{4R^2-1}\sqrt{\frac{4-xt}{xt}}
\sin\left[R\arccos\left(1-\frac{xt}{2}\right)\right]
\right.
\\
&\ \ \ \ \ \ \ \ \ \ \ \ \ \ \ \ \ \ \ \ 
\left.
-\frac{1}{4R^2-1}\cos\left[R\arccos\left(1-\frac{xt}{2}\right)\right]
\right\}
dt
\\
&
+\frac{1}{R}\sum_{k=0}^nf_k
\frac
{
\frac{(-R)_k\,(R)_k\,(3/2)_{v+k}}{(3/2)_k\,(1)_{v+k}}
\left(\frac x4 \right)^k
\frac{\Gamma(v+k+2)}{\Gamma(v+k+3/2)\Gamma(1/2)}
}
{
\frac{(-R)_{k+1}\,(R)_{k+1}}{(3/2)_{k+1}}
\left(\frac x4 \right)^{k+1}
}
\frac x4
\frac{(-R+k)(R+k)}{(3/2+k)(1+v+k)}
\\
&
\times
\int_0^1 t^{v+k+1/2}(1-t)^{-1/2}\frac{d^{k+1}}{dt^{k+1}}
\left\{
\frac{2R}{4R^2-1}\sqrt{\frac{4-xt}{xt}}
\sin\left[R\arccos\left(1-\frac{xt}{2}\right)\right]
\right.
\\
&\ \ \ \ \ \ \ \ \ \ \ \ \ \ \ \ \ \ \ \ 
\left.
-\frac{1}{4R^2-1}\cos\left[R\arccos\left(1-\frac{xt}{2}\right)\right]
\right\}
dt,
\endalign
$$
which when substituting $v=qR+c$ becomes after simplifications
$$
\align
&
{T'}^{(n)}(R,qR+c;x)=
\frac{4}{\pi(4R^2-1)}\sum_{k=0}^nf_k(qR+c+k+1/2)
\int_0^1 t^{qR}\frac{t^{k+c-1/2}}{(1-t)^{1/2}}\\
&\ \ \ \ \ \ \ \ \ 
\times
\frac{d^k}{dt^k}
\left\{
\sqrt{\frac{4-xt}{xt}}
\sin\left[R\arccos\left(1-\frac{xt}{2}\right)\right]
-\frac{1}{2R}\cos\left[R\arccos\left(1-\frac{xt}{2}\right)\right]
\right\}
dt
\\
&
+
\frac{4}{\pi}\frac{1}{4R^2-1}\sum_{k=0}^nf_k
\int_0^1 t^{qR}\frac{t^{k+c+1/2}}{(1-t)^{1/2}}\\
&\ \ \ \ \ \ \ \ \ 
\times
\frac{d^{k+1}}{dt^{k+1}}
\left\{\sqrt{\frac{4-xt}{xt}}
\sin\left[R\arccos\left(1-\frac{xt}{2}\right)\right]
-\frac{1}{2R}\cos\left[R\arccos\left(1-\frac{xt}{2}\right)\right]\right\}dt.
\\
\tag7.22
\endalign
$$
Let $F(t)=\sqrt{(4-xt)/(xt)}\sin(R\arccos(1-xt/2))-1/(2R)\cos(R\arccos(1-xt/2))$.
Lemma 7.4 implies that
$$
\align
F^{(k)}(t)=
\sqrt{\frac{4-xt}{xt}}
\cos\left[R\arccos\left(1-\frac{xt}{2}\right)+\frac{(n-1)\pi}{2}\right]
\left(R\sqrt{\frac{x}{4t-xt^2}}\,\right)^k+O(R^{k-1})
,\tag7.23
\endalign
$$
where each term of the omitted linear combination is of the form 
$R^jx^s\cos(R\arccos(1-xt/2))$ or $R^jx^s\sin(R\arccos(1-xt/2))$ times a function 
of type (7.4), with $j\leq k-1$ and $s\geq -1/2$. 

Since there are only a finite number of omitted terms in (7.23) for each $k=0,\dotsc,n$, 
we obtain by
(7.22) and Corollary 7.3 that their contribution to the first sum in (7.22) is
bounded in absolute value by $L'_k{R}^{k-3/2}/\sqrt{x}$ for all $R\geq R'_k$ and
$x\in(0,1]$, where $L'_k$ and $R'_k$, $k=0,\dotsc,n$, are some constants independent of 
$x\in(0,1]$. Therefore, the overall contribution of the non-leading terms in (7.23) to the
first of the two terms on the right hand side of (7.22) is bounded in absolute value by
$L'_1{R}^{n-5/2}/\sqrt{x}$ for all $R\geq \rho'$ and
$x\in(0,1]$, where $L'_1$ and $\rho'$ are constants independent of $x\in(0,1]$.
Similarly, the overall contribution of the non-leading terms in (7.23) to the second term
on the right hand side of (7.22) is seen to be bounded in absolute value by 
$L'_2{R}^{n-5/2}/\sqrt{x}$ for all $R\geq \rho''$ and
$x\in(0,1]$, with constants $L'_2$ and $\rho''$ independent of $x\in(0,1]$. Therefore the
total contribution of the non-leading terms in (7.23) to ${T'}^{(n)}(R,qR+c;x)$ is bounded
in absolute value by $L'{R}^{n-5/2}/\sqrt{x}$ for all $R\geq \rho'_0$ and
$x\in(0,1]$, for some constants $L'$ and $\rho'_0$ independent of $x\in(0,1]$.

On the other hand, by using Corollary 7.3 as in the previous paragraph, 
one sees that the combined
contribution of the leading terms of (7.23) for $k=0,\dotsc,n-1$ to ${T'}^{(n)}(R,qR+c;x)$ 
is again bounded in absolute value by $K'{R}^{n-5/2}/\sqrt{x}$ for all $R\geq \rho'_1$ and
$x\in(0,1]$, for some constants $K'$ and $\rho'_1$ independent of $x\in(0,1]$.

By the previous two paragraphs and (7.23) we obtain from the $k=n$ terms of (7.22)
that
$$
\align
&\!\!\!\!\!\!\!\!\!\!\!\!\!\!\!\!\!\!\!\!\!\!
\left|
{T'}^{(n)}(R,qR+c;x)-
\left\{
\frac{q}{\pi R}
\int_0^1 t^{qR}\frac{t^{c-1/2}}{(1-t)^{1/2}}\sqrt{\frac{4-xt}{xt}}
\left(R\sqrt{\frac{xt}{4-xt}}\,\right)^n
\right.
\right.
\\
&\ \ \ \ \ \ \ \ \ \ \ \ \ \ \ \ \ \ \ \ \ \ \ \ \ 
\times
\cos\left[R\arccos\left(1-\frac{xt}{2}\right)+\frac{(n-1)\pi}{2}\right]dt\\
&
+\frac{1}{\pi R^2}
\int_0^1 t^{qR}\frac{t^{c-1/2}}{(1-t)^{1/2}}\sqrt{\frac{4-xt}{xt}}
\left(R\sqrt{\frac{xt}{4-xt}}\,\right)^{n+1}\\
&\ \ \ \ \ \ \ \ \ \ \ \ \ \ \ \ \ \ \ \ \ \ \ \ \ 
\left.
\left.
\times
\cos\left[R\arccos\left(1-\frac{xt}{2}\right)+\frac{n\pi}{2}\right]dt
\right\}
\right|
\leq \frac{1}{\sqrt{x}}M'_0R^{n-5/2},\tag7.24
\endalign
$$
for all $R\geq \rho'_2$ and
$x\in(0,1]$, where the constants $M'_0$ and $\rho'_2$ are independent of $x\in(0,1]$.

Denote by $I_1$ and $I_2$ the first and second integrals on the right hand side of (7.24),
respectively. They clearly satisfy the hypothesis of Proposition 7.2. 
By (7.5) and (7.6) we obtain
$$
\align
&
\left|
\frac{q}{\pi R}I_1-\Rep
\left\{
\frac{q}{\sqrt{\pi R}}\left(q^2+\frac{x}{4-x}\,\right)^{-1/4}
\left(R\sqrt{\frac{x}{4-x}}\,\right)^{n-1}
\right.
\right.
\\
&\ \ \ \ \ \ \ \ \ \ \ \ \ \ \ \ \ 
\left.
\left.
\times
\exp\left(i\left[R\arccos\left(1-\frac{x}{2}\right)-\frac12\arctan\frac1q\sqrt{\frac{x}{4-x}}
+\frac{(n-1)\pi}{2}\right]\right)dt
\right\}
\right|
\\
&\ \ \ \ \ \ \ \ \ \ \ \ \ \ \ \ \ \ \ \ \ \ \ \ \ 
\leq M'_1R^{n-5/2}x^{(n-1)/2}\tag7.25
\endalign
$$
and
$$
\align
&
\left|
\frac{1}{\pi R^2}I_2
-\Rep
\left\{
\frac{1}{\sqrt{\pi}R^{3/2}}\left(q^2+\frac{x}{4-x}\,\right)^{-1/4}
\left(R\sqrt{\frac{x}{4-x}}\,\right)^{n}
\right.
\right.
\\
&\ \ \ \ \ \ \ \ \ \ \ \ \ \ \ \ \ 
\left.
\left.
\times
\exp\left(i\left[R\arccos\left(1-\frac{x}{2}\right)-\frac12\arctan\frac1q\sqrt{\frac{x}{4-x}}
+\frac{n\pi}{2}\right]\right)dt
\right\}
\right|
\\
&\ \ \ \ \ \ \ \ \ \ \ \ \ \ \ \ \ \ \ \ \ \ \ \ \
\leq M'_2R^{n-5/2}x^{n/2},\tag7.26
\endalign
$$
for $R\geq \rho'_3$, where $\rho'_3$, $M'_1$ and $M'_2$ are constants independent of 
$x\in(0,1]$.

Using the fact that 
$$
q+i\sqrt{\frac{x}{4-x}}=\sqrt{q^2+\frac{x}{4-x}}\exp\left(i\arctan\frac1q\sqrt{\frac{x}{4-x}}
\,\right),
$$
the two second terms on the left hand sides of (7.25) and (7.26) are readily seen to add up 
precisely to the second term on the left hand side of (7.2). Since in (7.25) and (7.26) the 
integer $n$ is nonnegative and since $|x|\leq1$, inequalities (7.25), (7.26) and (7.24) imply
(7.2). \endpf

\proclaim{Lemma 7.4} Let the functions $F_1(t)$, $F_2(t)$ and $F_3(t)$ be given by
$$
\align
F_1(t)&=\cos\left[R\arccos\left(1-\frac{xt}{2}\right)\right]\\
F_2(t)&=\sin\left[R\arccos\left(1-\frac{xt}{2}\right)\right]\\
F_3(t)&=\sqrt{\frac{4-xt}{xt}}\sin\left[R\arccos\left(1-\frac{xt}{2}\right)\right],
\endalign
$$
where $x\in(0,1]$ is a parameter. Then for any $n\geq0$ one has
$$
\align
&F_1^{(n)}(t)=\left(R\sqrt{\frac{x}{4t-xt^2}}\,\right)^n
\cos\left[R\arccos\left(1-\frac{xt}{2}\right)+\frac{n\pi}{2}\right]
+\sum_{\nu\in V_1}c^{(1)}_\nu R^{j^{(1)}_\nu}x^{l^{(1)}_\nu}Q^{(1)}_\nu\tag7.27\\
&F_2^{(n)}(t)=\left(R\sqrt{\frac{x}{4t-xt^2}}\,\right)^n
\sin\left[R\arccos\left(1-\frac{xt}{2}\right)+\frac{n\pi}{2}\right]
+\sum_{\nu\in V_2}c^{(2)}_\nu R^{j^{(2)}_\nu}x^{l^{(2)}_\nu}Q^{(2)}_\nu\tag7.28\\
&F_3^{(n)}(t)=\sqrt{\frac{4-xt}{xt}}\left(R\sqrt{\frac{x}{4t-xt^2}}\,\right)^n
\cos\left[R\arccos\left(1-\frac{xt}{2}\right)+\frac{(n-1)\pi}{2}\right]\\
&\ \ \ \ \ \ \ \ \ \ \ \ \ \ \ \ \ \ \ \ 
+\sum_{\nu\in V_3}c^{(3)}_\nu R^{j^{(3)}_\nu}x^{l^{(3)}_\nu}Q^{(3)}_\nu,\tag7.29
\endalign
$$
where $l^{(1)}_\nu,l^{(2)}_\nu\geq 0$, $l^{(3)}_\nu\geq -1/2$, for all $\nu$, 
and for $k=1,2,3$ and all $\nu\in V_k$ one has

$(i)$ $V_k$ is some finite set  

$(ii)$ $c^{(k)}_\nu,j^{(k)}_\nu,l^{(k)}_\nu\in\Q$ 

$(iii)$ $j^{(k)}_\nu\leq n-1$

$(iv)$ $Q^{(k)}_\nu$ is a function of type $(7.4)$ multiplied by either 
$\cos(R\arccos(1-xt/2))$ or $\sin(R\arccos(1-xt/2))$.
\endproclaim

\pf 
It readily follows by induction on $n$ that for $n\geq1$ one has
$$
\frac{d^n}{dt^n}e^{h(t)}=e^{h(t)}\sum_{k=1}^n\sum_{i_1,\dotsc,i_k\geq1\atop i_1+\cdots+i_k=n}
\alpha_I h^{(i_1)}(t)\cdots h^{(i_k)}(t),\tag7.30
$$
with coefficients $\alpha_I=\alpha_{(i_1,\dotsc,i_k)}\in\Z$ and $\alpha_{(1,\dotsc,1)}=1$.

A similar argument shows that for $n\geq1$ one has
$$
\frac{d^n}{dt^n}[g(t)]^{-1/2}=[g(t)]^{-1/2}\sum_{k=1}^n\sum_{i_0,i_1,\dotsc,i_k\geq1}
\beta_I [g(t)]^{-i_0} g^{(i_1)}(t)\cdots g^{(i_k)}(t),\tag7.31
$$
where only finitely many of the coefficients $\beta_I=\beta_{(i_0,i_1,\dotsc,i_k)}\in\Q$
are nonzero.

Choosing $h(t)=iR\arccos(1-xt/2)$, the derivatives on the left hand side of (7.27) can 
be expressed as
$$
F_1^{(n)}(t)=\Rep\frac{d^n}{dt^n}e^{h(t)}.\tag7.32
$$
For our choice of $h$ we obtain
$$
h'(t)=iR\sqrt{\frac{x}{4t-xt^2}}=iR\sqrt{x}[g(t)]^{-1/2},\tag7.33
$$
with $g(t)=4t-xt^2$. Since $g'(t)=4-2xt$, $g''(t)=-2x$, and the higher derivatives of $g$ are
zero, (7.33) and (7.31) imply that for $j\geq1$ we have
$$
\frac{d^j}{dt^j}h'(t)=iR\sqrt{x}\sum_\nu \beta_\nu
\frac{(4-2xt)^{a_\nu}(-2x)^{s_\nu}}{(4t-xt^2)^{1/2+b_\nu}}
,\tag7.34
$$
where the sum is finite, $\beta_\nu\in\Q$ and $a_\nu,b_\nu,s_\nu\geq0$ are integers. 
From the 
expression of $h'(t)$ it is apparent that (7.34) holds also for $j=0$.

Substituting (7.34) in (7.30) we obtain for $n\geq1$ that 
$$
\align
&\frac{d^n}{dt^n}e^{h(t)}=e^{h(t)}[h'(t)]^n + 
e^{h(t)}\sum_{k=1}^{n-1}\sum_{i_1,\dotsc,i_k\geq1\atop i_1+\cdots+i_k=n}
\alpha_I h^{(i_1)}(t)\cdots h^{(i_1)}(t)\\
&
=e^{h(t)}\left(iR\sqrt{\frac{x}{4t-xt^2}}\,\right)^n
+e^{h(t)}\sum_\nu \gamma_\nu(iR\sqrt{x})^{n_\nu}
\frac{(4-2xt)^{a_\nu}x^{s_\nu}}{(4t-xt^2)^{b_\nu}}
,\tag7.35
\endalign
$$
where the sum is finite, $n-1\geq n_\nu\in\Z$, $0\leq a_\nu,s_\nu\in\Z$, 
$0\leq b_\nu\in\frac12\Z$, and $\gamma_\nu\in\Q$. Relation (7.35) clearly holds also for
$n=0$. By (7.32), taking the real and imaginary parts in (7.35) one obtains (7.27) and
(7.28), respectively.

To prove (7.29),
apply Leibniz's formula for the derivatives of the product $f_1(t)f_2(t)$ 
defining $F_3(t)$, with $f_1(t)=\sqrt{(4-xt)/(xt)}$, $f_2(t)=\sin(R(\arccos(1-xt/2))$.
It is clear by (7.28) that the highest power of $R$ in the resulting terms is $n$, and it
occurs in only one of these terms, namely in $f_1(t)$ times the highest order term in
$f_2^{(n)}(t)$. By (7.28) it also follows that the leading term
in $R$ of $F_3^{(n)}(t)$ is the first term on the right hand side of (7.29).

Furthermore, (7.28) implies that the successive derivatives of 
the second factor $f_2(t)$ have the form of the summand on the right hand side of (7.29). 
Therefore, to finish the proof it is enough to show that the derivatives of $f_1(t)$ are 
also of this form.

We have
$$
\frac{d^n}{dt^n}\sqrt{\frac{4-xt}{xt}}=x^{-1/2}\frac{d^n}{dt^n}\sqrt{\frac4t-x}.\tag7.36
$$
One easily obtains by induction on $n$ that for $n\geq1$
$$
\frac{d^n}{dt^n}[g(t)]^{1/2}=[g(t)]^{1/2}\sum_{k=1}^n\sum_{i_0,i_1,\dotsc,i_k\geq1}
\delta_I [g(t)]^{-i_0} g^{(i_1)}(t)\cdots g^{(i_k)}(t),\tag7.37
$$
where the sum is finite and the coefficients $\delta_I=\beta_{(i_0,i_1,\dotsc,i_k)}\in\Q$.

Choose $g(t)=\frac4t-x$. Since $g^{(j)}(t)=4(-1)^{j}j!\,t^{-j-1}$, $j\geq1$, it follows by
(7.37) and (7.36) that for $n\geq1$
$$
\frac{d^n}{dt^n}\sqrt{\frac{4-xt}{xt}}=x^{-1/2}\sqrt{\frac4t-x}
\sum_{k=1}^n\sum_{\nu}
\delta_\nu \left(\frac4t-x\right)^{-s_\nu} t^{l_\nu},\tag7.38
$$
where $0\leq s_\nu,l_\nu\in\Z$. Since the terms in (7.38) have the form of the summand
on the right hand side of (7.29), and since this is also true of the left hand side of (7.38)
for $n=0$, the proof of (7.29) is complete. \epf

\bigskip
\centerline{\bf 8. Replacement of the $T^{(k)}$'s and ${T'}^{(k)}$'s by their 
asymptotics}
\centerline{\bf does not affect the asymptotics of the moments $M$}

\bigskip
Denote the approximants of the $T^{(k)}$'s and ${T'}^{(k)}$'s in Proposition 7.1 by
$$
\align
&F_k(R,q;x)=\frac{2}{\sqrt{\pi}}\frac{1}{\left(q^2+\frac{x}{4-x}\right)^{1/4}}
\frac{1}{R^{3/2}}\left(R\sqrt{\frac{x}{4-x}}\,\right)^k\\
&\ \ \ \ \ \ \ \ \ \ \ \ \ \ \ \ \ \
\times
\cos\left[R\arccos\left(1-\frac x2\right)-\frac12\arctan\frac1q\sqrt{\frac{x}{4-x}}+\frac{k\pi}{2}
\right]\tag8.1
\endalign
$$
and
$$
\align
&
F'_k(R,q;x)=\frac{1}{\sqrt{\pi}}\frac{\left(q^2+\frac{x}{4-x}\right)^{1/4}}
{\sqrt{\frac{x}{4-x}}}\frac{1}{R^{3/2}}\left(R\sqrt{\frac{x}{4-x}}\,\right)^k\\
&\ \ \ \ \ \ \ \ \ \ \ 
\times
\cos\left[R\arccos\left(1-\frac x2\right)+\frac12\arctan\frac1q\sqrt{\frac{x}{4-x}}
+\frac{(k-1)\pi}{2}
\right].\tag8.2
\endalign
$$
\proclaim{Proposition 8.1} Consider the moments 
$M_{\alpha_1,\beta_1,\dotsc,\alpha_m,\beta_m;\gamma_1,\delta_1,\dotsc,\gamma_n,\delta_n}$
defined by $(6.6)$. As the variables $R_i$, $v_i$, $R'_j$, $v'_j$, $i=1,\dotsc,m$, 
$j=1,\dotsc,n$ approach infinity as specified by $(\TwoTwo)$, we have 
$$
\align
&\!\!\!\!\!\!\!\!\!\!\!\
M_{\alpha_1,\beta_1,\dotsc,\alpha_m,\beta_m;\gamma_1,\delta_1,\dotsc,\gamma_n,\delta_n}=
\int_0^1\cdots\int_0^1 
\prod_{i=1}^m\prod_{j=1}^n(x_{ij}y_{ij}z_{ij}w_{ij})^{q_iR_i+q'_jR'_j+c_i+c'_j+1}\\
&
\times
F_{\alpha_1}(R_1,q_1;\prod_{j=1}^n x_{1j}y_{1j})\,
F_{\beta_1}(R_1,q_1;\prod_{j=1}^n z_{1j}w_{1j})
\cdots\\
&\ \ \ \ \ \ \ \ \ \ \ \ 
\cdots
F_{\alpha_m}(R_m,q_m;\prod_{j=1}^n x_{mj}y_{mj})\,
F_{\beta_m}(R_m,q_m;\prod_{j=1}^n z_{mj}w_{mj})\\
&
\times
F'_{\gamma_1}(R'_1,q'_1;\prod_{i=1}^m x_{i1}z_{i1})\,
F'_{\delta_1}(R'_1,q'_1;\prod_{i=1}^m y_{i1}w_{i1})
\cdots\\
&\ \ \ \ \ \ \ \ \ \ \ \ 
\cdots
F'_{\gamma_n}(R'_n,q'_n;\prod_{i=1}^m x_{in}z_{in})\,
F'_{\delta_n}(R'_n,q'_n;\prod_{i=1}^m y_{in}w_{in})\,dx_{11}\cdots dw_{mn}\\
&
+O\left(R^{\sum_{i=1}^m(\alpha_i+\beta_i)+\sum_{i=1}^n(\gamma_i+\delta_i)-3m-3n-1}\right),
\tag8.3
\endalign
$$
where the integration variables are $x_{ij}$, $y_{ij}$, $z_{ij}$, $w_{ij}$, 
$i=1,\dotsc,m$, $j=1,\dotsc,n$.
\endproclaim

\pf By Proposition 7.1, for any fixed $k\geq0$ we have
$$
\align
|T^{(k)}(R,qR+c;x)-F_k(R,q;x)|&\leq MR^{k-5/2}\tag8.4\\
|{T'}^{(k)}(R,qR+c;x)-F'_k(R,q;x)|&\leq x^{-1/2}M'R^{k-5/2}\tag8.5
\endalign
$$
for $R\geq R_0$, where the constants $R_0$, $M$ and $M'$ are independent of $x\in(0,1]$.

From (8.1) and (8.2) it is clearly seen that there exist constants $M_1$ and $M'_1$ so that
$$
\align
|F_k(R,q;x)|&\leq M_1R^{k-3/2}\tag8.6\\
|F'_k(R,q;x)|&\leq x^{-1/2}M'_1R^{k-3/2},\tag8.7
\endalign
$$
for all $k\geq0$, $R\geq0$ and $x\in(0,1]$.

By (8.4)--(8.7) it follows that for any fixed $k\geq0$
$$
\align
|T^{(k)}(R,qR+c;x)|&\leq M_2R^{k-3/2}\tag8.8\\
|{T'}^{(k)}(R,qR+c;x)|&\leq x^{-1/2}M'_2R^{k-3/2},\tag8.9
\endalign
$$
for $R\geq R'_0$, for some constants $R'_0$, $M_2$ and $M'_2$ that are independent of 
$x\in(0,1]$.

By (6.6) we have
$$
\align
&
\left|
M_{\alpha_1,\beta_1,\dotsc,\alpha_m,\beta_m;\gamma_1,\delta_1,\dotsc,\gamma_n,\delta_n}-
\int_0^1\cdots\int_0^1 
\prod_{i=1}^m\prod_{j=1}^n(x_{ij}y_{ij}z_{ij}w_{ij})^{q_iR_i+q'_jR'j+c_i+c'_j+1}
\right.
\\
&\ \ \ \ \ \ \ \ \ \ \ \ 
\times
F_{\alpha_1}(R_1,q_1;\prod_{j=1}^n x_{1j}y_{1j})\,
F_{\beta_1}(R_1,q_1;\prod_{j=1}^n z_{1j}w_{1j})
\cdots\\
&\ \ \ \ \ \ \ \ \ \ \ \ \ \ \ \ \ \ \ \ \ \ \ \ 
\cdots
F_{\alpha_m}(R_m,q_m;\prod_{j=1}^n x_{mj}y_{mj})\,
F_{\beta_m}(R_m,q_m;\prod_{j=1}^n z_{mj}w_{mj})\\
&\ \ \ \ \ \ \ \ \ \ \ \ 
\times
F'_{\gamma_1}(R'_1,q'_1;\prod_{i=1}^m x_{i1}z_{i1})\,
F'_{\delta_1}(R'_1,q'_1;\prod_{i=1}^m y_{i1}w_{i1})
\cdots\\
&\ \ \ \ \ \ \ \ \ \ \ \ \ \ \ \ \ \ \ \ \ \ \ \ 
\left.
\cdots
F'_{\gamma_n}(R'_n,q'_n;\prod_{i=1}^m x_{in}z_{in})\,
F'_{\delta_n}(R'_n,q'_n;\prod_{i=1}^m y_{in}w_{in})\,dx_{11}\cdots dw_{mn}
\right|
\endalign
$$
\vskip-0.2in
$$
\align
&=
\left|
\int_0^1\cdots\int_0^1 
\prod_{i=1}^m\prod_{j=1}^n(x_{ij}y_{ij}z_{ij}w_{ij})^{q_iR_i+q'jR'_j+c_i+c'_j+1}
\right.
\\
&\ \ \ \ \ \ \ \ \ \ \ \ 
\times
\{
T^{(\alpha_1)}(R_1,v_1;\prod_{j=1}^n x_{1j}y_{1j})\,
T^{(\beta_1)}(R_1,v_1;\prod_{j=1}^n z_{1j}w_{1j})
\cdots\\
&\ \ \ \ \ \ \ \ \ \ \ \ \ \ \ \ \ \ \ \ \ \ \ \ 
\cdots
T^{(\alpha_m)}(R_m,v_m;\prod_{j=1}^n x_{mj}y_{mj})\,
T^{(\beta_m)}(R_m,v_m;\prod_{j=1}^n z_{mj}w_{mj})\\
&\ \ \ \ \ \ \ \ \ \ \ \ 
\times
{T'}^{(\gamma_1)}(R'_1,v'_1;\prod_{i=1}^m x_{i1}z_{i1})\,
{T'}^{(\delta_1)}(R'_1,v'_1;\prod_{i=1}^m y_{i1}w_{i1})
\cdots\\
&\ \ \ \ \ \ \ \ \ \ \ \ \ \ \ \ \ \ \ \ \ \ \ \ 
\cdots
{T'}^{(\gamma_n)}(R'_n,v'_n;\prod_{i=1}^m x_{in}z_{in})\,
{T'}^{(\delta_n)}(R'_n,v'_n;\prod_{i=1}^m y_{in}w_{in})
\endalign
$$
\vskip-0.3in
$$
\align
&\ \ \ \ \ \ \ \ \ \ \ 
-
F_{\alpha_1}(R_1,q_1;\prod_{j=1}^n x_{1j}y_{1j})\,
F_{\beta_1}(R_1,q_1;\prod_{j=1}^n z_{1j}w_{1j})
\cdots\\
&\ \ \ \ \ \ \ \ \ \ \ \ \ \ \ \ \ \ \ \ \ \ \ \ 
\cdots
F_{\alpha_m}(R_m,q_m;\prod_{j=1}^n x_{mj}y_{mj})\,
F_{\beta_m}(R_m,q_m;\prod_{j=1}^n z_{mj}w_{mj})
\endalign
$$
$$
\align
&\ \ \ \ \ \ \ \ \ \ \ \ 
\times
F'_{\gamma_1}(R'_1,q'_1;\prod_{i=1}^m x_{i1}z_{i1})\,
F'_{\delta_1}(R'_1,q'_1;\prod_{i=1}^m y_{i1}w_{i1})
\cdots\\
&\ \ \ \ \ \ \ \ \ \ \ \ \ \ \ \ \ \ \ \ \ \ \ \ 
\left.
\cdots
F'_{\gamma_n}(R'_n,q'_n;\prod_{i=1}^m x_{in}z_{in})\,
F'_{\delta_n}(R'_n,q'_n;\prod_{i=1}^m y_{in}w_{in})
\}
\,dx_{11}\cdots dw_{mn}
\right|.
\\
\tag8.10
\endalign
$$
Clearly, for any $2l$ quantities $f_i$, $g_i$, $i=1,\dotsc,l$ one has
$$
\align
&
f_1\cdots f_l - g_1\cdots g_l=f_1\cdots f_{l-1}(f_l-g_l)
+f_1\cdots f_{l-2}(f_{l-1}-g_{l-1})g_l\\
&\ \ \ \ \ \ \ \ \ \ \ \ \ \ \ \ \ \ \ \ \ \ \ \ 
+f_1\cdots f_{l-3}(f_{l-2}-g_{l-2})g_{l-1}g_l+\cdots+(f_1-g_1)g_2\cdots g_l.\tag8.11
\endalign
$$
Apply (8.11) to the expression $E$ in the curly braces on the right hand side of (8.10): we
have $l=2m+2n$, the $f_i$'s become $T^{(k)}$'s or ${T'}^{(k)}$'s, and the $g_i$'s 
become $F_k$'s or $F'_k$'s. 

This results in expressing $E$ as a sum of $2m+2n$ products of the form 
$h_1\cdots h_{2m+2n}$, where exactly one of the $h_i$'s, say $h_{i_0}$, 
is a difference $T^{(k)}-F_k$
or ${T'}^{(k)}-F'_k$, all others being of the form $T^{(k)}$, ${T'}^{(k)}$, $F_k$ or 
$F'_k$. 

Furthermore, exactly $2n$ of the $h_i$'s are of the form ${T'}^{(k)}$, $F'_k$, or
${T'}^{(k)}-F'_k$, and the set of their $x$-arguments is always
$$
\left\{\prod_{i=1}^m x_{i1}z_{i1},\prod_{i=1}^m y_{i1}w_{i1},\,\dotsc,\,
\prod_{i=1}^m x_{in}z_{in},\prod_{i=1}^m y_{in}w_{in}\right\}.\tag8.12
$$
Since the $2n$ subsets of variables that are multiplied together in the elements of 
(8.12) form a partition
of the set of our $4mn$ variables $x_{ij}$, $y_{ij}$, $z_{ij}$, $w_{ij}$, 
$i=1,\dotsc,m$, $j=1,\dotsc,n$, inequalities (8.4)--(8.9), and the fact that $|h_{i_0}|$
can be bounded using the sharper (8.4) or (8.5), imply
$$
|h_1\cdots h_{2m+2n}|\leq 
\bar{M}^{2m+2n}R^{\sum_{i=1}^m(\alpha_i+\beta_i)+\sum_{i=1}^n(\gamma_i+\delta_i)-3m-3n-1}
\prod_{i=1}^m\prod_{j=1}^n(x_{ij}y_{ij}z_{ij}w_{ij})^{-1/2},\tag8.13
$$
for all $R\geq R''_0$, where the constants $R''_0$ and $\bar{M}$ are independent of 
$x\in(0,1]$.
Taking into account also the contribution of the double product in (8.10) to the integrand 
of (8.10), by (8.13) the absolute value of the latter is majorized by
$$
\align
&(2m+2n)\bar{M}^{2m+2n}R^{\sum_{i=1}^m(\alpha_i+\beta_i)+\sum_{i=1}^n(\gamma_i+\delta_i)-3m-3n-1}\\
&\ \ \ \ \ \ \ \ \ \ \ 
\times
\prod_{i=1}^m\prod_{j=1}^n(x_{ij}y_{ij}z_{ij}w_{ij})^{q_iR_i+q'_jR'_j+c_i+c'_j+1/2}.
\endalign
$$
Since by (\TwoTwo) the exponent of $x_{ij}y_{ij}z_{ij}w_{ij}$ is non-negative, this shows
that (8.3) holds with implicit constant $(2m+2n)\bar{M}$.~\epf

\mysec{9. Proof of Proposition 7.2}

For fixed $x\in[0,1]$, the asymptotics of the integral in (7.5) can be readily obtained by
Laplace's method for contour integrals, as it is presented for instance in \cite{\Ol,Ch.4,\,\S6.1}.
However, what (7.5) states is the existence of a {\it uniform error bound} for the Laplace 
approximation, for $x\in[0,1]$. We obtain this by extending the arguments in the proof of 
\cite{\Ol, Theorem 6.1,\,p.\,125} to the case when the integrand depends on a parameter.

For $t\in(0,1)$, write the factor $(1-t)^{1/2}$ in the denominator of $Q(t)$ in (7.4) as
$(1-t)^{1/2}=-i(t-1)^{1/2}$, where the square root on the right hand side has its principal
determination. The integral in (7.5) becomes
$$
I(R)=-i\int_1^0e^{-Rp(t)}\tilde{Q}(t)dt=:-i\tilde{I}(R),\tag9.1
$$
where $p(t)$ is given by (7.3) and
$$
\tilde{Q}(t)=\frac{t^l}{(t-1)^{1/2}}\frac{(4-2xt)^a}{(4-xt)^b},\ \ \ t\in(0,1),\tag9.2
$$
where as in (7.4) $0\leq a\in\Z$, $-1/2\leq b\in\frac12\Z$, $l\in\frac12\Z$, and
the square root on the right hand side of (9.2) has the principal determination.

Regard the integral $\tilde{I}(R)$ defined by (9.1) as a contour integral over the path
$\Cal P=[1,0]$ (a line segment) in the complex plane. 
Choose the principal determinations for all the 
multiple-valued maps in the expressions (7.3) and (9.2) defining $p(t)$ and $\tilde{Q}(P)$.
Note that for any fixed $x\in[0,1]$ the integral $\tilde{I}(R)$ satisfies the following properties:

\smallpagebreak
$(i)$ $p(t)$ and $\tilde{Q}(t)$ are independent of $R$, single-valued and holomorphic
in the pointed open disk $\dot{D}(1,1)=D(1,1)\setminus\{1\}$.

$(ii)$ $\Cal P$ is independent of $R$, and $\Cal P_{(1,0)}$ (i.e., the path $\Cal P$ less its endpoints)
is contained in $\dot{D}(1,1)$.

$(iii)$ For $t\in D(1,1)$, the functions $p(t)$ and $\tilde{Q}(t)$ can be expanded in 
convergent series as
$$
p(t)=p(1)+\sum_{s=0}^\infty p_s(t-1)^{s+1},
$$
where $p(1)=-i\arccos(1-x/2)$ and $p_0=-q-i\sqrt{x/(4-x)}$, and
$$
\tilde{Q}(t)=\sum_{s=0}^\infty q_s(t-1)^{s-1/2},
$$
where $q_0=(4-2x)^a/(4-x)^b$ and $(t-1)^{1/2}$ has its principal determination.

$(iv)$ $\tilde{I}(R)$ converges at 0 absolutely and uniformly with respect to $R\geq -l/q$
and $x\in[0,1]$.

$(v)$ $\Rep\{p(t)-p(1)\}$ is positive when $t\in(0,1)$, and is bounded away from 0
uniformly with respect to $x\in[0,1]$ as $t\to0$ along $\Cal P$.

\smallpagebreak
To avoid interruption in proving Proposition 7.2, we phrase three facts we need in the proof
as Lemmas 9.1--9.3, and include them at the end of this section.

Consider the map $t\mapsto v(t)$, $|t-1|<1$, given by
$$
v(t)=p(t)-p(1)=-q\ln t-i\arccos\left(1-\frac{xt}{2}\right)+i\arccos\left(1-\frac{x}{2}\right).
\tag9.3
$$

Let $U_x$ be a neighborhood of $t=1$ and $D$ a disk centered at $v=0$ satisfying the statement
of Lemma 9.1 (in particular, $D$ is independent of $x\in[0,1]$).
By Lemma 9.1(a), 
(9.3) maps 
$U_x$ conformally
onto 
$D$. Thus the inverse function 
$v\mapsto t(v)$ is also holomorphic, and hence $t-1$ can be expanded in a convergent series 
$$
t-1=\sum_{s=1}^\infty c_sv^s,\ \ \ v\in D,\tag9.4
$$
where the coefficients $c_s$ are expressible in terms of the $p_s$; in particular, $c_1=1/p_0$.

By Lemma 9.2, one can choose $k\in[0,1)\cap U$, with $U$ as in Lemma 9.1(b),
$k$ independent of $R$ and independent of 
$x\in[0,1]$, such that the disk $|v|\leq|p(k)-p(1)|$ is contained in $D$ for all $x\in[0,1]$.
Then the portion $[1,k]$ of $\Cal P$ may be deformed to make its $v$-map a straight line 
$[0,\Cal K]$, for all $x\in[0,1]$, without changing the value of the integral $\tilde{I}(R)$
(by (i) and (ii)). Making the change of variable $v=p(t)-p(1)$ on this deformed portion 
$[1,k]$ of $\Cal P$ we obtain
$$
\int_1^k e^{-Rp(t)}\tilde{Q}(t)dt=e^{-Rp(1)}\int_0^\Cal K e^{-Rv}f(v)dv,\tag9.5
$$
where
$$
\Cal K=p(k)-p(1),\ \ \ \ f(v)=\tilde{Q}(t)\frac{dt}{dv}=\frac{\tilde{Q}(t)}{p'(t)},\tag9.6
$$
and the path of integration on the right hand side of (9.5) is a straight line.

By (iii) and (9.4), for $v\in D$, $f(v)$ has a convergent expansion of the form
$$
f(v)=\frac{q_0\left(\sum_{s=1}^\infty c_sv^s\right)^{-1/2}
+q_1\left(\sum_{s=1}^\infty c_sv^s\right)^{1/2}+\cdots}
{p_0+2p_1\left(\sum_{s=1}^\infty c_sv^s\right)+3p_2\left(\sum_{s=1}^\infty c_sv^s\right)^{2}},
\tag9.7
$$
the branch of the square root being the principal one. By the binomial theorem one obtains
from (9.7) that $f(v)$ can be expressed as a convergent series
$$
f(v)=a_0v^{-1/2}+a_1v^{1/2}+a_2v^{3/2}+\cdots,
$$
where $v^{1/2}$ has its principal value and the coefficients $a_s$ can be expressed in terms
of the $p_s$'s and $q_s$'s as in \cite{\Ol, Ch.3,\,\S8.1}; in particular, 
$$
a_0=\frac{q_0}{p_0^{1/2}},\tag9.8
$$
the square root having its principal determination.

Define $f_1(v)$ by the relations $f_1(0)=a_1$ and
$$
f(v)=a_0v^{-1/2}+v^{1/2}f_1(v)\ \ \ (v\neq0).\tag9.9
$$

By Lemma 9.3, we may assume that
$$
|f_1(v)|\leq M_2,\ \ \ |v|\leq\Cal K,\ x\in[0,1],\tag9.10
$$
where the constant $M_2$ is independent of $x\in[0,1]$. 
(Indeed, by Lemma 9.2, $k$ can be chosen close enough to 1 so that in addition
$|\Cal K|=|p(k)-p(1)|<\rho$, $x\in[0,1]$, for the $\rho$ of Lemma 9.3.)

Using (9.9), rearrange the integral on the right hand side of (9.5) as
$$
\align
\int_0^\Cal K e^{-Rv}f(v)dv&=\frac{a_0}{R^{1/2}}\int_0^\infty e^{-y}y^{-1/2}dy
-\frac{a_0}{R^{1/2}}\int_{\Cal K R}^\infty e^{-y}y^{-1/2}dy
+\int_0^\Cal K e^{-Rv}v^{1/2}f_1(v)dv\\
&
=a_0R^{-1/2}\Gamma(1/2)-\epsilon_1(R)+\epsilon_2(R),\tag9.11
\endalign
$$
where 
$$
\align
\epsilon_1(R)&=a_0R^{-1/2}\Gamma(1/2,\Cal K R)\tag9.12\\
\epsilon_2(R)&=\int_0^\Cal K e^{-Rv}v^{1/2}f_1(v)dv.\tag9.13
\endalign
$$
Note that in (9.11) $y^{1/2}$ has its principal value (since $y=Rv$, and $v^{1/2}$ does so)
and $\Cal K R$ is not on the negative half-axis (since $\Rep \Cal K>0$ by (v)),
so the incomplete Gamma function in (9.12) also takes its principal value.

The absolute value of $\epsilon_1(R)$ can be bounded as follows. By 
\cite{\Ol, Ch.4,\,(2.02),\,(2.04)}, for real $\alpha$ the incomplete Gamma function 
$\Gamma(\alpha,z)$ satisfies
$$
\align
\Gamma(\alpha,z)&=e^{-z}z^{\alpha-1}\{1+\epsilon(z)\}\tag9.14\\
|\epsilon(z)|&\leq\frac{|\alpha-1|}{|z|\cos(\theta-\beta)-\sigma(\beta)},\tag9.15
\endalign
$$
where $\theta=\ph z$ (i.e., $z=re^{i\theta}$ for some $r\geq0$), $\beta\in(-\pi,\pi)$ is arbitrary, 
$$
\sigma(\beta)=\sup_{\ph t=-\beta}\frac{\alpha-2}{|t|}\ln|1+t|\tag9.16
$$
and $z$ is restricted by $|\theta-\beta|<\pi/2$, $|z|\cos(\theta-\beta)>\sigma(\beta)$.

Apply (9.14) and (9.15) for our case, $\alpha=1/2$, $z=\Cal K R$.
Choose $\beta=0$. We have $\theta=\ph \Cal K R=\ph \Cal K$, so the first
condition on $z$, $|\theta-\beta|<\pi/2$, is met by property (v). The second condition
on $z$ is also met, because by (9.16)
$$
\sigma(0)=\sup_{t>0}-\frac{3}{2t}\ln(1+t)\leq0.
$$
By (9.14) and (9.15) we obtain
$$
\Gamma(1/2,\Cal K R)=e^{-\Cal K R}(\Cal K R)^{-1/2}\{1+\epsilon\},
$$
with
$$
|\epsilon|\leq\frac{1/2}{R|\Cal K|\cos(\ph \Cal K)}=\frac{1}{2R(-q\ln k)},
$$
for all $R\geq0$. Since $\Rep\Cal K=-q\ln k>0$ is independent of $x$,
by the last two relations it follows that 
$$
|\Gamma(1/2,\Cal K R)|\leq M_1R^{-1},\ \ \ R\geq r_1,\ x\in[0,1],\tag9.17
$$
for some constants $M_1$ and $r_1$ independent of $x$.
On the other hand, by (9.8) and (iii) 
$$
a_0=\frac{(4-2x)^a}{(4-x)^b}\left(-q-i\sqrt{\frac{x}{4-x}}\,\right)^{-1/2},
$$
so $|a_0|$ can clearly be bounded above uniformly for $x\in[0,1]$. By (9.17) and (9.12) it
follows that
$$
|\epsilon_1(R)|\leq M'_1R^{-3/2},\ \ \ R\geq r_1,\tag9.18
$$
for all $x\in[0,1]$, where the constants $M'_1$ and $r_1$ are independent of $x$.

Next, we bound the absolute value of $\epsilon_2(R)$. The substitution $v=\Cal K\tau$
in (9.13) implies
$$
\epsilon_2(R)=\Cal K^{3/2}\int_0^1e^{-R\Cal K\tau}\tau^{1/2}f_1(\Cal K\tau)d\tau.\tag 9.19
$$
We have
$$
\Rep\{-R\Cal K\tau\}=-R\tau\Rep\{p(k)-p(1)\}=-R\tau\eta_k,
$$
where $\eta_k=-q\ln k>0$ is independent of $x\in[0,1]$. Using this and (9.10) we deduce
$$
|\epsilon_2(R)|\leq M_2|\Cal K|^{3/2}\int_0^\infty e^{-R\eta_k\tau}\tau^{1/2}d\tau
=M_2|\Cal K|^{3/2}(R\eta_k)^{-3/2}\Gamma(3/2).\tag9.20
$$
Since
$$
|\Cal K|=|p(k)-p(1)|\leq -q\ln k+|\arccos(1-xk/2)-\arccos(1-x/2)|
$$
and the two functions inside the absolute value sign are continuous in $x\in[0,1]$, $\Cal K$
can be bounded uniformly and (9.20) implies
$$
|\epsilon_2(R)|\leq M'_2R^{-3/2},\ \ \ R\geq0,\tag9.21
$$
for all $x\in[0,1]$, where $M'_2$ is independent of $x$.

Substituting (9.18) and (9.21) in (9.11) and using $|e^{-Rp(1)}|=1$ for all $R\in\R$ and $x\in[0,1]$, 
we obtain by (9.5)
$$
\left|\int_1^ke^{-Rp(t)}\tilde{Q}(t)dt-\sqrt{\pi}a_0R^{-1/2}e^{-Rp(1)}\right|\leq M_3R^{-3/2},
\ \ \ R\geq r_1,\tag9.22
$$
for all $x\in[0,1]$, where $M_3$ and $r_1$ are constants independent of $x$.

We now turn to bounding the absolute value of the tail of $\tilde{I}(R)$ omitted by the 
integral on the left hand sides of (9.5) and (9.22). We have
$$
\Rep\{p(t)-p(1)\}=-q\ln t\geq\eta>0,\ \ \ t\in[k,0),
$$
for all $x\in[0,1]$, where $\eta$ is independent of $x$. Therefore for $R\geq r_0$ one has
$$
\align
\Rep\{Rp(t)-Rp(1)\}&=\{(R-r_0)+r_0\}\Rep\{p(t)-p(1)\}\\
&\geq(R-r_0)\eta+\Rep\{r_0p(t)-r_0p(1)\}.
\endalign
$$
We obtain
$$
\align
\left|\int_k^0e^{-Rp(t)}\tilde{Q}(t)dt\right|&\leq |e^{-Rp(1)}|
\int_k^0e^{-\Rep\{Rp(t)-Rp(1)\}}
|\tilde{Q}(t)|dt\\
&\leq e^{(r_0-R)\eta}\int_k^0|e^{-r_0p(t)}||\tilde{Q}(t)|dt
\endalign
$$
(we also used $|e^{-Rp(1)}|=1$ and $|e^{r_0p(1)}|=1$). Choosing $r_0=-l/q$, since 
$|e^{-r_0p(t)}|=t^{qr_0}$, the last integral is uniformly bounded for $x\in[0,1]$ 
(see property (iv)). We deduce
$$
\left|\int_k^0e^{-Rp(t)}\tilde{Q}(t)dt\right|\leq M_4R^{-3/2},\ \ \ R\geq r_2\tag9.23
$$
for all $x\in[0,1]$, where the constants $M_4$ and $r_2$ do not depend on $x$. 

By (9.22) and (9.23) it follows that
$$
\left|\tilde{I}(R)-\sqrt{\pi}a_0R^{-1/2}e^{-Rp(1)}\right|\leq MR^{-3/2},\ \ \ R\geq R_0,
$$
uniformly for $x\in[0,1]$. This can be rewritten by (9.1) as
$$
\left|I(R)-(-i)\sqrt{\pi}a_0R^{-1/2}e^{-Rp(1)}\right|\leq MR^{-3/2},\ \ \ R\geq R_0,\tag9.24
$$
for all $x\in[0,1]$, where the constants $M$ and $R_0$ do not depend on $x$.

However, by (9.8) and (iii)
$$
a_0=\frac{(4-2x)^a}{(4-x)^b}\left(-q-i\sqrt{\frac{x}{4-x}}\,\right)^{-1/2},
$$
where the phase of the quantity in the large parentheses has its principal value
$-\pi+\arctan 1/q\sqrt{x/(4-x)}$. This implies
$$
a_0=\frac{(4-2x)^a}{(4-x)^b}\left(-q-i\sqrt{\frac{x}{4-x}}\,\right)^{-1/4}
e^{i\left[\frac{\pi}{2}-\frac12\arctan\frac1q\sqrt{\frac{x}{4-x}}\,\right]}.
$$
This shows that the approximant of $I(R)$ in (9.24) has precisely the expression (7.6), and
the proof of Proposition 7.2 is complete.

\proclaim{Lemma 9.1} Let $t\mapsto v_x(t)=v(t)$ be the map defined by $(9.3)$, all multivalued maps 
in $(9.3)$ taking their principal values. Let $x\in[0,1]$ be arbitrary. 

$(${\rm a}$)$. There exists a disk
$D$ centered at $0$ in the $v$-plane, independent of $x\in[0,1]$, and a neighborhood $U_x$ 
containing~$1$ in the $t$-plane, so that $t\mapsto v_x(t)$ maps $U_x$ univalently onto $D(0,\rho)$,
and $v'_x(t)\neq0$ on $U_x$. 

$(${\rm b}$)$. $D$ and $U_x$ can be chosen in part $(${\rm a}$)$ so that there exists a disk 
$U$ centered at $1$ in the $t$-plane, independent of $x$, with $U\subset U_x$, for all $x\in[0,1]$.
\endproclaim

\pf We have
$$
\frac{d}{dt}v_x(t)=-\frac{q}{t}-i\sqrt{\frac{x}{4t-xt^2}}\neq0,\ \ \ x\in[0,1],\ |t-1|<1,\tag9.25
$$
so the last condition in the statement of part (a) is met.
To complete the proof of part (a) it suffices to find $\delta,\rho>0$ independent of 
$x\in[0,1]$ so that\footnote{The fact that in (9.26) $\delta$ is independent of $x$ is not 
necessary here, but will be needed in the proof of Lemma 9.3.}
$$
\text{\rm $v_x(t)-v_0=0$ has a unique solution in the disk $|t-1|<\delta$, 
for any $v_0\in D=D(0,\rho)$.}
\tag9.26
$$
Suppose $\delta>0$ is independent of $x\in[0,1]$ and satisfies
$$
\text{\rm $v_x(t)=0$ has exactly one root in $|t-1|\leq\delta$},\ 
\text{\rm for all $x\in[0,1]$ (namely $t=1$)}.\tag9.27
$$
Since the map $(x,t)\mapsto|v_x(t)|$ is continuous and non-zero on the compact set 
$[0,1]\times\{|t-1|=\delta\}$, there exists $m>0$ independent of $x\in[0,1]$ so that
$$
|v_x(t)|\geq m>0,\ |t-1|=\delta,\ x\in[0,1].
$$
Choose $\rho$ so that $0<\rho<m$. Then for any $v_0\in D(0,\rho)$ we have 
$|v_0|<\rho<m\leq|v_x(t)|$, for $|t-1|=\delta$, and Rouch\'e's theorem (see e.g. \cite{\A})
implies that $v_x(t)$ and $v_x(t)-v_0$ have the same number of roots inside $|t-1|=\delta$. By
(9.27) we obtain that the $\delta$ of (9.27) and our choice of $\rho$ satisfy (9.26).

To finish the proof of part (a) we need to prove the existence of some $\delta$ satisfying (9.27). 
%

Let $x_0\in[0,1]$ be fixed. Since $v_{x_0}(1)=0$, and by (9.25) $v'_{x_0}(1)\neq0$, there
exists $\delta_{x_0}>0$ so that $v_{x_0}(t)$ has a unique root in $|t-1|\leq\delta_{x_0}$. 
Set $b_{x_0}=\min_{|t-1|=\delta_{x_0}}|v_{x_0}(t)|>0$. Write
$$
v_x(t)=v_{x_0}(t)+(v_x(t)-v_{x_0}(t))
.\tag9.28
$$
We have
$$
v_x(t)-v_{x_0}(t)=-i[\arccos(1-xt/2)-\arccos(1-x_0t/2)]
+i[\arccos(1-x/2)-\arccos(1-x_0/2)].\tag9.29
$$
Let $f(u)=\arccos(1-u/2)$.

Consider first the case $x_0\neq0$. Since $f'(u)=(4u-u^2)^{-1/2}$, its absolute value is
bounded as long as $u$ is bounded away from 0. By the mean value theorem, it follows from (9.29) 
that there exists
an open interval $U_{x_0}$ containing $x_0$ so that $|v_x(t)-v_{x_0}(t)|<b_{x_0}$ for 
$x\in U_{x_0}$ and $|t-1|=\delta_{x_0}$. Rouch\'e's theorem applied
to (9.28) shows then that $v_x(t)$ has a unique root in $|t-1|\leq\delta_{x_0}$ (namely, $t=1$) 
for all $x\in U_{x_0}$.

For $x_0=0$ we have
$$
v_x(t)-v_{x_0}(t)=-i[\arccos(1-xt/2)-\arccos(1-x/2)],
$$
and by the continuity of $f(u)$ at $u=0$ one obtains again that Rouch\'e's theorem is applicable, and
there exists a neighborhood 
$U_0$ of $x_0=0$ in [0,1] so that $v_x(t)$ has a unique root in $|t-1|\leq\delta_{x_0}=
\delta_{0}$ for all $x\in U_{0}$, namely the root $t=1$.

Since $[0,1]$ is a compact set, it is covered by a finite subcollection, say 
$U_{x_1},\dotsc,U_{x_n}$, of $(U_x)_{x\in[0,1]}$. Then $\delta=\min_{i=1}^n \delta_{x_i}$
satisfies (9.27). 

To prove part (b), consider the inverse function $v\mapsto t_x(v)$, $v\in D$. 
Let $\delta$ and $\rho$ be as in part (a). Since the original domain of (9.3) is the disk
$|t-1|<1$, we clearly have $\delta<1$. By (9.25),
$$
\frac{dt_x}{dv}(v_x(t))=\frac{1}{-\frac{q}{t}-i\sqrt{\frac{x}{4t-xt^2}}},\ \ \ x\in[0,1],
\ |t-1|<\delta.
$$ 
The denominator above is bounded for $x$ and $t$ as indicated. Furthermore, by part (a), for any fixed 
$x\in[0,1]$, the range of $v_x(t)$, $|t-1|<\delta$, contains the disk $|v|<\rho$. It follows
that there exists $\lambda>0$ independent of $x$ so that $|\frac{dt_x}{dt}(v)|\geq\lambda$,
for all $|v|<\rho$ and $x\in[0,1]$. 

For any $0<\eta<\rho$, consider the circle $C_\eta$ in 
the $v$-plane centered at $0$ and having radius $\eta$. Since $t_x(v)$ is a conformal map, 
it follows that $t_x(C_\eta)$ is a simple closed curve in the $t$-plane containing $t=1$ in its
interior. By the above lower bound on $|\frac{dt_x}{dt}(v)|$, all points of $t_x(C_\eta)$ are
at least at distance $\lambda\eta$ from $t=1$. Therefore, $t_x(C_\eta)$ must contain the disk
$|t-1|<\lambda\eta$ in its interior. Thus, $U_x=t_x(D)=t_x(\cup_{0\leq\eta<\rho}C_\eta)$ contains
the disk $|t-1|<\lambda\rho$, for all $x\in[0,1]$. \endpf

\proclaim{Lemma 9.2} Let the function $p(t)$ be given by $(7.3)$.
Then for any $\rho>0$ one can choose $k\in[0,1)$, $k$ not depending on $x\in[0,1]$,
so that the disk $|v|\leq|p(k)-p(1)|$ is contained in $D(0,\rho)$, for all $x\in[0,1]$. 
\endproclaim

\pf We have
$$
p(k)-p(1)=-q\ln t -i[\arccos(1-xk/2)-\arccos(1-x/2)],
$$
and since the first term on the right hand side is independent of $x$, it suffices to show 
that there exists some $k\in[0,1)$ so that
$$
|\arccos(1-xk/2)-\arccos(1-x/2)|<\rho/2,\ \ \ x\in[0,1].\tag9.30
$$
For $f(u)=\arccos(1-u/2)$ one has $f'(u)=(4u-u^2)^{-1/2}$, $u\in(0,1]$, and thus $f(u)$
is increasing on $[0,1]$. Therefore the absolute value in (9.30) equals in fact
$$
g(x):=\arccos(1-x/2)-\arccos(1-xk/2).
$$
One readily sees that $g'(x)=(4x-x^2)^{-1/2}-(4x/k-x^2)^{-1/2}>0$, $k\in(0,1)$, $x\in[0,1]$,
so (9.30) is equivalent to $g(1)=\arccos(1-1/2)-\arccos(1-k/2)<\rho/2$. Since one can clearly
choose $k$ to satisfy the latter condition, the proof of the Lemma is complete. \endpf

\proclaim{Lemma 9.3} Let $f_1(v)$ be defined by $(9.9)$, where $f(v)$ is given by $(9.6)$, 
$(9.2)$ and $(7.3)$, and all involved multiple valued maps have their principal determination.
Then there exists $\rho>0$ and a constant $M$, both independent
of $x\in[0,1]$, so that
$$
|f_1(v)|\leq M,\ \ \ \text{\rm all $|v|<\rho$,\ $x\in[0,1]$.}
$$
\endproclaim

\pf 
By (9.26) and the proof of Lemma 9.1, there exist $0<\delta<1$ and $\rho>0$, both independent of $x$,
so that for any fixed $x\in[0,1]$, all $z$'s in the $v$-plane with $|z|<\rho$ can be written as 
$z=v(t)$, with $|t-1|<\delta$.
Therefore, to prove the current Lemma it is enough to show that $|f_1(v)|=|f_1(v(t))|\leq M$, for 
$|t-1|<\delta$ and $x\in[0,1]$, for some constant $M$ independent of $x$.

By (9.9) and (9.8) we have
$$
\align
f_1(v)&=v^{-1/2}\left\{\frac{\tilde{Q}(t)}{p'(t)}-a_0v^{-1/2}\right\}\\
&=v^{-1/2}\left\{\frac{H(t)}{p'(t)}(t-1)^{-1/2}-\frac{H(1)}{p'(1)^{1/2}}v^{-1/2}\right\},\tag9.31
\endalign
$$
where 
$$
H(t)=(t-1)^{1/2}\tilde{Q}(t).\tag9.32
$$
Write
$$
v=\frac{p(t)-p(1)}{t-1}(t-1)=p'(\xi)(t-1),\tag9.33
$$
where $\xi$ is on the line segment $[1,t]$. Then (9.31) becomes
$$
\align
f_1(v)&=v^{-1/2}(t-1)^{-1/2}\left\{\frac{H(t)}{p'(t)}-\frac{H(1)}{p'(1)^{1/2}p'(\xi)^{1/2}}\right\}\\
&=\frac{1}{(t-1)p'(\xi)^{1/2}}
\left\{
\left(\frac{H(t)}{p'(t)}-\frac{H(1)}{p'(t)}\right)
+\left(\frac{H(1)}{p'(t)}-\frac{H(1)}{p'(t)^{1/2}p'(1)^{1/2}}\right)
\right.
\\
&\ \ \ \ \ \ \ \ \ \ \ \ \ \ \ \ \ \ \ \ \ \ \ \ \ \ \ \ \ \ \ \ 
\left.
+\left(\frac{H(1)}{p'(t)^{1/2}p'(1)^{1/2}}-\frac{H(1)}{p'(\xi)^{1/2}p'(1)^{1/2}}\right)
\right\}
\\
&=\frac{1}{(t-1)p'(\xi)^{1/2}}
\left\{
\frac{1}{p'(t)}[H(t)-H(1)]+\frac{H(1)}{p'(t)^{1/2}}[U(t)-U(1)]
\right.
\\
&\ \ \ \ \ \ \ \ \ \ \ \ \ \ \ \ \ \ \ \ \ \ \ \ \ \ \ \ \ \ \ \
\left.
+\frac{H(1)}{p'(1)^{1/2}}[U(t)-U(\xi)]
\right\},
\tag9.34
\endalign
$$
where 
$$
U(t)=\frac{1}{p'(t)^{1/2}}.\tag9.35
$$
We have
$$
\align
H(t)-H(1)&=(t-1)H'(\xi_1)\\
U(t)-U(1)&=(t-1)U'(\xi_2)\\
U(t)-U(\xi)&=(t-\xi)U'(\eta),
\endalign
$$
where $\xi_1$, $\xi_2$ and $\eta$ are on the line segment $[1,t]$ (the latter because $\xi$
is on this segment). Therefore (9.34) implies
$$
\align
|f_1(v)|&\leq\frac{1}{|t-1||p'(\xi)^{1/2}|}
\left\{
\frac{1}{|p'(t)|}|H(t)-H(1)|+\frac{|H(1)|}{|p'(t)^{1/2}|}|U(t)-U(1)|
\right.
\\
&\ \ \ \ \ \ \ \ \ \ \ \ \ \ \ \ \ \ \ \ \ \ \ \ \ \ \ \ \ \ \ \
\left.
+\frac{|H(1)|}{|p'(1)^{1/2}|}|U(t)-U(\xi)|
\right\}
\\
&=\frac{1}{|t-1||p'(\xi)^{1/2}|}
\left\{
\frac{1}{|p'(t)|}|H'(\xi_1)||t-1|+\frac{|H(1)|}{|p'(t)^{1/2}|}|U'(\xi_2)||t-1|
\right.
\\
&\ \ \ \ \ \ \ \ \ \ \ \ \ \ \ \ \ \ \ \ \ \ \ \ \ \ \ \ \ \ \ \
\left.
+\frac{|H(1)|}{|p'(1)^{1/2}|}|U'(\eta)||t-\xi|
\right\}
\\
&\leq\frac{1}{|p'(\xi)^{1/2}|}
\left\{
\frac{1}{|p'(t)|}|H'(\xi_1)|+\frac{|H(1)|}{|p'(t)^{1/2}|}|U'(\xi_2)|
+\frac{|H(1)|}{|p'(1)^{1/2}|}|U'(\eta)|
\right\}
\endalign
$$
(where at the last step we used that $\xi$ is on the line segment $[1,t]$). Therefore, to
finish the proof we need upper bounds that are uniform in $x\in[0,1]$ for $|H(1)|$ and for
$|p'(t)|^{-1}$, $|H'(t)|$, and $|U'(t)|$, for $|t-1|<\delta$. 

By (9.32) and (7.4), we have
$$
H(t)=t^l(4-2xt)^a/(4-xt)^b,\tag9.36
$$
where $0\leq a\in\Z$, $-1/2\leq b\in\frac12\Z$ and $l\in\frac12\Z$. Since the right hand
side of (9.36) is continuous in $(x,t)$ on the compact set
$[0,1]\times\{|t-1|\leq\delta\}$, $|H(t)|$, and in particular $|H(1)|$ can be majorized
uniformly with respect to $x\in[0,1]$.

Since by (9.35) $U'(t)=-1/2[p'(t)]^{-3/2}p''(t)$, 
all remaining uniform upper bounds will follow from such bounds for $|H'(t)|$,
$1/|p'(t)|$ and $|p''(t)|$. 

By (9.36) we have
$$
H'(t)=lt^{l-1}\frac{(4-2xt)^a}{(4-xt)^b}+
t^l\frac{-2ax(4-2xt)^{a-1}(4-xt)^b+bx(4-2xt)^{a}(4-xt)^{b-1}}{(4-xt)^{2b}}.\tag9.37
$$

It follows from (7.3) that
$$
\align
p'(t)&=-\frac{q}{t}-i\sqrt{\frac{x}{4t-xt^2}}\tag9.38\\
p''(t)&=\frac{q}{t^2}+i\frac{\sqrt{x}(2-xt)}{(4t-xt^2)^{3/2}}.\tag9.39
\endalign
$$
Since the right hand sides in (9.37) and (9.39) are continuous in $(x,t)$ on 
the compact set $[0,1]\times\{|t-1|\leq\delta\}$, it follows that $|H'(t)|$
and $|p''(t)|$ can be majorized as desired. Since $|t-1|\leq\delta$, and by (9.38) 
$|p'(t)|\geq q/|t|$, one has $1/|p'(t)|\leq (1+\delta)/q$, and the proof is complete. \epf 

\mysec{10. The asymptotics of a multidimensional Laplace integral}

By Proposition 8.1, the asymptotics of the moments
$M_{\alpha_1,\beta_1,\dotsc,\alpha_m,\beta_m;\gamma_1,\delta_1,\dotsc,\gamma_n,\delta_n}$
is given by the multiple integral in (8.3). In determining the asymptotics of the latter 
we will use the following result.

\proclaim{Proposition 10.1} Let $h,a:D\to\C$, $(0,1]^n\subset D\subset\R^n$, be two 
functions. Assume
that $\int_{[0,1]^n}|h|<\infty$ and that there exists a neighborhood $V$ of $(1,\dotsc,1)$
so that $h\in C^1(V)$ and $a\in C^2(V)$. Then for fixed $q_1,\dotsc,q_n>0$ we have
$$
\align
&I(R)=\int_0^1\cdots\int_0^1 x_1^{Rq_1}\cdots x_n^{Rq_n}h(x_1,\dotsc,x_n)
e^{iRa(x_1,\dotsc,x_n)}dx_1\cdots dx_n\\
&
=\frac{1}{R^n}\frac{h(1,\dotsc,1)e^{iRa(1,\dotsc,1)}}
{[q_1+ia_{x_1}(1,\dotsc,1)]\cdots[q_n+ia_{x_n}(1,\dotsc,1)]}+O(R^{-n-1}),\tag10.1
\endalign
$$
where $a_{x_i}$ is the partial derivative of $a$ with respect to the variable $x_i$.
\endproclaim

We deduce the above result from the following.

\proclaim{Lemma 10.2} Let $k\in(0,1)$ and let $V\subset\R^n$ be an open set containing 
$[k,1]^n$. 
Let $h,a:V\to\C$ be functions so that $h\in C^1(V)$ and $a\in C^2(V)$. 
Then for fixed $q_1,\dotsc,q_n>0$ we have
$$
\align
&I_k(R)=\int_k^1\cdots\int_k^1 x_1^{Rq_1}\cdots x_n^{Rq_n}h(x_1,\dotsc,x_n)
e^{iRa(x_1,\dotsc,x_n)}dx_1\cdots dx_n\\
&\ \ 
=\frac{1}{R}\int_k^1\cdots\int_k^1 
\frac{x_2^{Rq_2}\cdots x_n^{Rq_n}h(1,x_2\dotsc,x_n)}{q_1+ia_{x_1}(1,x_2,\dotsc,x_n)}
e^{iRa(1,x_2,\dotsc,x_n)}dx_2\cdots dx_n\\
&\ \ \ \ \ \ \ \ \ \ \ \ \ \ \ \ \ \ \ \ \ \ \ \ 
+O(R^{-n-1}).\tag10.2
\endalign
$$
\endproclaim

\pf Write $I_k(R)$ as
$$
I_k(R)=\int_k^1\cdots\int_k^1 h(x_1,\dotsc,x_n)
e^{Rb(x_1,\dotsc,x_n)}dx_1\cdots dx_n,\tag10.3
$$
where 
$$
b(x_1,\dotsc,x_n)=q_1\ln x_1+\cdots+q_n\ln x_n+ia(x_1,\dotsc,x_n).\tag10.4
$$
Apply integration by parts \footnote{The idea 
of using integration by parts was suggested to the author by Andr\'as Vasy.}
with respect to the variable $x_1$ in (10.3).
We have
$$
\frac{\partial}{\partial x_1}e^{Rb(x_1,\dotsc,x_n)}=
R\left(\frac{\partial}{\partial x_1}b(x_1,\dotsc,x_n)\right)e^{Rb(x_1,\dotsc,x_n)}.\tag10.5
$$
By (10.4),
$$
\frac{\partial}{\partial x_1}b(x_1,\dotsc,x_n)=\frac{q_1}{x_1}
+i\frac{\partial}{\partial x_1}a(x_1,\dotsc,x_n),\tag10.6
$$
so in particular $b_{x_1}(x_1,\dotsc,x_n)\neq0$ for all $(x_1,\dotsc,x_n)\in[k,1]^n$.
Therefore (10.5) can be rewritten (omitting the arguments for brevity) as
$$
e^{Rb}=\frac{1}{Rb_{x_1}}\frac{\partial}{\partial x_1}e^{Rb}.\tag10.7
$$  
Writing the second factor of the integrand in (10.3) as in (10.7) and applying integration
by parts with respect to the variable $x_1$ we obtain
$$
\align
I_k(R)&=\int_k^1\cdots\int_k^1
\left\{\int_k^1 h\frac{1}{Rb_{x_1}}\frac{\partial e^{Rb}}{\partial x_1}dx_1 \right\}dx_2\cdots dx_n\\
&=\int_k^1\cdots\int_k^1
\left\{\left.\frac{h}{Rb_{x_1}}e^{Rb}\right|^1_k
-\int_k^1 e^{Rb}\frac{\partial}{\partial x_1}
\left(\frac{h}{Rb_{x_1}}\right)dx_1 \right\}dx_2\cdots dx_n\\
&=\frac{1}{R}\int_k^1\cdots\int_k^1
\frac{h(1,x_2,\dotsc,x_n)}{b_{x_1}(1,x_2,\dotsc,x_n)}e^{Rb(1,x_2,\dotsc,x_n)}dx_2\cdots dx_n\\
&
-\frac{1}{R}\int_k^1\cdots\int_k^1
k^{Rq_1}x_2^{Rq_2}\cdots x_n^{Rq_n}
\frac{h(k,x_2,\dotsc,x_n)}{b_{x_1}(k,x_2,\dotsc,x_n)}e^{iRa(k,x_2,\dotsc,x_n)}dx_2\cdots dx_n\\
&-\frac{1}{R}\int_k^1\cdots\int_k^1\frac{h_{x_1}b_{x_1}-hb_{x_1x_1}}{(b_{x_1})^2}
e^{Rb}dx_1\cdots dx_n\tag10.8
\endalign
$$
(where at the last equality we used (10.4) for the second term on the right hand side). 

By (10.6), 
$$
|b_{x_1}(x_1,\dotsc,x_n)|\geq\frac{q_1}{x_1}\geq q_1,\ \ \ (x_1,\dotsc,x_n)\in[k,1]^n.\tag10.9
$$
Therefore, the absolute value of the second term on the right hand side of (10.8) is majorized
by 
$$
\frac{1}{Rq_1}k^{Rq_1}\int_k^1\cdots\int_k^1|h(k,x_2,\dotsc,x_n)|dx_2\cdots dx_n,
$$
which in turn, by the presence of the exponential and
since $h$ is continuous on $[k,1]^n$, is less than $M_1R^{-n-1}$ for
$R\geq r_1$, for some suitable constants $M_1$ and $r_1$.

On the other hand, the absolute value of the third term on the right hand side of (10.8) can
be majorized, according to (10.9), by
$$
\align
&\frac{1}{R(q_1)^2}\int_k^1\cdots\int_k^1|h_{x_1}b_{x_1}-hb_{x_1x_1}e^{Rb}|dx_1\cdots dx_n\\
&=\frac{1}{R(q_1)^2}\int_k^1\cdots\int_k^1 x_1^{Rq_1}\cdots x_n^{Rq_n}
|h_{x_1}b_{x_1}-hb_{x_1x_1}|dx_1\cdots dx_n.
\endalign
$$
Since $h_{x_1}b_{x_1}-hb_{x_1x_1}$ is continuous on $[k,1]^n$ and 
$\int_{[k,1]^n}x_1^{Rq_1}\!\!\cdots x_n^{Rq_n}\leq
\int_{[0,1]^n}x_1^{Rq_1}\!\!\cdots x_n^{Rq_n}=\prod_{i=1}^n(Rq_i+1)^{-1}$, 
there exist constants $M_2$ and $r_2$ so that the right hand side of the above equation is
majorized by $M_2R^{-n-1}$, for all $R\geq r_2$.

By (10.8), (10.6) and the last two paragraphs we obtain (10.3). \endpf

{\it Proof of Proposition 10.1.} Choose $k\in(0,1)$ so that $[k,1]^n\subset V$. We note first
that in order to prove the Proposition it suffices to show that $I_k(R)$ has the asymptotics 
given by the right hand side of (10.1), where $I_k(R)$ is defined by (10.2). Indeed,
$I(R)=I_k(R)+J(R)$, where
$$
J(R)=\int_{[0,1]^n\setminus[k,1]^n}x_1^{Rq_1}\cdots x_n^{Rq_n}
h(x_1,\dotsc,x_n)e^{iRa(x_1,\dotsc,x_n)}dx_1\cdots dx_n.
$$
Since throughout the integration range above one has $x_i<k$ for at least one index $i$, 
denoting $q_0=\min(q_1,\dotsc,q_n)>0$ we obtain
$$
\align
|J(R)|&\leq \int_{[0,1]^n\setminus[k,1]^n}k^{Rq_0}|h(x_1,\dotsc,x_n)|dx_1\cdots dx_n\\
&\leq k^{Rq_0}\int_{[0,1]^n}|h(x_1,\dotsc,x_n)|dx_1\cdots dx_n.
\endalign
$$
By hypothesis the integral above is finite, so there exist constants 
$M_0$ and $r_0$ so that $|J(R)|\leq M_0R_0^{-n-1}$ for all $R\geq r_0$. Therefore addition of 
$J(R)$ to $I_k(R)$ does not change its asymptotics.

By Lemma 10.2, the asymptotics of $I_k(R)$ is the same as $1/R$ times the asymptotics of the 
integral on
the right hand side of (10.2). In turn, the latter integral meets the hypotheses of 
Lemma 10.2. Applying Lemma 10.2 to it we obtain from (10.2) that
$$
\align
&I_k(R)=
\frac{1}{R^2}\int_k^1\cdots\int_k^1 
\frac{x_3^{Rq_3}\cdots x_n^{Rq_n}h(1,1,x_3\dotsc,x_n)}
{[q_1+ia_{x_1}(1,1,x_3,\dotsc,x_n)][q_2+ia_{x_2}(1,1,x_3,\dotsc,x_n)]}
\\
&\ \ \ \ \ \ \ \ \ \ \ \ \ \ \ \ \ \ \ \ \ \ \ \ 
\times
e^{iRa(1,1,x_3,\dotsc,x_n)}dx_3\cdots dx_n+O(R^{-n-1}).
\endalign
$$
Lemma 10.2 can be applied again to the integral on the right hand side above, yielding an
$(n-3)$-fold integral that meets its hypotheses. All the successive applications in this manner
of Lemma 10.2 lead to multiple integrals to which it is again applicable. After $n$ applications
one obtains (10.1). \endpf

By (8.1) and (8.2), the multiple integral (8.3) leads to integrals having the form of $I(R)$,
but with the exponential function replaced by a cosine. As indicated below, this type of 
integrals can easily be handled by Proposition 10.2.

\proclaim{Corollary 10.3} Let $h,a,c:D\to\R$, $(0,1]^n\subset D\subset\R^n$, 
be three functions.  Assume
that $\int_{[0,1]^n}|h|<\infty$ and that there exists a neighborhood $V$ of $(1,\dotsc,1)$
so that $h,c\in C^1(V)$ and $a\in C^2(V)$. Then for fixed $q_1,\dotsc,q_n>0$ we have
$$
\align
&K(R)=\int_0^1\!\!\cdots\!\int_0^1 x_1^{Rq_1}\!\cdots x_n^{Rq_n}h(x_1,\dotsc,x_n)
\cos[Ra(x_1,\dotsc,x_n)+c(x_1,\dotsc,x_n)]dx_1\cdots dx_n\\
&\ \ 
=\frac{1}{R^n}\frac{h(1,\dotsc,1)}
{\sqrt{(q_1)^2+(a_{x_1}(1,\dotsc,1))^2}\cdots\sqrt{(q_n)^2+(a_{x_n}(1,\dotsc,1))^2}}\\
&\ \ \ \ \ \ 
\times
\cos\left[Ra(1,\dotsc,1)+c(1,\dotsc,1)-\arctan\frac{a_{x_1}(1,\dotsc,1)}{q_1}-\cdots
-\arctan\frac{a_{x_n}(1,\dotsc,1)}{q_n}\right]\\
&\ \ \ \ \ \ \ \ \ \ \ \ \ \ \ \ \ \ \ \ \ \ \ \ 
+O(R^{-n-1}),\tag10.10
\endalign
$$
where $a_{x_i}$ is the partial derivative of $a$ with respect to the variable $x_i$.
\endproclaim

\pf Since $h$, $a$ and $c$ are real-valued, $\cos(Ra+c)=\Rep e^{iRa+ic}$, and thus
$$
K(R)=\Rep\int_0^1\cdots\int_0^1 x_1^{Rq_1}\cdots x_n^{Rq_n}h(x_1,\dotsc,x_n)
e^{ic(x_1,\dotsc,x_n)}e^{iRa(x_1,\dotsc,x_n)}dx_1\cdots dx_n.
$$
Therefore, by Proposition 10.1
$$
K(R)=\frac{1}{R^n}\Rep \frac{h(1,\dotsc,1)e^{i[Ra(1,\dotsc,1)+c(1,\dotsc,1)]}}
{[q_1+ia_{x_1}(1,\dotsc,1)]\cdots[q_n+ia_{x_n}(1,\dotsc,1)]}+O(R^{-n-1}).\tag10.11
$$
Let $a_j=a_{x_j}(1,\dotsc,1)$ and write 
$q_j+ia_j=\sqrt{q_j^2+a_j^2}\,(\cos\theta_j+i\sin\theta_j)$, with 
$\theta_j=\arctan\frac{a_j}{q_j}$,
for $j=1,\dotsc,n$. The contribution of the exponential and the denominator to the fraction in
(10.11) becomes
$$
\frac{e^{i[{Ra(1,\dotsc,1)+c(1,\dotsc,1)-\theta_1-\cdots-\theta_n]}}}
{\sqrt{q_1^2+a_1^2}\cdots\sqrt{q_n^2+a_n^2}}.
$$
Substituting this in (10.11) gives (10.10). \epf

\mysec{11. The asymptotics of $\omega_b$. Proof of Theorem 2.2}

In this section we use the results of Section 10 to deduce the asymptotics of $\omega_b$
from its expression (6.5). To this end, we need to determine first the asymptotics of the moments
$M_{\alpha_1,\beta_1,\dotsc,\alpha_m,\beta_m;\gamma_1,\delta_1,\dotsc,\gamma_n,\delta_n}$
defined by (6.6).

Recall that the arguments of $\omega_b$---and therefore those of the moments $M$---approach 
infinity as specified by (\TwoTwo). In particular\footnote
{
In this section doubly indexed variables appear alongside the complex number $i$. For this reason, we
use here $k$ and $j$ as opposed to the more familiar $i$ and $j$ as indices of these doubly
indexed variables.
}, 
$v_k=q_kR_k+c_k$, $k=1,\dotsc,m$ and 
$v'_j=q'_jR'_j+c'_j$, $j=1,\dotsc,n$.

Substituting (8.1) and (8.2) into (8.3) we obtain that
$$
\align
&M_{\alpha_1,\beta_1,\dotsc,\alpha_m,\beta_m;\gamma_1,\delta_1,\dotsc,\gamma_n,\delta_n}
=\frac{2^{2m}}{\pi^{m+n}}\frac{1}{\prod_{k=1}^mR_k^3\prod_{j=1}^n{R'_j}^3}\\
&
\times
\int_0^1\cdots\int_0^1
\prod_{k=1}^m
\frac{
1
}
{
{\root4\of{{q}_k^2+\frac{\prod_{j=1}^nx_{kj}y_{kj}}{4-\prod_{j=1}^nx_{kj}y_{kj}}}}
{\root4\of{{q}_k^2+\frac{\prod_{j=1}^nz_{kj}w_{kj}}{4-\prod_{j=1}^nz_{kj}w_{kj}}}}
}\\
&
\times
\prod_{j=1}^n
\frac{
{\root4\of{{q'_j}^2+\frac{\prod_{k=1}^mx_{kj}z_{kj}}{4-\prod_{k=1}^mx_{kj}z_{kj}}}}
{\root4\of{{q'_j}^2+\frac{\prod_{k=1}^my_{kj}w_{kj}}{4-\prod_{k=1}^my_{kj}w_{kj}}}}
}
{
\sqrt{\frac{\prod_{k=1}^mx_{kj}z_{kj}}{4-\prod_{k=1}^mx_{kj}z_{kj}}}
\sqrt{\frac{\prod_{k=1}^my_{kj}w_{kj}}{4-\prod_{k=1}^my_{kj}w_{kj}}}
}\\
&
\times
\prod_{k=1}^m
\left(R_k\sqrt{\frac{\prod_{j=1}^nx_{kj}y_{kj}}{4-\prod_{j=1}^nx_{kj}y_{kj}}}\,\right)^{\alpha_k}
\left(R_k\sqrt{\frac{\prod_{j=1}^nz_{kj}w_{kj}}{4-\prod_{j=1}^nz_{kj}w_{kj}}}\,\right)^{\beta_k}\\
&
\times
\prod_{j=1}^n
\left(R'_j\sqrt{\frac{\prod_{k=1}^mx_{kj}z_{kj}}{4-\prod_{k=1}^mx_{kj}z_{kj}}}\,\right)^{\gamma_j}
\left(R'_j\sqrt{\frac{\prod_{k=1}^my_{kj}w_{kj}}{4-\prod_{k=1}^my_{kj}w_{kj}}}\,\right)^{\delta_j}\\
&
\times
\prod_{k=1}^m
\left\{
\cos\left[R_k\arccos\left(1-\frac{\prod_{j=1}^nx_{kj}y_{kj}}{2}\right) 
-\frac12\arctan\frac{1}{q_k}\sqrt{\frac{\prod_{j=1}^nx_{kj}y_{kj}}{4-\prod_{j=1}^nx_{kj}y_{kj}}}
+\frac{\alpha_k\pi}{2}\right]
\right.
\\
&\ \ \ \ \ \,
\left.
\times
\cos\left[R_k\arccos\left(1-\frac{\prod_{j=1}^nz_{kj}w_{kj}}{2}\right) 
-\frac12\arctan\frac{1}{q_k}\sqrt{\frac{\prod_{j=1}^nz_{kj}w_{kj}}{4-\prod_{j=1}^nz_{kj}w_{kj}}}
+\frac{\beta_k\pi}{2}\right]
\right\}
\\
&\!\!\!
\times
\prod_{j=1}^n
\left\{
\cos\left[R'_j\arccos\left(1-\frac{\prod_{k=1}^mx_{kj}z_{kj}}{2}\right) 
+\frac12\arctan\frac{1}{q'_j}\sqrt{\frac{\prod_{k=1}^mx_{kj}z_{kj}}{4-\prod_{k=1}^mx_{kj}z_{kj}}}
+\frac{(\gamma_j-1)\pi}{2}\right]
\right.
\\
&\ \ 
\left.
\times
\cos\left[R'_j\arccos\left(1-\frac{\prod_{k=1}^my_{kj}w_{kj}}{2}\right) 
+\frac12\arctan\frac{1}{q'_j}\sqrt{\frac{\prod_{k=1}^my_{kj}w_{kj}}{4-\prod_{k=1}^my_{kj}w_{kj}}}
+\frac{(\delta_j-1)\pi}{2}\right]
\right\}
\\
&
\times
\prod_{k=1}^m\prod_{j=1}^n(x_{kj}y_{kj}z_{kj}w_{kj})^{q_kR_k+q'_jR'_j+c_k+c'_j+1}
dx_{11}\cdots dw_{mn}.
\tag11.1
\endalign
$$
One readily proves by induction on $s$ that
$$
\cos\theta_1\cdots\cos\theta_s=\frac{1}{2^s}\sum_{\epsilon_1,\dotsc,\epsilon_s=\pm1}
\cos(\epsilon_1\theta_1+\cdots+\epsilon_s\theta_s).
$$
Since by (\TwoTwo) we have $R_k=A_kR$, $k=1,\dotsc,m$ and $R'_j=B_jR$,  $j=1,\dotsc,n$, the product of
the $2m+2n$ cosines in the integrand of (11.1) becomes by the above formula
$$
\spreadlines{4\jot}
\align
&\frac{1}{2^{2m+2n}}\sum_{\epsilon_1,\dotsc,\epsilon_{2m+2n}=\pm1}
\cos
\left\{
\phantom{\sum_{k=1}^m}
\right.
\\
&
R
\left[
\sum_{k=1}^m\left(
\epsilon_{2k-1}A_k\arccos\left(1-\frac{\prod_{j=1}^nx_{kj}y_{kj}}{2}\right) 
+\epsilon_{2k}A_k\arccos\left(1-\frac{\prod_{j=1}^nz_{kj}w_{kj}}{2}\right)
\right)
\right.
\\
&\!\!
\left.
+\sum_{j=1}^n\left(
\epsilon_{2m+2j-1}B_j\arccos\left(1-\frac{\prod_{k=1}^mx_{kj}z_{kj}}{2}\right) 
+\epsilon_{2m+2j}B_j\arccos\left(1-\frac{\prod_{k=1}^my_{kj}w_{kj}}{2}\right)
\right)
\right]
\\
&
-\sum_{k=1}^m\left(
\frac{\epsilon_{2k-1}}{2}
\arctan\frac{1}{q_k}\sqrt{\frac{\prod_{j=1}^nx_{kj}y_{kj}}{4-\prod_{j=1}^nx_{kj}y_{kj}}}
+\frac{\epsilon_{2k}}{2}
\arctan\frac{1}{q_k}\sqrt{\frac{\prod_{j=1}^nz_{kj}w_{kj}}{4-\prod_{j=1}^nz_{kj}w_{kj}}}\right)\\
&
+\sum_{j=1}^n\left(
\frac{\epsilon_{2m+2j-1}}{2}
\arctan\frac{1}{q'_j}\sqrt{\frac{\prod_{k=1}^mx_{kj}z_{kj}}{4-\prod_{k=1}^mx_{kj}z_{kj}}}
+\frac{\epsilon_{2m+2j}}{2}
\arctan\frac{1}{q'_j}\sqrt{\frac{\prod_{k=1}^my_{kj}w_{kj}}{4-\prod_{k=1}^my_{kj}w_{kj}}}\right)\\
&\!\!
\left.
+\sum_{k=1}^m(\epsilon_{2k-1}\alpha_k+\epsilon_{2k}\beta_k)\frac{\pi}{2}
+\sum_{j=1}^n(\epsilon_{2m+2j-1}\gamma_j+\epsilon_{2m+2j}\delta_j)\frac{\pi}{2}
-\sum_{j=1}^n(\epsilon_{2m+2j-1}+\epsilon_{2m+2j})\frac{\pi}{2}
\right\}.
\\
\tag11.2
\endalign
$$

The $4mn$-fold integral in (11.1), with the product of cosines in the integrand written as in
(11.2), becomes the sum of $2m+2n$ integrals of
the form of the multiple integral in Corollary 10.3. Our current functions $h$ satisfy
$\int_{[0,1]^{4mn}}|h|<\infty$. Indeed, since the product of the $2n$ products (8.12) is
$\prod_{k=1}^m\prod_{j=1}^nx_{kj}y_{kj}z_{kj}w_{kj}$, one readily sees that $|h|$ can be majorized by
$K\prod_{k=1}^m\prod_{j=1}^n(x_{kj}y_{kj}z_{kj}w_{kj})^{c_k+c'_j+1/2}$, where $K$ is a constant
independent of $x_{kj},y_{kj},z_{kj},w_{kj}$. Since by (\TwoTwo) the exponent in the previous double 
product is non-negative, it follows that $\int_{[0,1]^{4mn}}|h|<\infty$.

The other conditions in the hypothesis of Corollary 10.3 are clearly 
satisfied. The $a$-functions of Corollary 10.3 are now the coefficients $a(x_{11},\dotsc,w_{mn})$ of
$R$ in the argument of the cosines in (11.2). 

Note that there are exactly two terms of each of our 
$a(x_{11},\dotsc,w_{mn})$'s containing any given $x_{kj}$, $y_{kj}$, $z_{kj}$ or $w_{kj}$, and that 
all variables are set to 1 on the right hand side of (10.10). Since  
$\partial/\partial x \arccos(1-xu/2)=(4x/u-x^2)^{-1/2}$, we obtain
$$
\align
\frac{\partial a}{\partial x_{kj}}(1,\dotsc,1)&=\frac{\epsilon_{2k-1}A_k+\epsilon_{2m+2j-1}B_j}{\sqrt{3}}
\\
\frac{\partial a}{\partial y_{kj}}(1,\dotsc,1)&=\frac{\epsilon_{2k-1}A_k+\epsilon_{2m+2j}B_j}{\sqrt{3}}
\\
\frac{\partial a}{\partial z_{kj}}(1,\dotsc,1)&=\frac{\epsilon_{2k}A_k+\epsilon_{2m+2j-1}B_j}{\sqrt{3}}
\\
\frac{\partial a}{\partial w_{kj}}(1,\dotsc,1)&=\frac{\epsilon_{2k}A_k+\epsilon_{2m+2j}B_j}{\sqrt{3}}
.\tag11.3
\endalign
$$

By (11.1)--(11.3), Corollary 10.3 implies
$$
\align
&M_{\alpha_1,\beta_1,\dotsc,\alpha_m,\beta_m;\gamma_1,\delta_1,\dotsc,\gamma_n,\delta_n}
=\frac{E}{R^{4mn}}\prod_{k=1}^m\left(\frac{R_k}{\sqrt{3}}\right)^{\alpha_k+\beta_k}
\prod_{j=1}^n\left(\frac{R'_j}{\sqrt{3}}\right)^{\gamma_j+\delta_j}\\
&
\times\!\!\!\!\! \!\!\!\!\! 
\sum_{\epsilon_1,\dotsc,\epsilon_{2m+2n}=\pm1}\frac{1}{D^{\epsilon_1,\dotsc,\epsilon_{2m+2n}}}\cos
\left\{
\frac{R\pi}{3}\left[\sum_{k=1}^mA_k(\epsilon_{2k-1}+\epsilon_{2k})+
\sum_{j=1}^nB_j(\epsilon_{2m+2j-1}+\epsilon_{2m+2j})\right]
\right.
\\
&\ \ \ 
-\sum_{k=1}^m\frac{\epsilon_{2k-1}+\epsilon_{2k}}{2}\arctan\frac{1}{q_k\sqrt{3}}
+\sum_{j=1}^n\frac{\epsilon_{2m+2j-1}+\epsilon_{2m+2j}}{2}\arctan\frac{1}{q'_j\sqrt{3}}\\
&\ \ \   
+\sum_{k=1}^m(\epsilon_{2k-1}\alpha_k+\epsilon_{2k}\beta_k)\frac{\pi}{2}
+\sum_{j=1}^n(\epsilon_{2m+2j-1}\gamma_j+\epsilon_{2m+2j}\delta_j)\frac{\pi}{2}
-\sum_{j=1}^n(\epsilon_{2m+2j-1}+\epsilon_{2m+2j})\frac{\pi}{2}\\
&\ \ \  
-\sum_{k=1}^m\sum_{j=1}^n
\left(
\arctan\frac{\epsilon_{2k-1}A_k+\epsilon_{2m+2j-1}B_j}{(q_kA_k+q'_jB_j)\sqrt{3}}
+\arctan\frac{\epsilon_{2k-1}A_k+\epsilon_{2m+2j}B_j}{(q_kA_k+q'_jB_j)\sqrt{3}}
\right.
\\
&\ \ \ \ \ \ \ \ \ \ \ \ \ \ \ \ \ \ \ \ 
\left.
\left.
+\arctan\frac{\epsilon_{2k}A_k+\epsilon_{2m+2j-1}B_j}{(q_kA_k+q'_jB_j)\sqrt{3}}
+\arctan\frac{\epsilon_{2k}A_k+\epsilon_{2m+2j}B_j}{(q_kA_k+q'_jB_j)\sqrt{3}}
\right)
\right\}
\\
&\ \ \ \ \ \ \ \ 
+O(R^{-4mn+\sum_{k=1}^m(\alpha_k+\beta_k)+\sum_{j=1}^n(\gamma_j+\delta_j)-3m-3n-1}),
\tag11.4
\endalign
$$
where
$$
E=\frac{3^n}{2^{2n}\pi^{m+n}}\frac{1}{\prod_{k=1}^mR_k^3\prod_{j=1}^n{R'_j}^3}
\frac{\prod_{j=1}^n \sqrt{{q'_j}^2+\frac13}}{\prod_{k=1}^m\sqrt{{q}_k^2+\frac13}}\tag11.5
$$
and
$$
\align
D^{\epsilon_1,\dotsc,\epsilon_{2m+2n}}=\prod_{k=1}^m\prod_{j=1}^n
&
\left\{
\sqrt{(q_kA_k+q'_jB_j)^2+\frac{(\epsilon_{2k-1}A_k+\epsilon_{2m+2j-1}B_j)^2}{3}}
\right.
\\
&
\times
\sqrt{(q_kA_k+q'_jB_j)^2+\frac{(\epsilon_{2k-1}A_k+\epsilon_{2m+2j}B_j)^2}{3}}\\
&
\times
\sqrt{(q_kA_k+q'_jB_j)^2+\frac{(\epsilon_{2k}A_k+\epsilon_{2m+2j-1}B_j)^2}{3}}
\\
&
\times
\left.
\sqrt{(q_kA_k+q'_jB_j)^2+\frac{(\epsilon_{2k}A_k+\epsilon_{2m+2j}B_j)^2}{3}}
\right\}.
\tag11.6
\endalign
$$
Define
$$
\align
&S_{\alpha_1,\beta_1,\dotsc,\alpha_m,\beta_m;\gamma_1,\delta_1,\dotsc,\gamma_n,\delta_n}^{
\epsilon_1,\dotsc,\epsilon_{2m+2n}}:=\\
&
\prod_{k=1}^m\left(\frac{R_k}{\sqrt{3}}\right)^{\alpha_k+\beta_k}
\prod_{j=1}^n\left(\frac{R'_j}{\sqrt{3}}\right)^{\gamma_j+\delta_j}
\times
(\text{\rm summand of the $(2m+2n)$-fold sum in (11.4)).}
\tag 11.7
\endalign
$$
We call $(\epsilon_1,\dotsc,\epsilon_{2m+2n})$ {\it balanced} if 
$\epsilon_{2j-1}+\epsilon_{2j}=0$, $j=1,\dotsc,2m+2n$.

\proclaim{Lemma 11.1} Fix $\epsilon_j\in\{-1,1\}$, $j=1,\dotsc,2m+2n$. As in Section $6$, 
let $\Cal C$ be the collection of terms obtained by expanding out the left hand side of $(6.2)$, and for any $C\in\Cal C$ let $e(C)$, $\alpha_k(C)$, $\beta_k(C)$, $\gamma_j(C)$ and
$\delta_j(C)$ have the same significance as in $(6.2)$.

Then unless 
$(\epsilon_1,\dotsc,\epsilon_{2m+2n})$ is balanced, one has
$$
\sum_{C\in\Cal C} e(C)
S_{\alpha_1(C),\beta_1(C),\dotsc,\alpha_m(C),\beta_m(C);
\gamma_1(C),\delta_1(C),\dotsc,\gamma_n(C),\delta_n(C)}^{
\epsilon_1,\dotsc,\epsilon_{2m+2n}}=0.\tag11.8
$$
\endproclaim

\pf Suppose there exists $j\in[1,m]$ with $\epsilon_{2j-1}=\epsilon_{2j}$. Without loss of 
generality we may assume $\epsilon_{1}=\epsilon_{2}$. Then it follows from (11.7) that
$$
\align
S_{\alpha_1,\beta_1,\alpha_2,\beta_2,\dotsc,\alpha_m,\beta_m;
\gamma_1,\delta_1,\dotsc,\gamma_n,\delta_n}^{
\epsilon_1,\dotsc,\epsilon_{2m+2n}}
=
S_{\alpha_1-1,\beta_1+1,\alpha_2,\beta_2,\dotsc,\alpha_m,\beta_m;
\gamma_1,\delta_1,\dotsc,\gamma_n,\delta_n}^{
\epsilon_1,\dotsc,\epsilon_{2m+2n}},\ \ \ \alpha_1\geq1.\tag11.9
\endalign
$$
Partition the terms in the expansion $\Cal C$ of the left hand side of (6.2) into four
classes ${\Cal C}_1$, $\Cal C_2$, $\Cal C_3$, $\Cal C_4$, according to which of the terms
of the factor $(a_1-b_1)^2=(a_1^2-a_1b_1-b_1a_1+b_1^2)$ (in order, from left to right) is
chosen when expanding. 

By (11.9), the restriction of the sum (11.8) to ${\Cal C}_1$ is canceled by its restriction
to ${\Cal C}_2$, and the restriction to ${\Cal C}_3$ is canceled by the restriction
to ${\Cal C}_4$. This proves (11.8).

The case when $\epsilon_{2j-1}=\epsilon_{2j}$, $j\in[m+1,m+n]$, is treated in a perfectly 
analogous manner. \endpf

Define
$M_{\alpha_1,\beta_1,\alpha_2,\beta_2,\dotsc,\alpha_m,\beta_m;
\gamma_1,\delta_1,\dotsc,\gamma_n,\delta_n}^0$
to be given by the expression on the right hand side of (11.4), when the
summation range is restricted to balanced $(\epsilon_1,\dotsc,\epsilon_{2m+2n})$'s. 

Clearly, for $(\epsilon_1,\dotsc,\epsilon_{2m+2n})$ balanced we have
$$
\align
&\{\epsilon_{2k-1}A_k+\epsilon_{2m+2j-1}B_j,\epsilon_{2k-1}A_k+\epsilon_{2m+2j}B_j,
\epsilon_{2k}A_k+\epsilon_{2m+2j-1}B_j,\epsilon_{2k}A_k+\epsilon_{2m+2j}B_j\}\\
&\ \ \ \ 
=\{A_k+B_j,A_k-B_j,-A_k+B_j,-A_k-B_j\},
\endalign
$$
for all $k=1,\dotsc,m$, $j=1,\dotsc,n$. It follows that for 
$(\epsilon_1,\dotsc,\epsilon_{2m+2n})$ balanced one has
\medskip
$(i)$ the double sum of arctangents in (11.4) is 0, and

$(ii)$ the double product $D^{\epsilon_1,\dotsc,\epsilon_{2m+2n}}$ in (11.6) equals 
$$
D=\prod_{k=1}^m\prod_{j=1}^n\left[(q_kA_k+q'_jB_j)^2+\frac13(A_k-B_j)^2\right]
\left[(q_kA_k+q'_jB_j)^2+\frac13(A_k+B_j)^2\right],\tag11.10
$$
an expression independent of $(\epsilon_1,\dotsc,\epsilon_{2m+2n})$.

By $(i)$ and $(ii)$ above we obtain from our definition of the $M_0$'s that
$$
\align
&\!\!\!\!\!\!\!\!\!\!\!
M_{\alpha_1,\beta_1,\alpha_2,\beta_2,\dotsc,\alpha_m,\beta_m;
\gamma_1,\delta_1,\dotsc,\gamma_n,\delta_n}^0=\frac{E}{DR^{4mn}}
\prod_{k=1}^m\left(\frac{R_k}{\sqrt{3}}\right)^{\alpha_k+\beta_k}
\prod_{j=1}^n\left(\frac{R'_j}{\sqrt{3}}\right)^{\gamma_j+\delta_j}\\
&
\times
\sum_{\epsilon_{2l-1}=\pm1,\ l=1,\dotsc,m+n}
\cos\left(\sum_{k=1}^m\frac{\epsilon_{2k-1}(\alpha_k-\beta_k)}{2}\pi
+\sum_{j=1}^n\frac{\epsilon_{2m+2j-1}(\gamma_j-\delta_j)}{2}\pi\right).\tag11.11
\endalign
$$

By (6.2), when our parameters depend on $R$ as in (\TwoTwo), the order in $R$ of $e(C)$ plus 
$\sum_{k=1}^m(\alpha_k(C)+\beta_k(C))+\sum_{j=1}^n(\gamma_j(C)+\delta_j(C))$ is always less or equal
than the number of factors of the product $\Cal E$ of (6.2), namely $2m+2n+4{m\choose2}+4{n\choose2}=
2m^2+2n^2$. Thus, the omitted part of the approximation of 
$e(C)M_{\alpha_1(C),\beta_1(C),\dotsc,\alpha_m(C),\beta_m(C);
\gamma_1(C),\delta_1(C),\dotsc,\gamma_n(C),\delta_n(C)}$ resulting from (11.4) has at most order
$2m^2+2n^2-4mn-3m-3n-1$ in $R$. Therefore, by Lemma 11.1 and our definition of 
the $M_0$'s, (6.5) implies
$$
\align
&\omega_b\left(\matrix{R_1}\\{v_1}\endmatrix
\cdots\matrix{R_m}\\{v_m}\endmatrix;
\matrix{R'_1}\\{v'_1}\endmatrix\cdots\matrix{R'_n}\\{v'_n}\endmatrix\right)=\\
&\ \ \ \ 
X\left|\sum_{C\in\Cal C} e(C)
M^0_{\alpha_1(C),\beta_1(C),\dotsc,\alpha_m(C),\beta_m(C);
\gamma_1(C),\delta_1(C),\dotsc,\gamma_n(C),\delta_n(C)}\right|\\
&\ \ \ \ \ \ \ \
+O(R^{2m^2+2n^2-4mn-2m-1}),
\tag11.12
\endalign
$$
where
$$
X=\chi_{2m,2n}\prod_{k=1}^mR_k\prod_{j=1}^nR'_j(R'_j-1/2)(R'_j+1/2),\tag11.13
$$
with $\chi$ given by (4.3).

\proclaim {Lemma 11.2} We have
$$
\align
&\!\!\!\!\!\!\!\!\!\!\!\!\!\!\!\!\!\!\!\!\!
M^0_{\alpha_1,\beta_1,\dotsc,\alpha_m,\beta_m;\gamma_1,\delta_1,\dotsc,\gamma_n,\delta_n}
\\
&
=\left\{ 
\aligned 
&Y\prod_{k=1}^m\left(\frac{iR_k}{\sqrt{3}}\right)^{\alpha_k}
\prod_{k=1}^m\left(\frac{-iR_k}{\sqrt{3}}\right)^{\beta_k}
\prod_{j=1}^n\left(\frac{iR'_j}{\sqrt{3}}\right)^{\gamma_j}
\prod_{j=1}^n\left(\frac{-iR'_j}{\sqrt{3}}\right)^{\delta_j},\\
&\ \ \ \ \ \ \ \ \ \ \ \ \ \ \ \ \ \ \ \ \ \ 
\text{\rm if $\alpha_1=\beta_1(\mo 2),\dotsc,\gamma_n=\delta_n(\mo 2)$}\\
&0,\ \ \ \ \ \ \ \ \ \ \ \ \ \ \ \ \text{\rm otherwise},
\endaligned
\right.\tag11.14
\endalign
$$
where 
$$
\align
&\!\!\!\!\!\!\!\!\!\!\!
Y=\frac{2^{m-n}3^n}{\pi^{m+n}R^{4mn}}\frac{1}{\prod_{k=1}^mR_k^3\prod_{j=1}^n(R'_j)^3}
\frac{\prod_{j=1}^n \sqrt{{q'_j}^2+\frac13}}{\prod_{k=1}^m\sqrt{{q}_k^2+\frac13}}\\
&\!\!\!\!\!\!\!\!\!
\times
\frac{1}{\prod_{k=1}^m\prod_{j=1}^n[(q_kA_k+q'_jB_j)^2+\frac13(A_k-B_j)^2]
[(q_kA_k+q'_jB_j)^2+\frac13(A_k+B_j)^2]}.\tag11.15
\endalign
$$
\endproclaim

\pf If $\alpha_1=\beta_1(\mo 2),\dotsc,\gamma_n=\delta_n(\mo 2)$, all the constants 
multiplying $\pi$ in the argument of the cosine in (11.11) are integers. Since 
$\cos(-n\pi+x)=\cos(n\pi+x)$ for all $n\in\Z$ and all $x$, it follows that all $2^{m+n}$
terms in the sum (11.11) are equal to the term corresponding to 
$(\epsilon_1,\epsilon_3\dotsc,\epsilon_{2m+2n-1})=(1,\dotsc,1)$, which is
$$
\align
\cos\left(\sum_{k=1}^m\frac{\alpha_k-\beta_k}{2}\pi
+\sum_{j=1}^n\frac{\gamma_j-\delta_j}{2}\pi\right)
&=(-1)^{\sum_{k=1}^m\frac{\alpha_k-\beta_k}{2}
+\sum_{j=1}^n\frac{\gamma_j-\delta_j}{2}}\\
&=\prod_{k=1}^m i^{\alpha_k}(-i)^{\beta_k}\prod_{j=1}^n i^{\gamma_j}(-i)^{\delta_j}.
\endalign
$$
Using this and substituting in (11.11) the expressions for $E$ and $D$ given by (11.5) and 
(11.10) we obtain (11.14).

Assume next that there is an index $k$ so that $\alpha_k$ and $\beta_k$ have opposite 
parity. Since
$\cos(-n\pi+x)=-\cos(n\pi+x)$ for all $n\in\frac12+\Z$ and all $x$, it follows that the
$\epsilon_{2k-1}=1$ part of the multiple sum in (11.11) cancels the $\epsilon_{2k-1}=-1$ part, so the
multiple sum is 0. The same argument applies if there is an index $j$ so that 
$\gamma_j$ and $\delta_j$ have opposite parity.~\endpf

\bigskip
By (11.12) and Lemma 11.2 we obtain
$$
\align
&\!\!\!\!\!\!\!\!\!\!\!\!\!\!
\omega_b\left(\matrix{R_1}\\{v_1}\endmatrix
\cdots\matrix{R_m}\\{v_m}\endmatrix;
\matrix{R'_1}\\{v'_1}\endmatrix\cdots\matrix{R'_n}\\{v'_n}\endmatrix\right)=XY\\
&\!\!\!\!\!\!\!\!\!\!\!\!\!\!
\times
\left|\sum_{C\in\Cal C} e(C)
\left<\prod_{k=1}^m\left(\frac{iR_k}{\sqrt{3}}\right)^{\alpha_k(C)}
\prod_{k=1}^m\left(\frac{-iR_k}{\sqrt{3}}\right)^{\beta_k(C)}
\prod_{j=1}^n\left(\frac{iR'_j}{\sqrt{3}}\right)^{\gamma_j(C)}
\prod_{j=1}^n\left(\frac{-iR'_j}{\sqrt{3}}\right)^{\delta_j(C)}\right>\right|\\
&\ \ \ \ \ \ \ \
+O(R^{2m^2+2n^2-4mn-2m-1}),
\tag11.16
\endalign
$$
where for any monomial $\mu=\prod_{k=1}^mx_k^{\alpha_k}y_k^{\beta_k}
\prod_{j=1}^nz_j^{\gamma_j}w_j^{\delta_j}$, the angular brackets denote
$\left<\mu\right>=\mu$ in case 
$\alpha_k=\beta_k(\mo 2)$, $\gamma_j=\delta_j(\mo 2)$, for all $k$ and $j$, and 
$\left<\mu\right>=0$ otherwise.

Note that if in (11.16) we had $\left<\mu\right>=\mu$ for {\it all} monomials $\mu$, 
the sum in (11.16) would evaluate to the specialization of the expression $\Cal E$ of (6.2)
when
$$
a_k=\frac{iR_k}{\sqrt{3}},\ b_k=-\frac{iR_k}{\sqrt{3}},\ 
c_j=\frac{iR'_j}{\sqrt{3}},\ d_j=-\frac{iR'_j}{\sqrt{3}},\tag11.17
$$
$k=1,\dotsc,m$, $j=1,\dotsc,n$. 

This observation, together with the fortunate situation presented in the result below, 
allows us to evaluate the sum (11.16).

\proclaim{Lemma 11.3} Under the substitutions $(11.17)$, all contributions to the expansion
of $\Cal E$ coming from monomials
$\mu_C=e(C)a_1^{\alpha_1(C)}b_1^{\beta_1(C)}\cdots a_m^{\alpha_m(C)}b_m^{\beta_m(C)}
c_1^{\gamma_1(C)}d_1^{\delta_1(C)}\cdots c_n^{\gamma_n(C)}d_n^{\delta_n(C)}$, $C\in\Cal C$,
for which not all pairs $(\alpha_k(C),\beta_k(C))$, $(\gamma_j(C),\delta_j(C))$
have components of the same parity, cancel out.
\endproclaim

\pf We define first the following map $C\mapsto C'$ on the subcollection $\Cal C_0$ 
consisting
of those $C\in\Cal C$ for which not all pairs $(\alpha_k(C),\beta_k(C))$, 
$(\gamma_j(C),\delta_j(C))$ have components of the same parity.

Totally order the disjoint union of the set of indices in the pairs 
$(\alpha_k(C),\beta_k(C))$ with the set of indices in the pairs 
$(\gamma_j(C),\delta_j(C))$.
Let $C\in\Cal C_0$, and consider the smallest pair index for which the components have
opposite parity. Assume that this smallest index $k_0$
occurs in a pair of the first type, $(\alpha_{k_0}(C),\beta_{k_0}(C))$. Our arguments
apply the same way to the case when
the smallest such index occurs in a pair of type $(\gamma_j(C),\delta_j(C))$.

By (6.2), $C$ is obtained by selecting a signed term from each of the 
$2m+2n+4{m\choose2}+4{n\choose2}$ factors of $\Cal E$. Define $C'$ as being obtained
by making the following selections:

\medskip
\flushpar
$(i)$ From the $m$ pairs of factors $(a_k-b_k)(a_k-b_k)$, $k=1,\dotsc,m$, 
make the selection as follows:

(a) if $l\neq k_0$, select in $C'$ the same signed terms as in $C$;

(b) if $k=k_0$, select in $C'$ the signed term of each factor of 
$(a_{k_0}-b_{k_0})(a_{k_0}-b_{k_0})$ that was {\it not} selected in $C$.

\smallpagebreak
\flushpar
$(ii)$ From the $n$ pairs of factors $(c_j-d_j)(c_j-d_j)$, $j=1,\dotsc,n$, 
select in $C'$ the same signed terms as in $C$. 

\smallpagebreak
\flushpar
$(iii)$ From the ${m\choose2}$ groups of four factors 
$$
((v_j-v_k)+a_j-a_k)((v_j-v_k)+a_j-b_k)((v_j-v_k)+b_j-a_k)((v_j-v_k)+b_j-b_k),
$$
$1\leq k<j\leq m$, make the selection as follows: 

(a) if $k_0\notin\{k,j\}$, make in $C'$ the same selection as in $C$;

(b) if $k=k_0$, and the selected terms in $C$ from the four factors
$$
((v_j-v_{k_0})+a_j-a_{k_0})((v_j-v_{k_0})+a_j-b_{k_0})((v_j-v_{k_0})+b_j-a_{k_0})
((v_j-v_{k_0})+b_j-b_{k_0})
$$
are the $k_1$th, $k_2$th, $k_3$th and $k_4$th, respectively, then select in $C'$ the
        $k_2$th, $k_1$th, $k_4$th and $k_3$th terms of the above factors, respectively;

(c) if $j=k_0$, and the selected terms in $C$ from the four factors
$$
((v_{k_0}-v_k)+a_{k_0}-a_k)((v_{k_0}-v_k)+a_{k_0}-b_k)((v_{k_0}-v_k)+b_{k_0}-a_k)
((v_{k_0}-v_k)+b_{k_0}-b_k)
$$
are the $k_1$th, $k_2$th, $k_3$th and $k_4$th, respectively, then select in $C'$ the
        $k_3$th, $k_4$th, $k_1$th and $k_2$th terms of the above factors, respectively.

\smallpagebreak
\flushpar
$(iv)$ From the final ${n\choose2}$ groups of four factors make the same selections in
$C'$ as in $C$.

\medskip
Let us now compare the monomials 
$$
\mu_C=e(C)\prod_{k=1}^ma_k^{\alpha_k(C)}b_k^{\beta_k(C)}
\prod_{j=1}^mc_j^{\gamma_j(C)}d_j^{\delta_j(C)}
$$ 
and 
$$
\mu_{C'}=e(C')\prod_{k=1}^ma_k^{\alpha_k(C')}b_k^{\beta_k(C')}
\prod_{j=1}^mc_j^{\gamma_j(C')}d_j^{\delta_j(C')}
$$ 
generated by the selections $C$ and $C'$.

It is clear from our construction that the portions of $C$ and $C'$ covered by step 
$(iii)$ produce contributions to $\mu_C$ and $\mu_{C'}$ whose only difference is that 
the roles of $a_{k_0}$ and $b_{k_0}$ are interchanged. 
 
The same is true for step $(i)$. Indeed, for $k=k_0$, to the four possible selections
$(a_{k_0})(a_{k_0})$, $(a_{k_0})(-b_{k_0})$, $(-b_{k_0})(a_{k_0})$, $(-b_{k_0})(-b_{k_0})$ 
in $C$ there correspond the four selections 
$(-b_{k_0})(-b_{k_0})$, $(-b_{k_0})(a_{k_0})$, $(a_{k_0})(-b_{k_0})$, $(a_{k_0})(a_{k_0})$
in $C'$, respectively. So the four possible contributions to $\mu_C$ and $\mu_{C'}$
are $a_{k_0}^2b_{k_0}^0$, $-a_{k_0}b_{k_0}$, $-a_{k_0}^2b_{k_0}$, $a_{k_0}^0b_{k_0}^2$ and
$a_{k_0}^0b_{k_0}^2$, $-a_{k_0}b_{k_0}$, $-a_{k_0}^2b_{k_0}$, $a_{k_0}^2b_{k_0}^0$, 
respectively.

Since steps $(ii)$ and $(iv)$ do not involve the index $k_0$, and the portions of 
$C$ and $C'$ covered by these steps coincide, it follows from the previous two paragraphs 
that 
$$
\align
&\mu_C(a_1,b_1,\dotsc,a_{k_0},b_{k_0},\dotsc,a_m,b_m,c_1,d_1,\dotsc,c_n,d_n)\\
&\ \ \ \ \ \ \ \ \ \ \ \ \ \ \ \ \ \ \ \ \ 
=
\mu_{C'}(a_1,b_1,\dotsc,b_{k_0},a_{k_0},\dotsc,a_m,b_m,c_1,d_1,\dotsc,c_n,d_n).
\endalign
$$
Since by
(11.17) $a_{k_0}$ and $b_{k_0}$ become the negatives of each other, and since 
$\alpha_{k_0}(C)$ and $\beta_{k_0}(C)$ have opposite parity, it follows that under
the specialization (11.17) the monomials $\mu_C$ and $\mu_{C'}$ cancel out.

Our map $C\mapsto C'$ clearly maps $\Cal C_0$ to itself, and sends $C'$ back to $C$. By the
previous paragraph, when making the specialization (11.17), all the monomials corresponding
to term selections in $\Cal C_0$ cancel out in pairs. This proves the Lemma. \epf

\proclaim{Corollary 11.4} We have 
$$
\align
&\!\!\!\!\!\!\!\!\!\!\!\!\!\!
\omega_b\left(\matrix{R_1}\\{v_1}\endmatrix
\cdots\matrix{R_m}\\{v_m}\endmatrix;
\matrix{R'_1}\\{v'_1}\endmatrix\cdots\matrix{R'_n}\\{v'_n}\endmatrix\right)=XY\\
&\!\!\!\!\!\!\!\!\!\!\!\!\!\!
\times
\left|\sum_{C\in\Cal C} e(C)
\prod_{k=1}^m\left(\frac{iR_k}{\sqrt{3}}\right)^{\alpha_k(C)}
\prod_{k=1}^m\left(\frac{-iR_k}{\sqrt{3}}\right)^{\beta_k(C)}
\prod_{j=1}^n\left(\frac{iR'_j}{\sqrt{3}}\right)^{\gamma_j(C)}
\prod_{j=1}^n\left(\frac{-iR'_j}{\sqrt{3}}\right)^{\delta_j(C)}\right|\\
&\ \ \ \ \ \ \ \
+O(R^{2m^2+2n^2-4mn-2m-1}),\tag11.18
\endalign
$$
where $X$ and $Y$ are given by $(11.13)$ and $(11.15)$.
\endproclaim

\pf This follows directly from (11.16) and Lemma 11.3. \endpf

Our results allow us now to obtain the asymptotics of $\omega_b$, thus proving Theorem 2.2.


\smallpagebreak
{\it Proof of Theorem 2.2.} By Corollary 11.4, it suffices to evaluate the sum in (11.18). 
However, by (6.2),
this sum is just the specialization (11.17) of the product $\Cal E$ in (6.2). The 
product of the first $2m+2n$ factors of $\Cal E$ specializes to
$$
\prod_{k=1}^m\left(\frac{2iR_k}{\sqrt{3}}\right)^2
\prod_{j=1}^n\left(\frac{2iR'_j}{\sqrt{3}}\right)^2
=\frac{(-4)^{m+n}}{3^{m+n}}\prod_{k=1}^mR_k^2\prod_{j=1}^n(R'_j)^2.\tag11.19
$$
The product of the next $4{m\choose2}$ factors of $\Cal E$ specializes to
$$
\align
\prod_{1\leq k<j\leq m}
&
\left\{
\left((v_j-v_k)+\left(\frac{iR_j}{\sqrt{3}}-\frac{iR_k}{\sqrt{3}}\right)\right)
\left((v_j-v_k)+\left(\frac{iR_j}{\sqrt{3}}+\frac{iR_k}{\sqrt{3}}\right)\right)
\right.
\\
\times
&
\left.
\left((v_j-v_k)+\left(-\frac{iR_j}{\sqrt{3}}-\frac{iR_k}{\sqrt{3}}\right)\right)
\left((v_j-v_k)+\left(-\frac{iR_j}{\sqrt{3}}+\frac{iR_k}{\sqrt{3}}\right)\right)
\right\}
\\
=
\prod_{1\leq k<j\leq m}&
\left[(v_j-v_k)^2+\frac13(R_j-R_k)^2\right]\left[(v_j-v_k)^2+\frac13(R_j+R_k)^2\right],\tag11.20
\endalign
$$
while the product of the last $4{n\choose2}$ factors of $\Cal E$ specializes to
$$
\align
\prod_{1\leq k<j\leq n}
&
\left\{
\left((v'_j-v'_k)+\left(\frac{iR'_j}{\sqrt{3}}-\frac{iR'_k}{\sqrt{3}}\right)\right)
\left((v'_j-v'_k)+\left(\frac{iR'_j}{\sqrt{3}}+\frac{iR'_k}{\sqrt{3}}\right)\right)
\right.
\\
\times
&
\left.
\left((v'_j-v'_k)+\left(-\frac{iR'_j}{\sqrt{3}}-\frac{iR'_k}{\sqrt{3}}\right)\right)
\left((v'_j-v'_k)+\left(-\frac{iR'_j}{\sqrt{3}}+\frac{iR'_k}{\sqrt{3}}\right)\right)
\right\}
\\
=
\prod_{1\leq k<j\leq n}&
\left[(v'_j-v'_k)^2+\frac13(R'_j-R'_k)^2\right]\left[(v'_j-v'_k)^2+\frac13(R'_j+R'_k)^2\right].\tag11.21
\endalign
$$
In the expression $Y$ given by (11.15) write, using (\TwoTwo),
$$
\sqrt{q_k^2+\frac13}=\frac1{R_k}\sqrt{(q_kR_k)^2+\frac13R_k^2}
=\frac1{R_k}\sqrt{(v_k-c_k)^2+\frac13R_k^2},
$$
$$
\sqrt{{q'_j}^2+\frac13}=\frac1{R'_j}\sqrt{(q'_jR'_j)^2+\frac13(R'_j)^2}
=\frac1{R'_j}\sqrt{(v'_j-c'_j)^2+\frac13(R'_j)^2},
$$
and
$$
\align
\left[(q_kA_k+q'_jB_j)^2+\frac13(A_k\pm B_j)^2\right]
&=\frac{1}{R^2}\left[(q_kA_kR+q'_jB_jR)^2+\frac13(A_kR\pm B_jR)^2\right]\\
&=\frac{1}{R^2}\left[(v_k+v'_j-c_k-c'_j)^2+\frac13(R_k\pm R'_j)^2\right].
\endalign
$$
Substituting the resulting expression for $Y$ and formula (11.13) for $X$ into (11.18), we obtain 
by Corollary 11.4 and (11.19)--(11.21) that
$$
\align
&\!\!\!\!
\omega_b\left(\matrix{R_1}\\{v_1}\endmatrix
\cdots\matrix{R_m}\\{v_m}\endmatrix;
\matrix{R'_1}\\{v'_1}\endmatrix\cdots\matrix{R'_n}\\{v'_n}\endmatrix\right)=
\chi_{2m,2n}\prod_{k=1}^mR_k\prod_{j=1}^nR'_j(R'_j-1/2)(R'_j+1/2)\\
&
\times
\frac{2^{m-n}3^n}{\pi^{m+n}}\frac{1}{\prod_{k=1}^mR_k^2\prod_{j=1}^n(R'_j)^4}
\frac{\prod_{j=1}^n \sqrt{(v'_j-c'_j)^2+\frac13(R'_j)^2}}
{\prod_{k=1}^m\sqrt{(v_k-c_k)^2+\frac13R_k^2}}\\
&
\times
\frac{1}{\prod_{k=1}^m\prod_{j=1}^n[(v_k+v'_j-c_k-c'_j)^2+\frac13(R_k- R'_j)^2]
[(v_k+v'_j-c_k-c'_j)^2+\frac13(R_k+ R'_j)^2]}\\
&
\times
\left|
\frac{(-4)^{m+n}}{3^{m+n}}\prod_{k=1}^mR_k^2\prod_{j=1}^n(R'_j)^2
\right.
\\
&
\times
\prod_{1\leq k<j\leq m}
\left[(v_j-v_k)^2+\frac13(R_j-R_k)^2\right]\left[(v_j-v_k)^2+\frac13(R_j+R_k)^2\right]\\
&
\left.
\times
\prod_{1\leq k<j\leq n}
\left[(v'_j-v'_k)^2+\frac13(R'_j-R'_k)^2\right]\left[(v'_j-v'_k)^2+\frac13(R'_j+R'_k)^2\right]
\right|
\\
&
+O(R^{2m^2+2n^2-4mn-2m-1})
.\tag11.22
\endalign
$$
By (\TwoTwo), the parameters $R_1,\dotsc,R_m$, $v_1,\dotsc,v_m$ and
$R'_1,\dotsc,R'_n$, $v'_1,\dotsc,v'_n$ approach infinity as $R\to\infty$, while
$c_1,\dotsc,c_m$ and $c'_1,\dotsc,c'_n$ are constant. It follows from this 
that the difference
between the product on the right hand side of (11.22) and what it becomes when one omits
the additive constants $\pm1/2$ on its first line and the constants $c_k$ 
and~$c'_j$ is $O(R^{2m^2+2n^2-4mn-2m-1})$. The latter is readily brought to the form 
given by (2.9) and (2.10). By the assumption on the distinctness of the pairs 
$(A_k,q_k)$, $k=1,\dotsc,m$, and $(B_j,q'_j)$, $j=1,\dotsc,n$ in (\TwoTwo), the expression
on the right hand side of (2.9) has degree
$2m^2+2n^2-4mn-2m$ in $R$. Therefore, (2.9) does indeed give the asymptotics of $\omega_b$.~\epf

\mysec{12. Another simple product formula for correlations along the boundary}

By Proposition 3.2, the asymptotics of the joint correlation at the center $\omega$ will follow
provided we also work out the asymptotics of the boundary-influenced correlation 
$\bar{\omega}_b$ defined by (3.3). We present this in Section 13. But first we need an analog of
Proposition 4.1 corresponding to the regions $E_N$ used to define $\bar{\omega}_b$.

Let $E$ be the region determined by the common outside boundary of the regions
$E_N\left(\matrix{R_1}\\{v_1}\endmatrix
\cdots\matrix{R_m}\\{v_m}\endmatrix;
\matrix{R'_1}\\{v'_1}\endmatrix\cdots\matrix{R'_n}\\{v'_n}\endmatrix\right)$ defined in 
Section 3, for
fixed $m$, $n$ and $N$. Then $E$ is the half-hexagonal lattice region with  
four straight sides---the southern side of length $N+2n+1$, 
southeastern of length $2N+4m-1$, northeastern of length $2N+4n$, and northern 
of length $N+2m$---followed by
$N+2n$ descending zig-zags to the lattice point $O$, two extra unit steps southeast of
$O$ followed by one step west, and $N+2m-1$ more descending zig-zags to close up the boundary 
(an example can be seen in Figure 12.1).
In addition, 
the $N+2m-1$ dimer positions weighted 1/2 in the regions $E_N$ are also weighted so in
$E$.

\topinsert
\twoline{\mypic{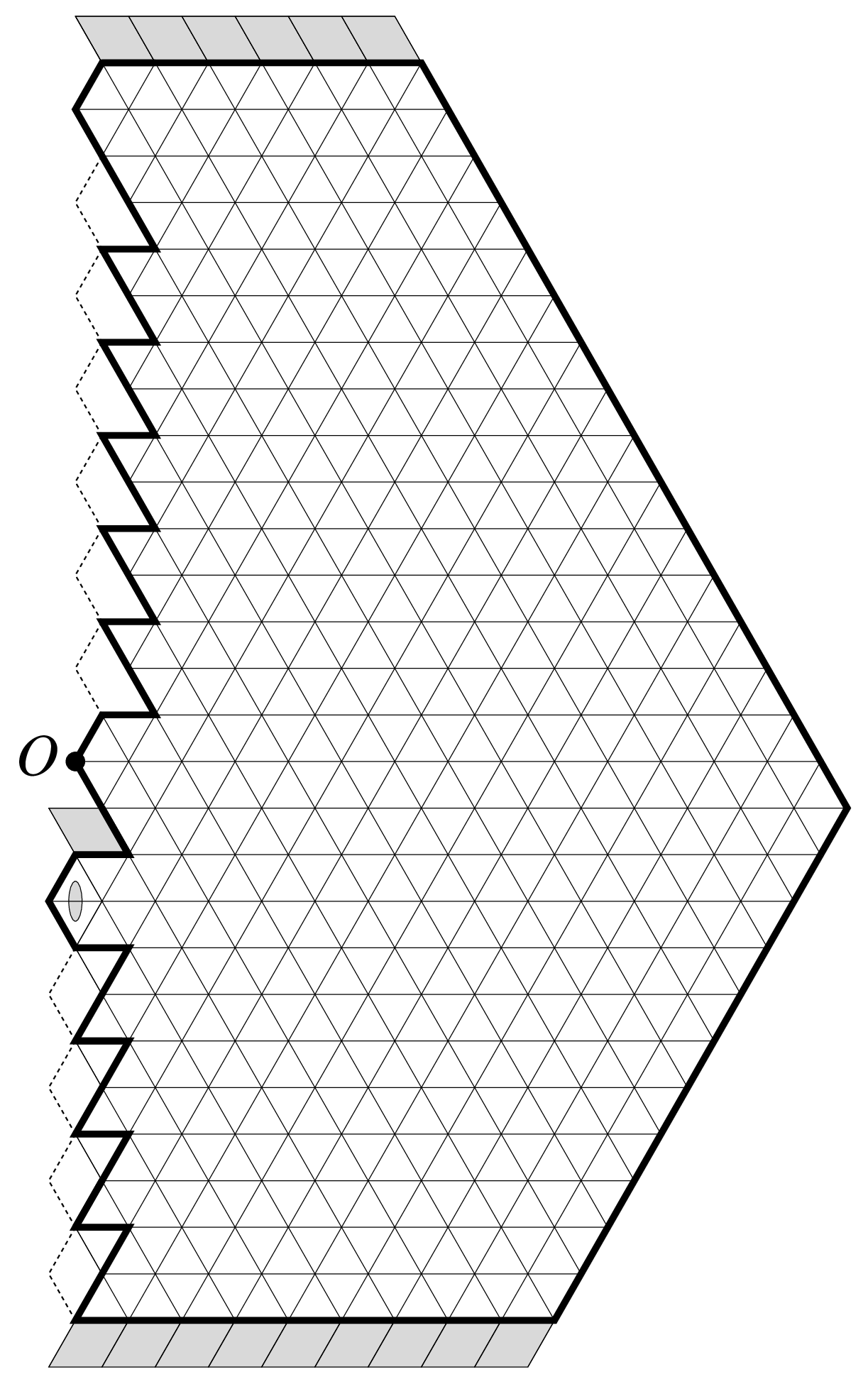}}{\mypic{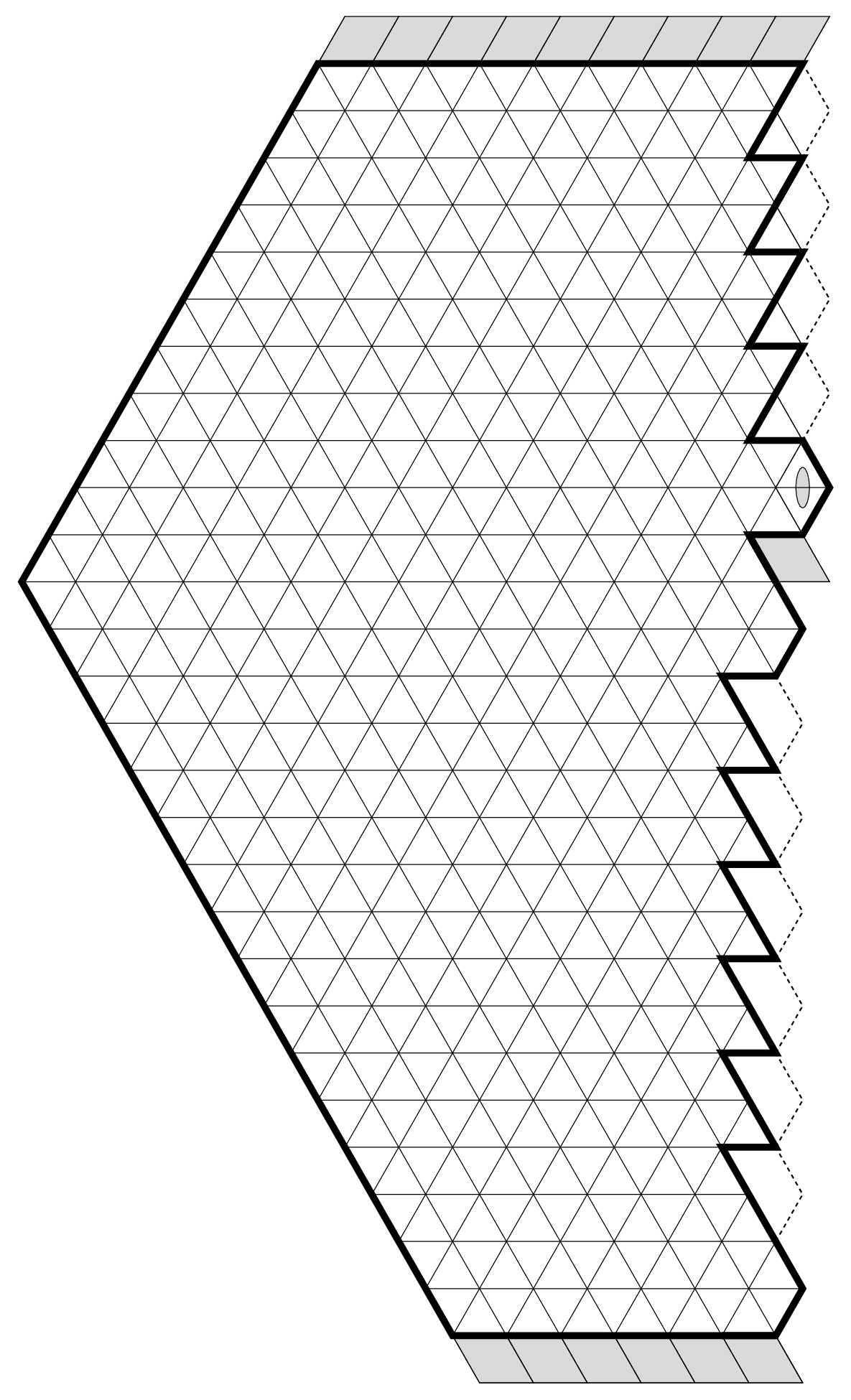}}
\twoline{Figure~12.1. {\rm $E_2[1,2,3,4;1,2,3,4,5,6]$.}}
{Figure~12.2. {\rm Region denoted }}
\twoline{\ \ \ \ \ \ \ \ \ \ \ \ \ \ \ \ \ \ \ \ \ \ \ \ \ \ \ \ \ \ \ \ \ \ \ \ \ \ \ \ \ \ \ \ }
{\rm $R_{[2,3,4,5,6,7][2,3,4,5]}(6)$ in \cite{\Cthree}.}
\endinsert

As in the case of 
the regions $W$ of Section 3, the vertical, jagged boundary of $E$ can be viewed as 
consisting of bumps---in the present case, pairs of adjacent
lattice segments forming an angle that opens to the {\it east}: $N+2m-1$ bumps below $O$, and
$N+2n$ above $O$. Label the former by $0,1,\dotsc,N+2m-2$ and the latter by 
$0,1,\dotsc,N+2n-1$, both labelings starting with the bumps closest to $O$ and then moving
successively outwards. 

In the description of $E$ the parameters $m$
and $n$ always appear with even coefficients. We re-denote, as in Section 3, $2m$ by $m$
and $2n$ by $n$, for notational simplicity. 
Therefore we consider the four straight sides of $E$ to have lengths 
$N+n$, $2N+2m-1$, $2N+2n$ and $N+m$, while the number of bumps below and above $O$ 
is $N+m-1$ and $N+n$, respectively. 

We allow removal of any bump exactly like in Section 3: above $O$, place an up-pointing 
quadromer across it and discard the three monomers of $E$ it covers; below $O$, use
down-pointing quadromers.

Define $E_N[k_1,\dotsc,k_{m};l_1,\dotsc,l_{n}]$ to be the region 
obtained from $E$ by removing the bumps below $O$ with labels 
$0\leq k_1<k_2<\cdots<k_{m}\leq N+m-2$, and the bumps above $O$ with labels 
$0\leq l_1<l_2<\cdots<l_{n}\leq N+n-1$. Figure 12.1 shows $E_2[1,2,3,4;1,2,3,4,5,6]$.

The correlations of the removed bumps on the boundary of the regions $E_N$ turn out to be
{\it exactly} the same as the corresponding correlations we found in Section 3
for the regions~$W_N$.

\proclaim{Proposition 12.1} For $m,n\geq0$ and fixed integers 
$0\leq k_1<k_2<\cdots<k_{m}$ and $0\leq l_1<l_2<\cdots<l_{n}$ we have
$$
\align
\lim_{N\to\infty}&\frac{\M\left(E_N[k_1,\dotsc,k_{m};l_1,\dotsc,l_{n}]\right)}
{\M\left(E_N[0,\dotsc,m-1;0,\dotsc,n-1]\right)}=\\
=
&
\chi_{m,n}
\prod_{i=1}^{m}\frac{(3/2)_{k_i}}{(2)_{k_i}}
\prod_{i=1}^{n}\frac{(3/2)_{l_i}}{(1)_{l_i}}
\frac{
{\displaystyle \prod_{1\leq i<j\leq m}}(k_j-k_i)
{\displaystyle \prod_{1\leq i<j\leq n}}(l_j-l_i)}
{{\displaystyle \prod_{i=1}^{m}\prod_{i=1}^{n}}(k_i+l_j+2)},\tag12.1
\endalign
$$
where $\chi_{m,n}$ is given by $(4.3)$. 
\endproclaim

\pf We proceed in analogy to the proof of Proposition 4.1. The results of \cite{\Cthree} provide
an explicit formula for $\M(E_N[k_1,\dotsc,k_{m};l_1,\dotsc,l_{n}])$ as well. Indeed, in
the notation of \cite{\Cthree} we have
$$
\align
&E_N[k_1,\dotsc,k_{m};l_1,\dotsc,l_{n}]=\\
&\ \ \ \ \ \ 
R_{[1,\dotsc,N+n]\setminus[l_1+1,\dotsc,l_{n}+1],
[1,\dotsc,N-1+m]\setminus[k_1+1,\dotsc,k_{m}+1]}(N+m)\tag12.2
\endalign
$$
(this is illustrated by Figure 12.2, which is just the $180^\circ$ rotation of Figure 12.1).

Proposition 2.1 of \cite{\Cthree} and formulas \cite{\Cthree,(1.1),\, (1.3),\, (1.5)} provide an 
explicit formula for $\M(R_{\bold p,\bold q}(x))$, for any pair of lists 
$\bold p=[p_1,\dotsc,p_s]$, $1\leq p_1<\cdots<p_s$ and $\bold q=[q_1,\dotsc,q_t]$, 
$1\leq q_1<\cdots<q_t$, and any nonnegative integer $x\geq q_t-p_s-t+s-1$. 

Written as a constant times a monic polynomial in $x$, this formula takes the form
$$
\M\left(R_{\bold p,\bold q}(x)\right)=
c_{\bold p,\bold q}G_{{\bold p},{\bold q}}(x),\tag12.3
$$ 
where by \cite{\Cthree,(1.3)} we have
$$
c_{\bold p,\bold q}=2^{{t-s\choose 2}-s}\prod_{i=1}^s\frac{1}{(2p_i)!}
\prod_{i=1}^t\frac{1}{(2q_i-1)!}
\frac{
{\prod_{1\leq i<j\leq s}}(p_j-p_i)
{\prod_{1\leq i<j\leq t}}(q_j-q_i)}
{{\prod_{i=1}^{s}\prod_{i=1}^{t}}(p_i+q_j)}.\tag12.4
$$
Furthermore, it is straightforward to check, using (1.1) and (1.5), that the 
polynomials $G_{{\bold p},{\bold q}}(x)$ satisfy 
$$
\align
\frac{G_{{\bold p}^{|i\rangle},{\bold q}}(x)}{G_{{\bold p},{\bold q}}(x)}
&=(x-p_i+p_s)(x+p_i+p_s-s+t+2),\ \ \ \ \text{\rm for $1\leq i<s$}\tag12.5\\
\frac{G_{{\bold p},{\bold q}^{|i\rangle}}(x)}{G_{{\bold p},{\bold q}}(x)}
&=(x+q_i+p_s+1)(x-q_i+p_s-s+t+1),\ \ \ \ \text{\rm for $1\leq i\leq t$}\tag12.6
\endalign
$$
(as in Section 4, ${\bold p}^{|i\rangle}$ denotes the list obtained from ${\bold p}$ by increasing its 
$i$-th element by 1, and is defined only if $l_{i+1}-l_i\geq2$).

Consider first the limit
$$
\lim_{N\to\infty}\frac{\M\left(E_N[k_1,\dotsc,k_{i-1},k_i+1,k_{i+1},\dotsc,k_{m};
l_1,\dotsc,l_{n}]\right)}
{\M\left(E_N[k_1,\dotsc,k_{i-1},k_i,k_{i+1},\dotsc,k_{m};
l_1,\dotsc,l_{n}]\right)},\tag12.7
$$
for $k_i+1<k_{i+1}$.

Use (12.2) to view the regions involved in this fraction as $R_{{\bold p},{\bold q}}(x)$'s. 
Comparing formulas (12.4) and (4.6), and using ${s-t\choose2}-t={t-s\choose2}-s$, one sees
that
$$
c_{\bold p,\bold q}=\bar{c}_{\bold q,\bold p}.
$$
For the lists on the right hand side of (12.2) this implies 
$$
\align
&c_{[1,\dotsc,N+n]\setminus[l_1+1,\dotsc,l_{n}+1],
[1,\dotsc,N-1+m]\setminus[k_1+1,\dotsc,k_{m}+1]}\\
&\ \ \ \ \ \ 
=
\bar{c}_{[1,\dotsc,N-1+m]\setminus[k_1+1,\dotsc,k_{m}+1],
[1,\dotsc,N+n]\setminus[l_1+1,\dotsc,l_{n}+1]}.\tag12.8
\endalign
$$
Therefore, the contribution to the fraction in (12.7) coming (via (12.2)) from the 
$c_{\bold p,\bold q}$-parts of (12.3) follows from our work in Section 4. The only difference
from that case is that in the $\bar{c}_{\bold p,\bold q}$ of (12.8) the largest element of 
${\bold p}$ is now $(N-1)+m$, as opposed to $N+m$. However, substitution of $N$ by $N-1$ in 
(4.10) makes no difference in the limit $N\to\infty$, so the contribution of the 
$c_{\bold p,\bold q}$-parts to the limit (12.7) is, asymptotically for $N\to\infty$, 
precisely the same as (4.10).

On the other hand, one readily sees by (12.2) and (12.6) that the contribution of the 
$G_{\bold p,\bold q}(x)$-parts to the fraction in (12.7) is
$$
\frac{1}{(2N+m+n+k_i+2)(2N+m+n-k_i-1)},
$$
a fraction whose asymptotics as $N\to\infty$ is clearly the same as that of (4.11).

It follows from the above two paragraphs that the limit (12.7) is equal to the product on the 
right hand side of (4.12), so it has exactly the same expression as in the case of the 
regions $W_N$ treated in Section 4.

A similar analysis shows that 
$$
\align
&\lim_{N\to\infty}\frac{\M\left(E_N[k_1,\dotsc,k_{m};
l_1,\dotsc,l_{i-1},l_i+1,l_{i+1},\dotsc,l_{n}]\right)}
{\M\left(E_N[k_1,\dotsc,k_{m};
l_1,\dotsc,l_{i-1},l_i,l_{i+1},\dotsc,l_{n}]\right)}\\
&\ \ \ \ \ \ 
=
\lim_{N\to\infty}\frac{\M\left(W_N[k_1,\dotsc,k_{m};
l_1,\dotsc,l_{i-1},l_i+1,l_{i+1},\dotsc,l_{n}]\right)}
{\M\left(W_N[k_1,\dotsc,k_{m};
l_1,\dotsc,l_{i-1},l_i,l_{i+1},\dotsc,l_{n}]\right)},
\endalign
$$
so that also decrementing a single element from the second index list has exactly the same effect
for the regions $E_N$ as for the regions $W_N$ of Section 4. Formula (12.1) follows therefore
by Proposition 4.1. \epf

\mysec{13. The asymptotics of $\bar{\omega}_b$. Proof of Theorem 2.1}


Our reasoning from Section 5 applies with no change to the regions $E_N$ defined in Section 3.
Each missing
quadromer can be placed back at the expense of performing a Laplace expansion in the 
corresponding Gessel-Viennot determinant. The resulting $2\times2$ cofactors have precisely 
the forms (5.5) and (5.13). The analog of (5.19) we get is
$$
\align
&\M\left(E_N\left(\matrix{R_1}\\{v_1}\endmatrix\cdots\matrix{R_m}\\{v_m}\endmatrix;
\matrix{R'_1}\\{v'_1}\endmatrix\cdots\matrix{R'_n}\\{v'_n}\endmatrix\right)\right)=
2^{m+n}\prod_{i=1}^mR_i(R_i-1/2)(R_i+1/2)\prod_{i=1}^nR'_i\\
&\ \  
\times
\left|
\sum_{0\leq a_1<b_1\leq R_1}\cdots\sum_{0\leq a_m<b_m\leq R_m}
\sum_{0\leq c_1<d_1\leq R'_1}\cdots\sum_{0\leq c_n<d_n\leq R'_n}
(-1)^{\sum_{i=1}^m (a_i+b_i) +\sum_{i=1}^n (c_i+d_i)}
\right.
\\
&\ \ 
\times
\prod_{i=1}^m
\frac{(b_i-a_i)(R_i+a_i-1)!\,(R_i+b_i-1)!}{(2a_i+1)!\,(R_i-a_i)!\,(2b_i+1)!\,(R_i-b_i)!}\\
&\ \ 
\times
\prod_{i=1}^n
\frac{(d_i-c_i)(R'_i+c_i-1)!\,(R'_i+d_i-1)!}{(2c_i)!\,(R'_i-c_i)!\,(2d_i)!\,(R'_i-d_i)!}\\
&\ \ 
\times
\sign(v'_1+c_1,v'_1+d_1,\dotsc,v'_n+c_n,v'_n+d_n)
\sign(v_1+a_1,v_1+b_1,\dotsc,v_m+a_m,v_m+b_m)\\
&\ \ 
\left.
\times
\M(E_N[\{v_1+a_1,\dotsc,v_m+b_m\}_{<};
\{v'_1+c_1,\dotsc,v'_n+d_n\}_{<}])
\right|,
\tag13.1
\endalign
$$
where the summation extends only over those summation indices for which both lists of 
arguments in the region on the right hand side have distinct elements, and $\sign$ denotes
the sign of such a list when regarded as a permutation.

After removing the forced dimers, the normalizing region at the denominator of (3.3),
$$
E_N\left(\matrix{1}&3\\{0}&0\endmatrix\cdots\matrix{2m-1}\\{0}\endmatrix;
\matrix{1}&3\\{0}&0\endmatrix\cdots\matrix{2n-1}\\{0}\endmatrix\right),\tag13.2
$$ 
is readily seen to be precisely $E_N[0,1,\dotsc,2m-1;0,1,\dotsc,2n-1]$.

Moreover, by the sameness of the right hand sides of (12.1) and (4.2), (5.21) implies
$$
\align
\lim_{N\to\infty}&\frac{\M\left(E_N[\{v_1+a_1,\dotsc,v_m+b_m\}_{<};
\{v'_1+c_1,\dotsc,v'_n+d_n\}_{<}]\right)}
{\M\left(E_N[0,\dotsc,2m-1;0,\dotsc,2n-1]\right)}=\chi_{2m,2n}\\
&\!\!\!\!\!\!\!\!\!\!\!\!
\times
\sign(v_1+a_1,v_1+b_1,\dotsc,v_m+a_m,v_m+b_m)
\sign(v'_1+c_1,v'_1+d_1,\dotsc,v'_n+c_n,v'_n+d_n)\\
\times
&\prod_{i=1}^m\frac{(3/2)_{v_i+a_i}\,(3/2)_{v_i+b_i}}
{(2)_{v_i+a_i}\,(2)_{v_i+b_i}}
\prod_{i=1}^n\frac{(3/2)_{v'_i+c_i}\,(3/2)_{v'_i+d_i}}
{(1)_{v'_i+c_i}\,(1)_{v'_i+d_i}}\\
\times
&\prod_{1\leq i<j\leq m}(v_j-v_i+a_j-a_i)(v_j-v_i+a_j-b_i)(v_j-v_i+b_j-a_i)
(v_j-v_i+b_j-b_i)\\
&\times\prod_{1\leq i<j\leq n}(v'_j-v'_i+c_j-c_i)(v'_j-v'_i+c_j-d_i)(v'_j-v'_i+d_j-c_i)
(v'_j-v'_i+d_j-d_i)\\
&
\times\frac{\prod_{i=1}^m(a_i-b_i)\prod_{i=1}^n(c_i-d_i)}
{\prod_{i=1}^m\prod_{j=1}^n(u_{ij}+a_i+c_j)(u_{ij}+a_i+d_j)
(u_{ij}+b_i+c_j)(u_{ij}+b_i+d_j)  },\tag13.3
\endalign
$$
where $\chi$ is given by (4.3) and $u_{ij}=v_i+v'_j+2$, $i=1,\dotsc,m$, $j=1,\dotsc,n$.

Divide (13.1) by the matching generation function of the region (13.2) and
let $N\to\infty$. By (3.3) and (13.3) we obtain
$$
\align
&\!\!\!\!\!\!\!\!\!\!\!
\bar{\omega}_b\left(\matrix{R_1}\\{v_1}\endmatrix
\cdots\matrix{R_m}\\{v_m}\endmatrix;
\matrix{R'_1}\\{v'_1}\endmatrix\cdots\matrix{R'_n}\\{v'_n}\endmatrix\right)
=2^{m+n}\chi_{m,n}\prod_{i=1}^mR_i(R_i-1/2)(R_i+1/2)
\prod_{i=1}^nR'_i
\\
&\!\!\!\!\!\!\!\!\!\!\!\!
\times
\left|
\sum_{0\leq a_1<b_1\leq R_1}\cdots\sum_{0\leq a_m<b_m\leq R_m}
\sum_{0\leq c_1<d_1\leq R'_1}\cdots\sum_{0\leq c_n<d_n\leq R'_n}
(-1)^{\sum_{i=1}^m (a_i+b_i) +\sum_{i=1}^n (c_i+d_i)}
\right.
\\
&\times\prod_{i=1}^m\frac{(R_i+a_i-1)!\,(R_i+b_i-1)!}
{(2a_i+1)!\,(R_i-a_i)!\,(2b_i+1)!\,(R_i-b_i)!}
\\
&\times\prod_{i=1}^n\frac{(R_i+c_i-1)!\,(R_i+d_i-1)!}
{(2c_i)!\,(R_i-c_i)!\,(2d_i)!\,(R_i-d_i)!}
\\
&\times\prod_{i=1}^m\frac{(3/2)_{v_i+a_i}\,(3/2)_{v_i+b_i}}
{(2)_{v_i+a_i}\,(2)_{v_i+b_i}}
\prod_{i=1}^n\frac{(3/2)_{v'_i+c_i}\,(3/2)_{v'_i+d_i}}
{(1)_{v'_i+c_i}\,(1)_{v'_i+d_i}}
\endalign
$$
\vskip-0.1in
$$
\align
\times
\prod_{1\leq i<j\leq m}&(v_j-v_i+a_j-a_i)(v_j-v_i+a_j-b_i)(v_j-v_i+b_j-a_i)(v_j-v_i+b_j-b_i)\\
\times\prod_{1\leq i<j\leq
n}&(v'_j-v'_i+c_j-c_i)(v'_j-v'_i+c_j-d_i)(v'_j-v'_i+d_j-c_i)(v'_j-v'_i+d_j-d_i)\\
&\!\!\!\!\!\!\!\!\!\!\!\!\!\!\!\!\!\!\!\!\!\!\!\!\!\!\!\!
\left.
\times\frac{\prod_{i=1}^m(a_i-b_i)^2\prod_{i=1}^n(c_i-d_i)^2}
{\prod_{i=1}^m\prod_{j=1}^n(u_{ij}+a_i+c_j)(u_{ij}+a_i+d_j)
(u_{ij}+b_i+c_j)(u_{ij}+b_i+d_j)  }
\right|,
\tag13.4
\endalign
$$
where the summation range is restricted to those summation variables for which
$v_1+a_1,v_1+b_1,\dotsc,v_m+a_m,v_m+b_m$,
as well as $v'_1+c_1,v'_1+d_1,\dotsc,v'_n+c_n,v'_n+d_n$, are distinct.

By the argument in the last paragraph of the proof of Lemma 5.1, the restrictions $a_i<b_i$,
$i=1,\dotsc,m$, and $c_i<d_i$, $i=1,\dotsc,n$ can be dropped at the expense of a multiplicative 
factor of $1/2^{m+n}$. We obtain the following result.

\proclaim{Lemma 13.1} For fixed $R_1,\dotsc,R_m,R'_1,\dotsc,R'_n\geq1$ and
$v_1,\dotsc,v_m,v'_1,\dotsc,v'_n\geq0$ we have
$$
\align
&\!\!\!\!\!\!\!\!\!\!\!\!\!\!\!\!\!\!\!\!\!\!\!\!
\bar{\omega}_b\left(\matrix{R_1}\\{v_1}\endmatrix
\cdots\matrix{R_m}\\{v_m}\endmatrix;
\matrix{R'_1}\\{v'_1}\endmatrix\cdots\matrix{R'_n}\\{v'_n}\endmatrix\right)
=\chi_{2m,2n}\prod_{i=1}^mR_i(R_i-1/2)(R_i+1/2)
\prod_{i=1}^nR'_i
\\
&\!\!\!\!\!\!\!\!\!\!\!\!\!\!\!\!\!\!\!\!\!\!\!\!\!\!\!\!\!\!\!\!\!\!\!\!
\times
\left|
\sum_{a_1,b_1=0}^{R_1}\cdots\sum_{a_m,b_m=0}^{R_m}
\sum_{c_1,d_1=0}^{R'_1}\cdots\sum_{c_n,d_n=0}^{R'_n}
(-1)^{\sum_{i=1}^m (a_i+b_i) +\sum_{i=1}^n (c_i+d_i)}
\right.
\\
&\times\prod_{i=1}^m\frac{(R_i+a_i-1)!\,(R_i+b_i-1)!}
{(2a_i+1)!\,(R_i-a_i)!\,(2b_i+1)!\,(R_i-b_i)!}
\frac{(3/2)_{v_i+a_i}\,(3/2)_{v_i+b_i}}{(2)_{v_i+a_i}\,(2)_{v_i+b_i}}
\\
&\times\prod_{i=1}^n\frac{(R_i+c_i-1)!\,(R_i+d_i-1)!}
{(2c_i)!\,(R_i-c_i)!\,(2d_i)!\,(R_i-d_i)!}
\frac{(3/2)_{v'_i+c_i}\,(3/2)_{v'_i+d_i}}{(1)_{v'_i+c_i}\,(1)_{v'_i+d_i}}
\\
\times
\prod_{1\leq i<j\leq m}&(v_j-v_i+a_j-a_i)(v_j-v_i+a_j-b_i)(v_j-v_i+b_j-a_i)(v_j-v_i+b_j-b_i)\\
\times\prod_{1\leq i<j\leq
n}&(v'_j-v'_i+c_j-c_i)(v'_j-v'_i+c_j-d_i)(v'_j-v'_i+d_j-c_i)(v'_j-v'_i+d_j-d_i)\\
&\!\!\!\!\!\!\!\!\!\!\!\!\!\!\!\!\!\!\!\!\!\!\!\!\!\!\!\!
\left.
\times\frac{\prod_{i=1}^m(a_i-b_i)^2\prod_{i=1}^n(c_i-d_i)^2}
{\prod_{i=1}^m\prod_{j=1}^n(u_{ij}+a_i+c_j)(u_{ij}+a_i+d_j)
(u_{ij}+b_i+c_j)(u_{ij}+b_i+d_j)  }
\right|,\tag13.5
\endalign
$$
where 
$$
u_{ij}=v_i+v'_j+2
$$
for $i=1,\dotsc,m$, $j=1,\dotsc,n$, and $\chi$ is given by $(4.3)$.
\endproclaim

Note that expressions (13.5) and (5.1) are nearly identical. There are only two differences: 
first, the roles of $R_i$ and $R'_i$ are interchanged in the products on the first lines of 
their right hand sides; and second, in
the denominators on the third and fourth lines of (13.5) one has the expressions $(2a_i+1)!$, 
$(2b_i+1)!$, $(2c_i)!$ and $(2d_i)!$, while in the corresponding positions in (5.1) one has
$(2a_i)!$, $(2b_i)!$, $(2c_i+1)!$ and $(2d_i+1)!$, respectively.

The fact that these are the only differences allows us to obtain the asymptotics of 
$\bar{\omega}_b$ from the asymptotics of $\omega_b$ worked out in Sections 6--12. 

The arguments of Section 6 apply equally well to yield an expression for $\bar{\omega}_b$
analogous to the expression (6.5) for $\omega_b$. 
By the the first two formulas in the proof of Proposition 6.1, besides interchanging the roles
of $R_i$ and $R'_i$ in the products on the first line of (5.1), 
the only effect of the difference between (13.5) and (5.1) is that the 
fractions $1/2$ and $3/2$ at the denominators of (6.7) and (6.8) swap places
for the sums $\bar{T}^{(n)}$ and ${\bar{T'}}^{(n)}$---the analogs of $T^{(n)}$ and ${T'}^{(n)}$.
More precisely, define
$$
\align
\bar{T}^{(n)}(R,v;x)&:=\frac{1}{R}\sum_{a=0}^R \frac{(-R)_a\,(R)_a\,(3/2)_{v+a}}
{(1)_a\,(3/2)_a\,(2)_{v+a}}\left(\frac{x}{4}\right)^a a^n\tag13.6\\
{\bar{T'}}^{(n)}(R,v;x)&:=\frac{1}{R}\sum_{c=0}^R \frac{(-R)_c\,(R)_c\,(3/2)_{v+c}}
{(1)_c\,(1/2)_c\,(1)_{v+c}}\left(\frac{x}{4}\right)^c c^n
.\tag13.7
\endalign
$$
Then by the arguments that proved Proposition 6.1 we obtain the following result.

\proclaim{Proposition 13.2} The boundary-influenced correlation $\bar{\omega}_b$ can be written 
as
$$
\align
&\bar{\omega}_b\left(\matrix{R_1}\\{v_1}\endmatrix
\cdots\matrix{R_m}\\{v_m}\endmatrix;
\matrix{R'_1}\\{v'_1}\endmatrix\cdots\matrix{R'_n}\\{v'_n}\endmatrix\right)
=\chi_{2m,2n}\prod_{i=1}^mR_i(R_i-1/2)(R_i+1/2)
\prod_{i=1}^nR'_i
\\
&\ \ \ \ \ \ \ \ \ \ \ \ 
\times
\left|\sum_{C\in\Cal C} e(C)
\bar{M}_{\alpha_1(C),\beta_1(C),\dotsc,\alpha_m(C),\beta_m(C);
\gamma_1(C),\delta_1(C),\dotsc,\gamma_n(C),\delta_n(C)}\right|,
\tag13.8
\endalign
$$
where $\chi$ is given by $(4.3)$, the collection $\Cal C$ and
$e(C)$, $\alpha_i(C)$, $\beta_i(C)$, $\gamma_j(C)$, $\delta_j(C)$
are as in $(6.2)$, and the ``moment'' 
$\bar{M}_{\alpha_1,\beta_1,\dotsc,\alpha_m,\beta_m;\gamma_1,\delta_1,\dotsc,\gamma_n,\delta_n}$
equals the $4mn$-fold integral
$$
\align
&\!\!\!\!\!\!\!\!\!\!\!\
\bar{M}_{\alpha_1,\beta_1,\dotsc,\alpha_m,\beta_m;\gamma_1,\delta_1,\dotsc,\gamma_n,\delta_n}=
\int_0^1\cdots\int_0^1 \prod_{i=1}^m\prod_{j=1}^n(x_{ij}y_{ij}z_{ij}w_{ij})^{v_i+v'_j+1}\\
&
\times
\bar{T}^{(\alpha_1)}(R_1,v_1;\prod_{j=1}^n x_{1j}y_{1j})\,
\bar{T}^{(\beta_1)}(R_1,v_1;\prod_{j=1}^n z_{1j}w_{1j})
\cdots\\
&\ \ \ \ \ \ \ \ \ \ \ \ 
\cdots
\bar{T}^{(\alpha_m)}(R_m,v_m;\prod_{j=1}^n x_{mj}y_{mj})\,
\bar{T}^{(\beta_m)}(R_m,v_m;\prod_{j=1}^n z_{mj}w_{mj})\\
&
\times
{\bar{T'}}^{(\gamma_1)}(R'_1,v'_1;\prod_{i=1}^m x_{i1}z_{i1})\,
{\bar{T'}}^{(\delta_1)}(R'_1,v'_1;\prod_{i=1}^m y_{i1}w_{i1})
\cdots\\
&\ \ \ \ \ \ \ \ \ \ \ \ 
\cdots
{\bar{T'}}^{(\gamma_n)}(R'_n,v'_n;\prod_{i=1}^m x_{in}z_{in})\,
{\bar{T'}}^{(\delta_n)}(R'_n,v'_n;\prod_{i=1}^m y_{in}w_{in})\,dx_{11}\cdots dw_{mn},\tag13.9
\endalign
$$
with $\bar{T}^{(n)}(R,v;x)$ and ${\bar{T'}}^{(n)}(R,v;x)$ defined by $(13.6)$ and $(13.7)$.
\endproclaim

Furthermore, the calculations that proved Lemma 6.2 show that
$$
\align
&\!\!\!\!\!\!\!\!\!\!\!\!
\bar{T}^{(n)}(R,v;x)=\\
&\!\!\!\!\!\!\!\!\!\!
\frac{1}{R}\frac{(3/2)_v}{(2)_v}
\sum_{k=0}^n f_k\frac{(-R)_k\,(R)_k\,(v+3/2)_{k}}{(3/2)_k\,(v+2)_{k}}\left(\frac{x}{4}\right)^{k}
{}_3F_2\!\left[\matrix{-R+k,\,R+k,\,\frac{3}{2}+v+k}\\{\frac32+k,2+v+k}\endmatrix;\frac x4\right]
\tag13.10\\
&\!\!\!\!\!\!\!\!\!\!\!\!
{\bar{T'}}^{(n)}(R,v;x)=\\
&\!\!\!\!\!\!\!\!\!\!
\frac{1}{R}\frac{(3/2)_v}{(1)_v}
\sum_{k=0}^n f_k\frac{(-R)_k\,(R)_k\,(v+3/2)_{k}}{(1/2)_k\,(v+1)_{k}}\left(\frac{x}{4}\right)^{k}
{}_3F_2\!\left[\matrix{-R+k,\,R+k,\,\frac{3}{2}+v+k}\\{\frac12+k,1+v+k}\endmatrix;\frac x4\right],
\tag13.11
\endalign
$$
where the $f_k$'s are as in $(6.9)$ $($in particular $f_n=1$$)$. These expressions are nearly
the same as (6.10) and (6.11): the only difference is that, compared to $T^{(n)}$ and 
${T'}^{(n)}$, the fractions $1/2$ and $3/2$ at the denominators (including the denominator 
parameters of the~${}_3F_2$'s) swap places in the ``barred'' 
versions $\bar{T}^{(n)}$ and $\bar{T'}^{(n)}$.

Using (7.8) to write the ${}_3F_2$'s of (13.10) as integrals of ${}_2F_1$'s, the derivation
rule for~${}_2F_1$'s displayed after (7.9), and the ${}_2F_1$ evaluation (7.21), we obtain 
after simplifications
$$
\align
&\bar{T}^{(n)}(R,qR+c;x)=
\frac{2}{\pi R}\sum_{k=0}^n f_k
\int_0^1 t^{qR+c+k+1/2}(1-t)^{-1/2}\\
&\ \ \ \ \ \ \ \ \ 
\times
\frac{d^k}{dt^k}
\left\{
\frac{2R}{4R^2-1}\sqrt{\frac{4-xt}{xt}}
\sin\left[R\arccos\left(1-\frac{xt}{2}\right)\right]
\right.
\\
&\ \ \ \ \ \ \ \ \ \ \ \ \ \ \ \ \ \ \ \ \ \ \ \ \ \ \ 
\left.
-\frac{1}{4R^2-1}\cos\left[R\arccos\left(1-\frac{xt}{2}\right)\right]
\right\}
dt.
\tag13.12
\endalign
$$
To obtain a similar expression for $\bar{T'}^{(n)}$ we need first an analog of (7.16). One
readily sees that the calculation that proved (7.16) also shows that
$$
\align
&
{}_3F_2\left[\matrix{-R+k,\,R+k,\,\frac32+v+k}\\{\frac12+k,\,1+v+k}\endmatrix;\frac x4
\right]={}_3F_2\left[\matrix{-R+k,\,R+k,\,\frac12+v+k}\\{\frac12+k,\,1+v+k}\endmatrix;\frac x4
\right]\\
&\ \ \ \ \ \ \ \ \ 
+\frac{x}{4}\frac{(-R+k)(R+k)}{(1/2+k)(1+v+k)}
{}_3F_2\left[\matrix{-R+k+1,\,R+k+1,\,\frac32+v+k}\\{\frac32+k,\,2+v+k}\endmatrix;\frac x4
\right].
\endalign
$$
Express the ${}_3F_2$'s of (13.11) using the above identity. Use (7.8) to write the resulting 
${}_3F_2$'s as integrals of ${}_2F_1$'s. In turn, express the latter, using the derivation
rule for ${}_2F_1$'s displayed after (7.9), in terms of the ${}_2F_1$ evaluation (7.11). 
We obtain after simplifications
$$
\align
&
\bar{T'}^{(n)}(R,qR+c;x)=\\
&
\frac{2}{\pi R}\sum_{k=0}^nf_k(qR+c+k+1/2)
\int_0^1 t^{qR}\frac{t^{k+c-1/2}}{(1-t)^{1/2}}\frac{d^k}{dt^k}
\cos\!\left[R\arccos\left(1-\frac{xt}{2}\right)\right]dt\\
&
+
\frac{2}{\pi R}\sum_{k=0}^nf_k
\int_0^1 t^{qR}\frac{t^{k+c+1/2}}{(1-t)^{1/2}}\frac{d^{k+1}}{dt^{k+1}}
\cos\!\left[R\arccos\left(1-\frac{xt}{2}\right)\right]dt.\tag13.13
\endalign
$$

The asymptotics of the $\bar{T}^{(n)}$'s and $\bar{T'}^{(n)}$'s can be obtained using the 
same approach we employed in Section 7 for the ${T}^{(n)}$'s and ${T'}^{(n)}$'s. Indeed, by 
Lemma 7.4, when (7.23) is substituted in (13.12) the omitted terms in (7.23) give rise to
integrals like in Proposition~7.2. By the argument in the proof of Proposition 7.1, we
deduce that the asymptotics of $\bar{T}^{(n)}(R,qR+c;x)$ is generated by the main term on the 
right hand side of (7.23) for $k=n$. More precisely, we obtain
$$
\align
&\!\!\!\!\!\!\!\!\!\!
\left|
\bar{T}^{(n)}(R,qR+c;x)-\frac{1}{\pi R^2}
\int_0^1 t^{qR}\frac{t^{n+c+1/2}}{(1-t)^{1/2}}\sqrt{\frac{4-xt}{xt}}
\left(R\sqrt{\frac{x}{4t-xt^2}}\,\right)^n
\right.
\\
&\ \ \ \ \ \ \ \ \ \ \ \ \ \ \ \ \ \ \ \ \ \ \ \ \ 
\left.
\times
\cos\left[R\arccos\left(1-\frac{xt}{2}\right)+\frac{(n-1)\pi}{2}\right]dt
\right|
\leq \bar{M}_0R^{n-7/2},\tag13.14
\endalign
$$
for $R\geq \bar{R}_0$, with $\bar{R}_0$ and $\bar{M}_0$ independent of $x\in[0,1]$. 
Approximating the integral in (13.14) by Proposition 7.1 we deduce the first part of the
following result.

\proclaim{Proposition 13.3} Let $q>0$ be a fixed rational number, and let $n\geq0$ and $c$ be 
fixed integers. Then for any real number $x\in(0,1]$, we have
$$
\align
&\!\!\!\!\!\!\!\!\!
\left|
\bar{T}^{(n)}(R,qR+c;x)-\frac{1}{\sqrt{\pi}}
\frac{1}{ \root 4\of{ q^2+\frac{x}{4-x} }\sqrt{\frac{x}{4-x}} }
\frac{1}{R^{5/2}}\left(R\sqrt{\frac{x}{4-x}}\,\right)^n
\right.
\\
&\ \ \ \ \ \ 
\left.
\times
\cos\left[R\arccos\left(1-\frac x2\right)-\frac12\arctan\frac1q\sqrt{\frac{x}{4-x}}
+\frac{(n-1)\pi}{2}\right]
\right|
\leq \frac{1}{\sqrt{x}}\bar{M}R^{n-7/2}\tag13.15
\\
&\!\!\!\!\!\!\!\!\!
\left|
\bar{T'}^{(n)}(R,qR+c;x)-\frac{2}{\sqrt{\pi}}{ \root 4\of{q^2+\frac{x}{4-x} } }
\frac{1}{R^{1/2}}\left(R\sqrt{\frac{x}{4-x}}\,\right)^n
\right.
\\
&
\left.
\times
\cos\left[R\arccos\left(1-\frac x2\right)+\frac12\arctan\frac1q\sqrt{\frac{x}{4-x}}
+\frac{n\pi}{2}
\right]
\right|
\leq \bar{M'}R^{n-3/2},\tag13.16
\endalign
$$
for $R\geq \bar{R}_0$, where $\bar{R}_0$, $\bar{M}$ and $\bar{M'}$ are independent of 
$x\in(0,1]$.
\endproclaim

\pf To prove (13.16), substitute (7.13) into (13.13). By Lemma 7.4, all omitted terms in 
(7.13) give rise to integrals of the type addressed by Proposition 7.2. By the arguments
in the proof of Proposition 7.1, the asymptotics of $\bar{T'}^{(n)}(R,qR+c;x)$ comes
from the term on the right hand side of (7.13), when $k=n$. More precisely, we obtain
$$
\align
&\!\!\!
\left|
\bar{T'}^{(n)}(R,qR+c;x)-
\right.
\\
&
\left\{
\frac{2q}{\pi}
\int_0^1 t^{qR}\frac{t^{c-1/2}}{(1-t)^{1/2}}
\left(R\sqrt{\frac{xt}{4-xt}}\,\right)^n
\cos\left[R\arccos\left(1-\frac{xt}{2}\right)+\frac{n\pi}{2}\right]dt
\right.
\\
&
\left.
\left.
+\frac{2}{\pi R}
\int_0^1 t^{qR}\frac{t^{c-1/2}}{(1-t)^{1/2}}
\left(R\sqrt{\frac{xt}{4-xt}}\,\right)^{n+1}
\cos\left[R\arccos\left(1-\frac{xt}{2}\right)+\frac{(n+1)\pi}{2}\right]dt
\right\}
\right|
\\
&\ \ \ \ \ \ \ \ \ \ \ \ \ \ \ \ \ \ \ \ \ \ \ \ \ \ \ \ 
\leq \bar{K}R^{n-3/2},\tag13.17
\endalign
$$
for all $R\geq {\bar{\rho}}$ and
$x\in(0,1]$, where the constants $\bar{K}$ and $\bar\rho$ are independent of $x\in(0,1]$.

The two integrals on the right hand side of (13.17) can be approximated by means of 
Proposition 7.1. Carrying this out and using the equation displayed after (7.26) one obtains 
(13.16). \endpf

By Proposition 13.2, the analysis of the asymptotics of $\bar\omega_b$ is a repeat
of the analysis of the asymptotics of $\omega_b$, with the $T^{(n)}$'s and ${T'}^{(n)}$'s
being replaced by the $\bar{T}^{(n)}$'s and $\bar{T'}^{(n)}$'s, respectively.

Comparing Proposition 13.3 with Proposition 7.1 shows that the only differences between
the formulas approximating the $\bar{T}^{(n)}$'s and $\bar{T'}^{(n)}$'s on the one hand and 
the ones approximating their unbarred versions on the other are:

\smallpagebreak
$(i)$ The numerators 2 and 1 of the first fractions in the approximations are swapped in 
(13.15)--(13.16) as compared to (7.1)--(7.2);

$(ii)$ The factor $\sqrt{\frac{x}{4-x}}$ in the denominator of the approximant of ${T'}^{(n)}$
is moved to the denominator of the approximant of the ``unprimed'' $\bar{T}^{(n)}$;

$(iii)$ The powers of $R$ in the approximants (13.15)--(13.16) are $R^{-5/2}$ and $R^{-1/2}$,
as opposed to both being $R^{-3/2}$ in (7.1)--(7.2);

$(iv)$ Decrementation of the argument of the cosine by $\pi/2$ occurs for  $\bar{T}^{(n)}$ in 
(13.15)--(13.16), as opposed to occuring for  ${T'}^{(n)}$ in (7.1)--(7.2).

\smallpagebreak
Find the asymptotics of the moments 
$\bar{M}_{\alpha_1,\beta_1,\dotsc,\alpha_m,\beta_m;\gamma_1,\delta_1,\dotsc,\gamma_n,\delta_n}$
by the same reasoning we used for the moments $M$ in Sections 8 and 11. 
Relations (11.1) and (11.2) change slightly for the present case, to reflect differences 
$(i)$--$(iv)$ above. Relation (11.3) stays unchanged. The resulting analog of (11.4) is
$$
\align
&\bar{M}_{\alpha_1,\beta_1,\dotsc,\alpha_m,\beta_m;\gamma_1,\delta_1,\dotsc,\gamma_n,\delta_n}
=\frac{\bar{E}}{R^{4mn}}\prod_{i=1}^m\left(\frac{R_i}{\sqrt{3}}\right)^{\alpha_i+\beta_i}
\prod_{j=1}^n\left(\frac{R'_j}{\sqrt{3}}\right)^{\gamma_j+\delta_j}\\
&
\times\!\!\!\!\! \!
\sum_{\epsilon_1,\dotsc,\epsilon_{2m+2n}=\pm1}
\frac{1}{D^{\epsilon_1,\dotsc,\epsilon_{2m+2n}}}
\cos
\left\{
\frac{R\pi}{3}\left[\sum_{i=1}^mA_i(\epsilon_{2i-1}+\epsilon_{2i})+
\sum_{j=1}^nB_j(\epsilon_{2m+2j-1}+\epsilon_{2m+2j})\right]
\right.
\\
&\ \ \ 
-\sum_{i=1}^m\frac{\epsilon_{2i-1}+\epsilon_{2i}}{2}\arctan\frac{1}{q_i\sqrt{3}}
+\sum_{j=1}^n\frac{\epsilon_{2m+2j-1}+\epsilon_{2m+2j}}{2}\arctan\frac{1}{q'_j\sqrt{3}}\\
&\ \ \   
+\sum_{i=1}^m(\epsilon_{2i-1}\alpha_i+\epsilon_{2i}\beta_i)\frac{\pi}{2}
+\sum_{j=1}^n(\epsilon_{2m+2j-1}\gamma_j+\epsilon_{2m+2j}\delta_j)\frac{\pi}{2}
-\sum_{j=1}^n(\epsilon_{2j-1}+\epsilon_{2j})\frac{\pi}{2}\\
&\ \ \  
-\sum_{i=1}^m\sum_{j=1}^n
\left(
\arctan\frac{\epsilon_{2i-1}A_i+\epsilon_{2m+2j-1}B_j}{q_iA_i+q'_jB_j}
+\arctan\frac{\epsilon_{2i-1}A_i+\epsilon_{2m+2j}B_j}{q_iA_i+q'_jB_j}
\right.
\\
&\ \ \ \ \ \ \ \ \ \ \ \ \ \ \ \ \ \ \ \ 
\left.
\left.
+\arctan\frac{\epsilon_{2i}A_i+\epsilon_{2m+2j-1}B_j}{q_iA_i+q'_jB_j}
+\arctan\frac{\epsilon_{2i}A_i+\epsilon_{2m+2j}B_j}{q_iA_i+q'_jB_j}
\right)
\right\}
\\
&\ \ \ \ \ \ \ \ 
+O(R^{-4mn+\sum_{i=1}^m(\alpha_i+\beta_i)+\sum_{j=1}^n(\gamma_j+\delta_j)-3m-3n-1}),
\tag13.18
\endalign
$$
where
$$
\bar{E}=\frac{3^m}{2^{2m}\pi^{m+n}}\frac{1}{\prod_{i=1}^mR_i^5\prod_{i=1}^nR'_i}
\frac{\prod_{j=1}^n \sqrt{{q'_j}^2+\frac13}}{\prod_{i=1}^m\sqrt{{q}_i^2+\frac13}}\tag13.19
$$
and $D^{\epsilon_1,\dotsc,\epsilon_{2m+2n}}$ is given by (11.6). 

All consequences of differences $(i)$ and $(ii)$ are reflected in the change of exponents of
2 and 3 in (13.19) versus (11.5). All consequences of difference $(iii)$ are reflected in the 
changed exponents of $R_i$ and $R'_i$ in (13.19) versus (11.5). The combined effect of 
differences $(i)$--$(iii)$ on (11.4) is thus a multiplicative factor independent of 
$\alpha_1,\beta_1,\dotsc,\gamma_n,\delta_n$. 

The only effect of difference $(iv)$ is that the last sum on the fourth line of (13.18) is
$\sum_{j=1}^n(\epsilon_{2j-1}+\epsilon_{2j})\frac{\pi}{2}$, as opposed to
$\sum_{j=1}^n(\epsilon_{2m+2j-1}+\epsilon_{2m+2j})\frac{\pi}{2}$ in (11.4). Furthermore, the 
statement of Lemma 11.1 is clearly valid also for the multiple sum on the right hand side of 
(13.18). Therefore, in the multiple sum of (13.18) summation can be restricted to balanced
$(\epsilon_1,\dotsc,\epsilon_{2m+2n})$'s without changing its value. However, this restriction
eliminates the only difference between the multiple sums in (13.18) and (11.4). By the
arguments that proved (11.22) we obtain
$$
\align
&\!\!\!\!
\bar{\omega}_b\left(\matrix{R_1}\\{v_1}\endmatrix
\cdots\matrix{R_m}\\{v_m}\endmatrix;
\matrix{R'_1}\\{v'_1}\endmatrix\cdots\matrix{R'_n}\\{v'_n}\endmatrix\right)=
\chi_{2m,2n}\prod_{i=1}^mR_i(R_i-1/2)(R_i+1/2)
\prod_{j=1}^nR'_j\\
&
\times
\frac{2^{n-m}3^m}{\pi^{m+n}}\frac{1}{\prod_{i=1}^mR_i^4\prod_{j=1}^n(R'_j)^2}
\frac{\prod_{j=1}^n \sqrt{(v'_j-c'_j)^2+\frac13(R'_j)^2}}
{\prod_{i=1}^m\sqrt{(v_i-c_i)^2+\frac13R_i^2}}\\
&
\times
\frac{1}{\prod_{i=1}^m\prod_{j=1}^n\left[(v_i+v'_j-c_i-c'_j)^2+\frac13(R_i- R'_j)^2\right]
\left[(v_i+v'_j-c_i-c'_j)^2+\frac13(R_i+ R'_j)^2\right]}
\endalign
$$
\vskip-0.3in
$$\align
&
\times
\frac{(-4)^{m+n}}{3^{m+n}}\prod_{i=1}^mR_i^2\prod_{j=1}^n(R'_j)^2\\
&
\times
\prod_{1\leq i<j\leq m}
\left[(v_j-v_i)^2+\frac13(R_j-R_i)^2\right]\left[(v_j-v_i)^2+\frac13(R_j+R_i)^2\right]\\
&
\times
\prod_{1\leq i<j\leq n}
\left[(v'_j-v'_i)^2+\frac13(R'_j-R'_i)^2\right]\left[(v'_j-v'_i)^2+\frac13(R'_j+R'_i)^2\right]\\
&
+O(R^{2m^2+2n^2-4mn-2m-1})
.\tag13.20
\endalign
$$
This leads to the following result.

\proclaim{Proposition 13.4}
 Let the parameters $R_1,\dotsc,R_m$, $v_1,\dotsc,v_m$ and
$R'_1,\dotsc,R'_n$, $v'_1,\dotsc,v'_n$ depend on $R$ as in $(\TwoTwo)$. 
Then as $R\to\infty$ the asymptotics of $\bar{\omega}_b$ is given by
$$
\align
&\!\!\!\!\!\!\!\!\!\!\!\!\!\!
\bar{\omega}_b\left(\matrix{R_1}\\{v_1}\endmatrix
\cdots\matrix{R_m}\\{v_m}\endmatrix;
\matrix{R'_1}\\{v'_1}\endmatrix\cdots\matrix{R'_n}\\{v'_n}\endmatrix\right)=
\bar{\phi}_{2m,2n}
\prod_{i=1}^m(2R_i)\prod_{j=1}^n(2R'_j)
\frac{\prod_{j=1}^n\sqrt{(R'_j)^2+3(v'_j)^2}}{\prod_{i=1}^m\sqrt{R_i^2+3v_i^2}}\\
&\ \ \ \ \ \ \ 
\times
\prod_{1\leq i<j\leq m}[(R_j-R_i)^2+3(v_j-v_i)^2]
[(R_j+R_i)^2+3(v_j-v_i)^2]\\
&\ \ \ \ \ \ \ 
\times
\frac{\prod_{1\leq i<j\leq n}[(R'_j-R'_i)^2+3(v'_j-v'_i)^2]
[(R'_j+R'_i)^2+3(v'_j-v'_i)^2]}
{\prod_{i=1}^m\prod_{j=1}^n[(R'_j-R_i)^2+3(v'_j+v_i)^2]\,
[(R'_j+R_i)^2+3(v'_j+v_i)^2]}
\\
&+O(R^{2m^2+2n^2-4mn-2m-1}),\tag13.21
\endalign
$$
where
$$
\bar{\phi}_{k,l}=\frac{2^{l}3^{-(k-l)^2/4+(3k-l)/4}}{\pi^{(k+l)/2}}
\prod_{j=0}^{k-1}\frac{(2)_j}{(1)_j(3/2)_j}
\prod_{j=0}^{l-1}\frac{(j+2)_k}{(3/2)_j}.\tag13.22
$$
\endproclaim

\pf Since by (\TwoTwo) the parameters $R_1,\dotsc,R_m$, $v_1,\dotsc,v_m$ and
$R'_1,\dotsc,R'_n$, $v'_1,\dotsc,v'_n$ approach infinity as $R\to\infty$, while
$c_1,\dotsc,c_m$ and $c'_1,\dotsc,c'_n$ are constant, it follows 
that the difference
between the product on the right hand side of (13.20) and what it becomes when one omits
the additive constants $\pm1/2$ on its first line and the constants $c_i$ and
$c'_j$ is $O(R^{2m^2+2n^2-4mn-2m-1})$. This proves (13.21).
Since by assumption the pairs $(A_i,q_i)$, $i=1,\dotsc,m$ of (\TwoTwo) are distinct, as well
as the pairs $(B_j,q'_j)$, $j=1,\dotsc,n$ of (\TwoTwo), the expression on the right hand
side of (13.21) has degree $2m^2+2n^2-4mn-2m$ in $R$. Therefore, (13.21) does indeed give
the asymptotics of $\bar{\omega}_b$. \epf

The proof of Theorem 2.1 now follows readily.

\smallpagebreak
{\it Proof of Theorem 2.1.} The statement follows directly from Proposition 3.2, using Theorem 2.2
and Proposition 13.4. \epf

\mysec{14. A conjectured general two dimensional Superposition Principle}

The choice of the side-lengths of the regions (\TwoZero) might seem unmotivated at first, 
but is in fact quite natural. Indeed, one readily sees that for any lattice hexagon
on the triangular lattice, the difference between the lengths of all 
pairs of opposite sides is the same. Furthermore, this common difference is equal to the 
difference between the number of up-pointing and down-pointing unit triangles enclosed by 
the hexagon. Therefore, if we want to choose a lattice hexagon $H$ to contain all our 
plurimers and possess dimer coverings after their removal, the difference between opposite
sides of $H$ has to match the difference $4n+1-4m$ between the number of up-pointing
and down-pointing unit triangles in the union of the plurimers. It follows that the sides
of $H$ must have the form indicated in Section 2.

By a result of Cohn, Larsen and Propp \cite{\CLP}, a dimer covering of a large regular 
hexagon sampled according to the uniform distribution on all dimer coverings has maximal 
entropy statistics
at its center (and only at its center). This suggests that it is natural to place
the hexagon $H$ enclosing the plurimers so that they stay at its center when one scales 
and lets the size of $H$ grow to infinity. By this observation and the previous paragraph,
it follows that from this point of view, up to a translation by an absolute constant 
(independent of the size of $H$), (\TwoZero) are the most natural regions to use in the
definition (\TwoOne) of the plurimer correlation. (We note that translations by a constant 
vector that keep the symmetry about $\ell$ can be handled by exactly the same approach we
presented in Sections 2--14.)

In the results of this paper we assumed that all plurimers are triangular. Furthermore,
we assumed that all plurimers, except for a single
monomer $u$, have side-length 2, are distributed symmetrically about the say vertical
symmetry axis $\ell$ of $u$, and that the two plurimer orientations are {\it separated} 
by any horizontal $h$ 
through $u$, i.e., all plurimers above $h$ point upward, and all below $h$ point downward.
However, we conjecture that the Superposition Principle (2.6) is valid for arbitrary
plurimers.

%

To make this conjecture precise,
fix a vertex $O$ of the triangular lattice. Consider a rectangular system of coordinates
centered at $O$, with the unit on the horizontal axis equal to the side of a unit triangle,
and the unit on the vertical axis equal to twice the height of a unit triangle (these units
are chosen so as to maintain consistency with Section~2). 

Let $P_1,\dotsc,P_n$ be
arbitrary plurimers on the triangular lattice (i.e., finite unions of unit triangles), and let
${\bold p}_i$ be a distinguished lattice point (a base point) of 
$P_i$, $i=1,\dotsc,n$. Write $P_i(R_i,v_i)$ for the translation of the plurimer $P_i$
that takes the base point ${\bold p}_i$ to the lattice point of coordinates $(R_i,v_i)$.

As in Section 2, let the charge $\ch(P)$ of the plurimer $P$ be the number of 
the up-pointing unit triangles of $P$ minus the number of its down-pointing ones.
Let $H_N$ be the lattice hexagon centered at $O$
and having side-lengths alternating between $N$ and $N-k$, where 
$k=\ch(P_1)+\cdots+\ch(P_n)$. 
Define the correlation $\omega(P_1(R_1,v_1),\dotsc,P_n(R_n,v_n))$ by 
$$
\omega(P_1(R_1,v_1),\dotsc,P_n(R_n,v_n))=\lim_{N\to\infty}
\frac{M(H_N\setminus P_1(R_1,v_1)\cup\cdots\cup P_n(R_n,v_n))}
{M(H_N\setminus P_1(a_1,b_1)\cup\cdots\cup P_n(a_n,b_n))},\tag14.1
$$
where $a_i$ and $b_i$, $i=1,\dotsc,n$, are some fixed integers specifying a reference position
of the plurimers.

We conjecture that the following generalization of Theorem 2.1 holds.

\proclaim{Conjecture 14.1} Suppose the coordinates $R_i$ and $v_i$, $i=1,\dotsc,n$ are expressed
in terms of the integer parameter $R$ as 
$$
\align
R_i&=A_iR\\
v_i&=q_iR_i\tag14.2
\endalign
$$
where $0<A_i\in\Q$, $0<q_i\in\Q$, $i=1,\dotsc,n$ are all 
fixed.

Then the asymptotics of the plurimer correlation is given by
$$
\align
\omega(P_1(R_1,v_1),\dotsc,P_n(R_n,v_n))&=c\prod_{1\leq i<j\leq n}
[(R_j-R_i)^2+3(v_j-v_i)^2]^{\ch(P_i)\ch(P_j)/4}\\
&\ \ \ \ \ \ \ \ \ \ \ \ \ \ \ \ \ \ \ \ 
+O(R^{\sum_{1\leq i<j\leq n}\ch(P_i)\ch(P_j)/2-1})\\
&=c\prod_{1\leq i<j\leq n}
\de(P_i(R_i,v_i),P_j(R_j,v_j))^{\ch(P_i)\ch(P_j)/2}\\
&\ \ \ \ \ \ \ \ \ \ \ \ \ \ \ \ \ \ \ \ 
+O(R^{\sum_{1\leq i<j\leq n}\ch(P_i)\ch(P_j)/2-1}),\tag14.3
\endalign
$$
where $\de$ is the Euclidean distance, and 
$c$ depends just on the shapes of the plurimers $P_1,\dotsc,P_n$, and not on their 
coordinates $(R_i,v_i)$.
\endproclaim

Writing, for the sake of notational brevity, $P_i=P_i(R_i,v_i)$,
it follows from (14.3) that the correlation $\omega$ satisfies
$$
\align
\omega(P_1,P_2)&=c\de(P_1,P_2)^{\ch(P_1)\ch(P_2)/2}
+O(R^{\ch(P_1)\ch(P_2)/2-1})\tag14.4\\
\omega(P_1,\dotsc,P_n)&=c'\prod_{1\leq i<j\leq n}
\omega(P_i,P_j)+O(R^{\sum_{1\leq i<j\leq n}\ch(P_i)\ch(P_j)/2-1}),\tag14.5
\endalign
$$
where $c'$ is a constant independent of $R$.
When taking the logarithm in (14.4), the contribution of the constant is
negligible as $R\to\infty$ and we obtain
$$
\ln\omega(P_1,P_2)\sim\frac{\ch(P_1)\ch(P_2)}{2}\de(P_1,P_2),
\ \ \ R\to\infty.\tag14.6
$$
Similarly, provided $\sum_{1\leq i<j\leq n}\ch(P_i)\ch(P_j)
\neq0$, when taking the logarithm in (14.5)
one can neglect the contribution of the constant and we obtain
$$
\ln\omega(P_1,\dotsc,P_n)\sim\sum_{1\leq i<j\leq n}\ln\omega(P_i,P_j), \ \ \ R\to\infty.
\tag14.7
$$

Equations (14.6) and (14.7) show that $\ln\omega$ satisfies the characterizing properties
of the two-dimensional electrostatic potential---Coulomb's law and the Superposition 
Principle. Since all classical electrostatics can be deduced from these two properties (see
e.g. \cite{\Ftwo,Ch. 4}), Conjecture 14.1 implies that our random tiling model 
described in Section 2 indeed models classical two-dimensional electrostatics.

We believe that in fact (14.3) holds in still larger generality. There are two ingredients to
this extension. First, allow the family of regions used to define the plurimer correlation to
be any family with the property that in the scaling limit the plurimers are situated in the
region where 
dimer coverings have maximal entropy statistics. We believe that the
correlation defined by means of any such family of regions satisfies (14.3). One instance of
this situation is presented in \cite{\Cfour}, where it is shown that (14.3) holds for two charges
of magnitudes 2 and $-2$. 
An extension found by the author of the result in \cite{\Cfour} to an arbitrary number of 
even-side plurimers will be presented in a sequel of the present paper.

Second, based on our result (14.9) below and the conjectured rotational invariance of 
monomer-monomer
correlations (see \cite{\FS}), we conjecture that the natural analog of Conjecture 14.1 on 
the {\it square} lattice $\Z^2$ also holds. 

More precisely, let $P_1,\dotsc,P_n$ be arbitrary plurimers on the square lattice (i.e., finite 
unions of unit squares). Fix a chessboard coloring of the square lattice, and define
the charge $\ch(P)$ of plurimer $P$ to be the difference between the number of its white and
black unit squares. As we did for the triangular lattice, consider a base lattice point 
${\bold p}_i$ in each $P_i$, and denote by $P_i(x_i,y_i)$ the  
translation of $P_i$ taking ${\bold p}_i$ to $(x_i,y_i)$ (in order for $P_i(x_i,y_i)$ to
``look like'' $P_i$, we assume all such translations to be color preserving).

Let $AR_N$ be the ``Aztec rectangle region'' of sides $N$ and $N+k$ centered at the origin 
(i.e., the lattice region dual to the corresponding Aztec rectangle graph defined in 
\cite{\Cone}), where $k=\ch(P_1)+\cdots+\ch(P_n)$. 
Define the correlation of the $n$ plurimers by
$$
\omega(P_1(x_1,y_1),\dotsc,P_n(x_n,y_n))=\lim_{N\to\infty}
\frac{M(AR_N\setminus P_1(x_1,y_1)\cup\cdots\cup P_n(x_n,y_n))}
{M(AR_N\setminus P_1(a_1,b_1)\cup\cdots\cup P_n(a_n,b_n))},
$$
where $a_i$ and $b_i$, $i=1,\dotsc,n$, are some fixed integers specifying a reference position
of the plurimers.

\proclaim{Conjecture 14.2} 
Suppose the coordinates $x_i$ and $y_i$, $i=1,\dotsc,n$ are expressed
in terms of the integer parameter $R$ as 
$$
\align
x_i&=A_iR\\
y_i&=q_ix_i,
\endalign
$$
where $0<A_i\in\Q$, $0<B_i\in\Q$, $i=1,\dotsc,n$ are all 
fixed.

Then the asymptotics of the plurimer correlation is given by
$$
\align
\omega(P_1(x_1,y_1),\dotsc,P_n(x_n,y_n))&=c\prod_{1\leq i<j\leq n}
[(x_j-x_i)^2+(y_j-y_i)^2]^{\ch(P_i)\ch(P_j)/4}\\
&\ \ \ \ \ \ \ \ \ \ \ \ \ \ \ \ \ \ \ \ 
+O(R^{\sum_{1\leq i<j\leq n}\ch(F_i)\ch(F_j)/2-1})\\
&=c\prod_{1\leq i<j\leq n}
\de(P_1(x_i,y_i),P_1(x_j,y_j))^{\ch(P_i)\ch(P_j)/2}\\
&\ \ \ \ \ \ \ \ \ \ \ \ \ \ \ \ \ \ \ \ 
+O(R^{\sum_{1\leq i<j\leq n}\ch(F_i)\ch(F_j)/2-1}),\tag14.8
\endalign
$$
where $c$ is a constant depending just on the types of the plurimers $P_1,\dotsc,P_n$.
\endproclaim

More generally, we believe that any bipartite planar, periodic (i.e., invariant under
translations by two non-collinear vectors) graph has an embedding so that joint correlations of 
plurimers
are given by Coulomb's law (using the Euclidean distance) and the Superposition Principle.

%
%

For the square lattice, the special case of (14.8) when $n=2$ and the two plurimers are in fact 
monomers of opposite color was suggested to hold by Fisher and Stephenson in \cite{\FS}. 
The further specialization when the second monomer is adjacent to a lattice diagonal through 
the first was proved by Hartwig \cite{\Har}. A natural extension of Hartwig's result would be to
study the correlation of an {\it arbitrary} collection of monomers along two consecutive 
lattice diagonals. We state below (see (14.9)) a result we found which is a 
close analog of this. The proof will appear elsewhere.

For this purpose, rather than phrasing everything in terms of lattice regions and their 
dimer (domino) coverings, it will be convenient to use the dual set-up of graphs (duals of regions) 
and their perfect matchings. 

Consider therefore the square lattice $\Z^2$ and regard it as a graph. 
It will be convenient to draw it so that the lattice lines form angles of $45^\circ$ 
with the horizontal. 

We say that a vertex $v$ of the square lattice has been {\it split} if $v$ is replaced
by two new vertices $v'$ and $v''$ slightly to the left and right of $v$,
respectively, and the four edges incident to $v$ are replaced by two edges joining
$v'$ to the former two neighbors of $v$ on its left, and two joining $v''$ to the 
former two neighbors of $v$ on its right. Figure 14.1 contains an example of a split
vertex.

Color the vertices of the square lattice $\Z^2$ black and white in a chessboard 
fashion. As above, regard removal of a vertex of
$\Z^2$ as creation of a unit charge, of sign determined by its color. In
addition, it is natural to consider the operation of splitting a vertex as creating a
unit charge {\it of opposite sign to the one that would be created by removing that
vertex} (this is readily seen to be justified if one views perfect matchings as being 
encoded by families of non-intersecting lattice paths). 

Let $m,n\geq0$ be integers, $\{v_1,\dotsc,v_m\}$ and $\{w_1,\dotsc,w_n\}$ two
disjoint sets of nonnegative integers, and $N$ an even nonnegative integer. We define
$AR_N(v_1,\dotsc,v_m;w_1,\dotsc,w_n)$ to be the subgraph of $\Z^2$ described 
as follows. 

Consider the {\it Aztec rectangle} (see \cite{\Cone}) of width $N$ and height $N+m-n$,
i.e., a subgraph of $\Z^2$ consisting of an $N$ by $N+m-n$ array of 4-cycles touching
only at vertices. Let $\ell$ be its vertical symmetry axis, and let $O$ be its vertex on
$\ell$ that is $N$ lattice segments away from its base. Label
the vertices on $\ell$ starting with 0 for $O$ and continuing with consecutive
integers as we proceed upward. We define $AR_N(v_1,\dotsc,v_m;w_1,\dotsc,w_n)$ to be
the graph obtained from our Aztec rectangle by removing the vertices labeled $v_i$,
$i=1,\dotsc,m$, and by splitting the vertices $w_i$, $i=1,\dotsc,n$ (an example is
shown in Figure 14.1).

\topinsert
\centerline{\mypic{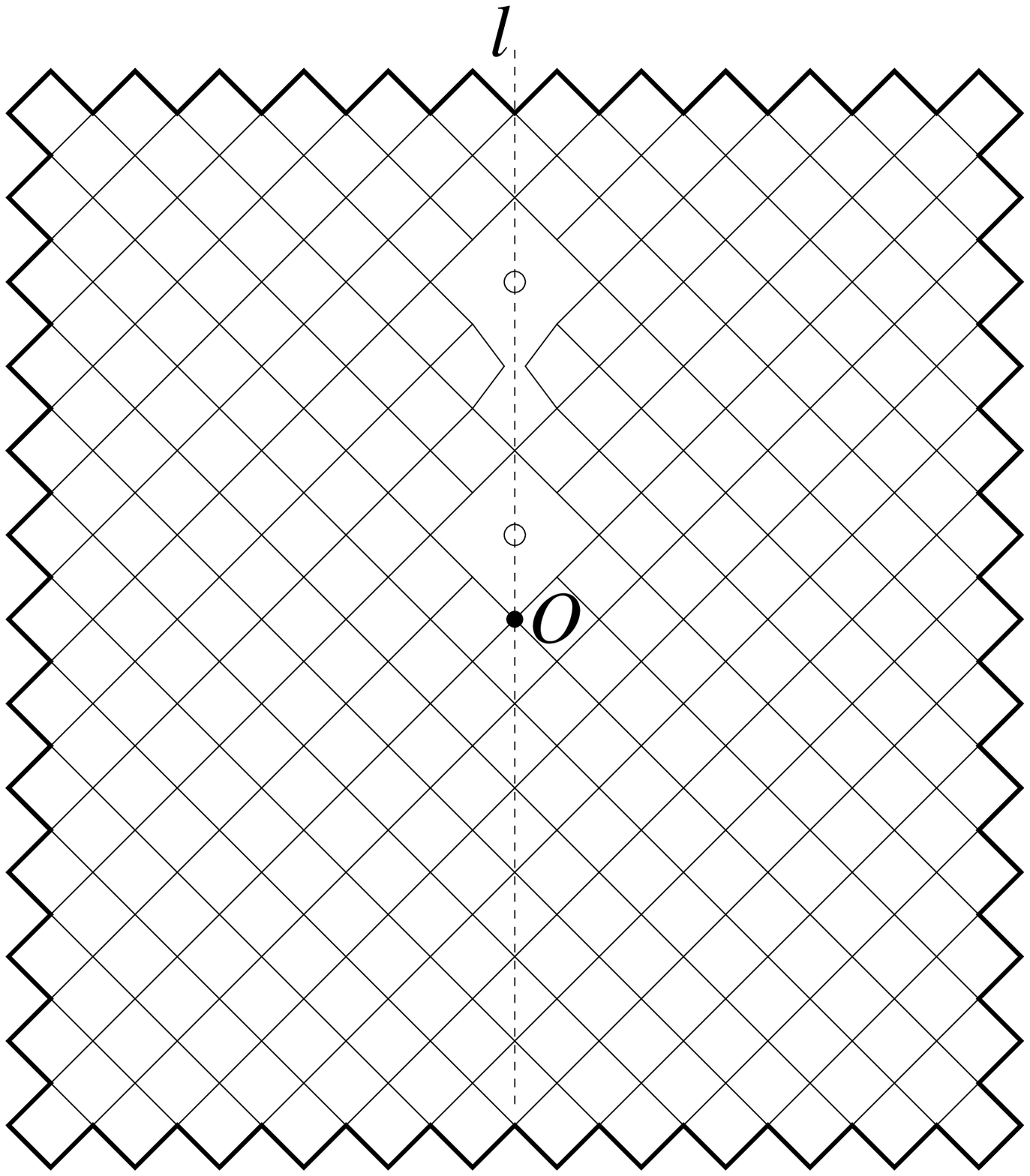}}
\centerline{Figure~14.1. $AR_{12}(1,4;3)$.{\rm }}
\endinsert

As mentioned above, we regard both kinds of altered vertices as unit charges.
We take the reference position of these charges to be when they are packed next to
$O$, and thus define the joint correlation of the charges as
$$\align
\omega(v_1,\dotsc,v_m;&w_1,\dotsc,w_n):=\\
&\lim_{N\to\infty}\frac{M(AR_N(v_1,\dotsc,v_m;w_1,\dotsc,w_n))}
{M(AR_N(0,1,\dotsc,m-1;m,m+1,\dotsc,m+n-1))}.
\endalign$$
Using results from \cite{\Cone} and doing the asymptotic analysis of the expressions
they lead to, we found that 
$$
\omega(v_1,\dotsc,v_m;w_1,\dotsc,w_n)\sim
c_{m,n}\frac{\prod_{1\leq i<j\leq m}(v_j-v_i)^{1/2}\prod_{1\leq i<j\leq n}(w_j-w_i)^{1/2}}
{\prod_{i=1}^m\prod_{i=1}^n|v_i-w_j|^{1/2}},\tag{14.9}
$$
as the distances between charges grow to infinity keeping constant mutual ratios,
where $c_{m,n}$ is some explicit constant depending only on $m$ and $n$. The above formula
shows that the Superposition Principle holds on the square lattice for {\it any}
distribution of unit charges along a lattice diagonal.

As a particular case of (14.9), we obtain, after completing the fairly laborious task
of working out the constant, that
$$
\omega(0,d)\sim \frac{\pi\sqrt{e}}{2^{1/3}A^6}\sqrt{d},\ \ \ d\to\infty\tag{14.10}$$
where $A=1.282427...$ is the Gleisher-Kinkelin constant \cite{\GR},
$$1^12^2\cdots n^n\sim An^{n^2/2+n/2+1/12}e^{-n^2/4}.
$$
Formula (14.10) is a counterpart of the result of Hartwig \cite{\Har}, which addresses the case 
of vertices of opposite colors removed from adjacent diagonals.

{\smc Remark 14.2.} After scaling, the above set-up places the momomers whose correlation is 
measured exactly in the center of the scaled Aztec rectangle, which in the limit approaches an
Aztec diamond. According to a result of Cohn, Elkies and Propp \cite{\CEP}, a uniformly sampled 
tiling of an Aztec diamond has maximal entropy statistics at its center, and only at its center. 
The above set-up seems therefore natural.

\mysec{15. Three dimensions and concluding remarks}

As far as the author knows, the question of working out the joint correlations of an arbitrary 
number of holes for the
dimer model on a bipartite lattice---and thereby establishing the emergence of electrostatics in 
this way---was not considered before in the literature.
Perhaps the case of removing several regions
from a lattice and study their joint correlations was not considered before due to the
complexity the problem presents already in the case of two removed monomers 
(see \cite{\FS}). This might have been also the reason for not considering the
case of two removed regions of arbitrary charges, a situation that already points to
the phenomena of electrostatics. Another possible reason is that the analysis of
\cite{\FS} was done on the square lattice, while the hexagonal lattice is the one on
which natural regions to remove---triangular regions---can have arbitrary charges.
(Indeed, on the square lattice to get a region of charge $s$ one needs for instance to 
line up $s$
diagonally adjacent monomers. Removing a whole dimer is natural on the square lattice,
and this was studied extensively in \cite{\FS}. However, since dimers have charge zero,
the study of their correlations does not detect the Superposition Principle!)

We mention that there are results in the physics literature outlining some connections of the
Ising model, a well studied statistical physical model, and electrostatics (e.g. Hurst and Green note 
in \cite{\HG} that the relations satisfied by certain matrices they employ to solve the Ising model 
``are similar to the relations satisfied by fermion emission and absorption matrices,'' and 
in \cite{\WMTB}, which presents exact
calculations of 2-point spin-spin correlations in the Ising model, Wu, McCoy, Tracy and Barouch
are led to certain integral equations that arose before in work of Myers \cite{\Myers} in the context 
of electromagnetic scattering from a strip; 
in \cite{\MPW} McCoy, Perk and Wu give recurrences for $n$-point spin correlations, but without 
analyzing their asymptotics).
Since the Ising model on a
lattice can be equivalently phrased as a dimer model on a modified, suitably weighted lattice (see e.g.
\cite{\Kastel}),
this
yields a connection between electrostatics and the dimer model. However, this turns out to lead
to a totally different object of study in the dimer model than the joint correlation of holes we 
considered in this paper: two spins being correlated in an Ising state
turns out to correspond to requiring two faces in the corresponding dimer lattice to be on the same
side of the union of certain polygons corresponding to the Ising state. Moreover, the fundamental
difference (4) mentioned in the Introduction between our model and the ones surveyed in \cite{\Nien} 
holds also between our model and the Ising model. 


One key feature of the work described above is that even though the quantities
that need to be studied asymptotically are not ``round'' (i.e., do not factor as 
products of small factors) in general, they can be expressed as multiple sums
of round terms, courtesy of the exact enumeration formulas from \cite{\Cthree} and
\cite{\Cone}. These multiple sums can then, in certain situations, be reduced to single 
or double sums that in turn lead to special functions whose asymptotics can be obtained
using specific techniques such as Laplace's method. This view supplies additional
motivation for the already well-represented study of regions whose tilings are
enumerated by simple product formulas.

The question of studying joint correlations of missing vertices in lattice graphs can 
naturally be phrased also in dimensions other than two. 
The really compelling case is that of three dimensions. 
{\it Does there exist a three dimensional
analog of our model that would model classical electrostatics in the physical, three
dimensional space?}

We present such a possibility in Question 15.1 below. In the case of an affirmative answer 
to this question, we show in Remark 15.2 how a natural new parameter could be introduced in
our model so that the parallel between the three dimensional analogs of (\TwoTwelve) and 
(\TwoEleven) holds for any temperature $T$.

Based on the two dimensional case, we are guided in the phrasing of (15.2)
by the assumption that in three dimensions
as well an analogous random tiling model would manifest the phenomena of electrostatics.

The square lattice version of our model is easiest to parallel in three dimensions. 
Consider $\R^3$ divided into unit cubes by the lattice $\Z^3$. Color the unit cubes black
and white so that adjacent cubes have different colors. Regard the unit cubes as monomers,
and their finite unions as plurimers. 

Consider two monomers $a=(0,0,0)$ and $b_r=(0,0,2r-1)$ of opposite color and include them in
a large cube $C_N$, where $N$ is even. Based on our guiding physical intuition, we wish to 
define their correlation $\omega(a,b_r)$ to be proportional to $\M(C_N\setminus\{a,b_r\})$,
for large $N$, in such a way that $-\ln\omega(a,b_r)$ behaves asymptotically like the 
potential energy of two unit charges of opposite sign at distance $2r$, i.e., like
a positive constant times $-1/(2r)$. In particular, we should have $\omega(a,b_r)\to1$,
as $r\to\infty$. 

This suggests that in the definition of $\omega(a,b_r)$ we should normalize by letting the
monomers be far apart. Based on this we define plurimer correlation as follows.

Fix a lattice point 
$O$, and consider a rectangular system of coordinates centered at $O$. 
Let $P_i$ be an arbitrary plurimer, and let ${\bold p}_i\in\Z^3$ be 
a base point of $P_i$, for $i=1,\dotsc,n$. 
For any integer $R\geq1$, define $RP_i$ to be the 
translation of $P_i$ that takes ${\bold p}_i$ to $R{\bold p}_i$.

Enclose the $n$ plurimers in a large cube $C_N$ of side $N$, centered at $O$.
For the sake of definition simplicity, assume that $\ch(P_1)+\cdots+\ch(P_n)=0$ (this
condition is not essential for defining the correlation below and phrasing Question 15.1---it
can be circumvented by replacing the enclosing cubes $C_N$ by a suitable family of regions $D_N$
with the property that $\ch(D_N)=\ch(P_1)+\cdots+\ch(P_n)$; for $\Z^2$, this was accomplished in 
Section 14 by choosing Aztec rectangles $AR_N$).

Assuming the 
following limit exists, define the plurimer correlation $\omega(P_1,\dotsc,P_n)$ by
$$
\omega(P_1,\dotsc,P_n)=\lim_{R\to\infty}\lim_{N\to\infty}
\frac{\M(C_N\setminus P_1\cup\cdots\cup P_n)}{\M(C_N\setminus RP_1\cup\cdots\cup RP_n)},
\tag15.1
$$
where $M(D)$ denotes the number of dimer coverings of the lattice region (on $\Z^3$) $D$.
(The inside limit should exist for any $R$---it is just the ratio of two joint correlations
of plurimers in a sea of dimers; we are assuming in addition that its value
approaches a limit as $R\to\infty$.)

The possibility we referred to above for modeling three dimensional electrostatics is 
phrased below. As in two dimensions, the charge $\ch(P)$ of a plurimer $P$ is defined to be
the difference between the numbers of its white and black unit cubes.

\proclaim{Question 15.1} Is it true that 
$$
\omega(RP_1,\dotsc,RP_n)=1-k_3\sum_{1\leq i<j\leq n}
\frac{\ch(P_i)\ch(P_j)}{\de(R{\bold p}_i,R{\bold p}_j)}+O(R^{-2})
\tag15.2
$$
as $R\to\infty$, where $k_3>0$ is an absolute constant?
\endproclaim

If the answer to this question is affirmative, taking the logarithm in (15.2) we obtain
$$
\ln\omega(RP_1,\dotsc,RP_n)\sim -k_3\sum_{1\leq i<j\leq n}
\frac{\ch(P_i)\ch(P_j)}{\de(R{\bold p}_i,R{\bold p}_j)},\ \ \ 
R\to\infty.\tag15.3
$$
This would show then that indeed $\ln\omega$ behaves like the electrostatic potential
in three dimensions.

\medskip
\flushpar
{\smc Remark 15.2.} Assuming the answer to Question 15.1 is affirmative, we can introduce
a new parameter in our model that allows it to parallel electrostatics for any 
temperature.

The new parameter, denoted by $x$, ranges over the odd positive integers. For any such
$x$, refine each unit cube $c$ of $\Z^3$ into $x^3$ equal smaller cubes, and properly color 
the latter black and white so that the smaller cubes fitting in the corners of $c$ have the
same color as $c$ (this can clearly be done, since $x$ is odd). Moreover, if we regard
the subdivided cube $c$ as a plurimer on the lattice $(\frac1x\Z)^3$, its charge is readily
seen to agree with the charge of $c$, regarded as a monomer on $\Z^3$ (again, this is due to $x$
being odd). 

Let $P_i$ and ${\bold p}_i$, $i=1,\dotsc,n$, be as above. Define the correlation
$\omega_x(P_1,\dotsc,P_n)$ by
$$
\omega_x(P_1,\dotsc,P_n)=\lim_{R\to\infty}\lim_{N\to\infty}
\frac{\M_x(C_N\setminus P_1\cup\cdots\cup P_n)}{\M_x(C_N\setminus RP_1\cup\cdots\cup RP_n)},
\tag15.4
$$
where for a lattice region $H$ in $\Z^3$, $M_x(H)$ denotes the number of its dimer coverings
when regarded as a lattice region in $(\frac1x\Z)^3$. (Since as noted in the previous
paragraph our lattice refinement preserves charge, $H$ has the same number of black and 
white fundamental regions when considered in $(\frac1x\Z)^3$, provided it does so when
considered in $\Z^3$.) Clearly, existence of the limit (15.1) implies existence of the 
above limit.

The asymptotics of $\omega_x(P_1,\dotsc,P_n)$ can be deduced from (15.1) as follows. Scale 
down the lattice $\Z^3$ by a linear factor of $x$, keeping our lattice point $O$ fixed, and
view the plurimers and their base points as being in the scaled down lattice: the plurimers
become $P_i^x$, $i=1,\dotsc,n$, and their base points ${\bold p}_i^x=x{\bold p}_i$, 
$i=1,\dotsc,n$. 
It follows from the definition that
$$
\omega_x(P_1,\dotsc,P_n)=\omega(P_1^x,\dotsc,P_n^x).
$$
Clearly, we also have 
$\de(R{\bold p}_i^x,R{\bold p}_j^x)=
x\de(R{\bold p}_i,R{\bold p}_j)$, for all $1\leq i<j\leq n$. 
Furthermore, the plurimer $(RP_i)^x$ is the same as the plurimer 
$RP_i^x$.
Therefore, we obtain by (15.2) that
$$
\omega_x(RP_1,\dotsc,RP_n)=1-\frac{k_3}{x}\sum_{1\leq i<j\leq n}
\frac{\ch(P_i)\ch(P_j)}{\de(R{\bold p}_i,R{\bold p}_j)}+O(R^{-2}).
\tag15.5
$$
Taking the logarithm in both sides we deduce from (15.5) that
$$
\ln\omega_x(RP_1,\dotsc,RP_n)\sim -\frac{k_3}{x}\sum_{1\leq i<j\leq n}
\frac{\ch(P_i)\ch(P_j)}{\de(R{\bold p}_i,R{\bold p}_j)},\ \ \ 
R\to\infty.\tag15.6
$$
Repeating the analysis at the end of Section 2, (15.6) implies that in the limit $N\to\infty$,
when one samples uniformly at
random from all dimer coverings of the $x$-subdivided $C_N\setminus P_1\cup\cdots\cup P_n$, 
the relative probability of having the plurimers at preassigned distances $d_{ij}$ versus
$d'_{ij}$
is, by the fundamental theorem of statistical mechanics (see e.g. \cite{\Fone, \S40)}), 
{\it exactly the same} as the relative probability of having $n$ corresponding 
electrical charges at
those distances, at temperature 
$$
T=xq_e^2/(4\pi\epsilon_0kk_3L_P),\tag15.7
$$ 
where $q_e$ is the charge of the electron, $k$ is Boltzmann's constant, $L_P$ is the
Planck length ($\sim10^{-35}$, smallest non-zero length that ``makes sense'' 
physically; in our analysis above, 
we express all physical distances using $L_P$ as a unit\footnote{Strictly speaking, one needs to
express physical distances in terms of $L_P$ also in (2.13), but since (2.13) is invariant under
changing the unit for distance, this was not necessary in the two dimensional case.}) 
and $\epsilon_0$ is the permittivity of empty space.
 


The free energy per unit volume in this model equals
the free energy per $\Z^3$-site in the dimer model on the 
lattice $(\frac1x\Z)^3$. This energy is $F=l_3x^3$, where $l_3\sim0.446\dots$ is the 
three-dimensional dimer constant (see e.g. \cite{\Ham}\cite{\Ctwo}). 

As mentioned in the footnote at the end of Section 2, there is an extra ``calibration'' parameter
one can consider in our model---a fixed positive integer $a$ so that a physical elementary charge
corresponds to a plurimer of charge $a$ in our model. The effect of this extra parameter in 
(15.7) is to multiply its right hand side by $1/a^2$. Expressing $x$ from the resulting relation
and substituting into $F=l_3x^3$ we obtain
$$
F=\left(\frac{4\pi\epsilon_0kk_3L_Pa^2}{q_e^2}\right)^3l_3T^3.\tag15.8
$$
This is similar in form to the formula for the free energy per unit volume, for high temperature $T$, 
obtained in quantum field theory.
Note that the constant of proportionality in (15.8) is
$\sim(10^{-30}a^2)^3$, so the value of $F$ depends crucially on $a$; for small $a$ the value is
negligible even for very large temperatures $T$, while for $a\sim10^{15}$---a conceivable value,
given that $L_P\sim10^{-35}$---it becomes quite significant. The free energy per unit volume $F$ 
in our model is reminiscent of the cosmological constant, which also encodes in some sense the 
overall energy stored in the vacuum of space. The counterpart in our model of the fact that the 
cosmological constant is believed to be positive but too small to be detected by current experiments
would then be that the calibration parameter $a$ satisfies $a\ll 10^{15}$.  

It is important to remark that while the Superposition Principle seems to hold independently of
the background bipartite lattice, the effect of lattice refinement on both (\TwoTwelve) and its 
three dimensional analog that (15.2) would imply depends crucially on the lattice. For instance,
(14.8) would imply that (\TwoTwelve) holds also on the square lattice. However, in this case 
$x$-fold refinement does not provide an extra parameter: for even $x$ the plurimers become neutral
on the refined lattice, while for odd $x$ the plurimer charges are invariant under refinement, 
a fact which
together with the invariance of (\TwoTwelve) under distance scaling shows that (\TwoTwelve) does
not change in this case under refinement. Similarly, for bipartite lattices in three dimensions 
the charge is not invariant in general under refinement, so different lattices lead to different 
analogs of (15.7) and (15.8).

An interesting question is the following: if bipartite lattices lead to electrostatic effects, 
what do non-bipartite lattices lead to? Since the latter do not have a black and white coloring,
it is natural to expect uniform behavior---either universal attraction, or universal repelling.
Is one perhaps led to the effects of some other fundamental physical force? 
The most natural non-bipartite plane lattice, the triangular lattice, is considered
in \cite{\FMS}, and an analysis of the monomer-monomer correlations on it, paralleling that of 
Fisher and Stephenson \cite{\FS} on the square lattice, is carried out. The first
15 correlations of two monomers in adjacent lattice lines are computed, and their values 
\cite{\FMS,Table III} provide convincing evidence for an exponential decay to a limiting positive
constant. Changing the normalization in the definition of correlation in \cite{\FMS,p.8} so that 
one normalizes by the reference position of two infinitely separated monomers (as in
(15.4)), this is equivalent to the modified correlation exponentially approaching zero. 
Then the ``potential'' would be $\ln\omega\sim -kr$, with $k>0$ a constant and $r$ the separation
between monomers, and the ``field'' $\frac{d}{dr}\ln\omega$---or more directly, according to
``discrete calculus'' (see e.g. \cite{\GKP}), $\omega(r+1)/\omega(r)-1$---would approach $-k$.
This is similar to the case of inter-quark force, which approaches a constant as the separation
between the quarks approaches infinity.
It would be very interesting to study the joint correlation of several plurimers on the 
triangular lattice in more 
detail, in particular to determine whether it is a function depending just on the pair 
correlations (as it is the case in the presence of a Superposition Principle), and to find out 
how it depends on the plurimer sizes. 

We conclude by mentioning that since the set of all tilings of a region is 
equivalent to the set of all families of non-intersecting lattice paths with certain starting and
ending points---which is in turn equinumerous, by the Gessel-Viennot theorem, with the set of 
all (appropriately signed) families of lattice paths with these starting and ending points---
averaging over all such tilings is reminiscent of the ``sum-over-paths'' interpretation of particles 
in quantum mechanics due to Feynman. The results in this paper show that if the different tilings of
a region with holes are associated with different ways for pairs of virtual particles and their 
antimatter companions to annihilate, then the microscopic frenzy of quantum-mechanical fluctuations
in the vacuum of empty space that Feynman once 
jokingly described as ``Created and annihilated, created and annihilated---what a waste of time,'' is 
seen to actually generate the phenomena of electrostatics\footnote{A consequence of quantum mechanics
is that the quantum fluctuations of the vacuum drive the intrinsic strength of the electric field of
charged particles to get larger when examined on short distance scales. This offers the opportunity to
perform a test for the parallel between our model and electrostatics. Explicit numerical calculations
based on results of this paper and \cite{\Cfour} confirm that the quantity $\omega(r+1)/\omega(r)-1$
(where $\omega(r)$ is the correlation between a fixed hole and another hole of fixed shape at distance 
$r$), the analog in our model of the electric field, when divided by the two dimensional electric field 
intensity $1/r$, does indeed yield a quantity that grows as $r\to0$. Details will appear in a sequel
to this paper.}.

\bigskip
\bigskip
{\bf Acknowledgments.} I thank Richard Stanley for his continuing interest in this work, his 
encouragements and many helpful discussions. I thank Timothy Chow for many interesting conversations and
my brother Alexandru Ciucu for following enthusiastically the development of the ideas presented in this 
paper. I also thank the anonymous referee for the thorough review and for helpful suggestions.




\mysec{References}
{\openup 1\jot \frenchspacing\raggedbottom
\roster

\myref{\A}
  L. V. Ahlfors, ``Complex analysis; an introduction to the theory of analytic functions of one 
complex variable,'' 2nd ed., McGraw-Hill, New York, 1966.

\myref{\Ca}
  P. G. Carter, ``An empirical equation for the resonance energy of polycyclic
aromatic hydrocarbons,'' {\it Trans. Faraday. Soc.} {\bf 45} (1949), 597--602.

\myref{\Cone}
  M. Ciucu, Enumeration of perfect matchings in graphs with reflective
symmetry, {\it J. Comb. Theory Ser. A} {\bf 77} (1997), 67--97.

\myref{\Ctwo}
  M. Ciucu, An improved upper bound for the three dimensional dimer problem,
{\it Duke Math. J.} {\bf 94} (1998), 1--11.

\myref{\Cthree} 
  M. Ciucu, Plane partitions I: A generalization of MacMahon's
formula, preprint (available at the Los Alamos 
archive, at http://arxiv.org/ps/math.CO/9808017).

\myref{\Cfour}
  M. Ciucu, Rotational invariance of quadromer correlations on the hexagonal lattice,
{\it Adv. in Math.}, in press.

\myref{\CEP}
  H. Cohn, N. Elkies, J. Propp, Local statistics for random domino tilings of the 
Aztec diamond, {\it Duke Math. J.} {\bf 85} (1996), 117--166.

\myref{\CLP}
  H. Cohn, M. Larsen, J. Propp, The shape of a typical boxed plane partition, 
{\it New York J. of Math.} {\bf 4} (1998), 137--165.

\myref{\FMS}
  P. Fendley, R. Moessner and S. L. Sondhi, The classical dimer model on the triangular lattice,
Electronic preprint dated June, 2002, available at arXiv:cond-mat/0206159.

\myref{\Fone}
  R. P. Feynman, ``The Feynman Lectures on Physics,'' vol. I, Addison-Wesley, Reading, 
Massachusetts, 1963.

\myref{\Ftwo}
  R. P. Feynman, ``The Feynman Lectures on Physics,'' vol. II, Addison-Wesley, Reading, 
Massachusetts, 1964.

\myref{\FS} 
  M. E. Fisher and J. Stephenson, Statistical mechanics of dimers on a plane 
lattice. II. Dimer correlations and monomers, {\it Phys. Rev. (2)} {\bf 132} (1963),
1411--1431.

\myref{\GV}
  I. M. Gessel and X. Viennot, Binomial determinants, paths, and hook length formulae,
{\it Adv. in Math.} {\bf 58} (1985), 300--321.

\myref{\GR} 
  I. S. Gradshtein and I. M. Ryzhik, ``Table of integrals, series, and 
products,'' Academic Press, New York, 1980.

\myref{\GKP}
  R. L. Graham, D. E. Knuth, O. Patashnik, ``Concrete Mathematics,'' Addison-Wesley, Reading, Mass., 1989.

\myref{\Ham}
  J. M. Hammersley, An improved lower bound for the multidimensional dimer problem,
{\it Proc. Cambridge Philos. Soc.} {\bf 64} (1968), 455--463.

\myref{\Har} 
  R. E. Hartwig, Monomer pair correlations, {\it J. Mathematical Phys.} {\bf 7}
(1966), 286--299.

\myref{\HG}
  C. A. Hurst and H. S. Green, New solution of the Ising problem for a rectangular lattice,
{\it J. Chem. Phys.} {\bf 33} (1960), 1059--1062.

\myref{\Kastel}
  P. W. Kasteleyn, Dimer statistics and phase transitions, {\it J. Math. Phys.} {\bf 4} (1963), 287--293.

\myref{\Kone}
  R. Kenyon, ``Local statistics of lattice dimers,'' {\it Ann. Inst. H. Poincar\'e Probab.
  Statist.} {\bf 33} (1997), 591--618. 

\myref{\Ktwo}
  R. Kenyon, ``Long-range properties of spanning trees. Probabilistic
  techniques in equilibrium and nonequilibrium statistical physics,'' J. Math. Phys. {\bf 41} (2000), 
 1338--1363.


\myref{\KM}
  Werner Krauth and  R. Moessner, ``Pocket Monte Carlo algorithm for classical doped dimer models,''
arXiv.org/cond-mat/0206177, June 2002.

\myref{\LP}
  L. Lov\'asz and M. D. Plummer, ``Matching theory,'' Elsevier Science Publishers, 
Amsterdam, 1986.

\myref{\Lu}
  Y. L. Luke, ``The special functions and their approximations,'' Academic Press, New York, 1969.

\myref{\MPW}
  B. M. McCoy, J. H. H. Perk and T. T. Wu, Ising field theory: Quadratic difference equations for the
n-point Green's functions on the lattice, {\it Phys. Rev. Lett.} {\bf 46} (1981), 757--760.

\myref{\Myers}
  J. Myers, Wave scattering and the geometry of a strip,
{\it J. Math. Phys.} {\bf 6} (1965), 1839--1846.

\myref{\Nien}
  B. Nienhuis, Coulomb gas formulation of two-dimensional phase transitions, ``Phase Transitions,''
vol. 11, 1--53, Academic Press, London, 1987. 

\myref{\Ol}
  F. W. J. Olver, Asymptotics and special functions, Academic Press, New York, 1974.


\myref{\Ste}  
  J. R. Stembridge, Nonintersecting paths, Pfaffians and plane partitions, 
{\it Adv. in Math.} {\bf 83} (1990), 96--131.

\myref{\WMTB}
  T. T. Wu, B. M. McCoy, C. A. Tracy and E. Barouch, Spin-spin correlation functions for the 
two-dimensional Ising model: Exact theory in the scaling region, {\it Phys. Rev. B} {\bf 13} (1976),
316--374.

\endroster\par}

\enddocument